\documentclass[journal]{IEEEtran}

\usepackage{times}
\usepackage{graphicx}

\usepackage{caption}
\usepackage{subcaption}
\usepackage{epsfig}
\usepackage{amsmath}
\usepackage{amssymb}
\usepackage{booktabs}
\usepackage{rotating}
\usepackage{algorithm}
\usepackage{algpseudocode}
\usepackage{multirow}
\usepackage{color}
\usepackage{xcolor}

\usepackage{enumitem}
\usepackage{enumerate}
\definecolor{dgreen}{rgb}{0,0,0}
\definecolor{dblue}{rgb}{0.220,0.325,0.639}
\definecolor{dred}{rgb}{0.933,0.122,0.137}

\definecolor{ddgreen}{HTML}{78c679}
\definecolor{ddyellow}{HTML}{fe9929}

\usepackage{mathtools}

\newcommand{\ie}{\textit{i}.\textit{e}.}

\usepackage[pagebackref=false,breaklinks=true,letterpaper=true,colorlinks,citecolor=black,linkcolor=black,bookmarks=false]{hyperref}

\begin{document}
\title{Boosting RGB-D Saliency Detection by Leveraging Unlabeled RGB Images}

\author{
    Xiaoqiang Wang,
    Lei Zhu,
    Siliang Tang,
    Huazhu Fu,~\IEEEmembership{Senior Member,~IEEE,} 
    Ping Li,\\
    Fei Wu,~\IEEEmembership{Senior Member,~IEEE,}
    Yi Yang,~\IEEEmembership{Senior Member,~IEEE,}
    Yueting Zhuang,~\IEEEmembership{Senior Member,~IEEE}

	\IEEEcompsocitemizethanks{
		\IEEEcompsocthanksitem X. Wang, S. Tang, F. Wu, Y. Yang and Y. Zhuang are with the College of Computer Science and Technology, Zhejiang University, China (E-mail: xq.wang@zju.edu.cn; siliang@zju.edu.cn;  wufei@zju.edu.cn; Yi.Yang@uts.edu.au; yzhuang@zju.edu.cn).
		\IEEEcompsocthanksitem L. Zhu is with ROAS Thrust, System Hub, Hong Kong University of Science and Technology (GZ) (E-mail: leizhu@ust.hk).
		\IEEEcompsocthanksitem H. Fu is with the Institute of High Performance Computing (IHPC), Agency for Science, Technology and Research (A*STAR), Singapore 138632. (E-mail: hzfu@ieee.org).
		\IEEEcompsocthanksitem P. Li is with the Department of Computing, The Hong Kong Polytechnic University, Kowloon, Hong Kong SAR, China. (Email: p.li@polyu.edu.hk).
		\IEEEcompsocthanksitem S. Tang is the corresponding author of this work.
	}
}

	\markboth{IEEE Transactions on Image Processing}%
    {Shell \MakeLowercase{\textit{et al.}}: Bare Demo of IEEEtran.cls for IEEE Journals}

\maketitle

\begin{abstract}
Training deep models for RGB-D salient object detection (SOD) often requires a large number of labeled RGB-D images. However, RGB-D data is not easily acquired, which limits the development of RGB-D SOD techniques. 
To alleviate this issue, we present a Dual-Semi RGB-D Salient Object Detection Network (DS-Net) to leverage unlabeled RGB images for boosting RGB-D saliency detection. 
We first devise a depth decoupling convolutional neural network (DDCNN), which contains a depth estimation branch and a saliency detection branch. The depth estimation branch is trained with RGB-D images and then used to estimate the pseudo depth maps for all unlabeled RGB images to form the paired data. The saliency detection branch is used to fuse the RGB feature and depth feature to predict the RGB-D saliency.
Then, the whole DDCNN is assigned as the backbone in a teacher-student framework for semi-supervised learning. Moreover, we also introduce a consistency loss on the intermediate attention and saliency maps for the unlabeled data,  as well as a supervised depth and saliency loss for labeled data.
Experimental results on seven widely-used benchmark datasets demonstrate that our DDCNN outperforms state-of-the-art methods both quantitatively and qualitatively. We also demonstrate that our semi-supervised DS-Net can further improve the performance, even when using an RGB image with the pseudo depth map.
\end{abstract}




\section{Introduction}
\label{sec:introduction}

RGB-D salient object detection (SOD) has attracted a surge in interest recently~\cite{cong2018review,wang2019salient,fan2020rethinking}.
Early RGB-D detectors~\cite{feng2016local,song2017depth,cong2017co,cong2019going} mainly examined the handcrafted priors, which degrades the detection performance since the assumptions of these heuristic priors are not always correct.
More recently, RGB-D SOD detectors~\cite{chen2018progressively,zhao2019contrast,piao2019depth,fu2020jl,zhang2020uc,zhang2020select,liu2020learning, Li2020_TC} based on convolutional neural networks (CNNs) have been developed by learning the features from RGB images and depth maps and exploring the complementary information between them.
Although these models have achieved impressive performances on the benchmark datasets, there are still several issues limiting the development of RGB-D SOD techniques: 1) pixel-level annotation for supervised learning is expensive and time-consuming, and 2) compared to RGB images, paired RGB-D images are more difficult to collect. Fortunately, it is easy to collect a large number of unlabeled RGB images. \textit{Thus, how to leverage the unlabeled RGB images to assist the RGB-D SOD methods is a desirable direction to explore.}

%

\begin{figure*}[!t]
	\centering
	\includegraphics[width=.95\linewidth]{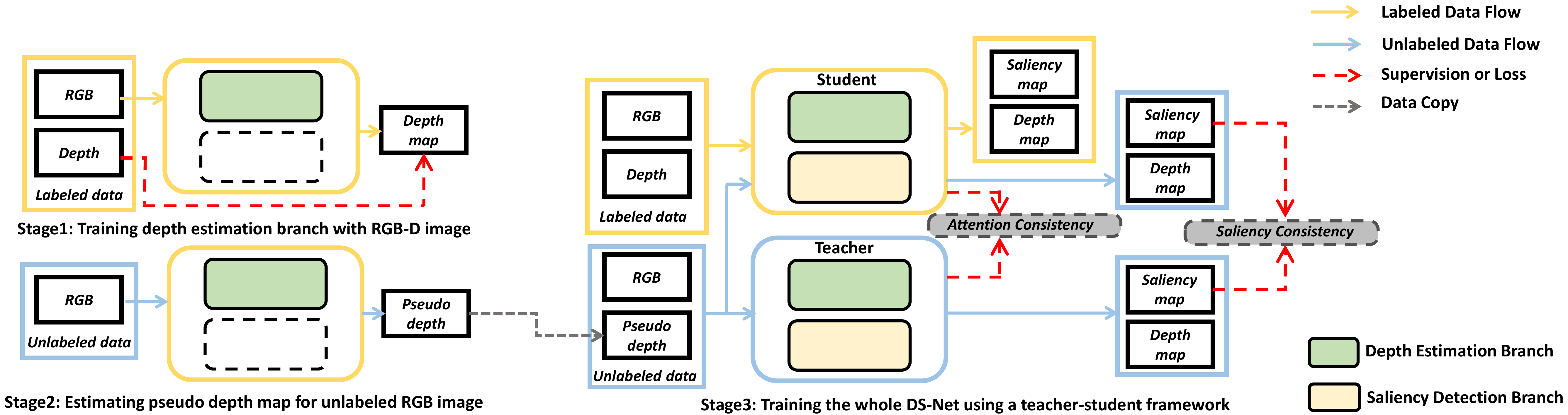} 
	\caption{
	Illustration of our DS-Net, which is trained in three main stages.
	1) Training depth estimation branch based on labeled RGB-D data. 2) Estimating the pseudo depth maps for all unlabeled RGB images. 3) Learning the whole network in a semi-supervised manner based on labeled RGB-D data and unlabeled RGB images with pseudo depth maps. 
}
\vskip -10pt
	\label{fig:arc} 
\end{figure*}

In this paper, we present a novel dual-semi RGB-D salient object detection network (DS-Net) for boosting the RGB-D saliency detection by leveraging unlabeled RGB data, as shown in Fig.~\ref{fig:arc}. 
Our DS-Net has a two-fold motivation: semi-supervised learning with unlabeled data and semi-paired data including RGB images without depth maps.
Specifically, we devise a depth decoupling convolutional neural network (DDCNN) to estimate depth maps of RGB images and detect RGB-D saliency maps \textbf{\textit{jointly}}. 
Our DDCNN disentangles two types of features from the RGB image, \ie, depth-aware features and depth-dispelled features.
The depth-aware features are used to estimate the pseudo depth maps for RGB images, while the depth-dispelled features are extracted from the input RGB image and then fused with the depth map features to predict the RGB-D saliency. 
Finally, we embed DDCNN as the backbone into a teacher-student framework to provide semi-supervision for training the whole DS-Net based on labeled RGB-D data and unlabeled RGB images with their pseudo depth maps.
Moreover, we also introduce a consistency loss to constrain the attention and saliency maps on unlabeled data, assisting the supervised loss for saliency and depth predictions on labeled data. In summary, the main contributions are:
\begin{itemize}
    \item
    A dual-semi RGB-D salient object detection network (DS-Net) is proposed for leveraging RGB images as unlabeled data to assist the RGB-D SOD task in a semi-supervised manner. We show an effective solution to improve the RGB-D task performance by incorporating RGB images with the pseudo depth maps.
    \item A depth decoupling convolutional neural network (DDCNN) is designed to \textbf{\textit{jointly}} estimate depth maps for RGB images and predict saliency maps for RGB-D images. Two types of features are disentangled from RGB image, \ie, depth-aware features and depth-dispelled features, enabling the network to identify the latent features specific to each modality and task.
    \item For enhancing the semi-supervised consistency, 
    a consistency loss is introduced in the teacher-student network to constrain the intermediate attention and saliency maps on the unlabeled data, assisting with the supervised depth and saliency loss on labeled data.
    \item Last but not least, experimental results on seven widely-used RGB-D SOD datasets show that our DDCNN outperforms state-of-the-art methods. We also demonstrate that the semi-supervised DS-Net can further improve the performance, even when using an RGB image with the pseudo depth map. 
\end{itemize}

Our code, the trained models, and the predicted saliency maps on all seven benchmark datasets are released at: \url{https://github.com/Robert-xiaoqiang/DS-Net}.


\section{Related Work}
\label{sec:related}




\subsection{RGB-D Salient Object Detection} \ 
RGB-D SOD methods based on deep learning can be roughly grouped into three categories: early fusion,  middle fusion, and later fusion.
Early fusion concatenates the RGB and depth images as a four-channel input and then passes this into CNNs for saliency detection.
Liu et al.~\cite{liu2019salient}, and Huang et al.~\cite{huang2018rgbd} developed a single-stream convolutional neural network and a fully convolutional network (FCN) with short connections to detect salient regions from the concatenated four-channel input, respectively.
Late fusion employs two separate backbone networks for RGB and depth to generate individual features which are fused together for final prediction.
{\color{dgreen}
Han et al.~\cite{han2017cnns} transferred the structure of the RGB-based CNN to be applicable for the depth view and fused the deep representations of both views automatically to obtain the final saliency map.
Wang et al.~\cite{wang2019adaptive} designed a two-streamed CNN to predict a saliency map from each modality separately and fuse the predicted saliency maps adaptively by learning a switch map.
}

%
As the most popular CNN framework for RGB-D saliency detection, middle fusion typically integrates multi-scale intermediate features from input RGB and depth modalities in different manners.
Chen et al.~\cite{chen2018progressively} developed complementarity-aware fusion (CA-Fuse) modules to progressively integrate RGB features and depth features.
Piao et al.~\cite{piao2019depth} devised depth refinement blocks to extract and
fuse multi-level paired complementary RGB and depth cues.
Fan et al.~\cite{fan2020rethinking} built a depth-depurator network to filter out noise in the depth map for better fusing cross-modal features.
Fu et al.~\cite{fu2020jl} utilized a shared backbone to extract hierarchical features from RGB and depth inputs simultaneously for a multi-scale cross-module fusion.
Zhang et al.~\cite{zhang2020uc} presented a probabilistic RGB-D saliency detection network via conditional variational autoencoders to approximate human annotation uncertainty and produce multiple saliency maps for each input image.
Zhang et al.~\cite{zhang2020select} included complementary interaction models, consisting of a cross-modal attention unit and a boundary supplement unit, to select useful RGB and depth features for salient object location and boundary detail refinement.
Li et al.~\cite{li2020cross} adopted a three-level Siamese encoder-decoder structure to develop three modules to fuse low-level, middle-level, and high-level RGB and depth features, respectively, for cross-modal and cross-scale RGB-depth interactions.


Although existing CNN-based methods have achieved more accurate results than traditional RGB-D saliency detectors, their network training requires a large amount of data with pixel-level saliency annotations.
Moreover, annotated training data are collected from limited scenarios, causing the networks to suffer from degraded performance on unseen photos.  
As such, this work presents a semi-supervised network to fuse unlabeled data with the labeled data for boosting RGB-D saliency detection. \textit{More importantly, rather than relying on unlabeled RGB-D paired images, our unlabeled data consists of only RGB images, which are much easier to collect in our daily life.} 

 \begin{figure*} [!t]
	\centering
	\includegraphics[width=0.95\linewidth]{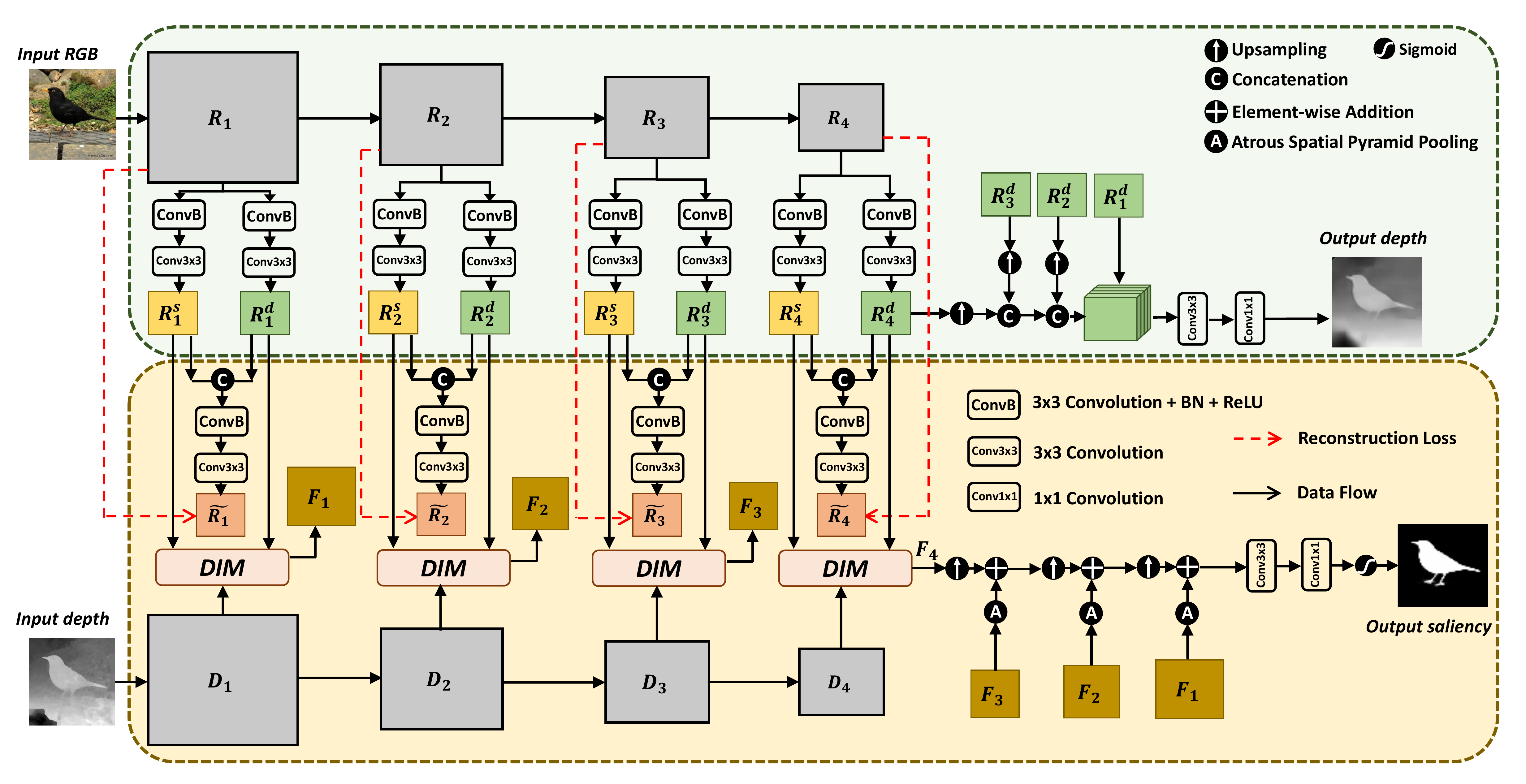}
	\caption{Illustration of the proposed DDCNN, which includes a depth estimation branch (\textcolor{ddgreen}{green block}) and a saliency detection branch (\textcolor{ddyellow}{yellow block}). 
	{\color{dgreen}
    Stage 1 and stage 2 use the green block only, while stage 3 uses both green and yellow blocks.
	}
	The depth estimation branch disentangles the image features $R_i$ into depth-aware features $R^d_i$ for estimating depth maps and depth-dispelled features $R^s_i$ for predicting saliency. 
	The saliency detection branch uses a depth-induced fusion module (DIM) to fuse depth-dispelled features $R_i^s$ and depth features. Finally, the saliency map is predicted by fusing all features $\{F_i\}_{i=1}^4$ produced by DIMs.
}
	\label{fig:DDCNN}
	\vskip -10pt
\end{figure*}

\subsection{Semi-Supervised Learning}

By integrating labeled and unlabeled data for network training, semi-supervised learning (SSL) has achieved remarkable results in many computer vision tasks~\cite{chen2020multi, InfNet2020}.
As a typical kind of SSL technique, self-ensembling usually devises a consistency loss on the unlabeled data to guarantee invariant predictions for perturbations of unlabeled data.
For example, the $\Pi$-model~\cite{laine2016temporal} devised
consistency constraints between the current network prediction and the temporal average of network predictions for unlabeled data.
The mean teacher (MT) framework~\cite{tarvainen2017mean} proposed to ensemble the network parameters to replace the network predictions of the $\Pi$-model~\cite{laine2016temporal}, achieving improved performance in the semi-supervised learning.
Developing semi-supervised CNNs for RGB-D saliency detection usually requires numerous paired RGB-D images, which are not easy to collect.
RGB images, however, are much easier to collect.
{\color{dgreen}
\textit{Hence, our model is devised to leverage unpaired RGB images to formulate semi-supervised RGB-D saliency detection.}
}



\section{Methodology}
\label{sec:method}

Fig.~\ref{fig:arc} shows an illustration of our dual-semi RGB-D salient object detection network (DS-Net), which integrates the labeled RGB-D data and unlabeled RGB images. A depth decoupling convolutional neural network (DDCNN) is utilized as the backbone, which contains two components: a depth estimation branch and a saliency detection branch for estimating the depth maps of RGB images and predicting the saliency maps of RGB-D images, respectively. 
To train the whole DS-Net with labeled RGB-D data and unlabeled RGB images, we first train the depth estimation branch of DDCNN using labeled RGB-D data to learn the mapping from an RGB image to its depth map. Then, we estimate the pseudo depth maps for all unlabeled RGB images to form the paired data. Finally, we utilize DDCNN as the backbone in a teacher-student framework. 
To train the whole DS-Net in a semi-supervised manner, we utilize a supervised loss on depth and saliency predictions for labeled RGB-D data and a consistency loss on intermediate attention maps and saliency predictions for unlabeled RGB images with their pseudo depth maps.

\begin{figure*}[!t]
	\centering
	\includegraphics[width=0.95\linewidth]{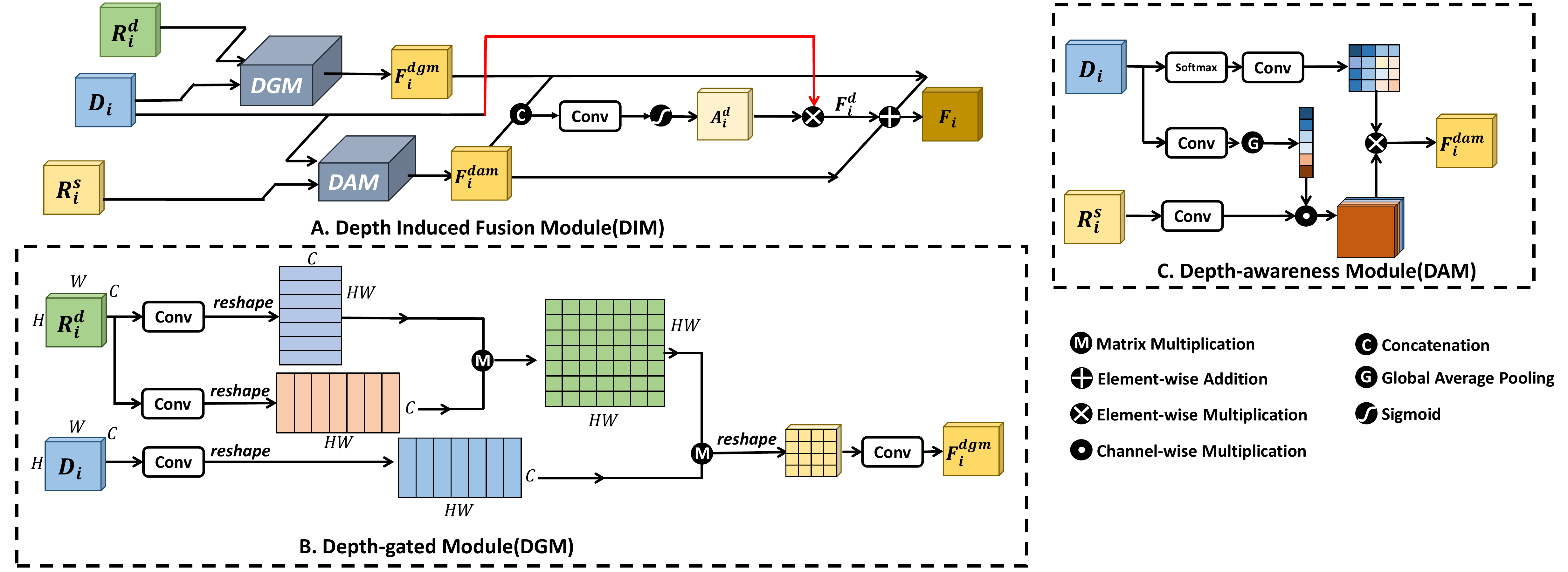}
	\caption{The illustration of DIM, which contains the Depth-awareness Module (DAM) and Depth-gated Module (DGM).}
	\label{fig:DIM}
	\vskip -10pt
\end{figure*}

{\color{dgreen}
\subsection{DDCNN}
}

Fig.~\ref{fig:DDCNN} illustrates the architecture of our DDCNN, which is a two-branch structure including a depth estimation branch (green block) and saliency detection branch (yellow block). 
Given a pair of \textcolor{dgreen}{input} RGB-D images, the RGB image is passed to an encoder to generate RGB features (\ie, $\{R_i\}_{i=1}^4$), while the depth map is fed to another encoder to extract depth features (\ie, $\{D_i\}_{i=1}^4$).  

In the depth estimation branch of DDCNN, each RGB feature $R_i$ is  disentangled into a depth-aware feature $R^d_i$ for estimating the depth map and a depth-dispelled feature $R^s_i$ for predicting saliency,
using a convolutional block with
``Conv(3$\times$3)
$\rightarrow$
BN
$\rightarrow$
ReLU
$\rightarrow$
Conv(3$\times$3)".
The  depth-aware features $R^d_i$ are then up-sampled to the same spatial resolution of $R_1^d$ and concatenated together to predict a depth map by applying a
``Conv(3$\times$3)
$\rightarrow$
Conv(1$\times$1)" convolutional block.
%
Moreover, we fuse $R_i^d$ and $R_i^s$ to reconstruct RGB features $\tilde{R_i}$ and compute a reconstruction loss $\mathcal{L}_{i}^{r}$ to regularize the decoupling process:
\vskip -5pt
\begin{equation} \label{eq:reconstruction_loss}
\small
\mathcal{L}_{i}^{r}  = L_{MSE}(\tilde{R_i}, R_i), \; \text{with} \; 
\tilde{R_i} = \textit{CB} \left( Cat(R_i^d, R_i^s) \right)  ,
\end{equation}
where $\textit{CB}(\cdot)$ is a convolutional block with 
``Conv(3$\times$3)
$\rightarrow$
BN
$\rightarrow$
ReLU
$\rightarrow$
Conv(3$\times$3)".
$Cat(\cdot)$ denotes a feature concatenation operation, and $L_{MSE}$ represents the mean square error (MSE) loss.

In the saliency detection branch of DDCNN, we devise a depth-induced fusion module (DIM) to fuse depth-dispelled features $R_i^s$ with two depth features ($R_i^d$ and $D_i$) at each CNN layer. The fused features from DIMs \textcolor{dgreen}{at the four levels of feature extraction layers} are denoted as $\{F_i\}_{i=1}^4$, as shown in the yellow block of Fig.~\ref{fig:DDCNN}.
Finally, we iteratively merge the four obtained features $\{F_i\}_{i=1}^4$ 
to predict a saliency map with a $3$$\times$$3$ convolution, a $1$$\times$$1$ convolution, and a sigmoid function. 
When merging features ($F_i$ and $F_{i+1}$) from two adjacent CNN layers, we up-sample the low-resolution features $F_{i+1}$ to the same resolution as the high-resolution features $F_{i}$, and then refine the feature map $F_{i}$ from the shallow layer by passing it to an atrous spatial pyramid pooling (ASPP) module~\cite{chen2017deeplab} with four dilated convolutional layers (dilation rates: $1$, $6$, $12$, $18$), followed by a $1$$\times$$1$ convolutional layer on the four dilated features concatenated together.
We then fuse the refined features of ASPP with up-sampled $F_{i+1}$ to produce the merged features of $F_i$ and $F_{i+1}$.

{\color{dgreen}
\vspace{2mm}
\noindent
\textbf{Depth-Induced Fusion Module.}
}
As shown in Fig.~\ref{fig:DDCNN}, a DIM at each CNN layer merges three feature maps (\ie $R_i^d$, $R_i^s$ and $D_i$) to leverage the complementary information of RGB and depth modalities.
Fig.~\ref{fig:DIM} gives an illustration of the DIM at the $i$-th CNN layer, which takes depth-aware feature $R_i^d$, depth-dispelled feature $R_i^s$ and depth feature $D_i$ as inputs and produces a fused feature $F_i$.
To be specific, we first devise a depth-gated module (DGM) to fuse depth features $D_i$ from the input depth map and $R_i^d$ from the depth estimation branch. The fused features of DGM are denoted as $F_i^{dgm}$.
Then, we adopt a depth-awareness module (DAM) to fuse $R_i^s$, and depth features $D_i$ from the input depth map to obtain new features $F_i^{dam}$.
Moreover, we concatenate $F_i^{dgm}$  and $F_i^{dam}$, and apply a $3$$\times$$3$ convolutional layer and a sigmoid function on the concatenated features to learn an attention map $A_i^d$ for weighting depth features $D_i$, thereby generating a new feature map $F_i^{d}$.
Finally, the output of DIM (denoted as $F_i$) is computed by adding $F_i^{dam}$, $F_i^{dgm}$, and $F_i^{d}$:
\vskip -5pt
\begin{equation}
	F_i = F_i^{dam} + F_i^{dgm} + F_i^{d} \ ,
\end{equation}

\vspace{2mm}
\noindent
\textbf{Depth-Awareness Module.} Considering the redundancy and noise in low-quality depth maps and the intrinsic difference between RGB and depth features, we design an effective fusion method to suppress noise and utilize the complementary information from the features $R_i^s$ and $D_i$ of the two modalities. 
Inspired by CBAM~\cite{woo2018cbam}, we design a DAM equipped with a channel attention and a spatial attention operation to merge RGB features $R_i^s$, and depth features $D_i$.
%
Fig.~\ref{fig:DIM} shows the workflow of DAM, which outputs a new feature map $F_i^{dam}$. 
This is achieved by applying a channel attention operation on $D_i$ to weight different channels of $R_i^s$, and then a spatial attention is computed on $D_i$ to recalibrate pixel-wise saliency cues of $R_i^s$ to obtain $F_i^{dam}$:
\vskip -5pt
\begin{equation}
    F_i^{dam} = S_{att}(D_i) \otimes (C_{att}(D_i) \odot R_i^s) \ ,  
\end{equation} 
where the channel attention $C_{att}(\cdot)$ includes a $3$$\times$$3$ convolution and a global average pooling.
The spatial attention $S_{att}(\cdot)$ consists of a $3$$\times$$3$ convolution with a softmax function.
$\odot$ denotes a channel-wise multiplication while $\otimes$ represents an element-wise multiplication.

\vspace{2mm}
\noindent
\textbf{Depth-Gated Module.} $R_i^d$ are the depth signals from RGB image, hence combining them and depth features $D_i$ from the input depth image enriches the depth representation from different depth modalities.
As such, we devise a DGM to merge $R_i^d$ (size: $H$$\times$$W$$\times$$C$) and $D_i$ (size: $H$$\times$$W$$\times$$C$) by considering long-range pixel dependencies for learning saliency cues.
First, we first apply a $3$$\times$$3$ convolutional layer on $R_i^d$, reshape the resultant features as an intermediate $C$$\times$$HW$ feature map, apply another $3$$\times$$3$ convolutional layer on $R_i^d$, reshape the resultant features as another intermediate $HW$$\times$$C$ feature map, and then multiply two intermediate features to generate a non-local similarity matrix \textcolor{dgreen}{with size} $HW$$\times$$HW$. 
Furthermore, we apply a $3$$\times$$3$ convolutional layer on $D_i$ and reshape the resultant features as an intermediate $HW$$\times$$C$ feature map, which is then multiplied with the non-local similarity matrix.
After that, we reshape the features $HW$$\times$$C$ resulting from the multiplication to a feature map \textcolor{dgreen}{with size} $H$$\times$$W$$\times$$C$, which undergoes a $3$$\times$$3$ convolution to obtain the features $F_i^{dgm}$ of DGM.

\begin{table*}[!t]
	\setlength\tabcolsep{2pt}
    \caption {Comparisons our network against state-of-the-art detectors on seven benchmark datasets. The best results are highlighted in \textbf{bold}. DS-Net (HRNet as backbone) is marked in \textbf{bold} when it outperforms all other baselines.}
    \label{table:state-of-the-art-part1}
    \resizebox{1.0\textwidth}{!}{%
\begin{tabular}{c|c|c|cccccccccccccccc|ccccc|cccc|ccc}
\hline
\multirow{2}{*}{Dataset}   &\multicolumn{1}{c|}{\multirow{2}{*}{Metric}}
&LBE &DF  &CTMF &PCF  &TANet &CPFP
&SSF  &UCNet  &JLDCF &\textcolor{dgreen}{JLDCF[J]} &HDF-Net
&PGA-Net &DANet &cmMS &Cas-Gnn &CMWNet &VGG-16
&DMRA &ATSA &\textcolor{dgreen}{SSDP} &\textcolor{dgreen}{DSA$^2$F} &VGG-19
&D$^3$Net  &BBS-Net &CoNet &ResNet-50
&DDCNN-semi &HRNet &HRNet                  \\
& \multicolumn{1}{c|}{}         &~\cite{feng2016local}          &~\cite{qu2017rgbd}  &~\cite{han2017cnns}      &~\cite{chen2018progressively}     &~\cite{chen2019three}   &~\cite{zhao2019contrast}

&~\cite{zhang2020select}  &~\cite{zhang2020uc} &~\cite{fu2020jl} &~\cite{fu2021siamese} &~\cite{pang2020hierarchical}

&~\cite{chen2020progressively} &~\cite{zhao2020single} &~\cite{li2020rgb} &~\cite{luo2020cascade}   &~\cite{li2020cross} &DS-Net

&~\cite{piao2019depth} &~\cite{zhang2020asymmetric} &~\cite{wang2020synergistic} &~\cite{sun2021deep} &DS-Net

&~\cite{fan2020rethinking} &~\cite{fan2020bbs} &~\cite{ji2020accurate} &DS-Net
&ourSplit &DDCNN  &DS-Net \\ \hline
    \multirow{4}{*}{NJU2K}
    & $S_m \; \; \; \; \uparrow$
    &0.695      &0.763      &0.849      &0.877      &0.878          &0.879
    &0.899      &0.897      &0.903      &\textcolor{dgreen}{0.911}  &0.908
    &0.906      &0.901      &0.904      &0.911          &0.903     &{\color{black}\bf 0.945}
    &0.886      &0.899      &\textcolor{dgreen}{0.878}	&\textcolor{dgreen}{0.903}  &{\color{black}\bf 0.944}          
    &0.895      &0.921    &0.894      &{\color{black}\bf 0.946}
    &0.922      &0.936           &{\color{black}\bf 0.950}
    \\
    & $F_\beta^{max}\uparrow$   
    &0.748      &0.804      &0.845      &0.872      &0.874          &0.877
    &0.886      &0.886      &0.903      &\textcolor{dgreen}{0.913}  &0.922    
    &0.883      &0.893      &0.914      &0.903          &0.902     &{\color{black}\bf 0.961}
    &0.886      &0.910      &\textcolor{dgreen}{0.852}	&\textcolor{dgreen}{0.901}  &{\color{black}\bf 0.964}          
    &0.889      &0.920      &0.872      &{\color{black}\bf 0.964}
    &0.943      &0.958           &{\color{black}\bf 0.965}
    \\
    & $E_\phi^{max}\uparrow$
    &0.803      &0.864      &0.913      &0.924      &0.925          &0.926 
    &-          &0.930      &0.944      &\textcolor{dgreen}{0.948}  &0.932      
    &0.914      &0.921      &-          &0.936          &0.933      &{\color{black}\bf 0.962}
    &0.927      &0.922      &\textcolor{dgreen}{0.909}	&\textcolor{dgreen}{0.923}  &{\color{black}\bf 0.964}          
    &0.932      &0.949    &0.912      &{\color{black}\bf 0.964}
    &0.944      &0.954           &{\color{black}\bf 0.966}
    \\
    &  $MAE\downarrow$    
    &0.153      &0.141      &0.085      &0.059      &0.060          &0.053
    &0.043      &0.043      &0.043      &\textcolor{dgreen}{0.040}  &0.038      
    &0.045      &0.040      &0.044      &0.035    &0.046  &{\color{black}\bf 0.026}     
    &0.051      &0.045      &\textcolor{dgreen}{0.055}	&\textcolor{dgreen}{0.039}  &{\color{black}\bf 0.026}          
    &0.051      &0.035    &0.047  &{\color{black}\bf 0.025}
    &0.037      &0.033           &{\color{black}\bf 0.024}
    \\ \hline
    
    \multirow{4}{*}{NLPR}
    & $S_m \; \; \; \; \uparrow$     
    &0.762      &0.802      &0.860      &0.874      &0.886      &0.888
    &0.914      &0.920      &0.925      &\textcolor{dgreen}{0.926}  &0.923      
    &0.918      &0.907      &0.900      &0.919          &0.917      &{\color{black}\bf 0.949}
    &0.899      &0.915      &\textcolor{dgreen}{0.875}	&\textcolor{dgreen}{0.918}  &{\color{black}\bf 0.950}          
    &0.905      &0.930    &0.907  &{\color{black}\bf 0.951}
    &0.922      &0.939           &{\color{black}\bf 0.952}
    \\
    & $F_\beta^{max}\uparrow$   
    &0.745      &0.778      &0.825      &0.841      &0.863          &0.867
    &0.875      &0.891      &0.916      &\textcolor{dgreen}{0.917}  &0.927    
    &0.871      &0.876      &0.914      &0.904          &0.903      &{\color{black}\bf 0.948}
    &0.879      &0.916      &\textcolor{dgreen}{0.809}	&\textcolor{dgreen}{0.897}  &{\color{black}\bf 0.948}          
    &0.885      &0.918      &0.848      &{\color{black}\bf 0.951}
    &0.919      &0.938           &{\color{black}\bf 0.953}
    \\
    & $E_\phi^{max}\uparrow$   
    &0.855      &0.880      &0.929      &0.925      &0.941          &0.932
    &-          &0.951      &0.962      &\textcolor{dgreen}{0.964}  &0.957      
    &0.948      &0.945      &-          &0.952          &0.951      &{\color{black}\bf 0.965}
    &0.947      &0.949      &\textcolor{dgreen}{0.915}	&\textcolor{dgreen}{0.950}  &{\color{black}\bf 0.966}          
    &0.946      &0.961   &0.936      &{\color{black}\bf 0.968}
    &0.942      &0.958            &{\color{black}\bf 0.970}
    \\
    &  $MAE\downarrow$    
    &0.081      &0.085      &0.056      &0.044      &0.041          &0.036
    &0.026      &0.025      &0.023      &\textcolor{dgreen}{0.023}  &0.023       
    &0.028      &0.028      &0.273      &0.025          &0.029      &{\color{black}\bf 0.020}
    &0.031      &0.028      &\textcolor{dgreen}{0.044}	&\textcolor{dgreen}{0.024}  &{\color{black}\bf 0.020}          
    &0.034      &0.023   &0.031      &{\color{black}\bf 0.018}
    &0.032      &0.024            &{\color{black}\bf 0.018}
    \\ \hline
    
    \multirow{4}{*}{STERE}
    & $S_m \; \; \; \; \uparrow$     
    &0.660      &0.757      &0.848      &0.875      &0.871          &0.879
    &0.893      &0.903      &0.905      &\textcolor{dgreen}{0.907}  &0.900      
    &0.897      &-          &0.889      &0.899          &0.905      &{\color{black}\bf 0.909}
    &0.886      &0.903      &\textcolor{dgreen}{0.893}	&\textcolor{dgreen}{-}  &{\color{black}\bf 0.910}          
    &0.891      &0.908   &0.908   &{\color{black}\bf 0.910}
    &0.892      &0.907            &{\color{black}\bf 0.914}
    \\
    & $F_\beta^{max}\uparrow$   
    &0.633      &0.757      &0.831      &0.860      &0.861          &0.874
    &0.880      &0.884      &0.901      &\textcolor{dgreen}{0.907}  &0.910       
    &0.884      &-          &0.908    &0.901      &0.901      &{\color{black}\bf 0.912}
    &0.886      &0.872      &\textcolor{dgreen}{0.878}	&\textcolor{dgreen}{-}  &{\color{black}\bf 0.914}          
    &0.881      &0.903      &0.885  &{\color{black}\bf 0.914}
    &0.896      &0.910            &{\color{black}\bf 0.915}
    \\
    & $E_\phi^{max}\uparrow$   
    &0.787      &0.847      &0.912      &0.925      &0.923          &0.925
    &-          &0.935      &{\color{black}\bf 0.936}   &\textcolor{dgreen}{0.935}  &0.931      
    &0.921      &-          &-          &0.930          &0.934  &0.935  
    &{\color{black}\bf 0.937}      &0.914      &\textcolor{dgreen}{0.936}	&\textcolor{dgreen}{-}  &{\color{black}\bf 0.937}          
    &0.930      &0.932      &0.923  &{\color{black}\bf 0.937}
    &0.911      &0.937      &{\color{black}\bf 0.947}
    \\
    &  $MAE\downarrow$    
    &0.250      &0.141      &0.086      &0.064      &0.060          &0.051
    &0.044      &{\color{black}\bf 0.039}   &0.042      &\textcolor{dgreen}{0.039}  &0.041        
    &{\color{black}\bf 0.039}   &-      &0.042          &{\color{black}\bf 0.039}   &0.043   &{\color{black}\bf 0.039}
    &0.047      &0.044      &\textcolor{dgreen}{0.045}	&\textcolor{dgreen}{-}  &{\color{black}\bf 0.039}          
    &0.054      &0.041   &0.041    &{\color{black}\bf 0.039}
    &0.041      &0.040      &{\color{black}\bf 0.037}
    \\ \hline
    
    \multirow{4}{*}{RGBD135}
    & $S_m \; \; \; \; \uparrow$     
    &0.703      &0.752      &0.863      &0.842      &0.858          &0.872
    &0.905      &{\color{black}\bf 0.934}   &0.929      &\textcolor{dgreen}{0.931}  &0.926      
    &0.894      &0.907      &-          &0.905          &{\color{black}\bf 0.934}   &{\color{black}\bf 0.934}
    &0.900      &0.924      &\textcolor{dgreen}{0.890}	&\textcolor{dgreen}{0.904}  &{\color{black}\bf 0.935}          
    &0.904      &0.933    &0.910  &{\color{black}\bf 0.934}
    &0.902      &0.925      &{\color{black}\bf 0.936}
    \\
    & $F_\beta^{max}\uparrow$   
    &0.788      &0.766      &0.844      &0.804      &0.827          &0.846
    &0.876      &0.919      &0.919      &\textcolor{dgreen}{0.929}  &{\color{black}\bf 0.932}     
    &0.870      &0.885      &-              &0.906      &0.930   &{\color{black}\bf 0.932}
    &0.888      &0.928    &\textcolor{dgreen}{0.864}	&\textcolor{dgreen}{0.898}  &{\color{black}\bf 0.932}      
    &0.885  &0.927  &0.861  &{\color{black}\bf 0.932}
    &0.908      &0.929          &{\color{black}\bf 0.933}
    \\
    & $E_\phi^{max}\uparrow$   
    &0.890      &0.870      &0.932      &0.893      &0.910          &0.923
    &-          &0.967      &0.968      &\textcolor{dgreen}{0.955}  &{\color{black}\bf 0.971}     
    &0.935  &0.952      &-              &0.947      &0.969   &0.959
    &0.943      &{\color{black}\bf 0.968}    &\textcolor{dgreen}{0.927}	&\textcolor{dgreen}{0.933}  &0.960      
    &0.946  &{\color{black}\bf 0.966}  &0.945  &0.960
    &0.913      &0.955      &0.961
    \\
    &  $MAE\downarrow$    
    &0.208      &0.093      &0.055      &0.049      &0.041          &0.038
    &0.025      &{\color{black}\bf 0.019}     &0.022    &\textcolor{dgreen}{0.022}  &0.021      
    &0.032      &0.024      &-          &0.028          &0.022  &0.021  
    &0.030      &0.023      &\textcolor{dgreen}{0.031}	&\textcolor{dgreen}{0.036}  &{\color{black}\bf 0.021}          
    &0.030      &{\color{black}\bf 0.021}   &0.027      &{\color{black}\bf 0.021}
    &0.032      &0.025      &0.021
    \\ \hline
    
    \multirow{4}{*}{LFSD}
    & $S_m \; \; \; \; \uparrow$     
    &0.729      &0.783      &0.788      &0.786      &0.794          &0.820
    &0.859      &0.854      &0.854      &\textcolor{dgreen}{0.863}  &0.854      
    &0.855   &-      &0.860          &0.849      &0.856      &{\color{black}\bf 0.866}
    &0.839      &0.833      &\textcolor{dgreen}{0.830}	&\textcolor{dgreen}{0.920}  &{\color{black}\bf 0.869}          
    &0.824      &0.854      &0.862  &{\color{black}\bf 0.872}
    &0.842      &0.862      &{\color{black}\bf 0.878}
    \\
    & $F_\beta^{max}\uparrow$   
    &0.722      &0.813      &0.787      &0.775          &0.792      &0.821
    &0.867      &0.855      &0.862      &\textcolor{dgreen}{0.862}  &0.883       
    &0.862      &-          &0.883   &0.864      &0.883   &{\color{black}\bf 0.884}
    &0.852      &0.830      &\textcolor{dgreen}{0.823}	&\textcolor{dgreen}{0.896}  &{\color{black}\bf 0.885}          
    &0.815      &0.858      &0.848  &{\color{black}\bf 0.884}
    &0.863      &0.883               &{\color{black}\bf 0.885}
    \\
    & $E_\phi^{max}\uparrow$   
    &0.797      &0.857      &0.857      &0.827      &0.840      &0.864
    &-          &0.901      &0.893      &\textcolor{dgreen}{0.900}  &0.891      
    &0.900  &-      &-          &0.877          &{\color{black}\bf 0.902}     &{\color{black}\bf 0.902}
    &0.893      &0.869      &\textcolor{dgreen}{0.879}	&\textcolor{dgreen}{0.962}  &{\color{black}\bf 0.903}          
    &0.856      &0.901    &0.897      &{\color{black}\bf 0.905}
    &0.874      &0.902      &{\color{black}\bf 0.905}
    \\
    &  $MAE\downarrow$    
    &0.214      &0.146      &0.127      &0.119      &0.118      &0.095
    &0.086      &0.086      &0.078      &\textcolor{dgreen}{0.079}  &{\color{black}\bf 0.076}      
    &0.086  &-          &0.082      &0.083          &0.086   &{\color{black}\bf 0.076}
    &0.083      &0.093      &\textcolor{dgreen}{0.090}	&\textcolor{dgreen}{0.021}  &{\color{black}\bf 0.073}          
    &0.106      &0.072  &0.071      &{\color{black}\bf 0.069}
    &0.082      &0.075      &{\color{black}\bf 0.064}
    \\ \hline
    
    \multirow{4}{*}{SIP}
    & $S_m \; \; \; \; \uparrow$     
    &0.727      &0.653      &0.716      &0.842      &0.835      &0.850
    &-          &0.875     &0.879  &\textcolor{dgreen}{0.882}   &{\color{black}\bf 0.886}     
    &0.875    &0.875    &-      &-          &0.867      &0.881
    &0.806      &-          &\textcolor{dgreen}{0.880}	&\textcolor{dgreen}{0.882}  &{\color{black}\bf 0.883}          
    &0.864      &0.879   &0.858      &{\color{black}\bf 0.883}
    &0.856      &0.881      &{\color{black}\bf 0.886}
    \\
    & $F_\beta^{max}\uparrow$   
    &0.751      &0.657      &0.694      &0.838      &0.830      &0.851
    &-          &0.867      &0.885      &\textcolor{dgreen}{0.900}  &0.901       
    &0.892    &0.848  &-          &-      &0.874      &{\color{black}\bf 0.909}
    &0.821      &-          &\textcolor{dgreen}{0.856}	&\textcolor{dgreen}{0.882}  &{\color{black}\bf 0.909}          
    &0.862      &0.883      &0.842  &{\color{black}\bf 0.912}
    &0.877      &0.902    &{\color{black}\bf 0.915}
    \\
    & $E_\phi^{max}\uparrow$   
    &0.853      &0.759      &0.829      &0.901      &0.895      &0.903
    &-          &0.914      &0.923      &\textcolor{dgreen}{0.919}  &0.922     
    &0.915    &0.908  &-          &-      &0.913      &{\color{black}\bf 0.924}
    &0.875      &-          &\textcolor{dgreen}{0.922}	&\textcolor{dgreen}{0.923}  &{\color{black}\bf 0.925}          
    &0.910      &0.922      &0.909  &{\color{black}\bf 0.927}
    &0.886      &0.923      &{\color{black}\bf 0.933}
    \\
    &  $MAE\downarrow$    
    &0.200      &0.185      &0.139      &0.071      &0.075      &0.064
    &-          &{\color{black}\bf 0.051}       &{\color{black}\bf 0.051}           &\textcolor{dgreen}{0.058}  &0.057     
    &0.054      &0.059      &-          &-          &0.062      &0.053
    &0.085      &-          &\textcolor{dgreen}{0.052}	&\textcolor{dgreen}{0.054}  &{\color{black}\bf 0.055}          
    &0.063      &0.055      &0.063  &{\color{black}\bf 0.051}
    &0.070      &0.052      &{\color{black}\bf 0.051}
    \\ \hline
    
    \multirow{4}{*}{DUTD}
    & $S_m \; \; \; \; \uparrow$     
    &-          &0.695      &0.499      &-          &0.526      &0.736
    &0.791      &0.801      &0.808      &\textcolor{dgreen}{-}  &0.818      
    &-          &0.899      &0.903      &-          &-      &{\color{black}\bf 0.913}
    &0.702      &0.889      &\textcolor{dgreen}{0.885}	&\textcolor{dgreen}{0.921}  &{\color{black}\bf 0.916}          
    &0.831      &-      &0.898  &{\color{black}\bf 0.917}
    &0.912          &0.923      &{\color{black}\bf 0.928}
    \\
    & $F_\beta^{max}\uparrow$   
    &-          &0.692      &0.411      &-          &0.458      &0.740
    &0.767      &0.771      &0.790      &\textcolor{dgreen}{-}  &0.898      
    &-          &0.918      &0.901      &-          &-      &{\color{black}\bf 0.919}
    &0.659      &0.795      &\textcolor{dgreen}{0.878}	&\textcolor{dgreen}{0.926}  &{\color{black}\bf 0.920}          
    &0.823      &-      &0.903  &{\color{black}\bf 0.924}
    &0.915          &0.929      &{\color{black}\bf 0.932}
    \\
    & $E_\phi^{max}\uparrow$   
    &-          &0.800      &0.654      &-          &0.709      &0.823
    &0.859      &0.856      &0.861      &\textcolor{dgreen}{-}  &0.859      
    &-          &0.937      &0.937      &-          &-      &{\color{black}\bf 0.938}
    &0.796      &0.933      &\textcolor{dgreen}{0.935}	&\textcolor{dgreen}{0.950}  &{\color{black}\bf 0.940}          
    &0.899      &-      &0.931  &{\color{black}\bf 0.940}
    &0.927          &0.939      &{\color{black}\bf 0.942}
    \\
    &  $MAE\downarrow$    
    &-          &0.220      &0.243      &-          &0.201      &0.144
    &0.113      &0.100      &0.093      &\textcolor{dgreen}{-}  &0.076      
    &-          &0.043      &0.043      &-          &-      &{\color{black}\bf 0.036}
    &0.122      &0.048      &\textcolor{dgreen}{0.057}	&\textcolor{dgreen}{0.030} &{\color{black}\bf 0.035}          
    &0.097      &-      &0.045  &{\color{black}\bf 0.036}
    &0.045          &0.042      &{\color{black}\bf 0.035}
    \\ \hline

    \end{tabular}
}
\end{table*}

\begin{table*}[!t]
	\setlength\tabcolsep{1pt}
    \caption {Quantitative results of our method and baseline networks for the ablation experiments on NJU2K~\cite{ju2014depth}, NLPR~\cite{peng2014rgbd}, STERE~\cite{niu2012leveraging}, RGBD135~\cite{cheng2014depth}, LFSD~\cite{li2014saliency}, and SIP~\cite{fan2020rethinking}. ``SL'' denotes supervised learning, while ``SSL'' represents semi-supervised learning.}
    \label{tab:ablation1}
    \resizebox{1.0\textwidth}{!}{
    \begin{tabular}{c|c|c|c|cc|cc|cc|cc|cc|cc}
    \toprule
    \multirow{2}{*}{\textbf{Name}} &
    \multirow{2}{*}{\textbf{Networks}} &
    \multirow{2}{*}{\textbf{SL}} & \multirow{2}{*}{\textbf{SSL}} & \multicolumn{2}{c|}{\textbf{NJU2K~\cite{ju2014depth}}} & \multicolumn{2}{c|}{\textbf{NLPR~\cite{peng2014rgbd}}} & \multicolumn{2}{c|}{\textbf{STERE~\cite{niu2012leveraging}}} &
    \multicolumn{2}{c|}{\textbf{RGBD135~\cite{cheng2014depth}}} & \multicolumn{2}{c|}{\textbf{LFSD~\cite{li2014saliency}}} & \multicolumn{2}{c}{\textbf{SIP~\cite{fan2020rethinking}}}  \\
    & & &
    & $S_m\uparrow$ &$MAE\downarrow$
    & $S_m\uparrow$ &$MAE\downarrow$
    & $S_m\uparrow$ &$MAE\downarrow$
    & $S_m\uparrow$ &$MAE\downarrow$
    & $S_m\uparrow$ &$MAE\downarrow$
    & $S_m\uparrow$ &$MAE\downarrow$ \\
    
    \midrule
    \midrule
    $M_1$ & ``DDCNN-w/o-DAM''  &$\surd$ & 
    &0.881 &0.065 	
    &0.887 &0.038
    &0.837 &0.088 	
    &0.834 &0.060 	
    &0.750 &0.157 	
    &0.755 &0.125 \\
    
    $M_2$ & ``DDCNN-w/o-DGM''  &$\surd$ & 
    &0.893 &0.054 	
    &0.888 &0.036
    &0.848 &0.071 	
    &0.859 &0.056 	
    &0.765 &0.148 	
    &0.762 &0.115 \\
    
    $M_3$ & ``DDCNN-w/o-DIM''  &$\surd$ & 
    &0.835 &0.089 	
    &0.878 &0.042
    &0.659 &0.176 	
    &0.811 &0.062 	
    &0.652 &0.201 	
    &0.672 &0.172 \\
    
    $M_4$ & ``DDCNN-w/o-depth''  &$\surd$ & 
    & 0.855 	&0.079 	
    & 0.888  	&0.037
    & 0.698  	&0.162
    & 0.829  	&0.056 
    & 0.669  	&0.194 
    & 0.685  	&0.164  \\
    
    $M_5$ & ``DDCNN-w/o-reconstr-loss''  & $\surd$ & 
    &0.894 	 	&0.051 	
    &0.908  	&0.038
    &0.835 	 	&0.081
    &0.899  	&0.042 
    &0.744  	&0.144 
    &0.766  	&0.110 \\
    
    $M_6$ & ``DDCNN-w/o-pretrain'' & $\surd$ &   
    &0.934 	 	&0.033 	
    &0.939 	 	&0.023
    &0.902 	 	&0.045
    &0.919  	&0.027
    &0.855  	&0.079
    &0.866  	&0.061 \\

    $M_7$ & DDCNN & $\surd$ & 
    &0.936 	 	&0.033 	
    &0.939 	 	&0.024
    &0.907 	 	&0.040
    &0.925  	&0.025
    &0.862  	&0.075
    &0.881  	&0.052 \\
    \hline
     
    $M_8$ & ``DDCNN-semi'' &  & $\surd$ 
    &0.941 	 	&0.025 	
    &0.942 	 	&0.020
    &0.911   	&0.038
    &0.934  	&0.024
    &0.869  	&0.071 
    &0.883  	&0.052  \\

    $M_9$ & ``DDCNN-semi-ourSplit''  & &$\surd$
    &0.922 &0.037 	
    &0.922 &0.032
    &0.892 &0.041 	
    &0.902 &0.032 	
    &0.842 &0.082 	
    &0.856 &0.070 \\

    $M_{10}$ & ``DDCNN-semi-depthEst''  & &$\surd$
    &0.938 &0.031 	
    &0.941 &0.022
    &0.906 &0.039 	
    &0.926 &0.025 	
    &0.866 &0.074 	
    &0.881 &0.053 \\
    \hline
    
    & \textbf{Our DS-Net} &  & $\surd$
    & \textbf{0.950} 	&\textbf{0.024}
    & \textbf{0.952} 	&\textbf{0.018}
    & \textbf{0.914} 	&\textbf{0.037}
    & \textbf{0.936} 	&\textbf{0.021}
    & \textbf{0.878} 	&\textbf{0.064}
    & \textbf{0.886} 	&\textbf{0.051} \\
    \bottomrule
    \end{tabular}
    }
\end{table*}

{\color{dgreen}
\subsection{Training Strategies}
\vspace{2mm}
\noindent
\textbf{Supervised loss for labeled data.}
}
For labeled data, we have a pair of input RGB and depth images with the corresponding annotated saliency mask.
It is natural to take the annotated saliency mask as the ground truth ($G_{s}$) for RGB-D saliency detection.
On the other hand, we also have the ground truth ($G_d$) for the depth estimation task, which is the input depth map.
With the two ground truths ($G_{s}$ and $G_{d}$), the supervised loss (denoted as $\mathcal{L}^{s}$) for a labeled image ($x$) is computed as the summation of the saliency detection loss and depth estimation loss:
\vskip -10pt
\begin{equation}
	\label{eq:supervised_loss}
    \mathcal{L}^s(x)= L_{BCE} \left( P_{s}, G_{s} \right) + \alpha \ L_{MSE} \left( P_{d}, G_{d} \right) \ ,
\end{equation}
where $P_{s}$ and $P_{d}$ denote the predicted saliency map and depth map, respectively.
$L_{BCE}$ and $L_{MSE}$ are the binary cross-entropy (BCE) loss and MSE loss functions, respectively.
We empirically set the weight $\alpha$$=$$1.0$ during the network training.

{\color{dgreen}
\vspace{2mm}
\noindent
\textbf{Consistency loss for unlabeled data}.
}
As a well-known semi-supervised manner, the teacher-student approach ensembles parameters of the network at different training processes as teacher network's parameters and devises a  consistency loss to make the student network learn from the teacher network for improving the quality of the network predictions.
In detail, we feed the unlabeled RGB data into the depth estimation branch of DDCNN to predict a depth map and take this prediction as a pseudo label of the depth map for the unlabeled RGB data. 
After that, we pass the unlabeled RGB image and corresponding pseudo depth image into the student and teacher networks to obtain two groups of depth map and saliency map prediction results.
We then force the predicted saliency map \textcolor{dgreen}{and the four learned attention maps} of DIMs to be consistent for the student and teacher networks. A consistency loss $\mathcal{L}^c$ on unlabeled data $y$ is defined as:
\vskip -10pt
\begin{equation}
\small
	\label{eq:consistency_loss}
	\mathcal{L}^c(y) = L_{MSE} \left( S_{s}, T_{s} \right) + \gamma \sum_{l=1}^{4}L_{MSE}(S_A^{l}, T_A^{l}) \ ,
\end{equation}
where $S_{s}$ and $T_{s}$ denote two detected saliency maps of the student and the teacher networks.
$S_{A}^{l}$ and $T_{A}^{l}$ represent the learned attention map of DIM at the $l$-th CNN layer from the student and the teacher networks.
We empirically set $\gamma=0.1$ in this work.

\begin{figure*}[!t]
	\centering
    \vspace*{0.5mm}
	\begin{subfigure}{0.071\textwidth}
	\includegraphics[width=\textwidth]{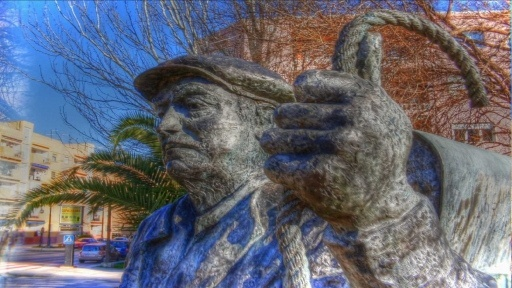}
	\end{subfigure}
	\begin{subfigure}{0.071\textwidth}
    \includegraphics[width=\textwidth]{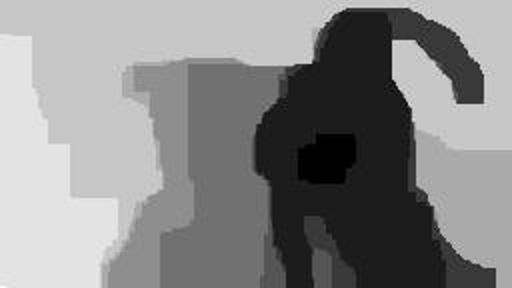}
	\end{subfigure}
    \begin{subfigure}{0.071\textwidth}
		\includegraphics[width=\textwidth]{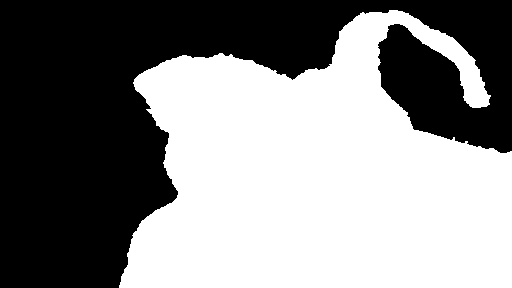}
	\end{subfigure}
	\begin{subfigure}{0.071\textwidth}
		\includegraphics[width=\textwidth]{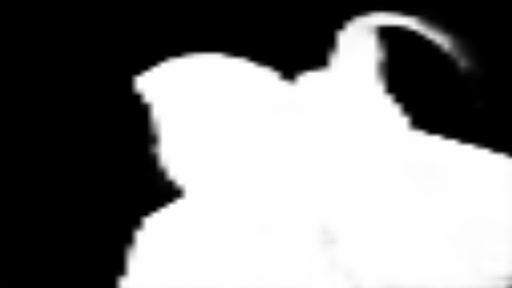}
	\end{subfigure}
	\begin{subfigure}{0.071\textwidth}
		\includegraphics[width=\textwidth]{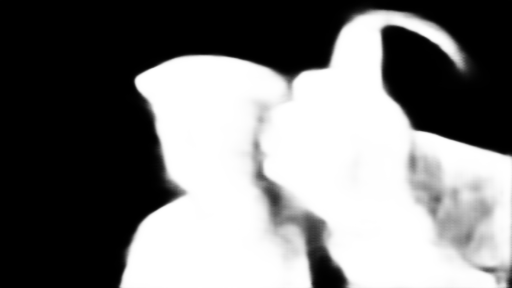}
	\end{subfigure}
	\begin{subfigure}{0.071\textwidth}
		\includegraphics[width=\textwidth]{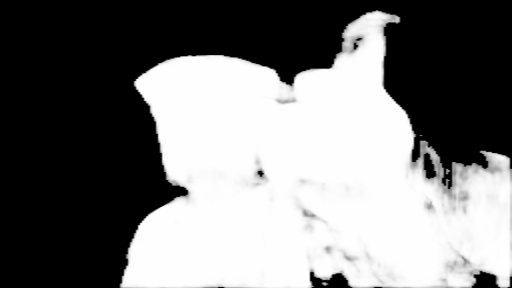}
	\end{subfigure}
	\begin{subfigure}{0.071\textwidth}
		\includegraphics[width=\textwidth]{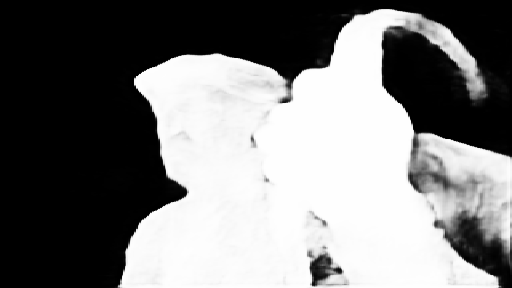}
	\end{subfigure}
	\begin{subfigure}{0.071\textwidth}
		\includegraphics[width=\textwidth]{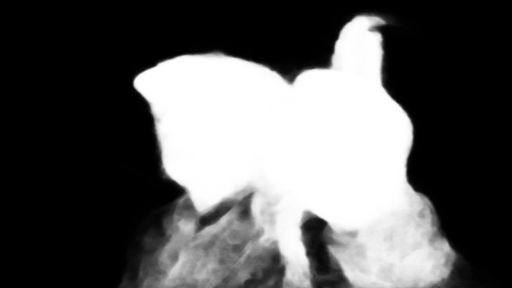}
	\end{subfigure}
    \begin{subfigure}{0.071\textwidth}
		\includegraphics[width=\textwidth]{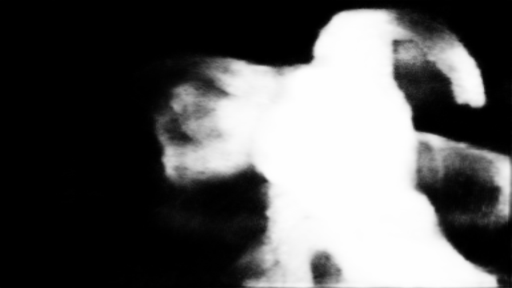}
	\end{subfigure}
	\begin{subfigure}{0.071\textwidth}
		\includegraphics[width=\textwidth]{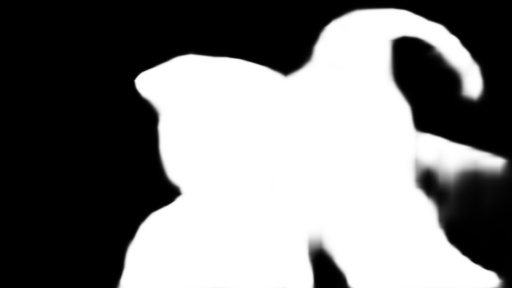}
	\end{subfigure}
	\begin{subfigure}{0.071\textwidth}
		\includegraphics[width=\textwidth]{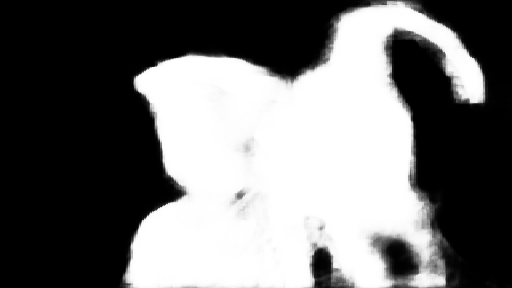}
	\end{subfigure}
    \begin{subfigure}{0.071\textwidth}
		\includegraphics[width=\textwidth]{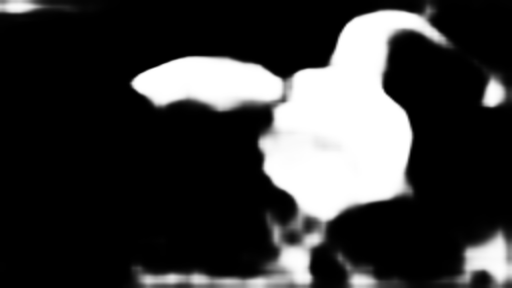}
	\end{subfigure}
	\begin{subfigure}{0.071\textwidth}
		\includegraphics[width=\textwidth]{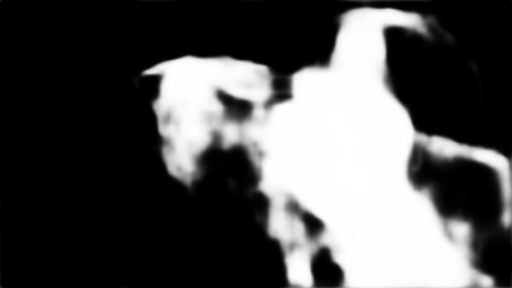}
	\end{subfigure}
	\ \\	
    \vspace*{0.5mm}
	\begin{subfigure}{0.071\textwidth}
	\includegraphics[width=\textwidth]{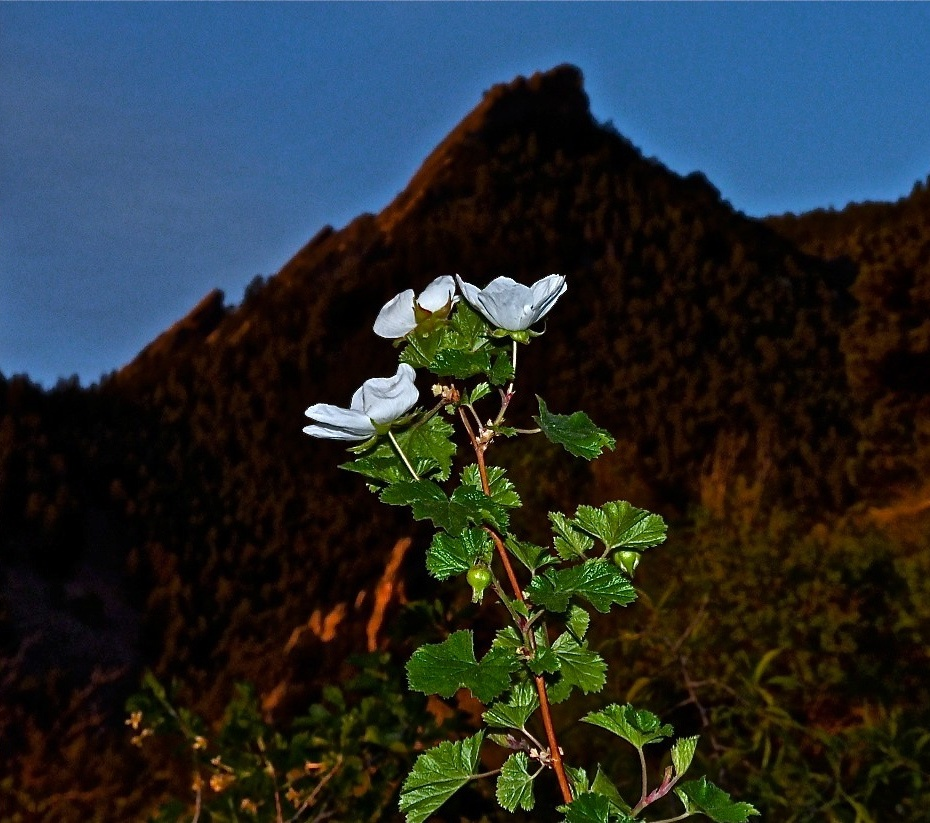}
	\end{subfigure}
	\begin{subfigure}{0.071\textwidth}
    \includegraphics[width=\textwidth]{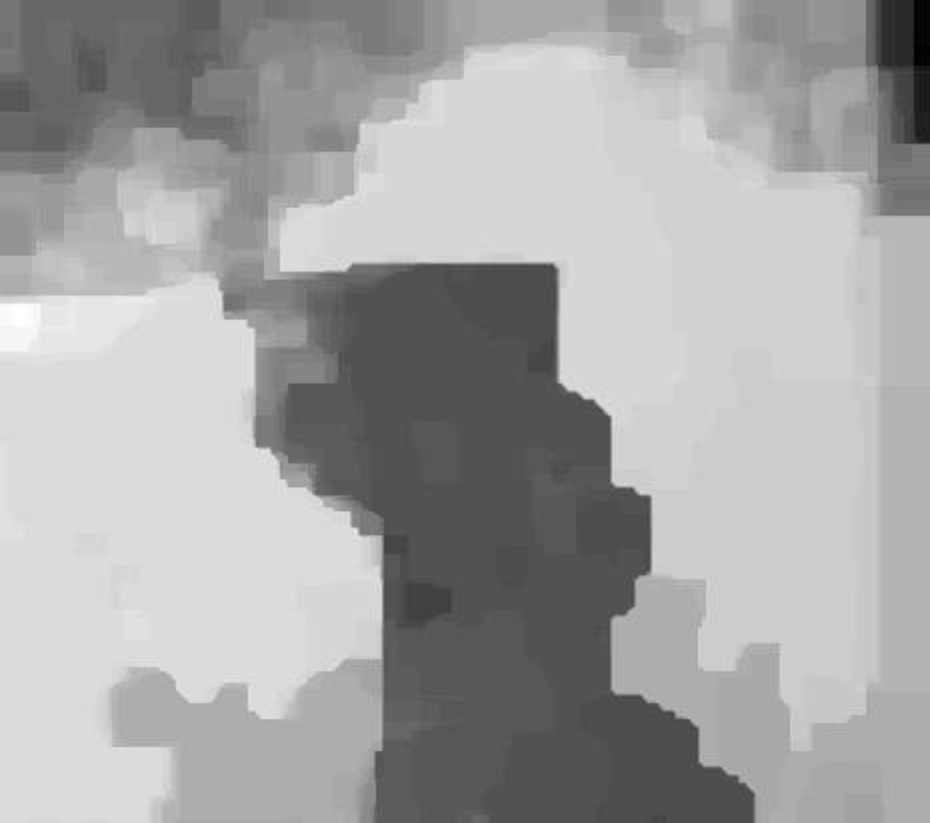}
	\end{subfigure}
    \begin{subfigure}{0.071\textwidth}
		\includegraphics[width=\textwidth]{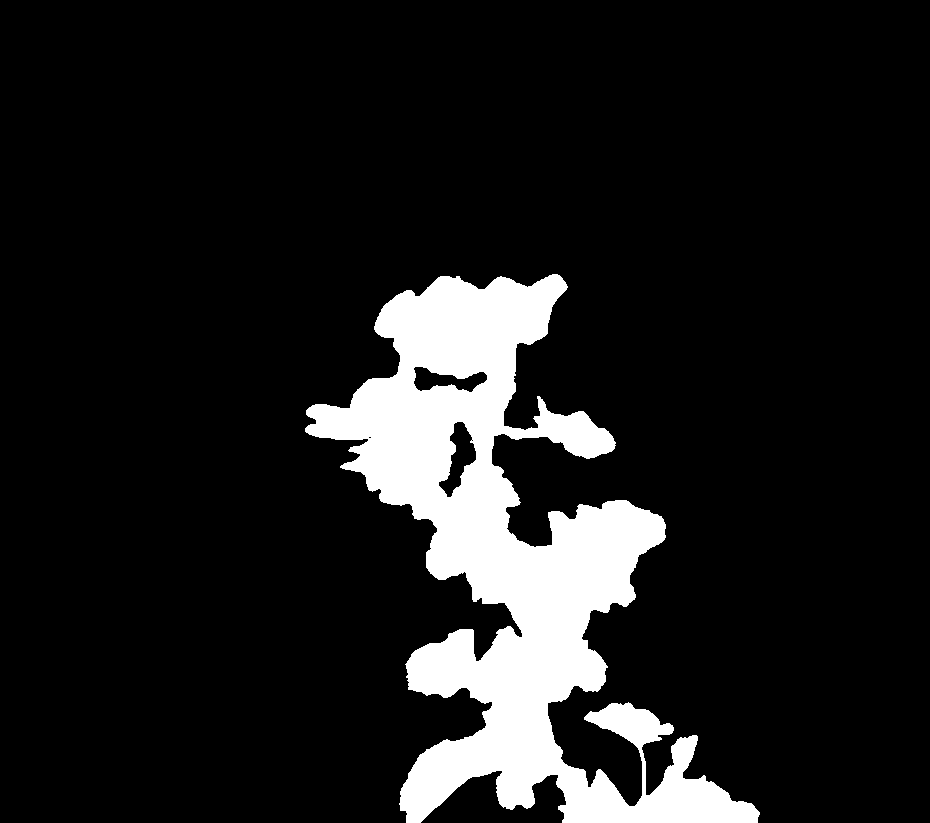}
	\end{subfigure}
	\begin{subfigure}{0.071\textwidth}
		\includegraphics[width=\textwidth]{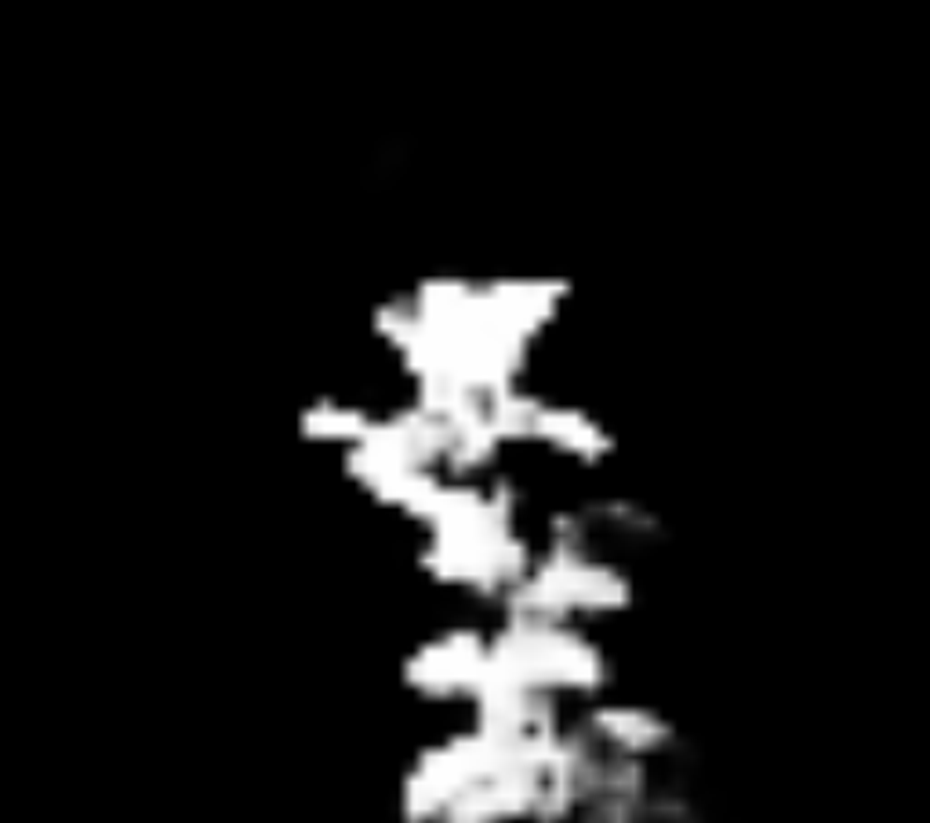}
	\end{subfigure}
	\begin{subfigure}{0.071\textwidth}
		\includegraphics[width=\textwidth]{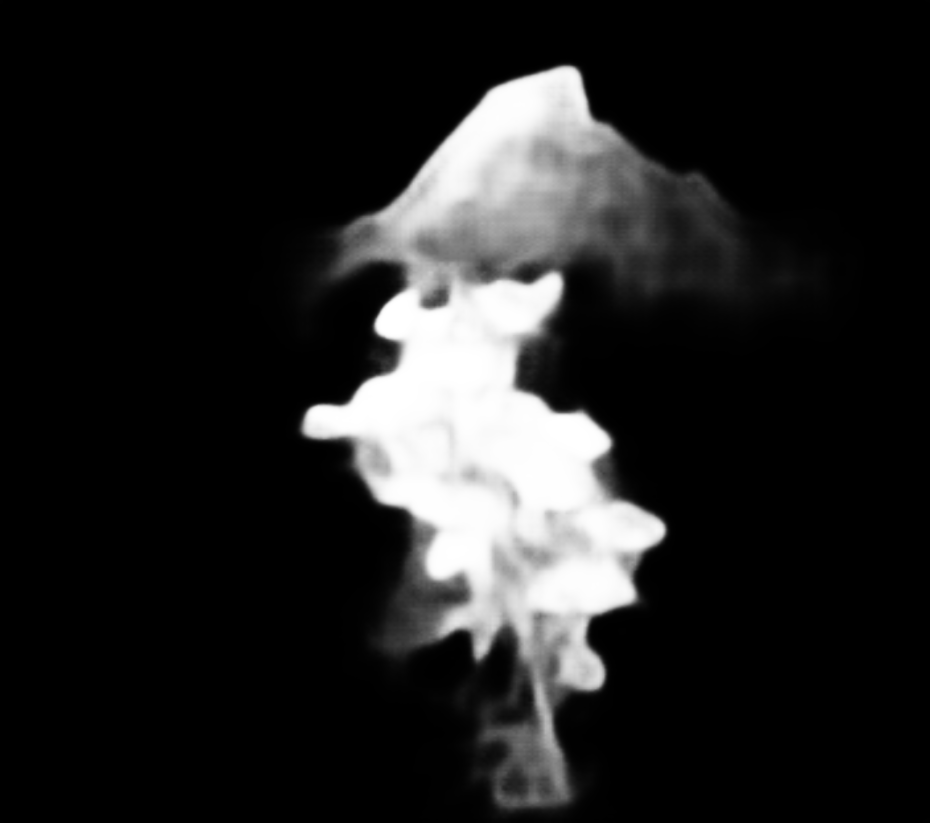}
	\end{subfigure}
	\begin{subfigure}{0.071\textwidth}
		\includegraphics[width=\textwidth]{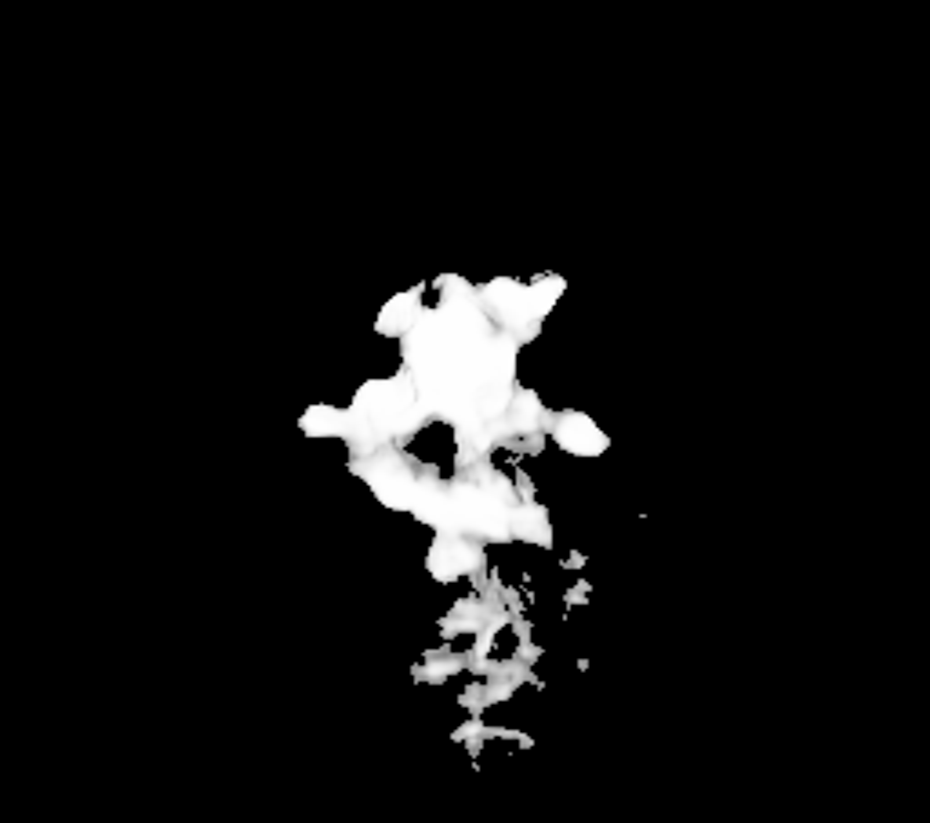}
	\end{subfigure}
	\begin{subfigure}{0.071\textwidth}
		\includegraphics[width=\textwidth]{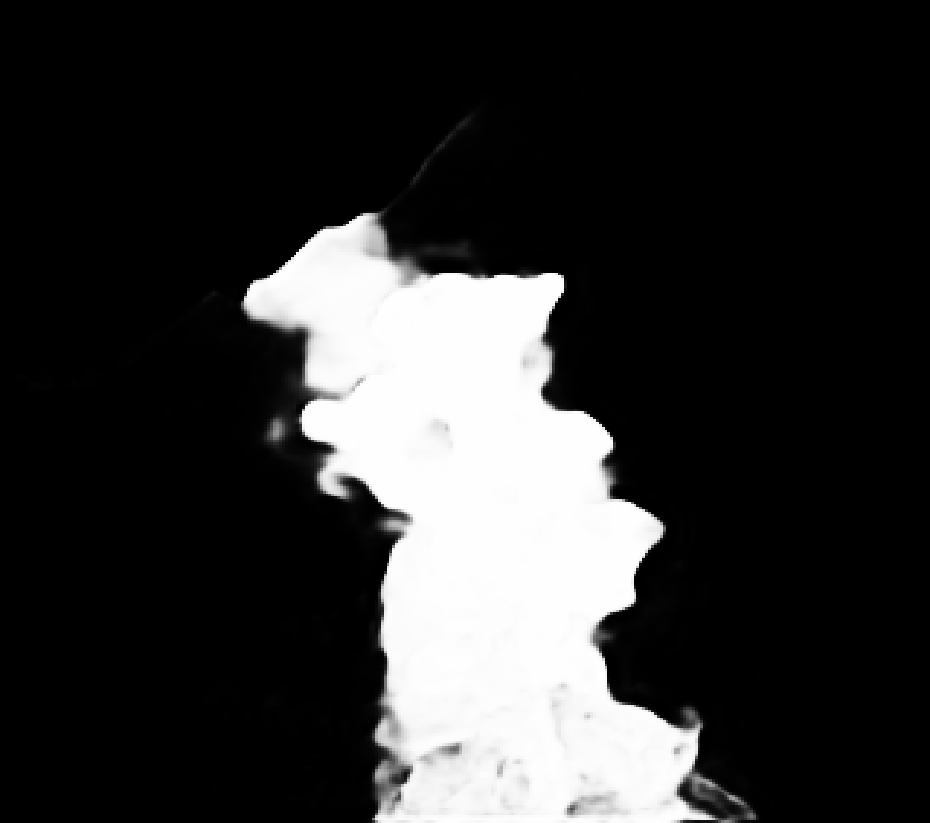}
	\end{subfigure}
	\begin{subfigure}{0.071\textwidth}
		\includegraphics[width=\textwidth]{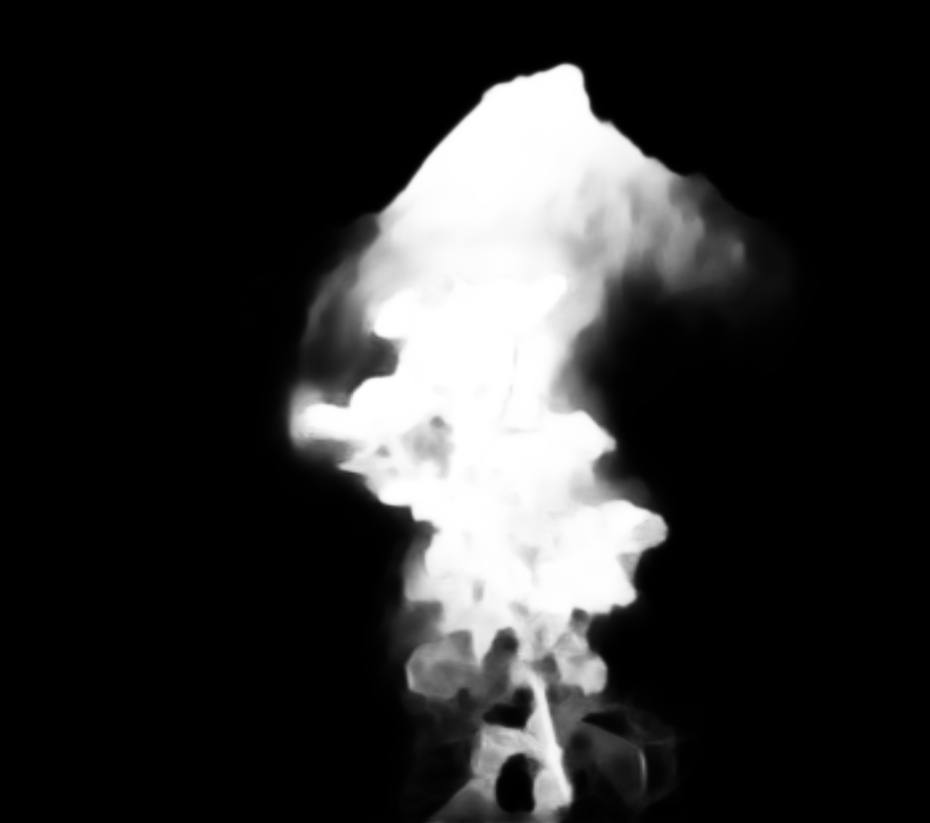}
	\end{subfigure}
    \begin{subfigure}{0.071\textwidth}
		\includegraphics[width=\textwidth]{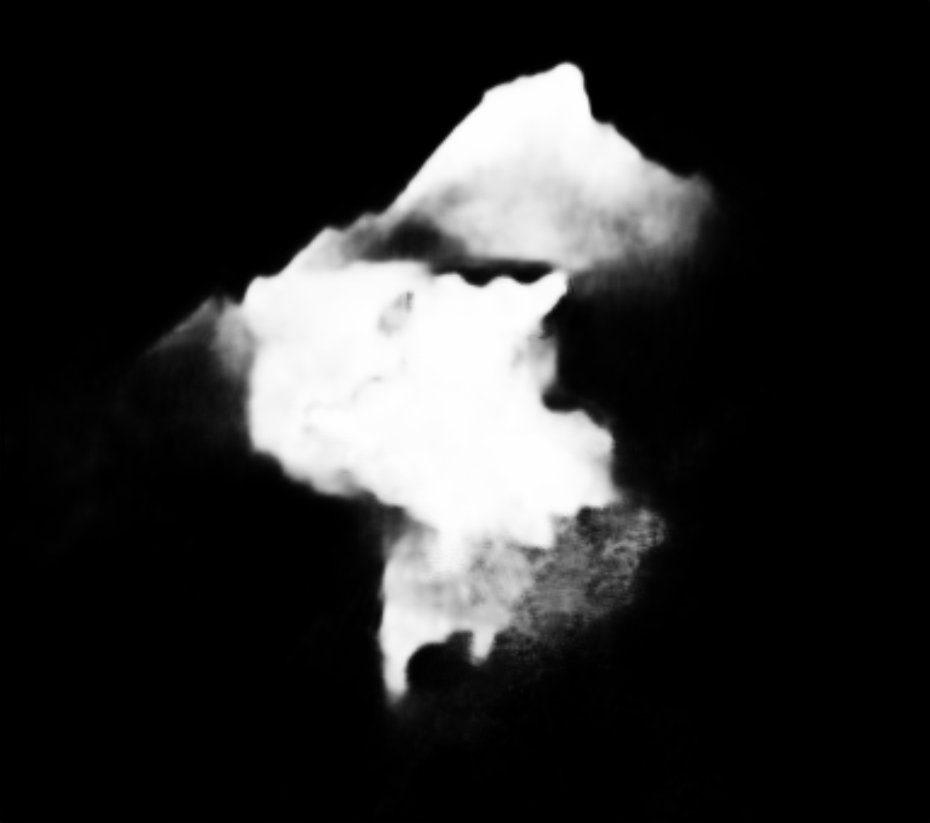}
	\end{subfigure}
	\begin{subfigure}{0.071\textwidth}
		\includegraphics[width=\textwidth]{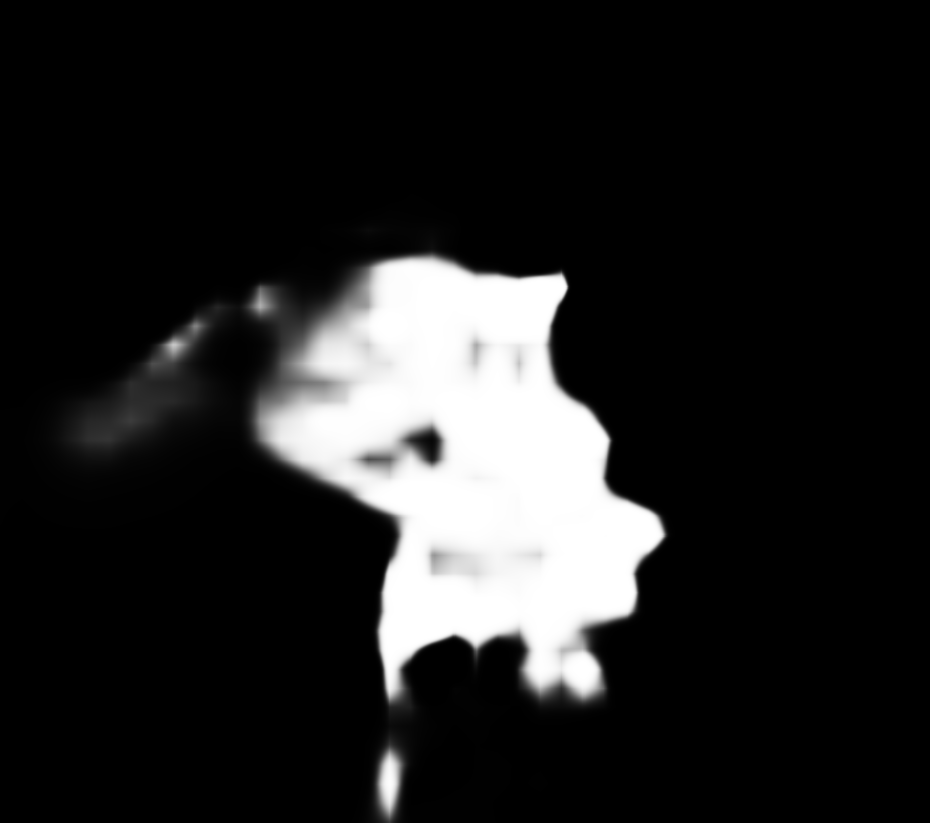}
	\end{subfigure}
	\begin{subfigure}{0.071\textwidth}
		\includegraphics[width=\textwidth]{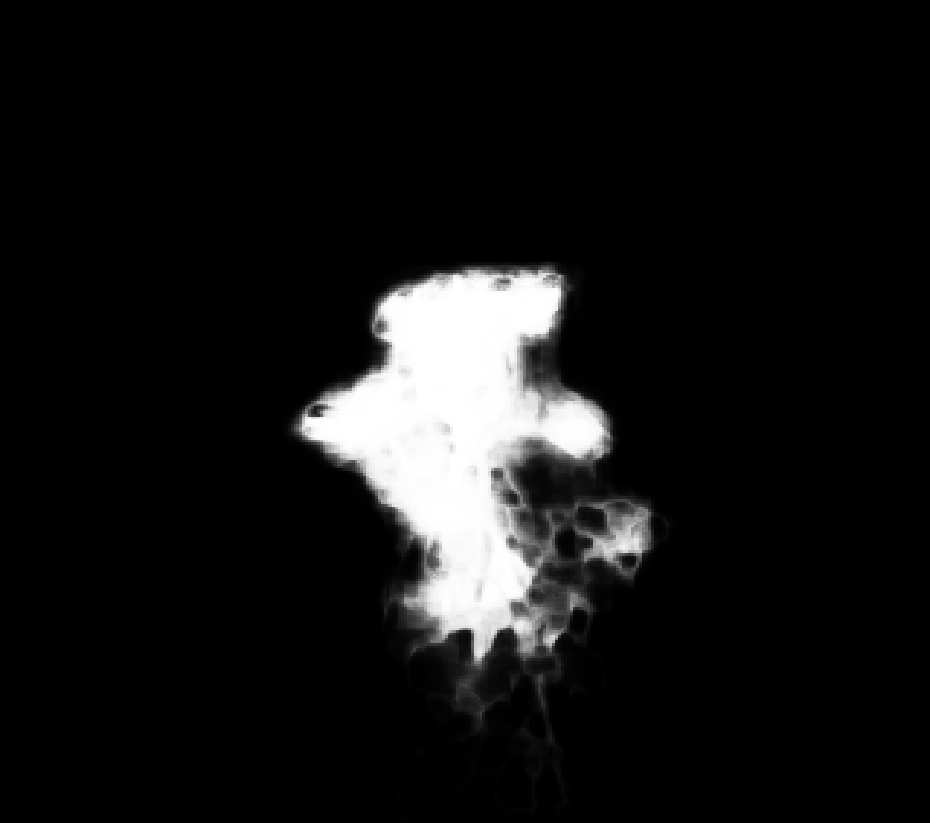}
	\end{subfigure}
    \begin{subfigure}{0.071\textwidth}
		\includegraphics[width=\textwidth]{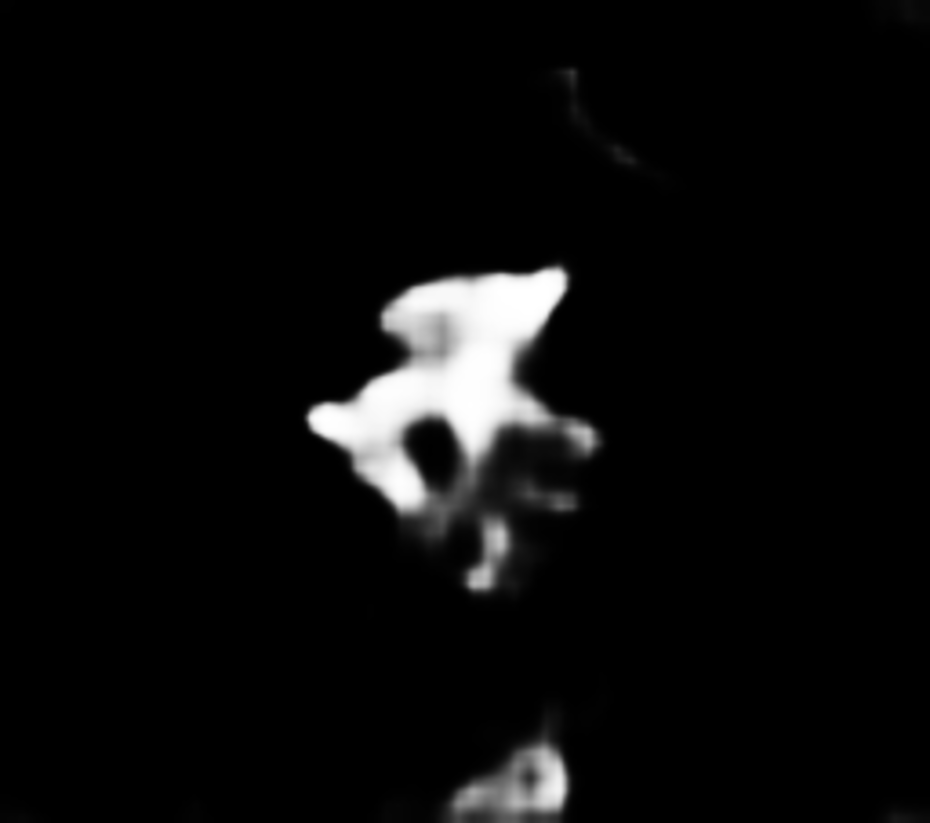}
	\end{subfigure}
	\begin{subfigure}{0.071\textwidth}
		\includegraphics[width=\textwidth]{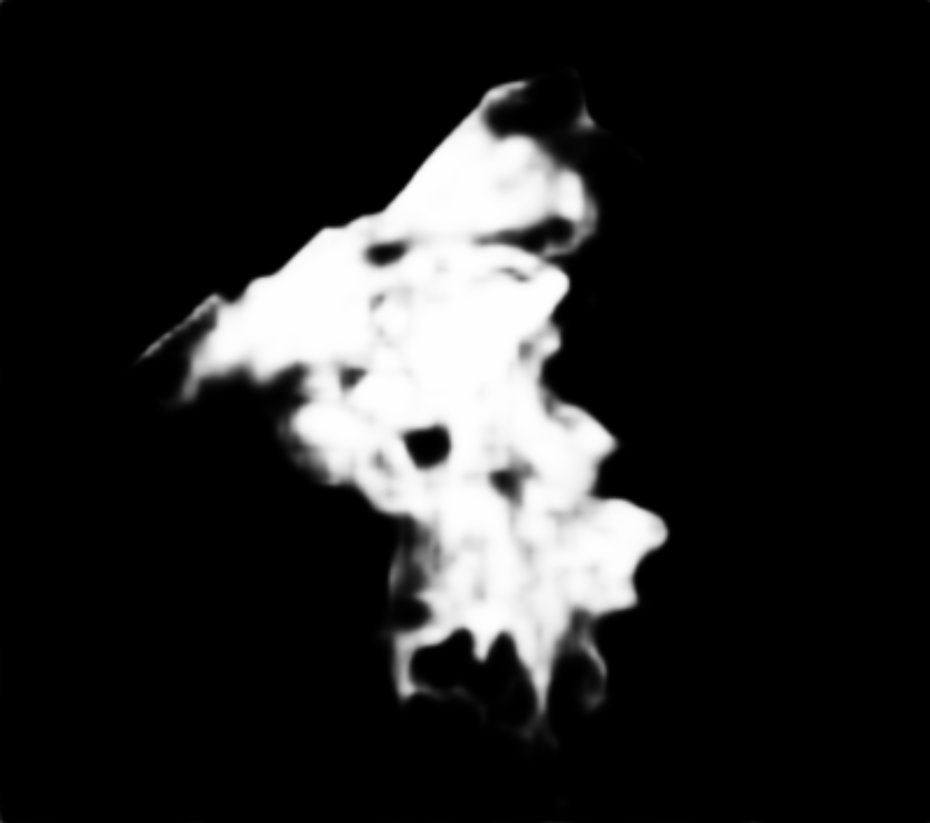}
	\end{subfigure}
	\ \\	
    \vspace*{0.5mm}
	\begin{subfigure}{0.071\textwidth}
	\includegraphics[width=\textwidth]{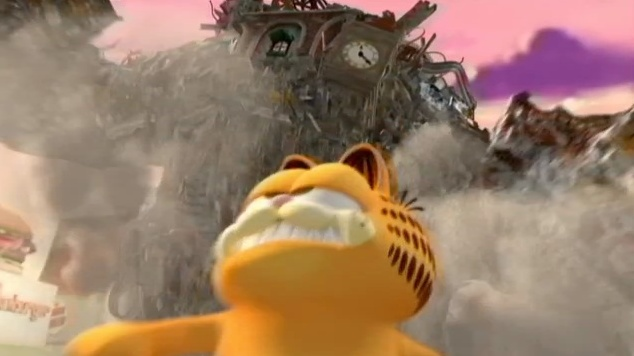}
	\end{subfigure}
	\begin{subfigure}{0.071\textwidth}
    \includegraphics[width=\textwidth]{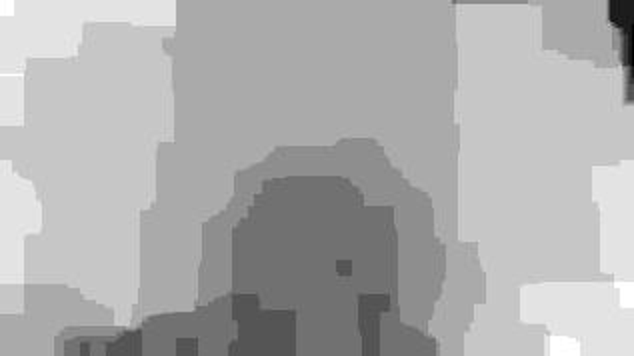}
	\end{subfigure}
    \begin{subfigure}{0.071\textwidth}
		\includegraphics[width=\textwidth]{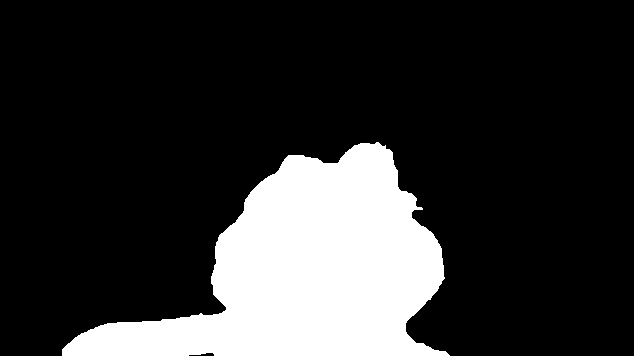}
	\end{subfigure}
	\begin{subfigure}{0.071\textwidth}
		\includegraphics[width=\textwidth]{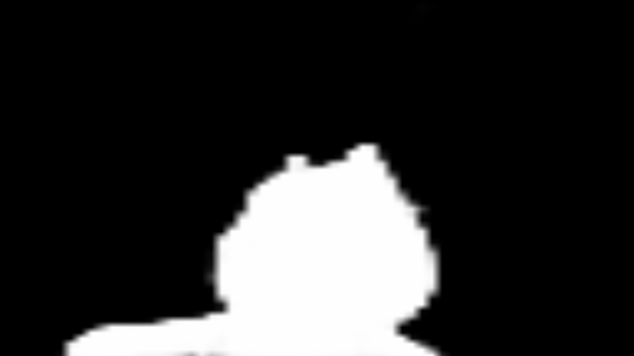}
	\end{subfigure}
	\begin{subfigure}{0.071\textwidth}
		\includegraphics[width=\textwidth]{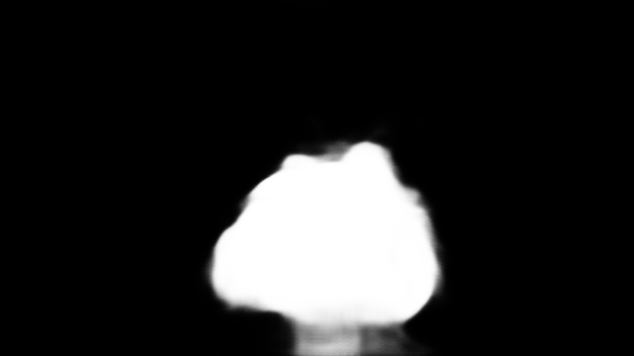}
	\end{subfigure}
	\begin{subfigure}{0.071\textwidth}
		\includegraphics[width=\textwidth]{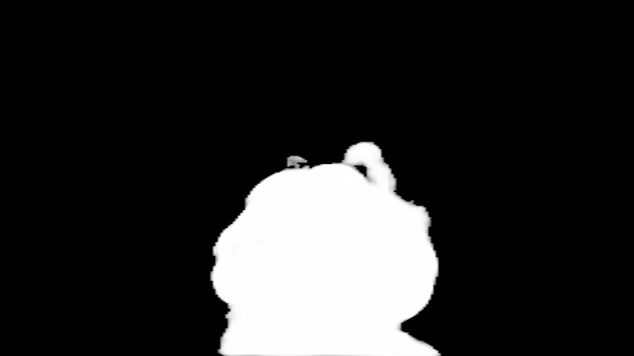}
	\end{subfigure}
	\begin{subfigure}{0.071\textwidth}
		\includegraphics[width=\textwidth]{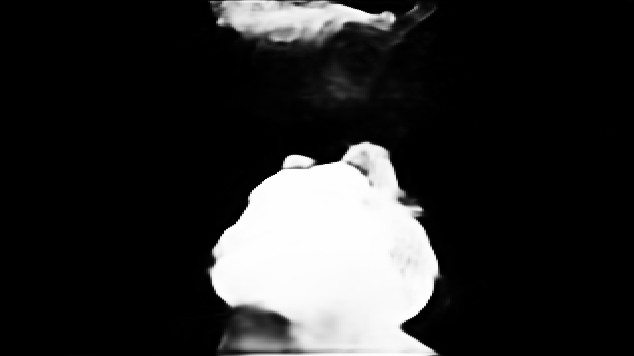}
	\end{subfigure}
	\begin{subfigure}{0.071\textwidth}
		\includegraphics[width=\textwidth]{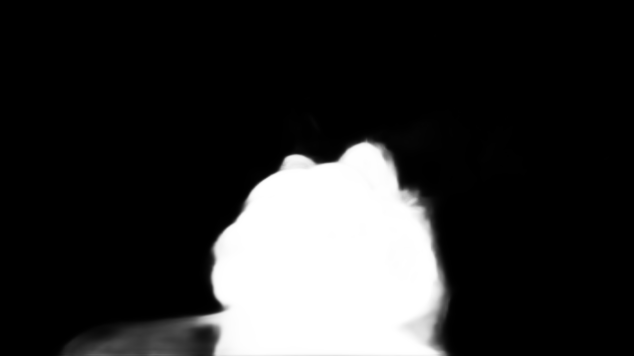}
	\end{subfigure}
    \begin{subfigure}{0.071\textwidth}
		\includegraphics[width=\textwidth]{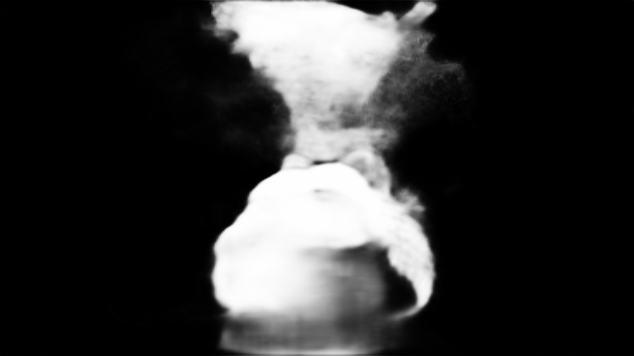}
	\end{subfigure}
	\begin{subfigure}{0.071\textwidth}
		\includegraphics[width=\textwidth]{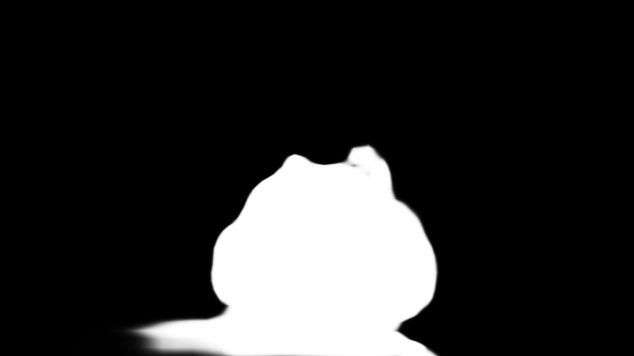}
	\end{subfigure}
	\begin{subfigure}{0.071\textwidth}
		\includegraphics[width=\textwidth]{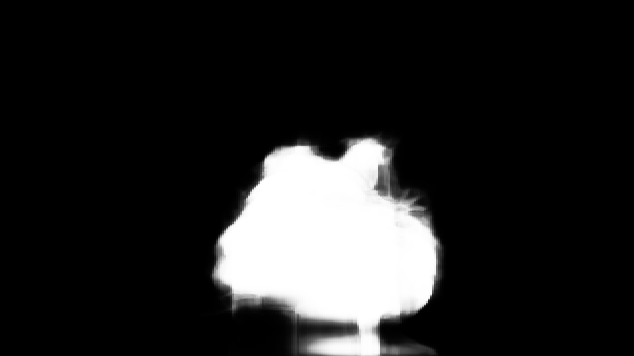}
	\end{subfigure}
    \begin{subfigure}{0.071\textwidth}
		\includegraphics[width=\textwidth]{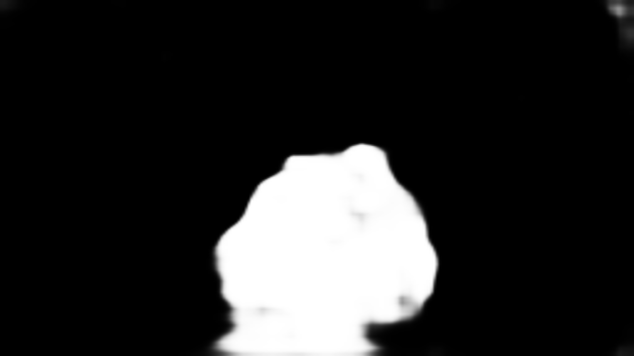}
	\end{subfigure}
	\begin{subfigure}{0.071\textwidth}
		\includegraphics[width=\textwidth]{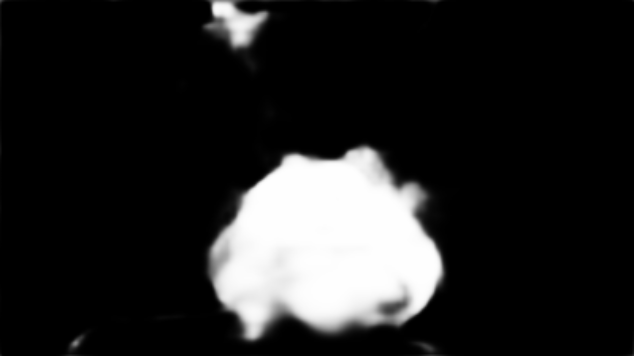}
	\end{subfigure}
 	\ \\
    \vspace*{0.5mm}
	\begin{subfigure}{0.071\textwidth}
	\includegraphics[width=\textwidth]{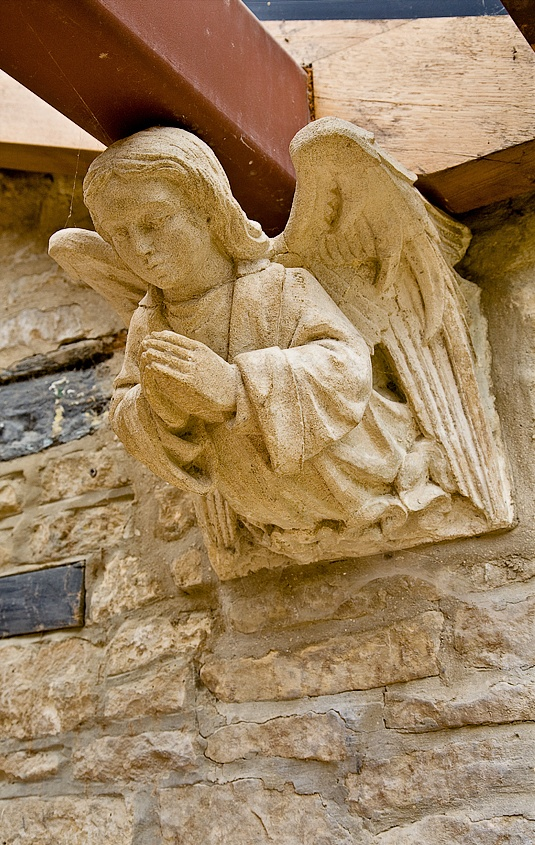}
	\end{subfigure}
	\begin{subfigure}{0.071\textwidth}
    \includegraphics[width=\textwidth]{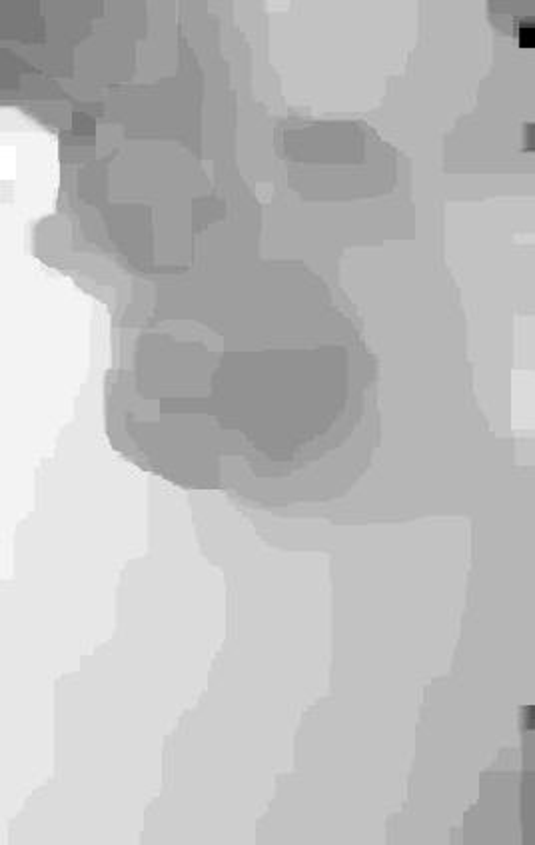}
	\end{subfigure}
    \begin{subfigure}{0.071\textwidth}
		\includegraphics[width=\textwidth]{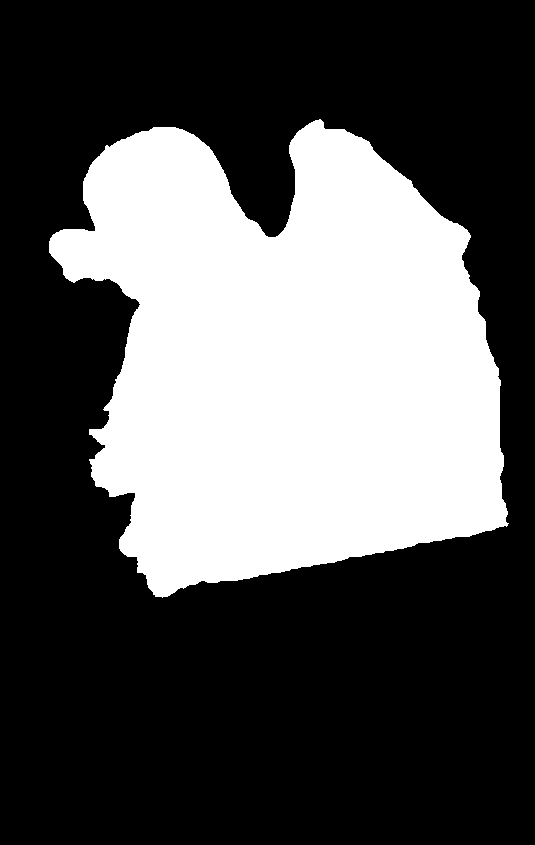}
	\end{subfigure}
	\begin{subfigure}{0.071\textwidth}
		\includegraphics[width=\textwidth]{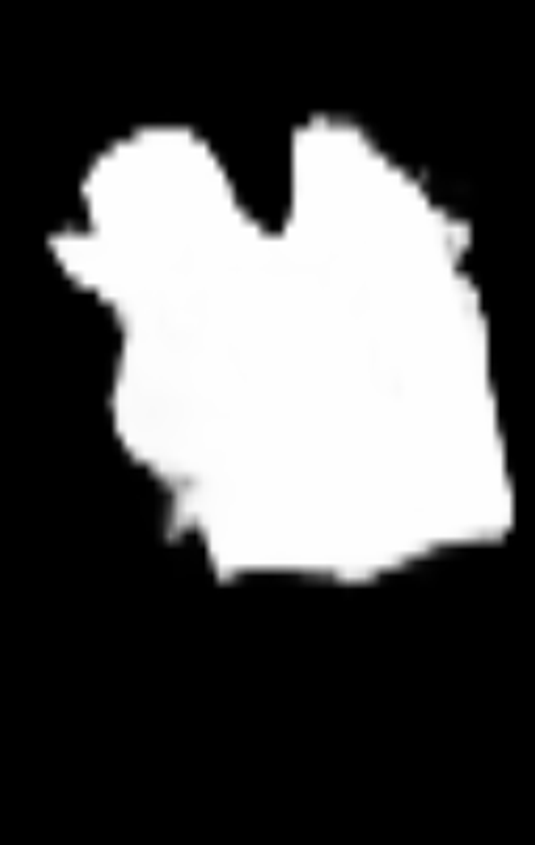}
	\end{subfigure}
	\begin{subfigure}{0.071\textwidth}
		\includegraphics[width=\textwidth]{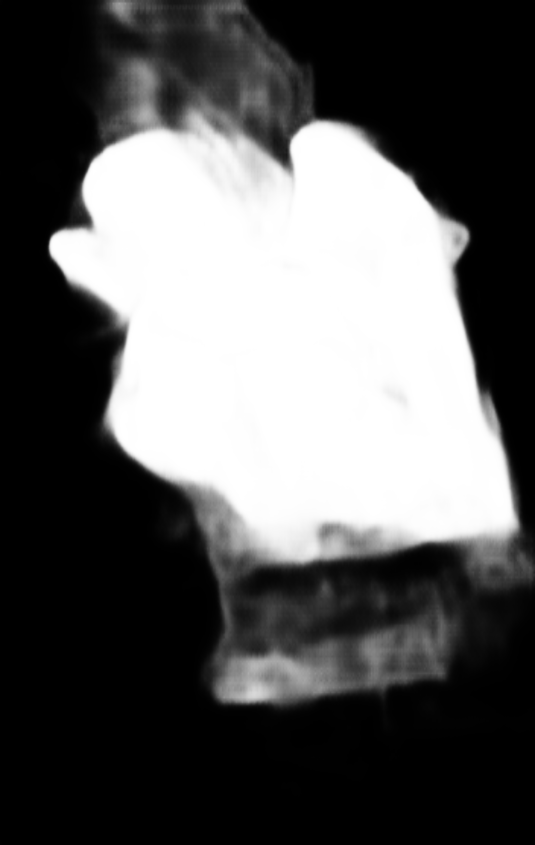}
	\end{subfigure}
	\begin{subfigure}{0.071\textwidth}
		\includegraphics[width=\textwidth]{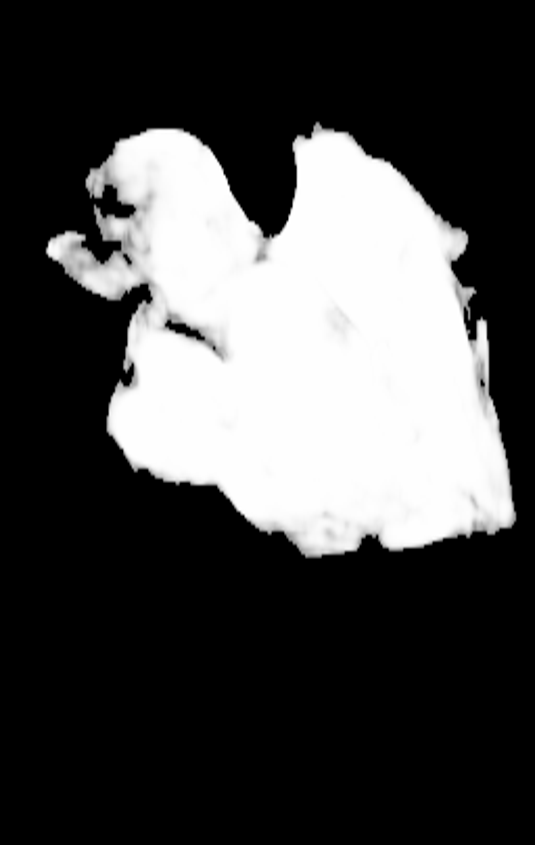}
	\end{subfigure}
	\begin{subfigure}{0.071\textwidth}
		\includegraphics[width=\textwidth]{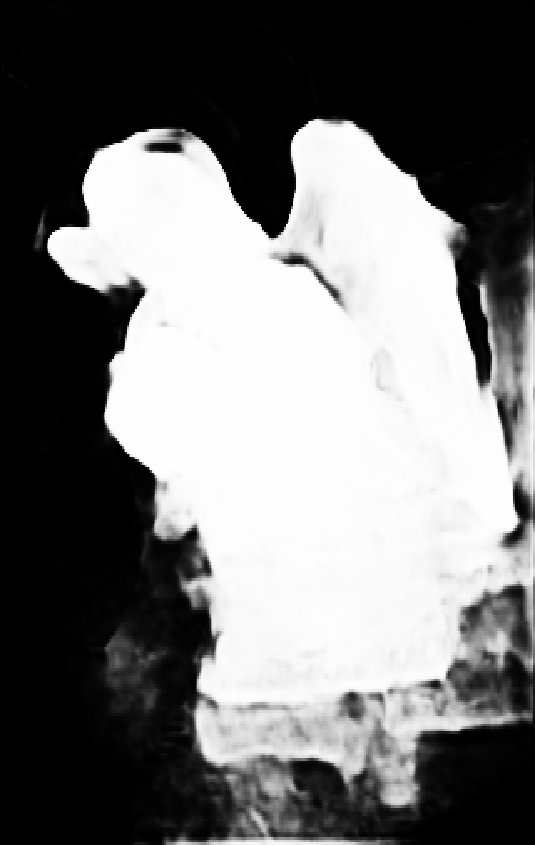}
	\end{subfigure}
	\begin{subfigure}{0.071\textwidth}
		\includegraphics[width=\textwidth]{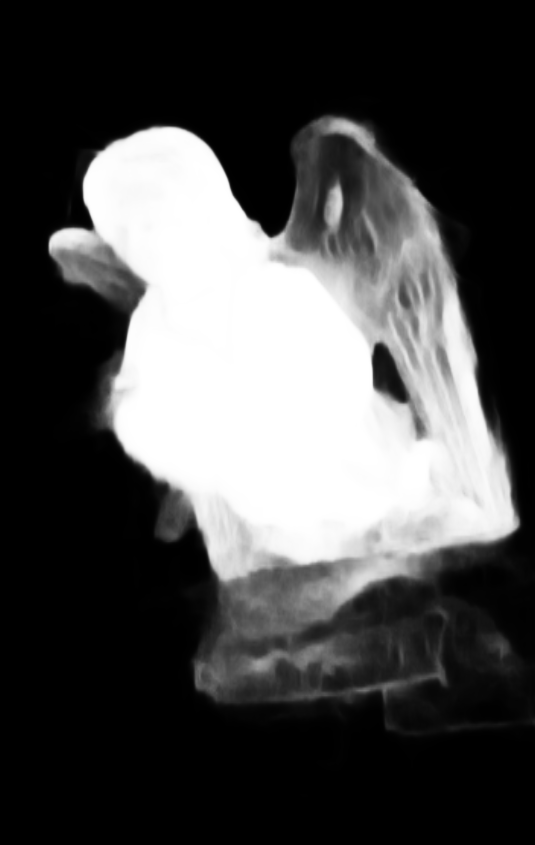}
	\end{subfigure}
    \begin{subfigure}{0.071\textwidth}
		\includegraphics[width=\textwidth]{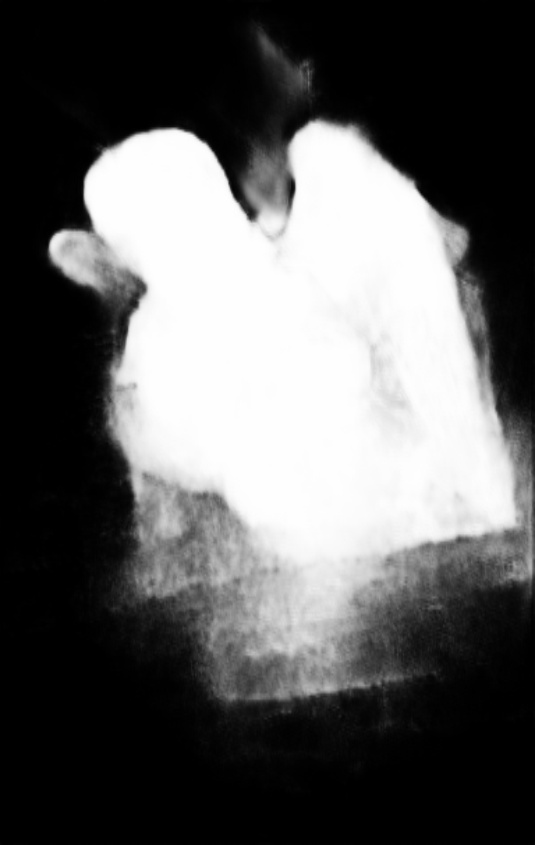}
	\end{subfigure}
    \begin{subfigure}{0.071\textwidth}
		\includegraphics[width=\textwidth]{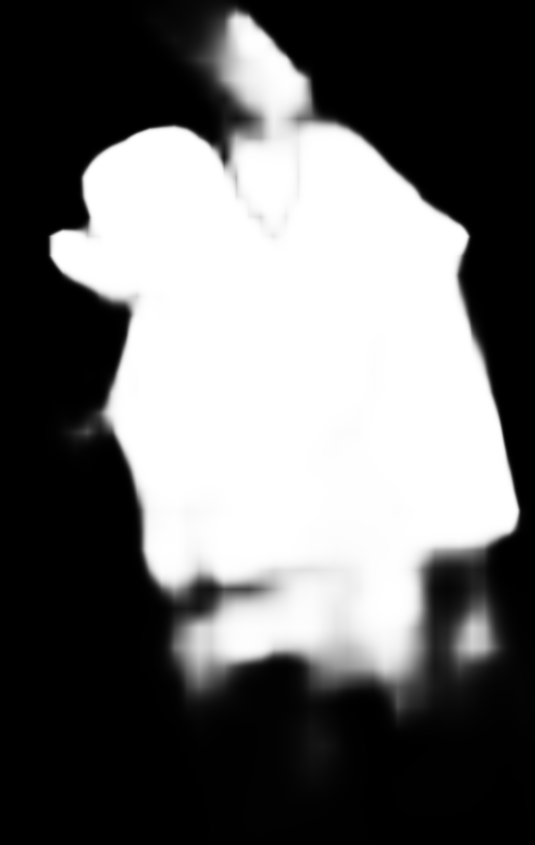}
	\end{subfigure}
	\begin{subfigure}{0.071\textwidth}
		\includegraphics[width=\textwidth]{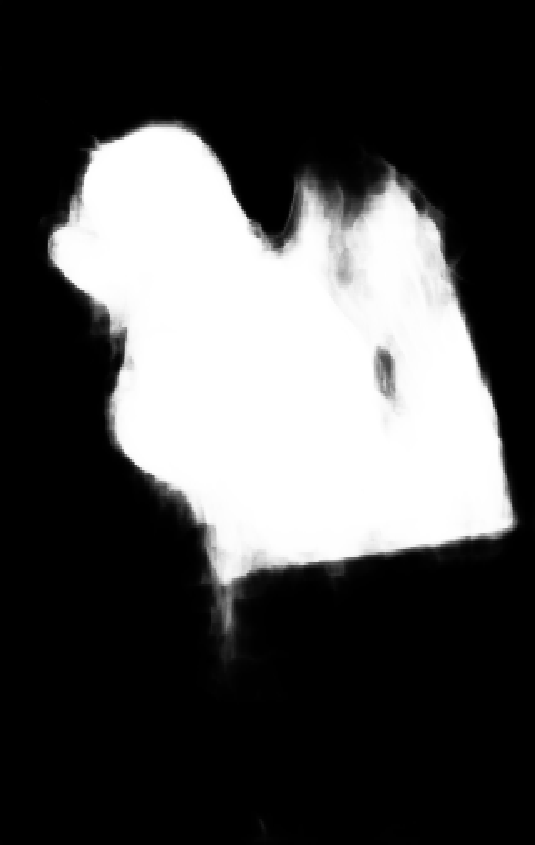}
	\end{subfigure}
    \begin{subfigure}{0.071\textwidth}
		\includegraphics[width=\textwidth]{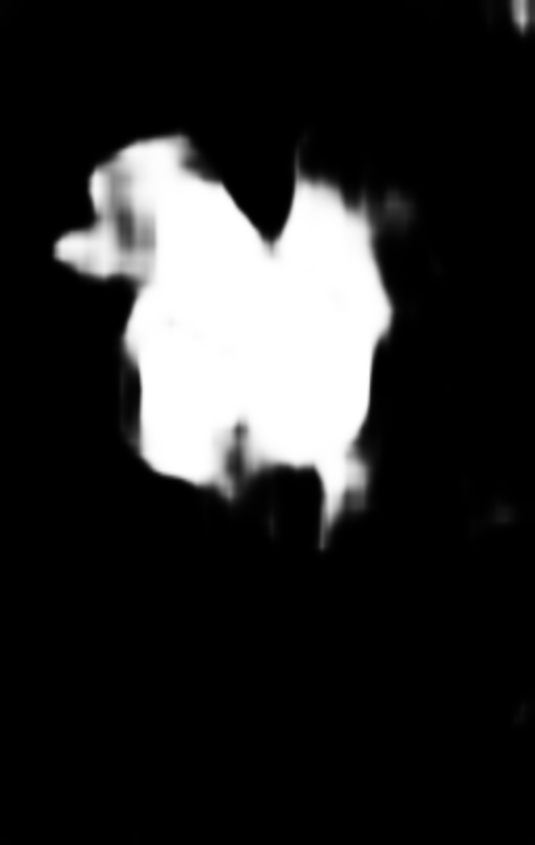}
	\end{subfigure}
	\begin{subfigure}{0.071\textwidth}
		\includegraphics[width=\textwidth]{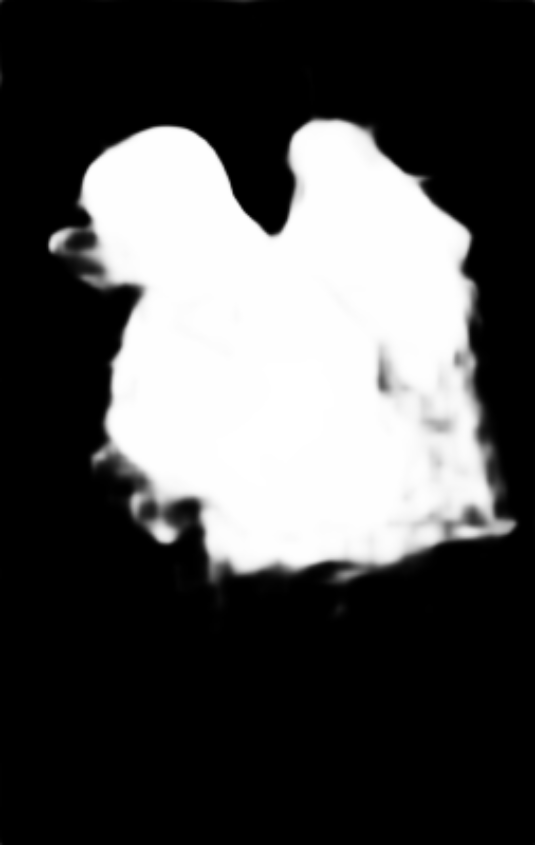}
	\end{subfigure}
	\ \\	
    \vspace*{0.5mm}
	\begin{subfigure}{0.071\textwidth}
		\includegraphics[width=\textwidth]{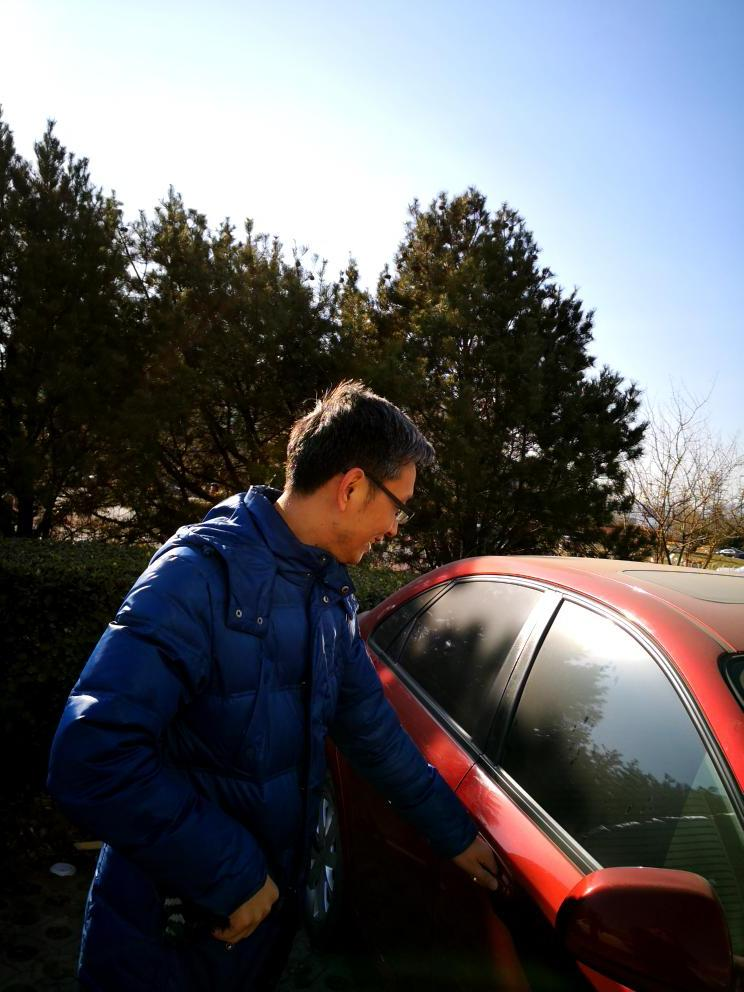}
	\end{subfigure}
	\begin{subfigure}{0.071\textwidth}
		\includegraphics[width=\textwidth]{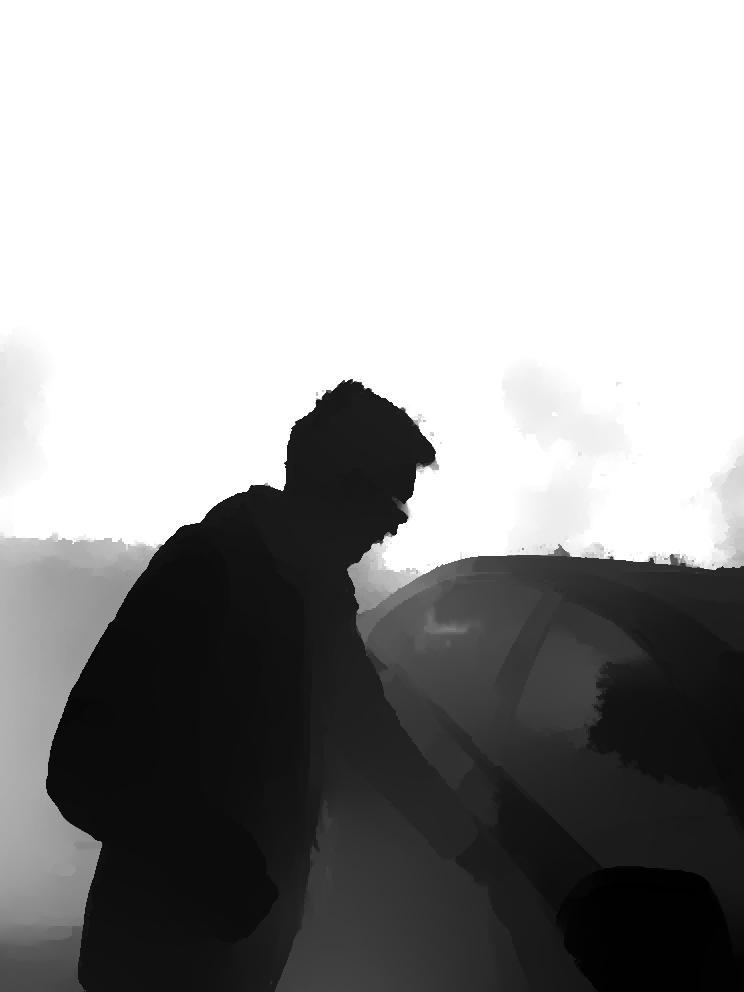}
	\end{subfigure}
    \begin{subfigure}{0.071\textwidth}
		\includegraphics[width=\textwidth]{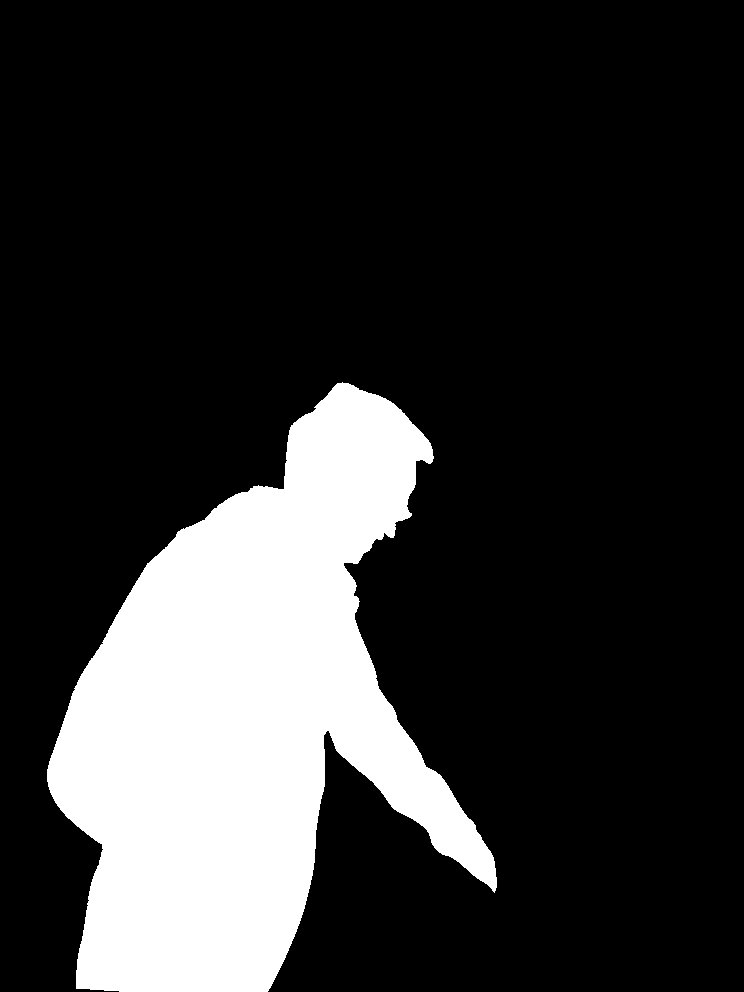}
	\end{subfigure}
	\begin{subfigure}{0.071\textwidth}
		\includegraphics[width=\textwidth]{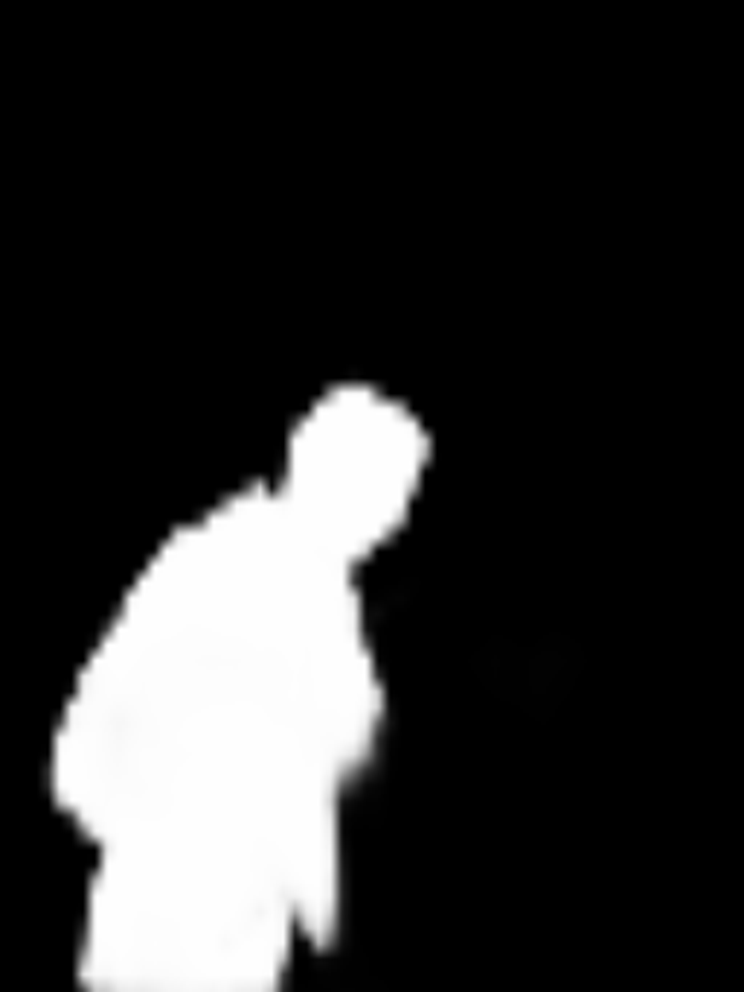}
	\end{subfigure}
	\begin{subfigure}{0.071\textwidth}
		\includegraphics[width=\textwidth]{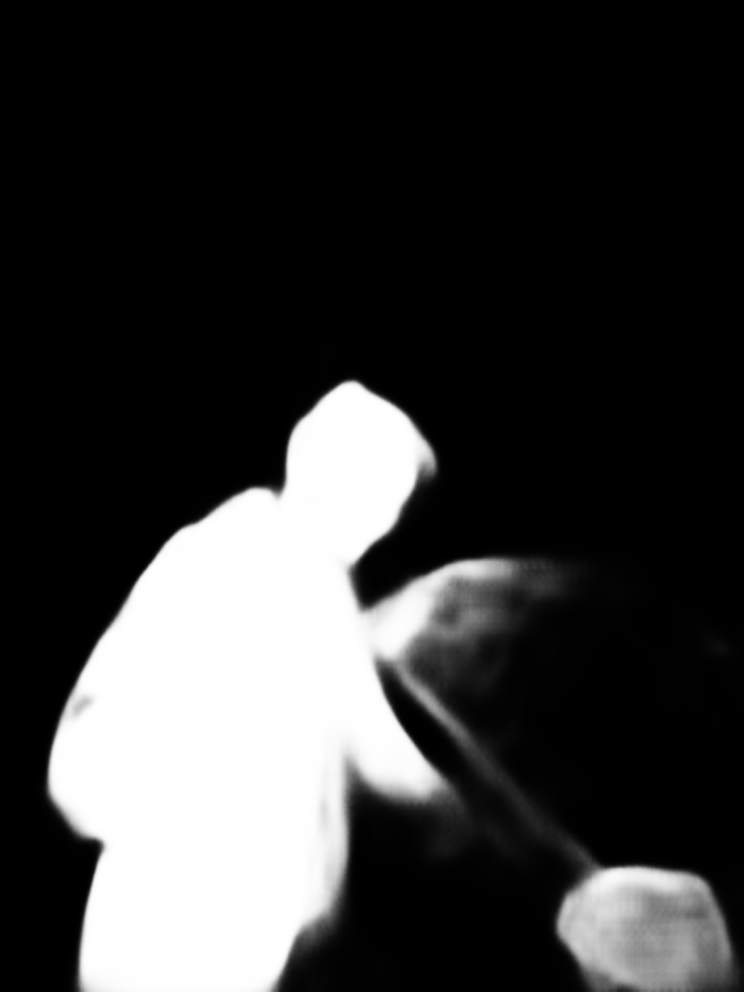}
	\end{subfigure}
	\begin{subfigure}{0.071\textwidth}
		\includegraphics[width=\textwidth]{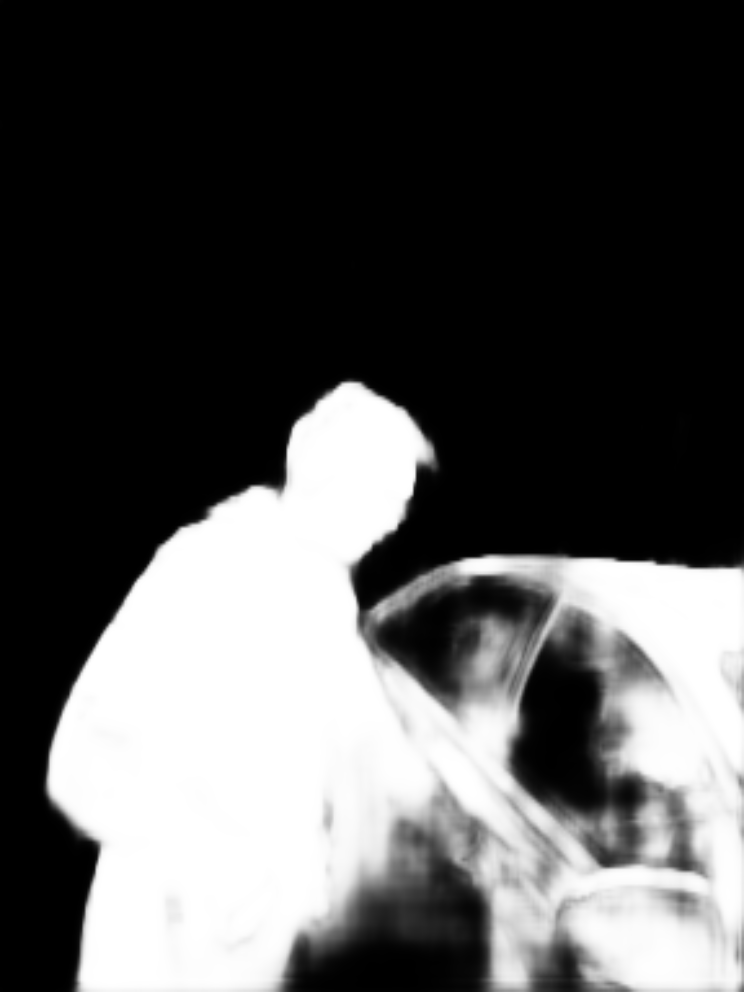}
	\end{subfigure}
	\begin{subfigure}{0.071\textwidth}
		\includegraphics[width=\textwidth]{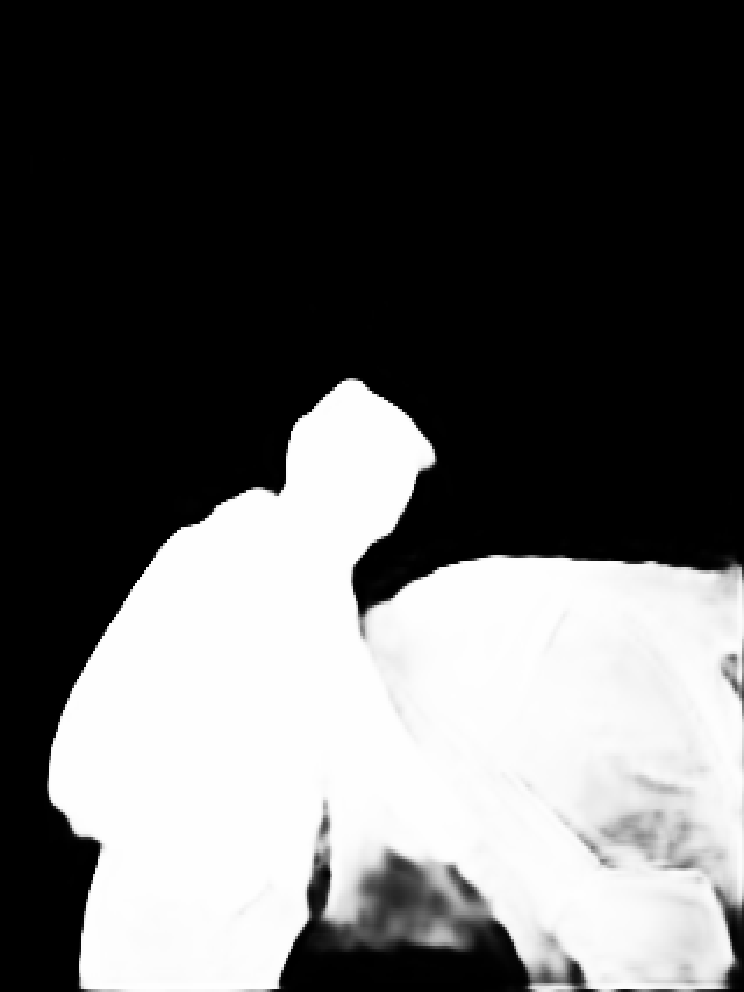}
	\end{subfigure}
	\begin{subfigure}{0.071\textwidth}
		\includegraphics[width=\textwidth]{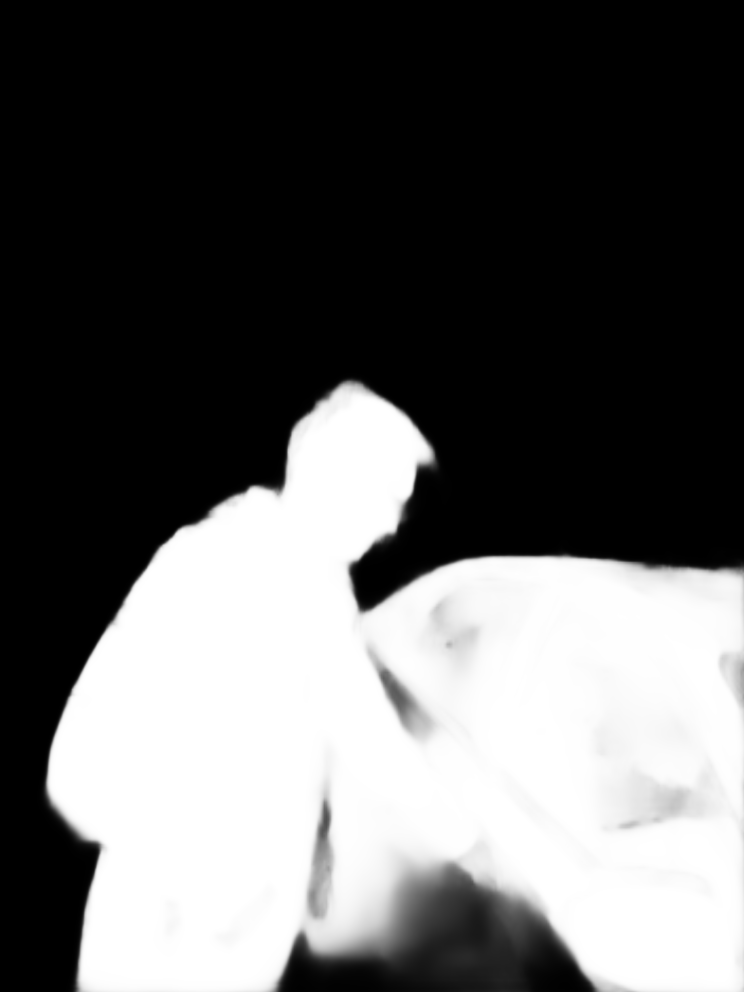}
	\end{subfigure}
    \begin{subfigure}{0.071\textwidth}
		\includegraphics[width=\textwidth]{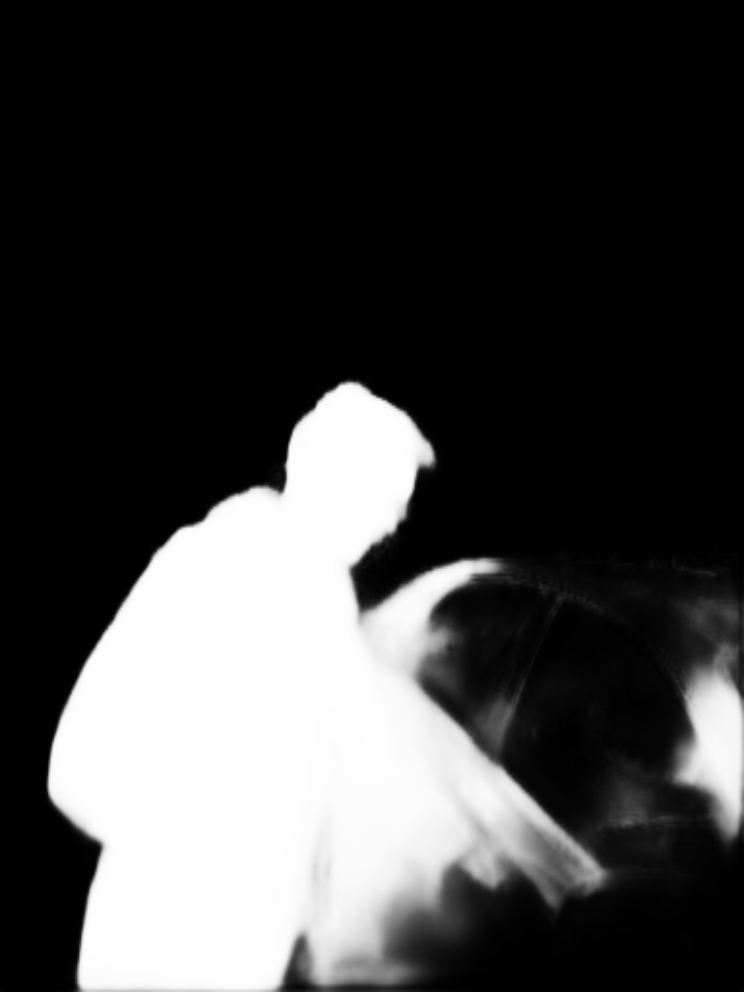}
	\end{subfigure}
    \begin{subfigure}{0.071\textwidth}
		\includegraphics[width=\textwidth]{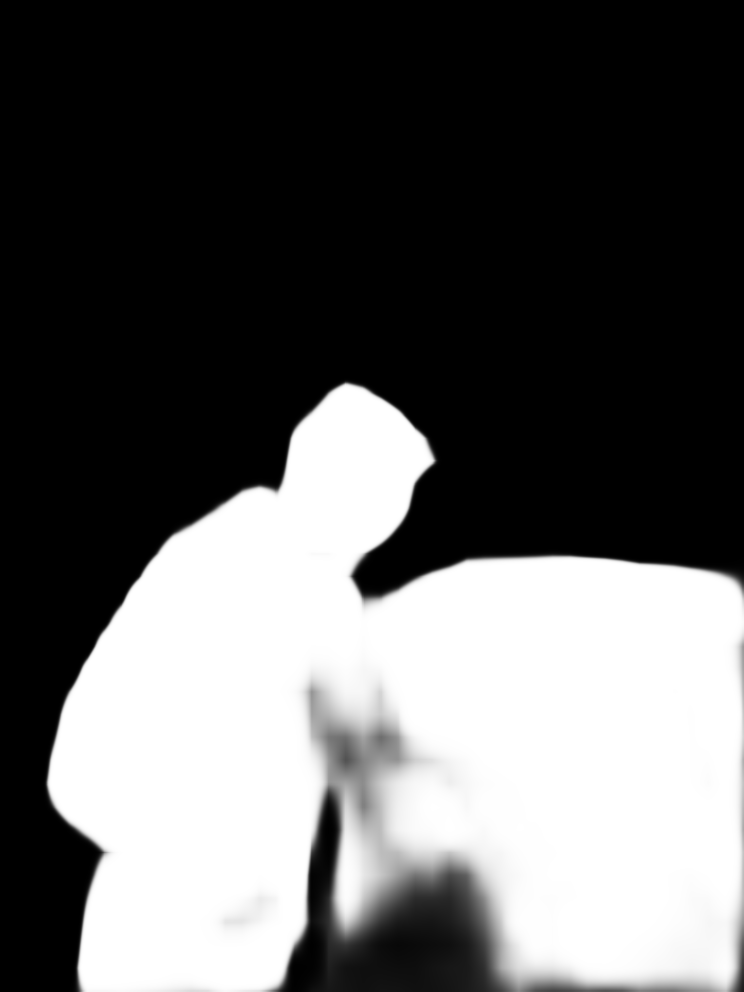}
	\end{subfigure}
	\begin{subfigure}{0.071\textwidth}
		\includegraphics[width=\textwidth]{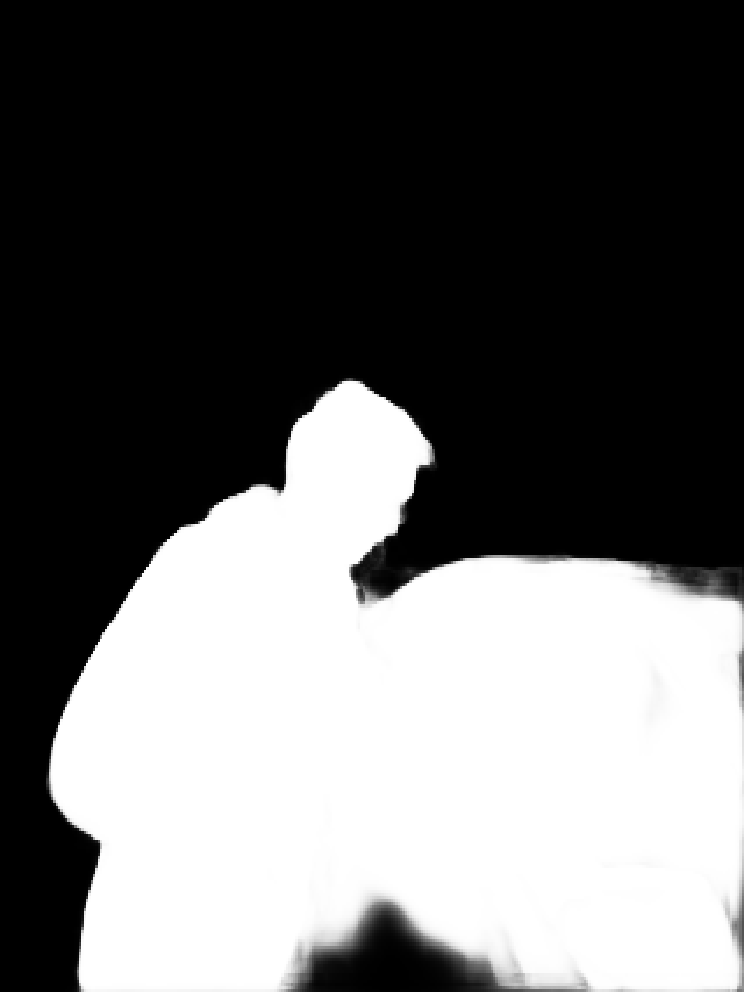}
	\end{subfigure}
    \begin{subfigure}{0.071\textwidth}
		\includegraphics[width=\textwidth]{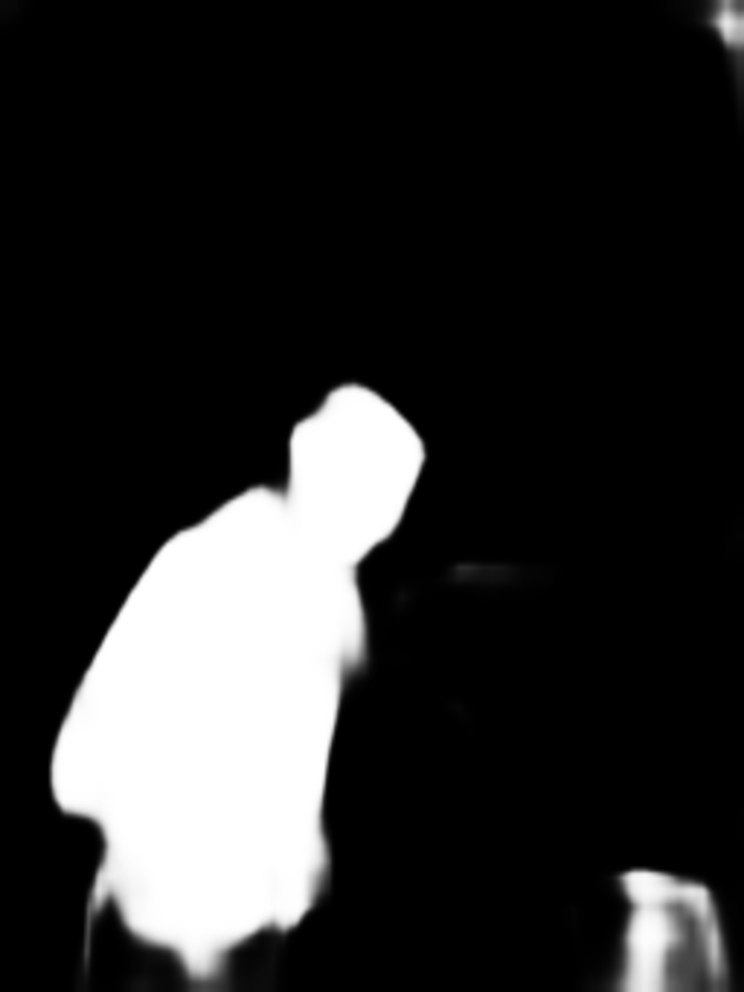}
	\end{subfigure}
	\begin{subfigure}{0.071\textwidth}
		\includegraphics[width=\textwidth]{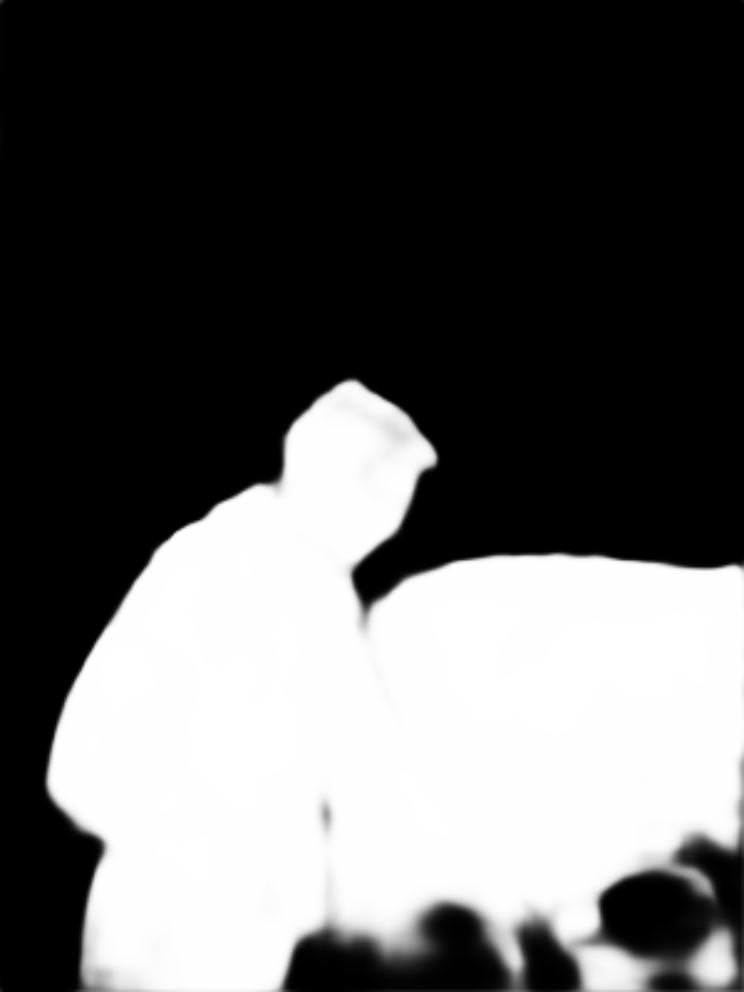}
	\end{subfigure}
	\ \\	
    \vspace*{0.5mm}
	\begin{subfigure}{0.071\textwidth}
		\includegraphics[width=\textwidth]{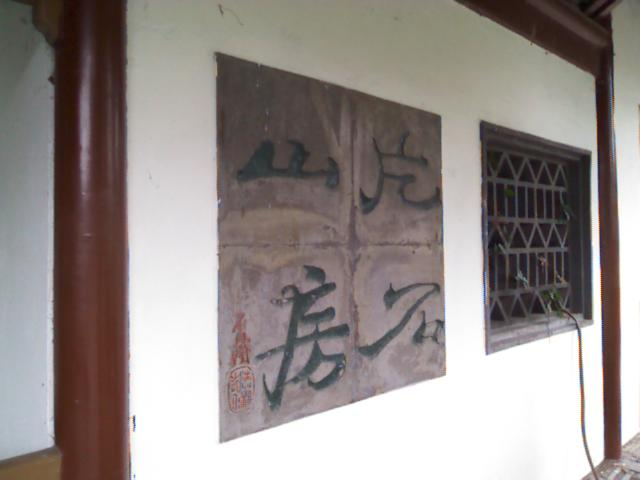}
	\end{subfigure}
	\begin{subfigure}{0.071\textwidth}
		\includegraphics[width=\textwidth]{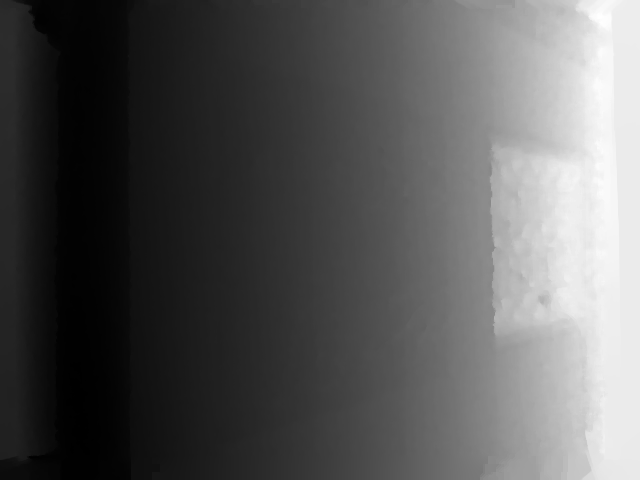}
	\end{subfigure}
    \begin{subfigure}{0.071\textwidth}
		\includegraphics[width=\textwidth]{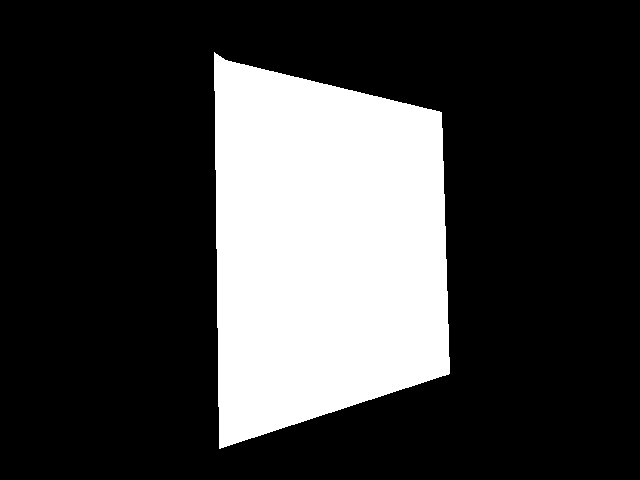}
	\end{subfigure}
	\begin{subfigure}{0.071\textwidth}
		\includegraphics[width=\textwidth]{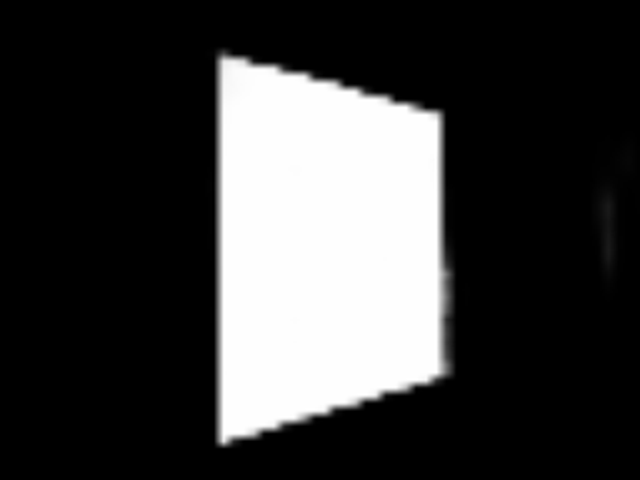}
	\end{subfigure}
	\begin{subfigure}{0.071\textwidth}
		\includegraphics[width=\textwidth]{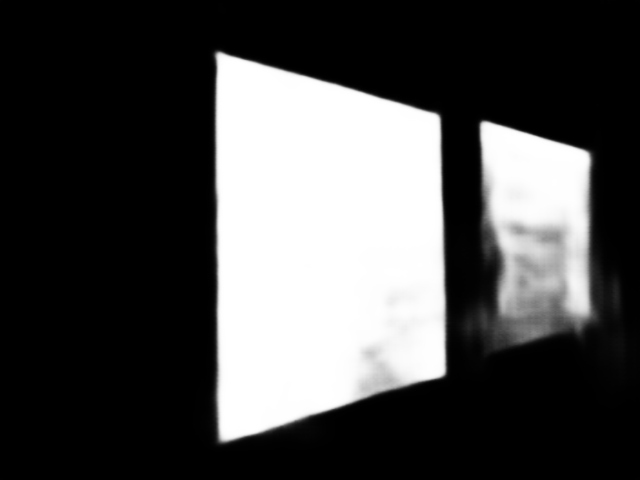}
	\end{subfigure}
	\begin{subfigure}{0.071\textwidth}
		\includegraphics[width=\textwidth]{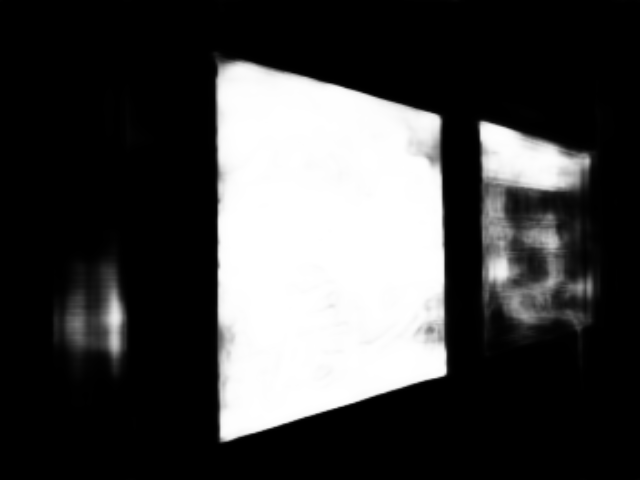}
	\end{subfigure}
	\begin{subfigure}{0.071\textwidth}
		\includegraphics[width=\textwidth]{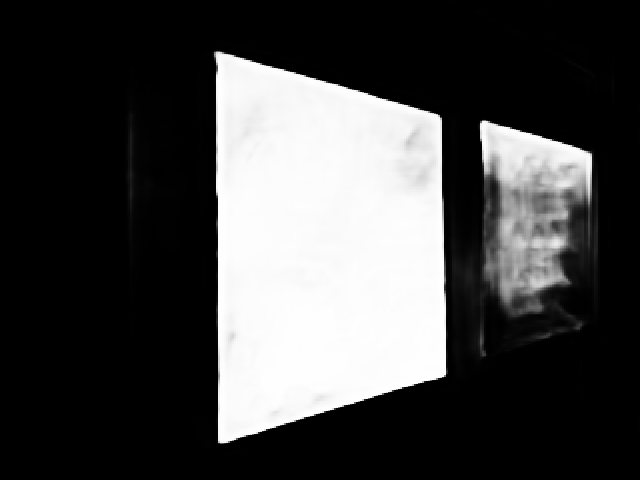}
	\end{subfigure}
	\begin{subfigure}{0.071\textwidth}
		\includegraphics[width=\textwidth]{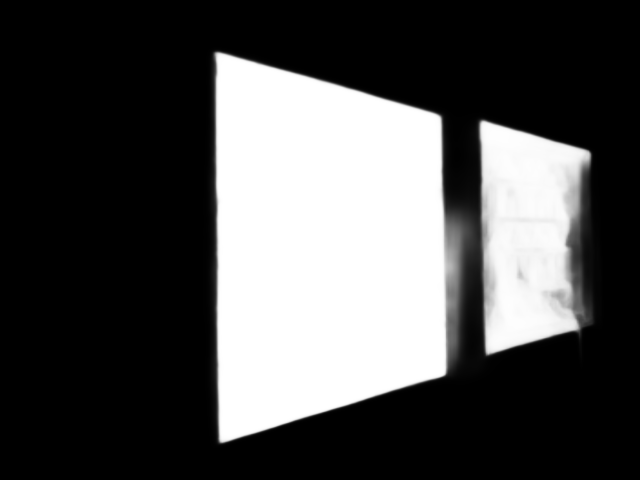}
	\end{subfigure}
    \begin{subfigure}{0.071\textwidth}
		\includegraphics[width=\textwidth]{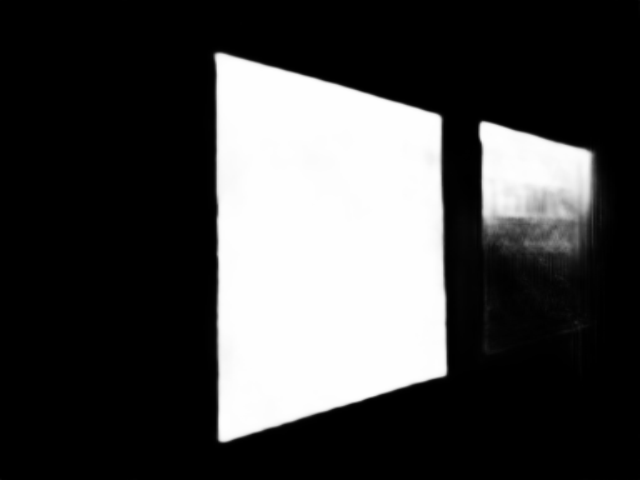}
	\end{subfigure}
    \begin{subfigure}{0.071\textwidth}
		\includegraphics[width=\textwidth]{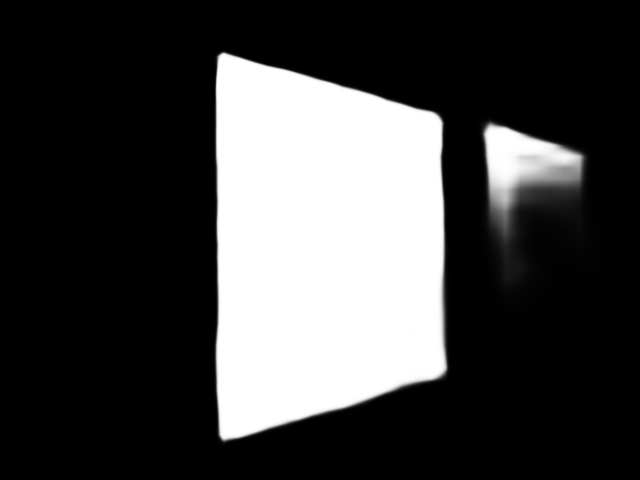}
	\end{subfigure}
	\begin{subfigure}{0.071\textwidth}
		\includegraphics[width=\textwidth]{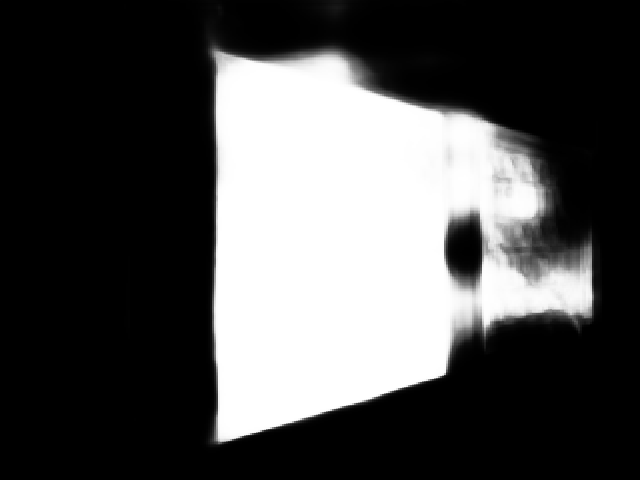}
	\end{subfigure}
    \begin{subfigure}{0.071\textwidth}
		\includegraphics[width=\textwidth]{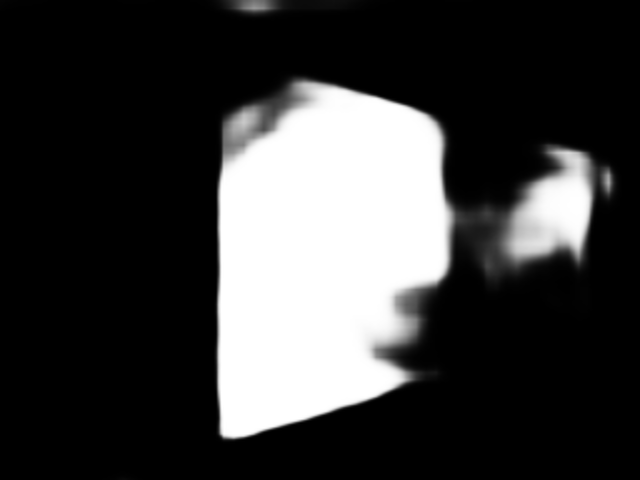}
	\end{subfigure}
	\begin{subfigure}{0.071\textwidth}
		\includegraphics[width=\textwidth]{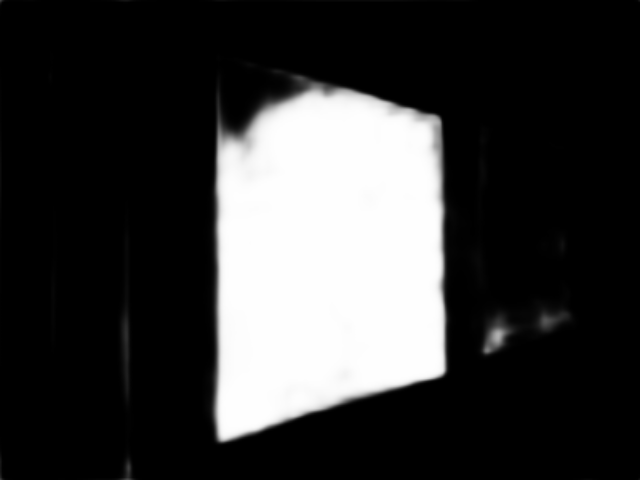}
	\end{subfigure}
	\ \\	
    \vspace*{0.5mm}
	\begin{subfigure}{0.071\textwidth}
		\includegraphics[width=\textwidth]{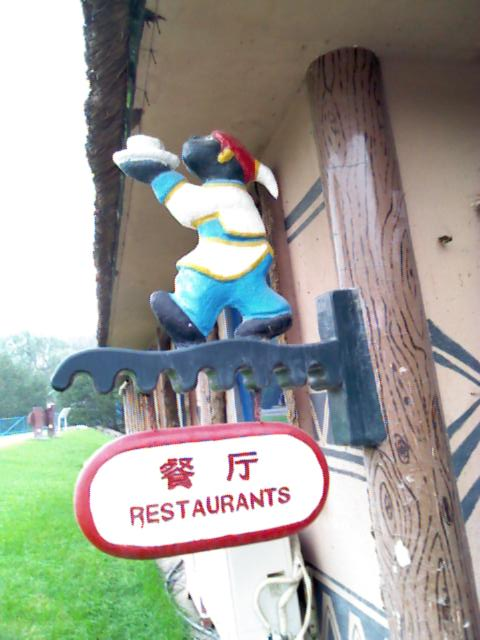}
	\end{subfigure}
	\begin{subfigure}{0.071\textwidth}
		\includegraphics[width=\textwidth]{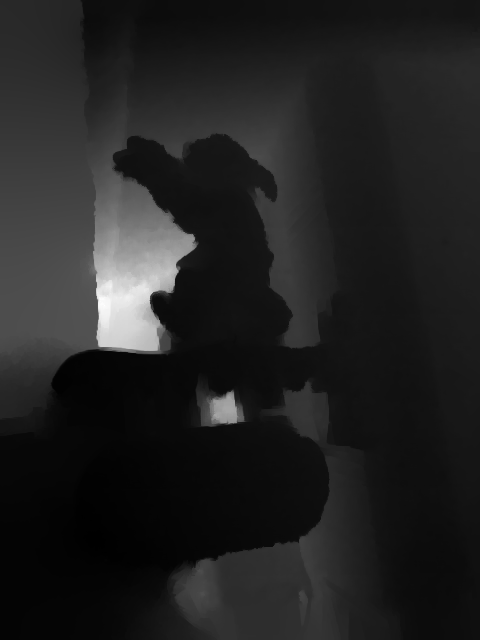}
	\end{subfigure}
    \begin{subfigure}{0.071\textwidth}
		\includegraphics[width=\textwidth]{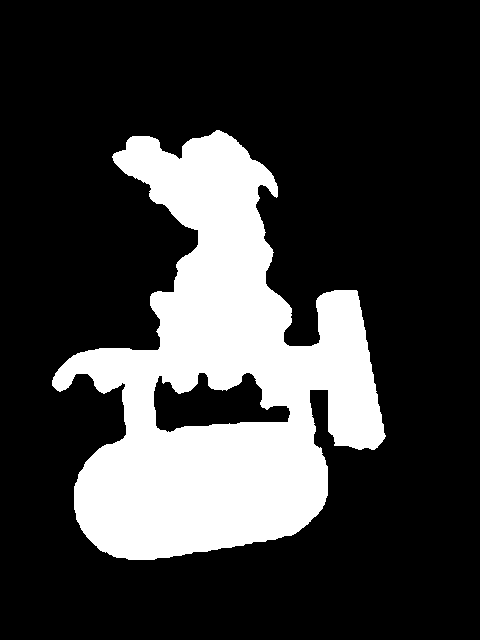}
	\end{subfigure}
	\begin{subfigure}{0.071\textwidth}
		\includegraphics[width=\textwidth]{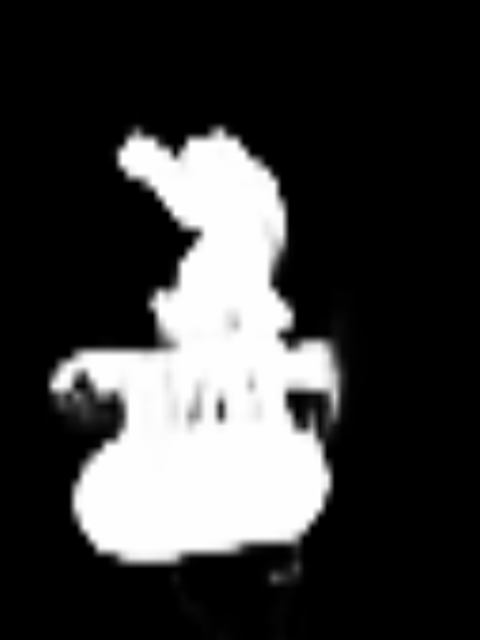}
	\end{subfigure}
	\begin{subfigure}{0.071\textwidth}
		\includegraphics[width=\textwidth]{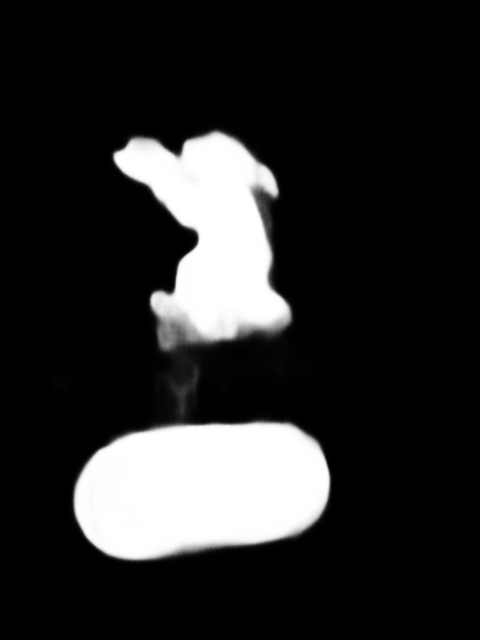}
	\end{subfigure}
	\begin{subfigure}{0.071\textwidth}
		\includegraphics[width=\textwidth]{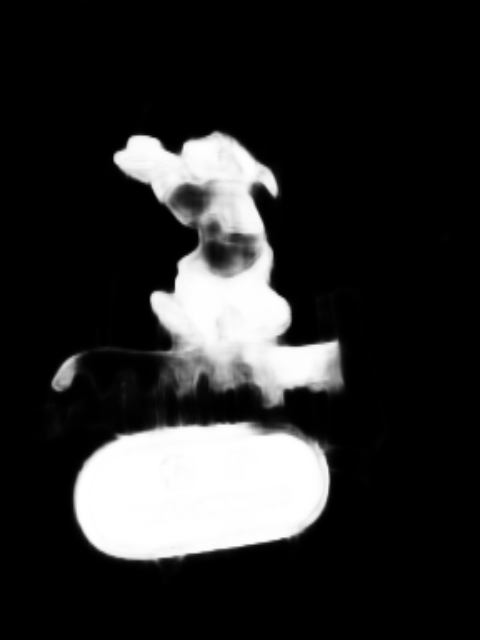}
	\end{subfigure}
	\begin{subfigure}{0.071\textwidth}
		\includegraphics[width=\textwidth]{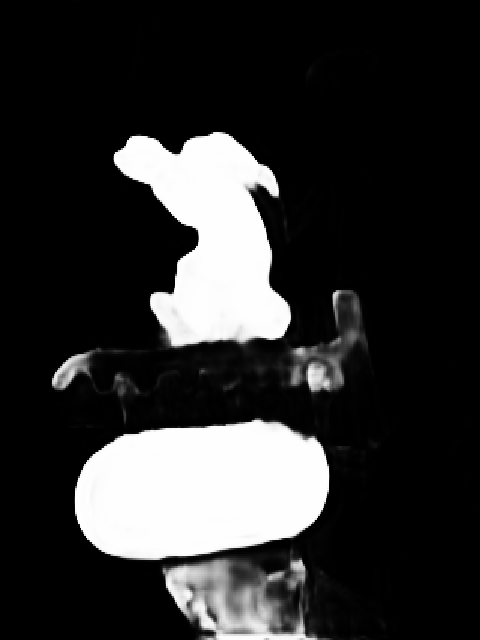}
	\end{subfigure}
	\begin{subfigure}{0.071\textwidth}
		\includegraphics[width=\textwidth]{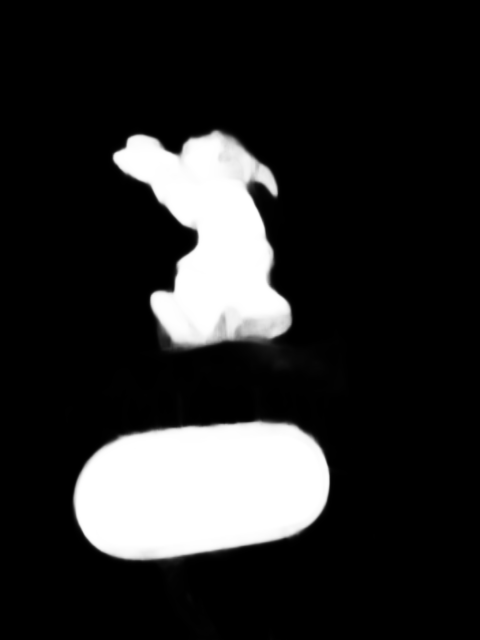}
	\end{subfigure}
    \begin{subfigure}{0.071\textwidth}
		\includegraphics[width=\textwidth]{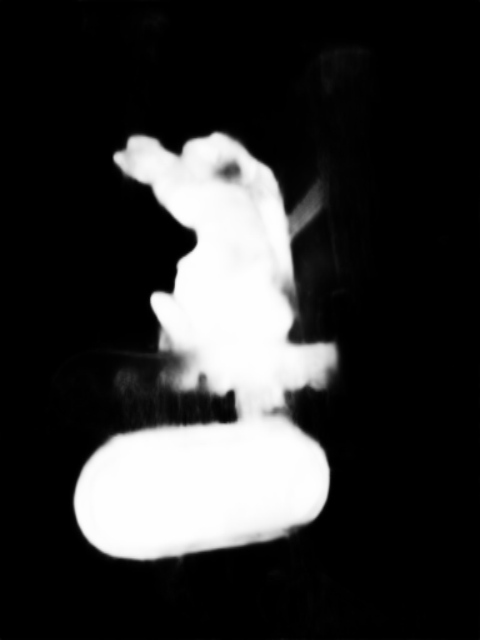}
	\end{subfigure}
    \begin{subfigure}{0.071\textwidth}
		\includegraphics[width=\textwidth]{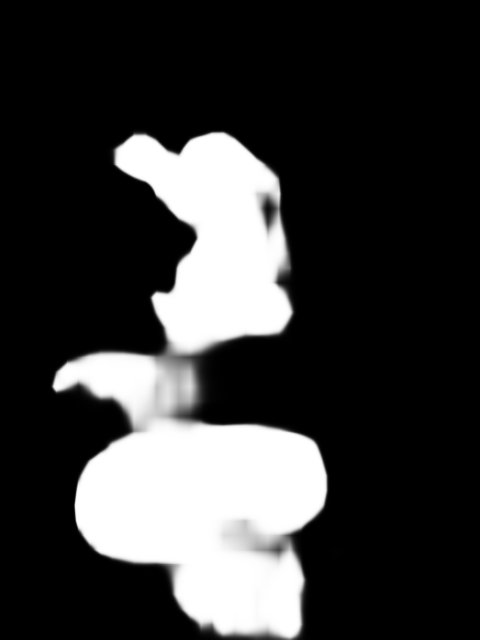}
	\end{subfigure}
	\begin{subfigure}{0.071\textwidth}
		\includegraphics[width=\textwidth]{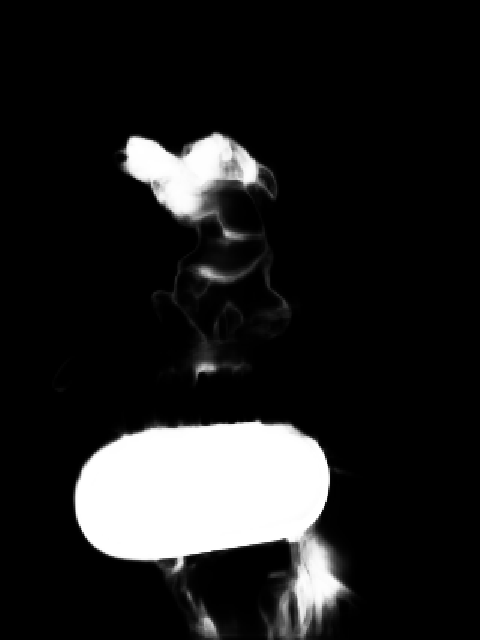}
	\end{subfigure}
    \begin{subfigure}{0.071\textwidth}
		\includegraphics[width=\textwidth]{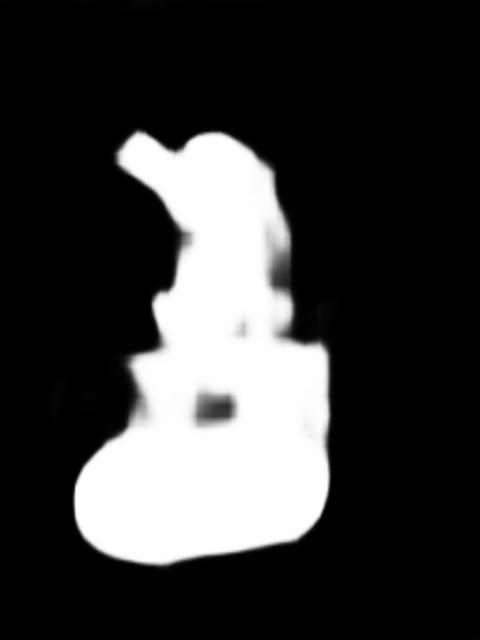}
	\end{subfigure}
	\begin{subfigure}{0.071\textwidth}
		\includegraphics[width=\textwidth]{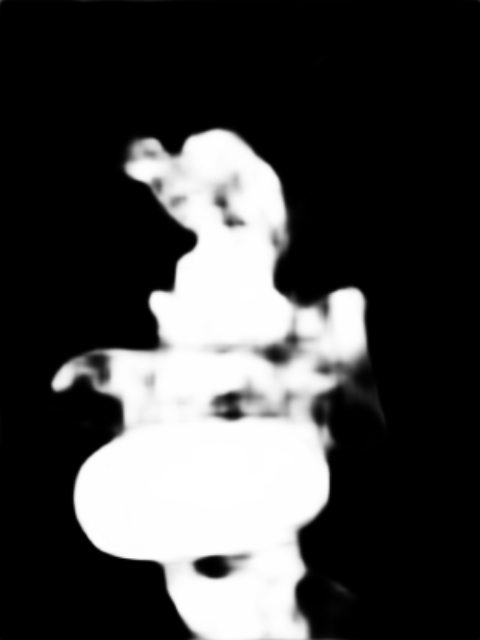}
	\end{subfigure}
	\ \\
   \vspace*{0.5mm}
   \begin{subfigure}{0.071\textwidth}
   	\includegraphics[width=\textwidth]{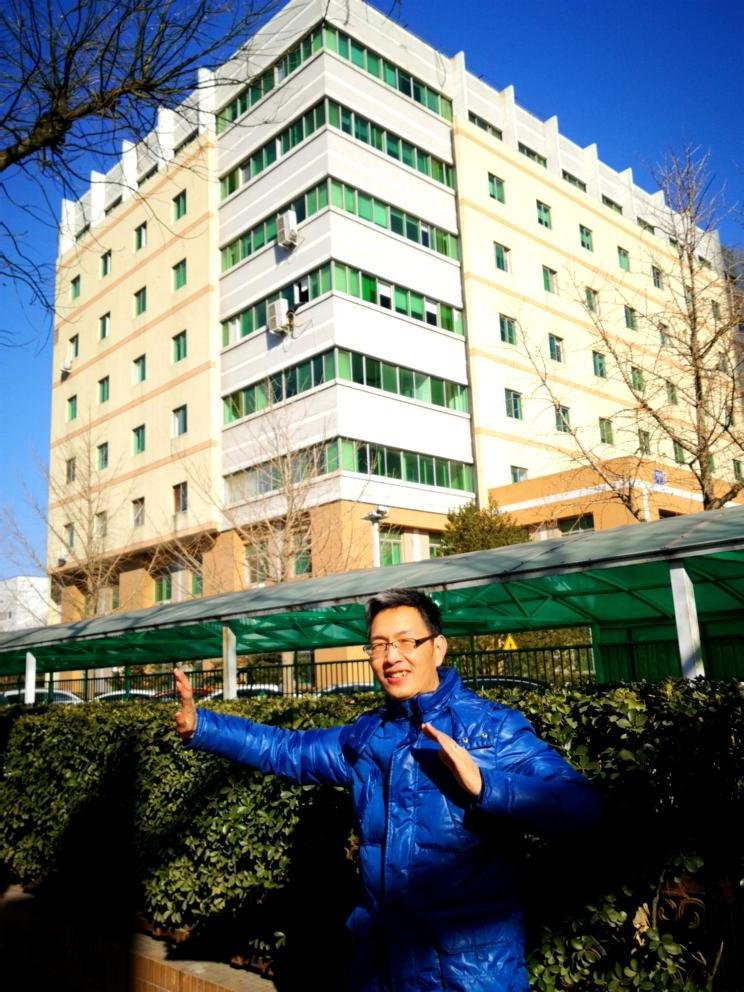}
    \vspace{-5.5mm} \caption{\footnotesize{Input}}
    \vspace{-2.5mm} \caption*{\footnotesize{RGB}}
   \end{subfigure}
   \begin{subfigure}{0.071\textwidth}
   	\includegraphics[width=\textwidth]{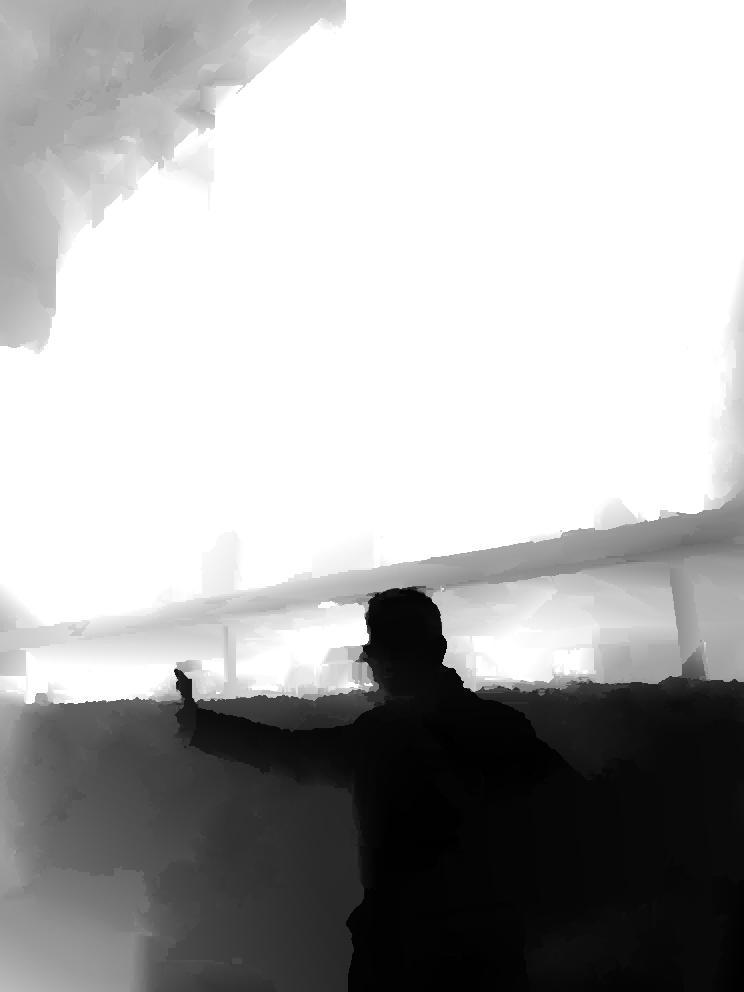}
    \vspace{-5.5mm} \caption{\footnotesize{Input}}
    \vspace{-2.5mm} \caption*{\footnotesize{depth}}
   \end{subfigure}
   \begin{subfigure}{0.071\textwidth}
   	\includegraphics[width=\textwidth]{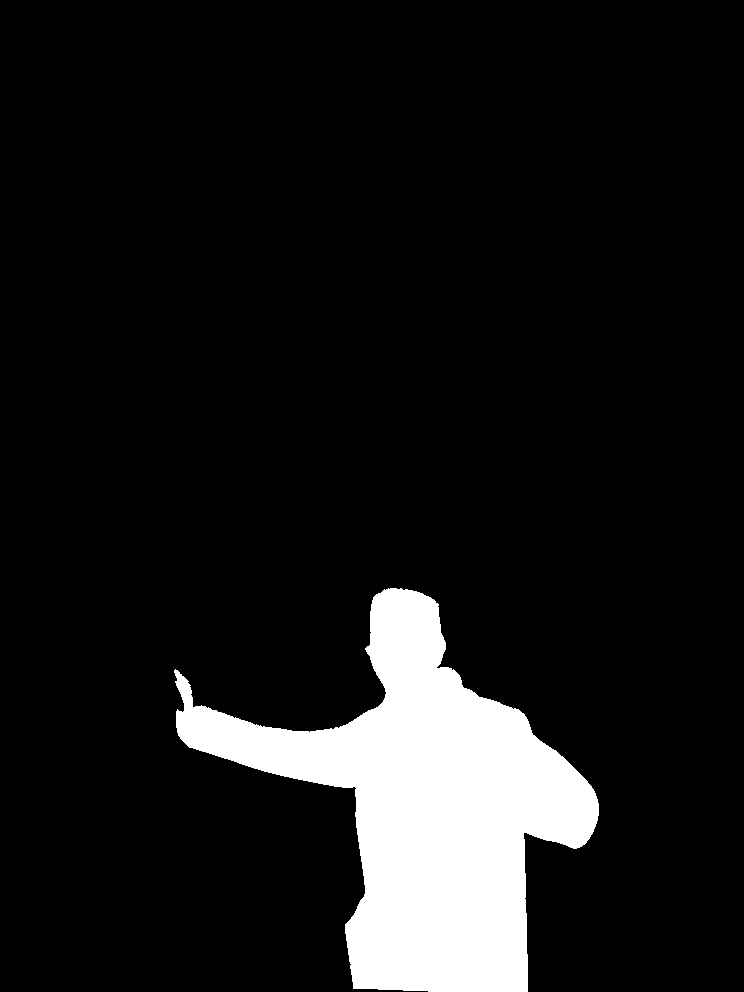}
    \vspace{-5.5mm} \caption{\footnotesize{Groun}}
   	\vspace{-2.5mm} \caption*{\footnotesize{d truth}}
   \end{subfigure}
   \begin{subfigure}{0.071\textwidth}
   	\includegraphics[width=\textwidth]{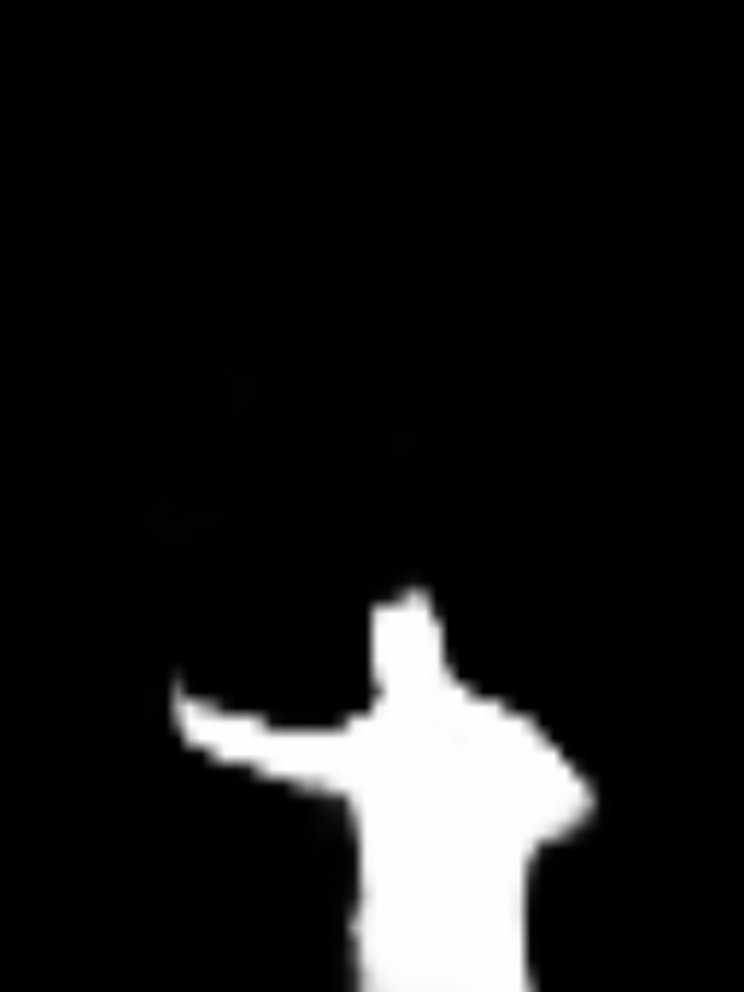}
    \vspace{-5.5mm} \caption{\footnotesize{Our}}
    \vspace{-2.5mm} \caption*{\footnotesize{method}}
   \end{subfigure}
   \begin{subfigure}{0.071\textwidth}
   	\includegraphics[width=\textwidth]{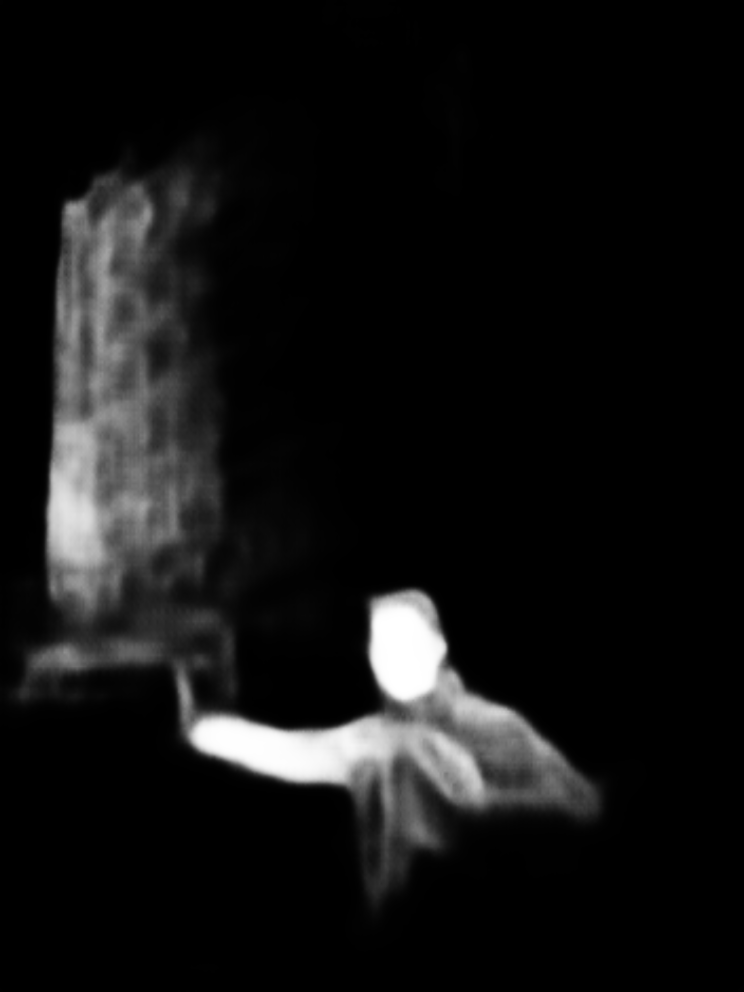}
    \vspace{-5.5mm} \caption{\footnotesize{BBS-}}
    \vspace{-2.5mm} \caption*{\footnotesize{Net~\cite{fan2020bbs}}}
   \end{subfigure}
   \begin{subfigure}{0.071\textwidth}
   	\includegraphics[width=\textwidth]{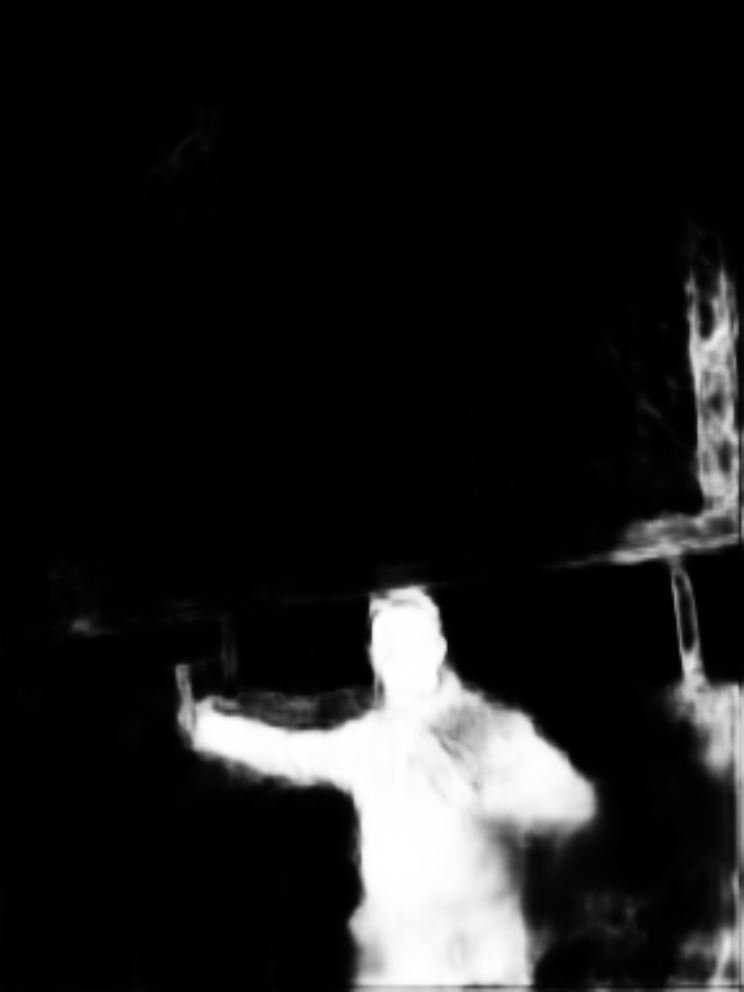}
    \vspace{-5.5mm} \caption{\footnotesize{CMW}}
   	\vspace{-2.5mm} \caption*{\footnotesize{Net~\cite{li2020cross}}}
   \end{subfigure}
   \begin{subfigure}{0.071\textwidth}
   	\includegraphics[width=\textwidth]{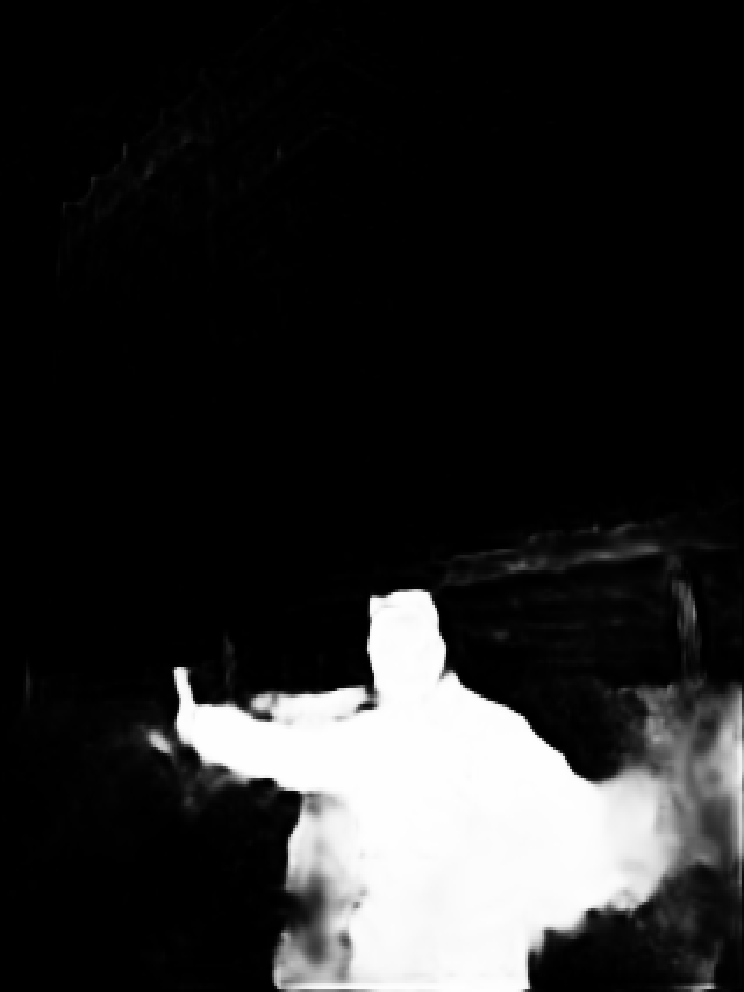}
    \vspace{-5.5mm} \caption{\footnotesize{HDF-}}
    \vspace{-2.5mm} \caption*{\footnotesize{Net~\cite{pang2020hierarchical}}}
   \end{subfigure}
   \begin{subfigure}{0.071\textwidth}
   	\includegraphics[width=\textwidth]{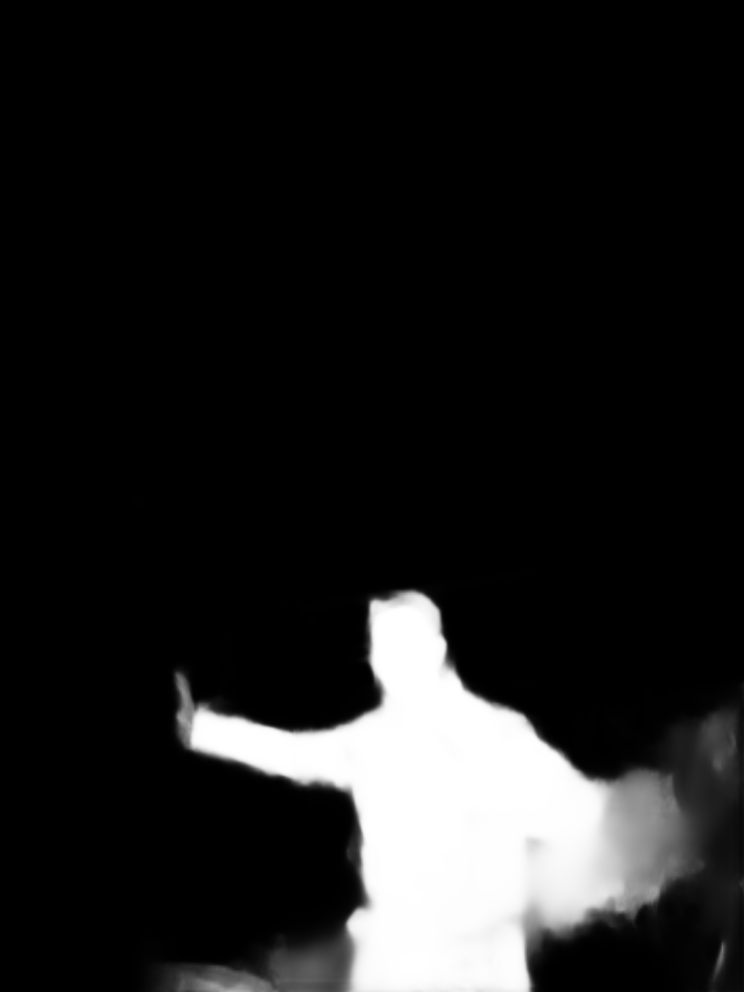}
    \vspace{-5.5mm} \caption{\footnotesize{PGA-}}
    \vspace{-2.5mm} \caption*{\footnotesize{Net~{\cite{chen2020progressively}}}}
   \end{subfigure}
   \begin{subfigure}{0.071\textwidth}
   	\includegraphics[width=\textwidth]{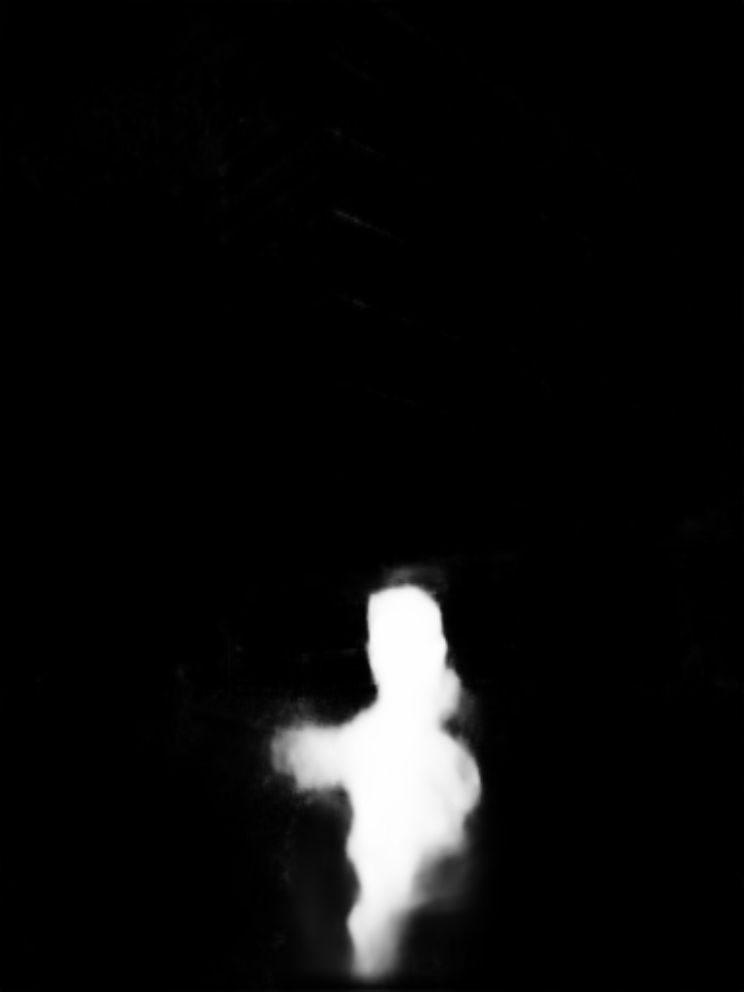}
    \vspace{-5.5mm} \caption{\footnotesize{DANet}}
    \vspace{-2.5mm} \caption*{\footnotesize{~\cite{zhao2020single}}}
   \end{subfigure}
   \begin{subfigure}{0.071\textwidth}
   	\includegraphics[width=\textwidth]{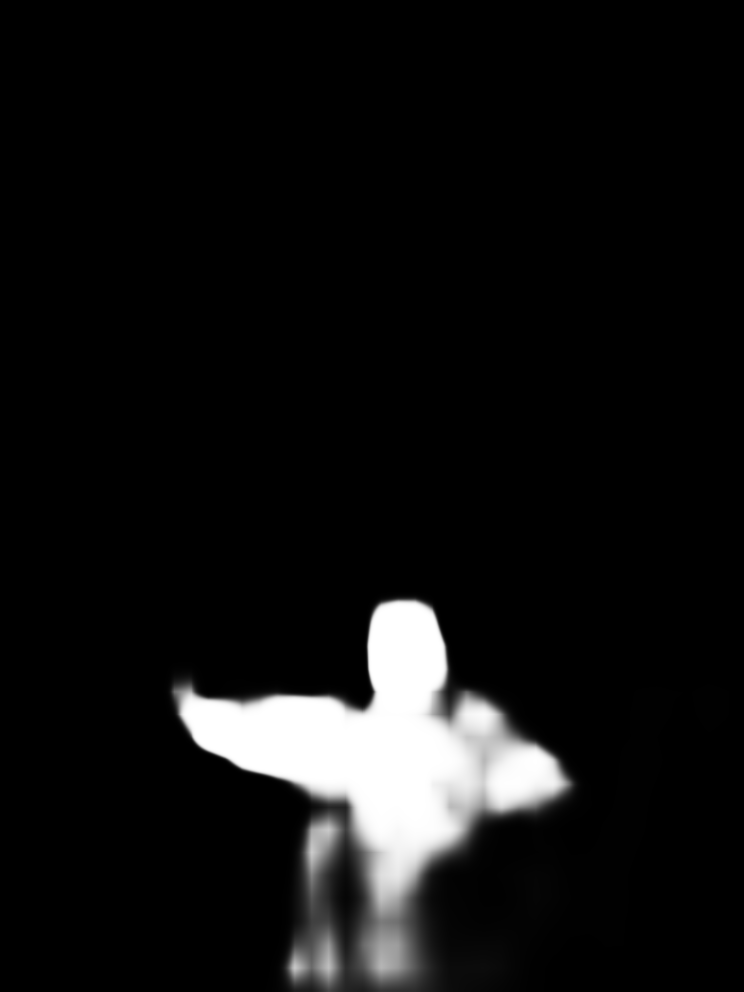}
    \vspace{-5.5mm} \caption{\footnotesize{UCNet}}
    \vspace{-2.5mm} \caption*{\footnotesize{~\cite{zhang2020uc}}}
   \end{subfigure}
   \begin{subfigure}{0.071\textwidth}
   	\includegraphics[width=\textwidth]{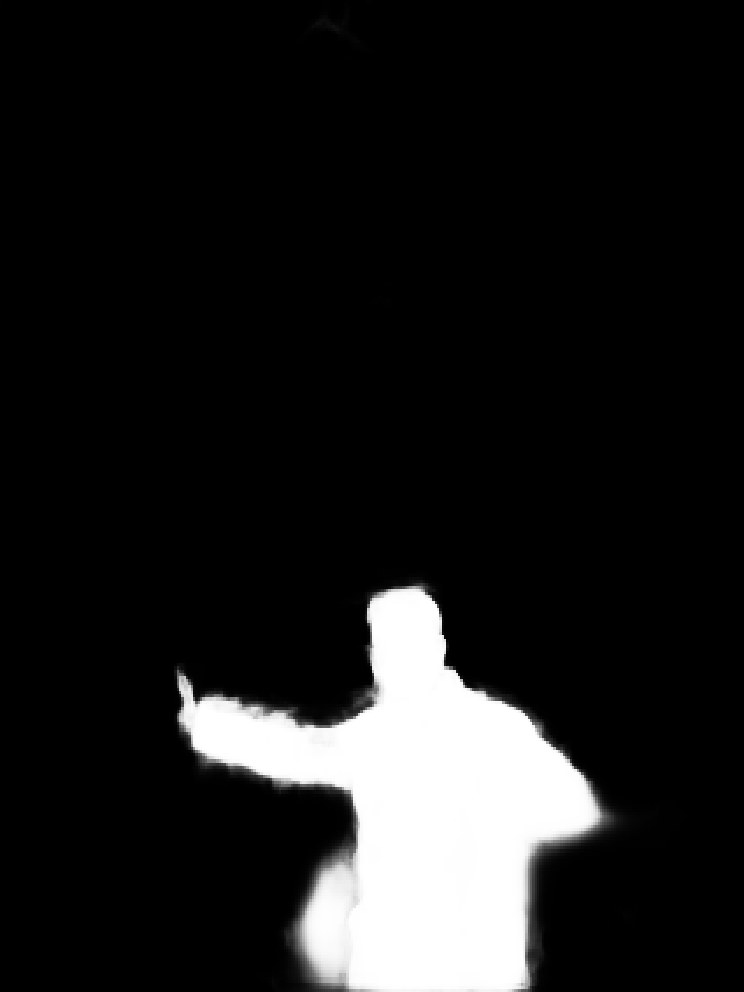}
    \vspace{-5.5mm} \caption{\footnotesize{JLD}}
   	\vspace{-2.5mm} \caption*{\footnotesize{CF~\cite{fu2020jl}}}
   \end{subfigure}
   \begin{subfigure}{0.071\textwidth}
   	\includegraphics[width=\textwidth]{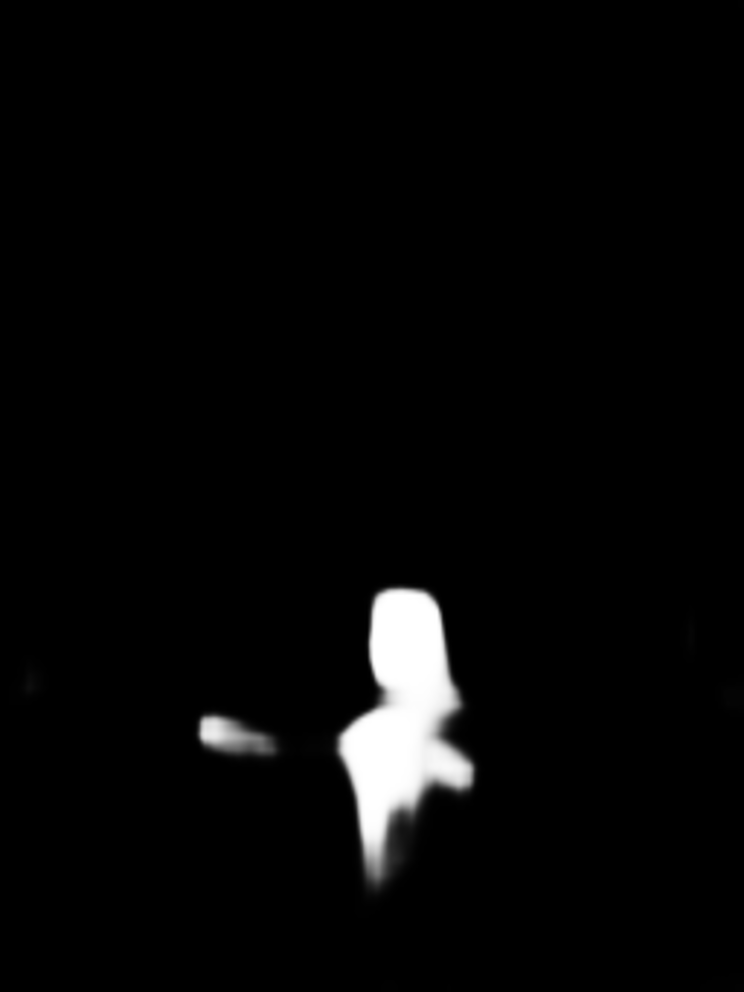}
    \vspace{-5.5mm} \caption{\footnotesize{DMRA}}
    \vspace{-2.5mm} \caption*{~{\cite{piao2019depth}}}
   \end{subfigure}
   \begin{subfigure}{0.071\textwidth}
   	\includegraphics[width=\textwidth]{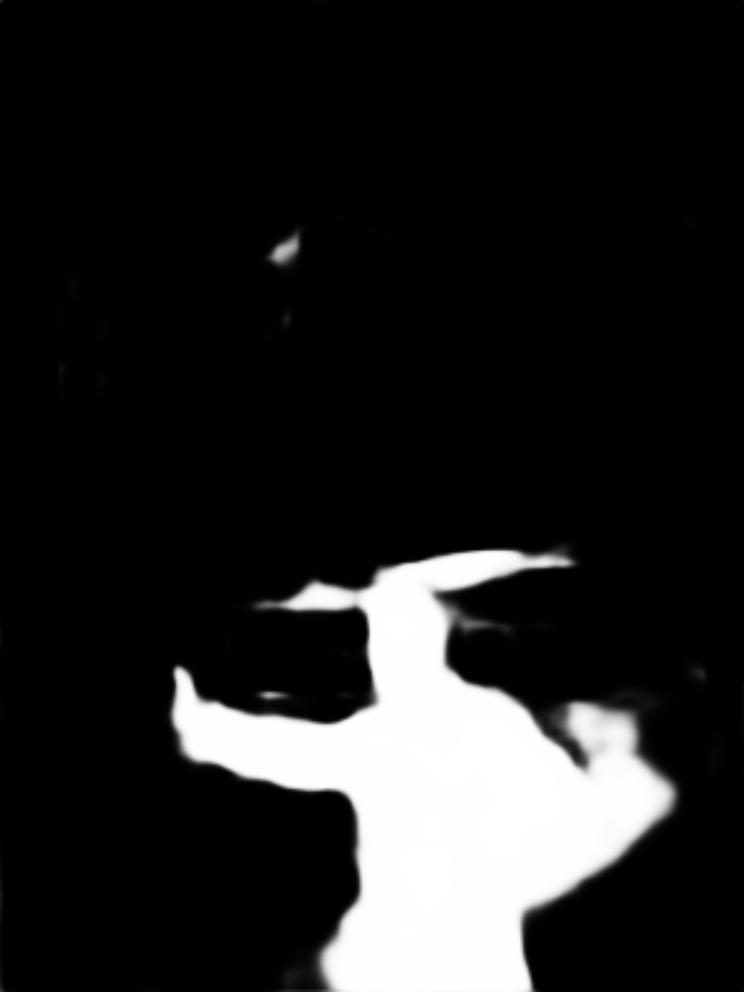}
    \vspace{-5.5mm} \caption{\footnotesize{CPFP}}
    \vspace{-2.5mm} \caption*{~\cite{zhao2019contrast}}
   \end{subfigure}
	\caption{Visual comparison of saliency map results produced by different methods.
	As can be seen, our network produces more accurate saliency maps than compared methods.
    }
	\label{fig:comparison_real_photos_part1}
	\vskip -10pt
\end{figure*}

{\color{dgreen}
\vspace{2mm}
\noindent
\textbf{Overall loss for the student network.}
}
The total loss of our network is computed as:
\vskip -10pt
\begin{equation}
\small
\begin{split}
\label{eq:total_loss}
	\mathcal{L}_{total} = \frac{1}{N_1} \sum_{i=1}^{N_1}{\left( \mathcal{L}^s(x_i) + \beta_1 \sum_{l=1}^{4}{\mathcal{L}_l^r(x_i)} \right)} & \\
	+ \lambda  \frac{1}{N_2} \sum_{j=1}^{N_2}{\left( \mathcal{L}^c(y_j) + \beta_2 \sum_{l=1}^{4}{\mathcal{L}_l^r(y_j)} \right)} & \ ,
\end{split}
\end{equation}
where $N_1$ and $N_2$ are the number of labeled RGB-D image pairs and unlabeled RGB images in our training set.
$\mathcal{L}^s(x_i)$ denotes the supervised loss (Eq.~\eqref{eq:supervised_loss}) for the $i$-th labeled image $x_i$ while $\mathcal{L}^c(y_j)$ is the consistency loss (Eq.~\eqref{eq:consistency_loss}) for the $j$-th unlabeled image $y_j$. $\mathcal{L}_l^r(x_i)$ and $\mathcal{L}_l^r(y_j)$ represent the reconstruction loss (Eq.~\eqref{eq:reconstruction_loss}) for labeled $x_i$ and unlabeled $y_j$, respectively.
%
We empirically set the weights $\beta_1 = 0.01$ and $\beta_2 = 1.0$.
The weight $\lambda$  balances the loss between the labeled and unlabeled data.
Following~\cite{chen2020multi}, we use a time dependent Gaussian warm-up function to update $\lambda$: $\lambda(t) = \lambda_{max} e^{(-5{(1-t/t_{max})}^2)}$, where $t$ denotes the current training iteration and $t_{max}$ is the maximum training iteration. 
In our experiments, we empirically set $\lambda_{max}=1.0$.

Following existing self-ensembling frameworks~\cite{tarvainen2017mean}, we minimize the total loss $\mathcal{L}_{total}$ of Eq.~\eqref{eq:total_loss} to train the student network. The parameters of the teacher network are computed as the exponential moving average (EMA) of the parameters of the student network to ensemble the information in different training steps.
{\color{dgreen} Following~\cite{chen2020multi}, during the inference stage, given an input RGB image and an input depth image, we pass them into the student network to predict a saliency detection map, which is then taken as the final result of our semi-supervised RGB-D saliency detection network, and the teacher network is not involved in the inference stage. 
}

{\color{dgreen}
\vspace{2mm}
\noindent
\textbf{Training configurations.}
We use the same HRNet architecture as the backbone for both the RGB image and the depth map. Given a 256x256 input image, the HRNet outputs multi-scale features at four convolutional blocks (layer 8, 27, 63, and 104) with a resolution of 64x64, 32x32, 16x16, and 8x8, respectively. These features are fed straight into the following network pipeline. As for the VGG-16, VGG-19 and ResNet-50 backbones, we employ the outputs from the layer (4, 7, 10, 13), (4, 8, 12, 16), and (11, 23, 41, 50) with the same resolution as the HRNet, respectively.
}
\section{Experimental Results}
\label{sec:experiment}

\begin{figure*}[t]
	\centering
	
	\vspace*{0.5mm}
	\begin{subfigure}{0.065\textwidth} 
		\includegraphics[width=\textwidth]{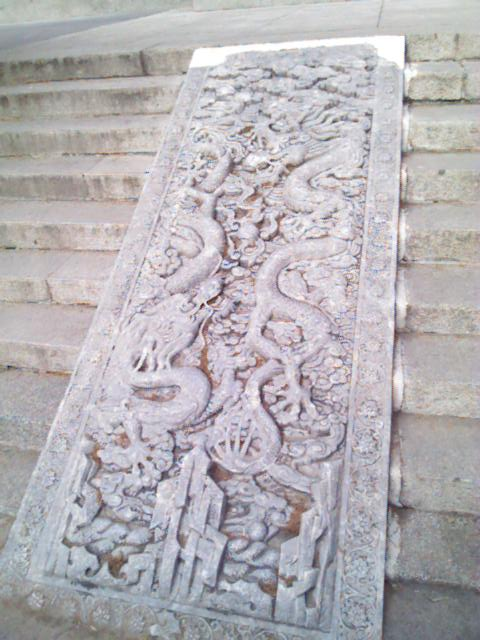}
    \end{subfigure}
	\begin{subfigure}{0.065\textwidth} 
		\includegraphics[width=\textwidth]{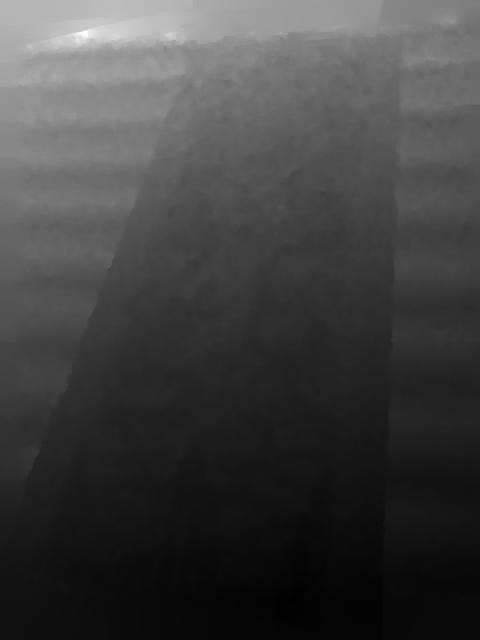}
	\end{subfigure}
	\begin{subfigure}{0.065\textwidth}
		\includegraphics[width=\textwidth]{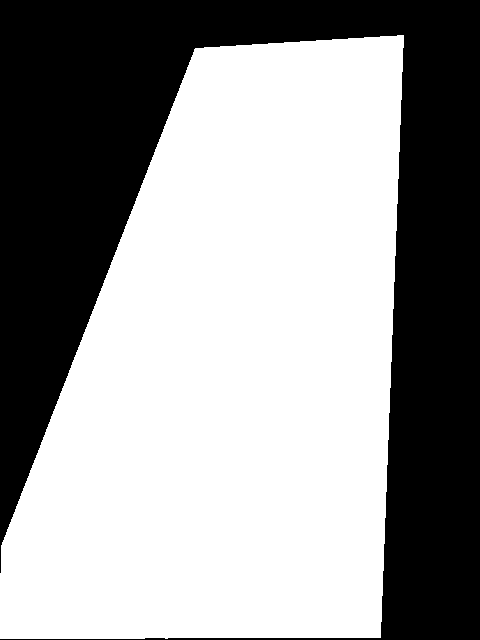}
	\end{subfigure}
    \begin{subfigure}{0.065\textwidth}
		\includegraphics[width=\textwidth]{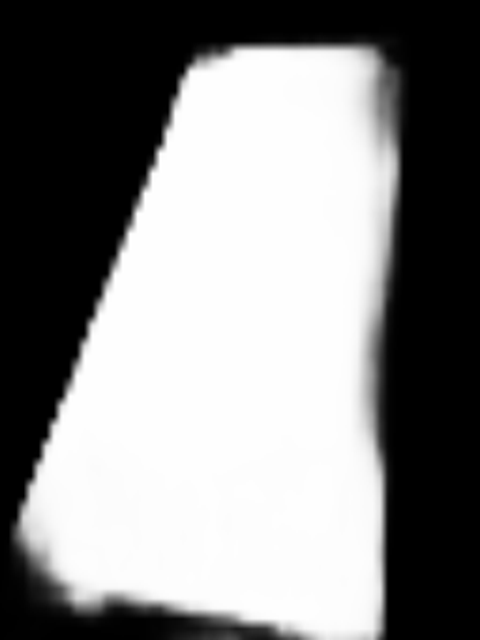}
	\end{subfigure}
    \begin{subfigure}{0.065\textwidth}
		\includegraphics[width=\textwidth]{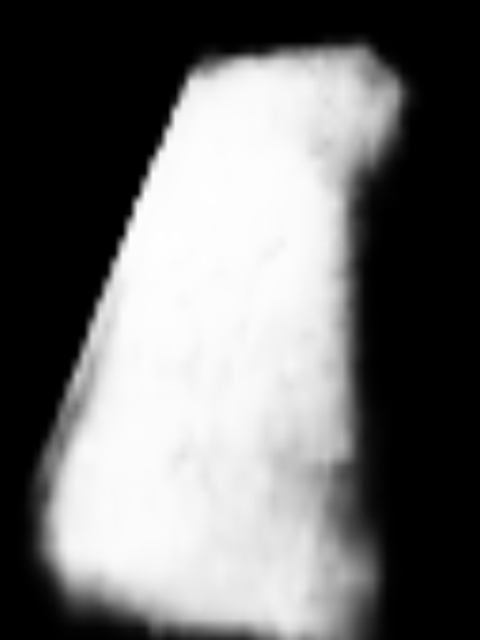}
	\end{subfigure}
    \begin{subfigure}{0.065\textwidth}
		\includegraphics[width=\textwidth]{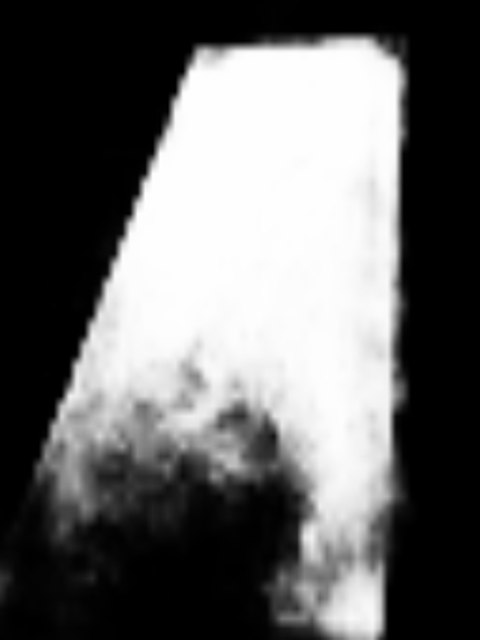}
	\end{subfigure}
	\begin{subfigure}{0.065\textwidth}
		\includegraphics[width=\textwidth]{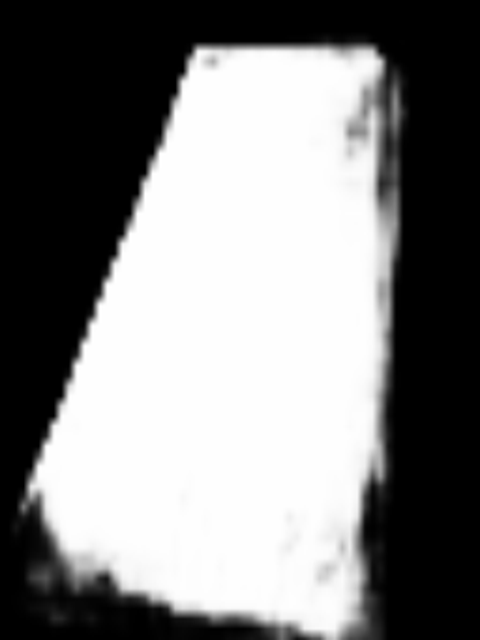}
	\end{subfigure}
	\begin{subfigure}{0.065\textwidth}
		\includegraphics[width=\textwidth]{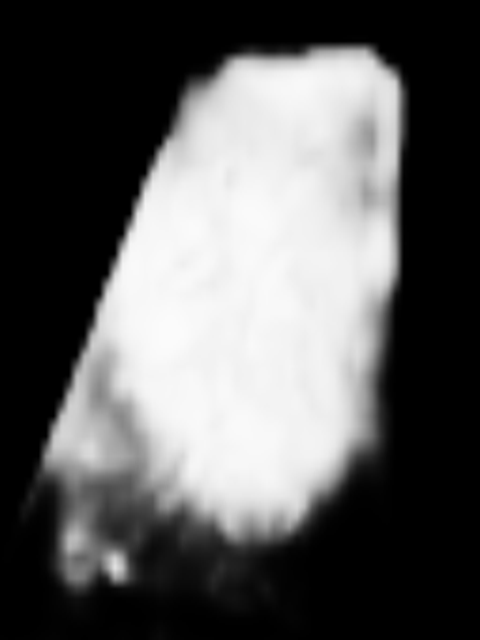}
	\end{subfigure}
    \begin{subfigure}{0.065\textwidth}
		\includegraphics[width=\textwidth]{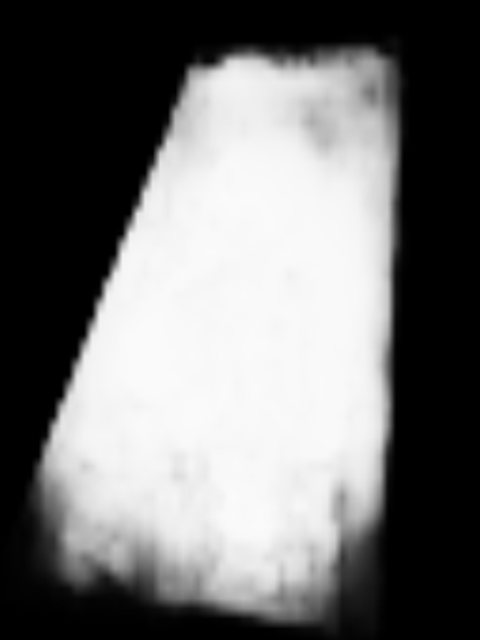}
	\end{subfigure}
	\begin{subfigure}{0.065\textwidth}
		\includegraphics[width=\textwidth]{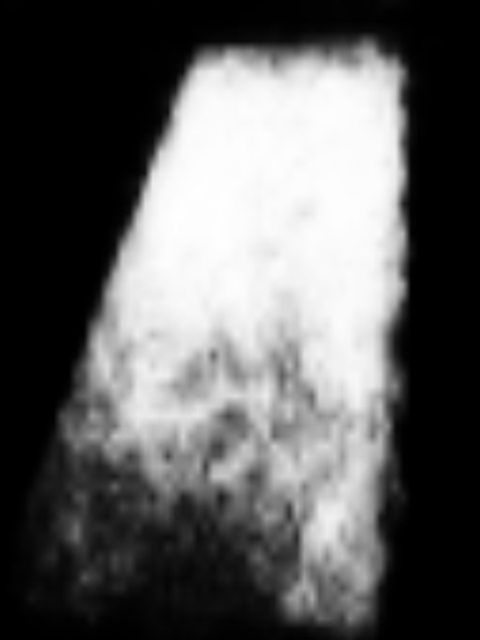}
	\end{subfigure}
	\begin{subfigure}{0.065\textwidth}
		\includegraphics[width=\textwidth]{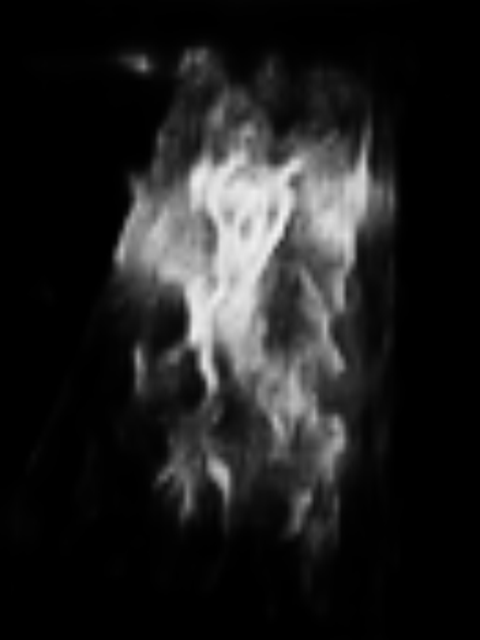}
	\end{subfigure}
	\begin{subfigure}{0.065\textwidth}
		\includegraphics[width=\textwidth]{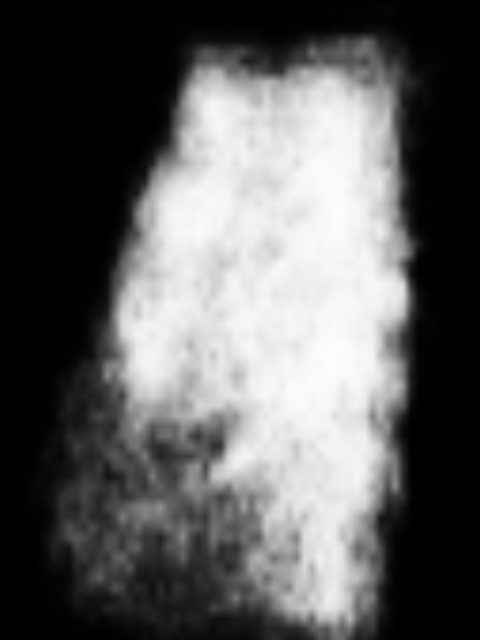}
	\end{subfigure}
	\begin{subfigure}{0.065\textwidth}
		\includegraphics[width=\textwidth]{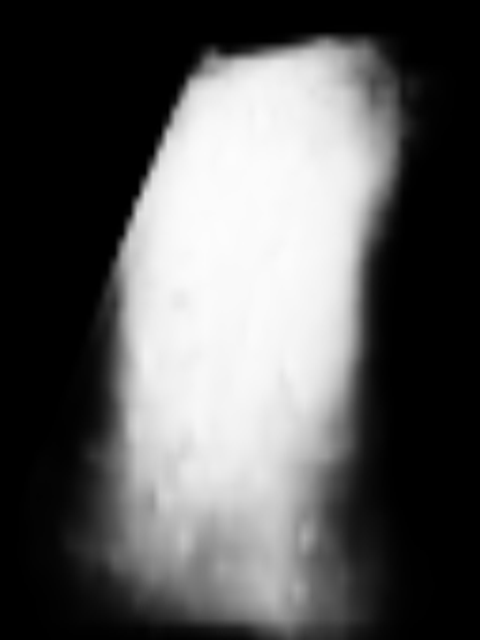}
	\end{subfigure}
	\begin{subfigure}{0.065\textwidth}
		\includegraphics[width=\textwidth]{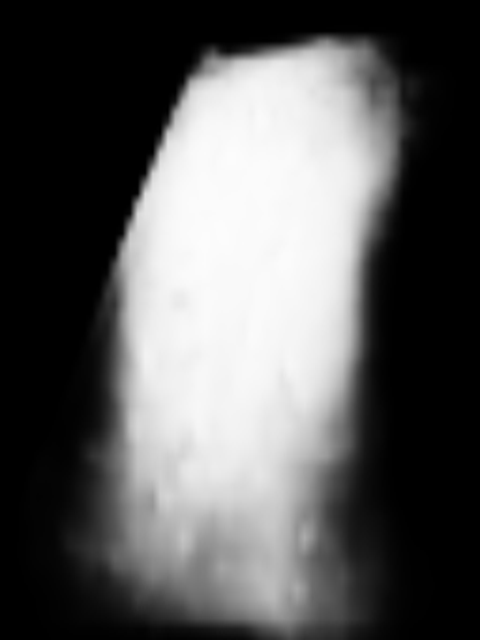}
	\end{subfigure}
	\ \\
	\vspace*{0.5mm}
	\begin{subfigure}{0.065\textwidth} 
		\includegraphics[width=\textwidth]{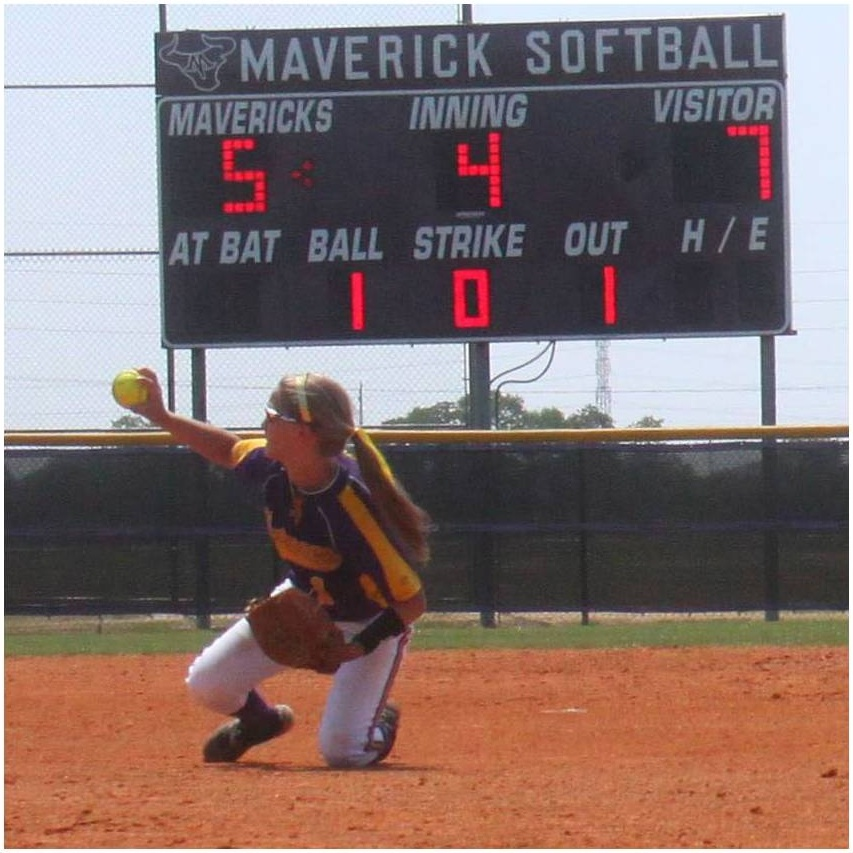}
    \end{subfigure}
	\begin{subfigure}{0.065\textwidth} 
		\includegraphics[width=\textwidth]{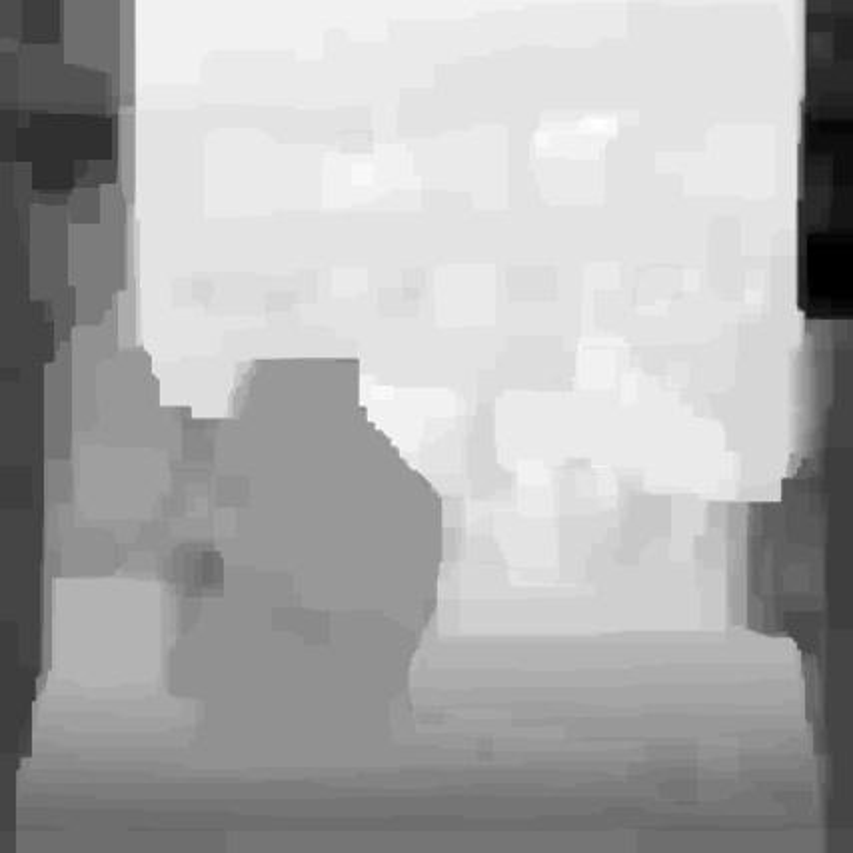}
	\end{subfigure}
	\begin{subfigure}{0.065\textwidth}
		\includegraphics[width=\textwidth]{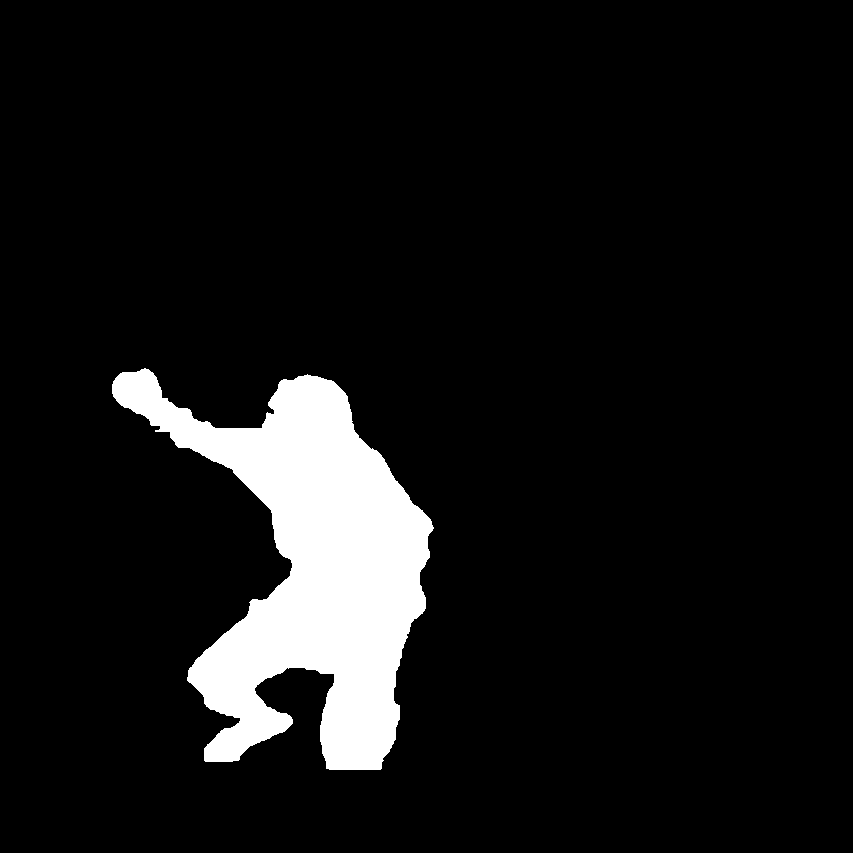}
	\end{subfigure}
    \begin{subfigure}{0.065\textwidth}
		\includegraphics[width=\textwidth]{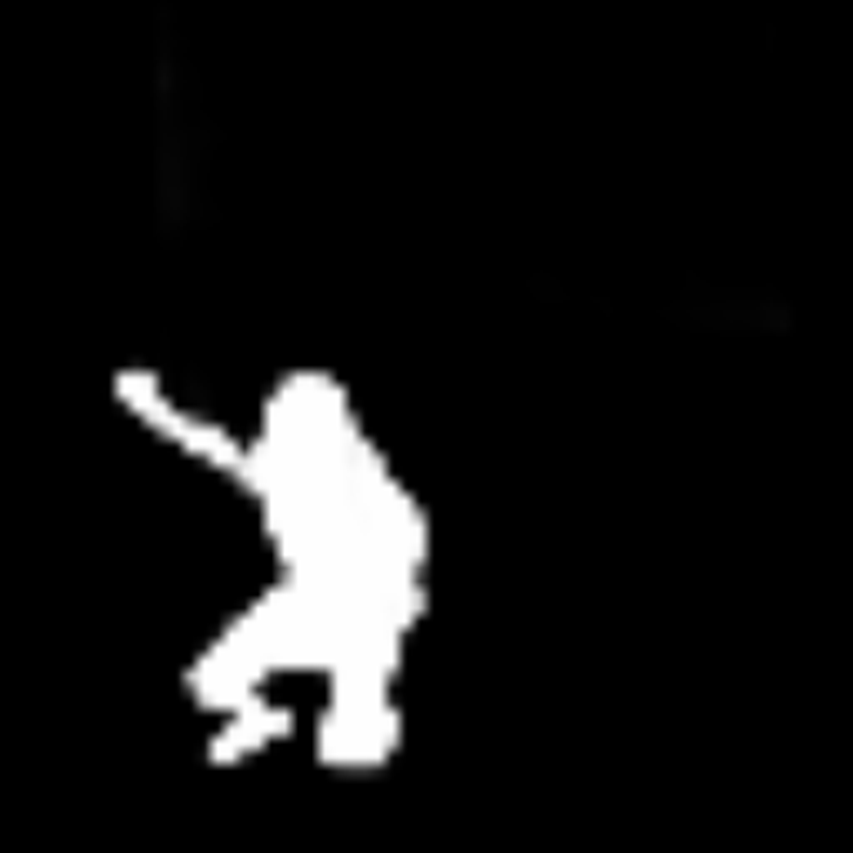}
	\end{subfigure}
    \begin{subfigure}{0.065\textwidth}
		\includegraphics[width=\textwidth]{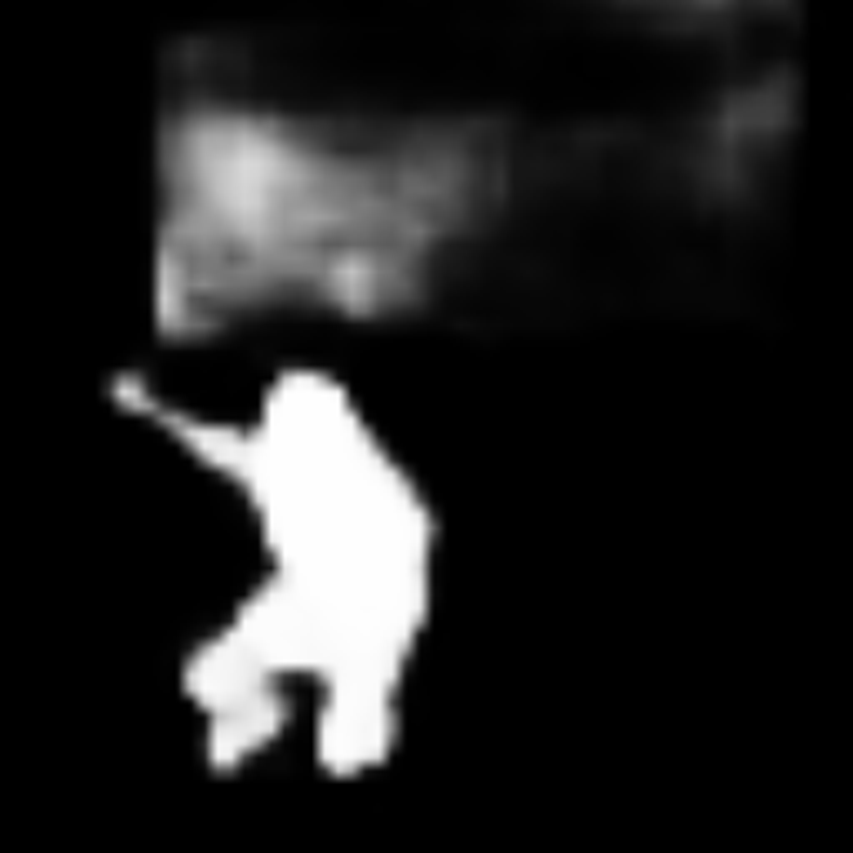}
	\end{subfigure}
    \begin{subfigure}{0.065\textwidth}
		\includegraphics[width=\textwidth]{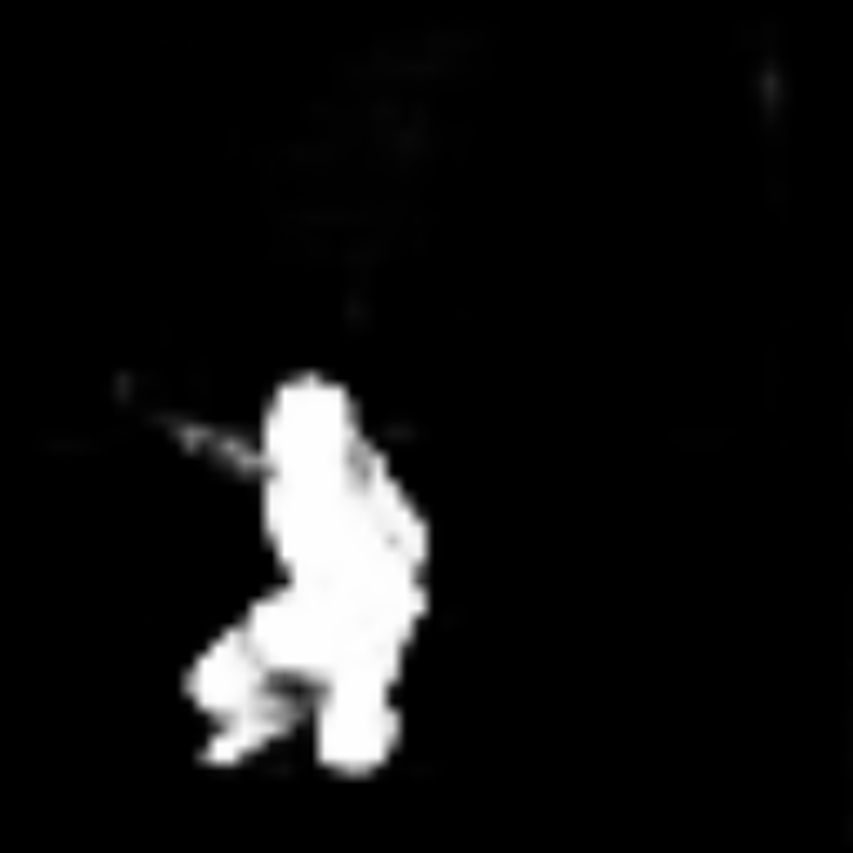}
	\end{subfigure}
	\begin{subfigure}{0.065\textwidth}
		\includegraphics[width=\textwidth]{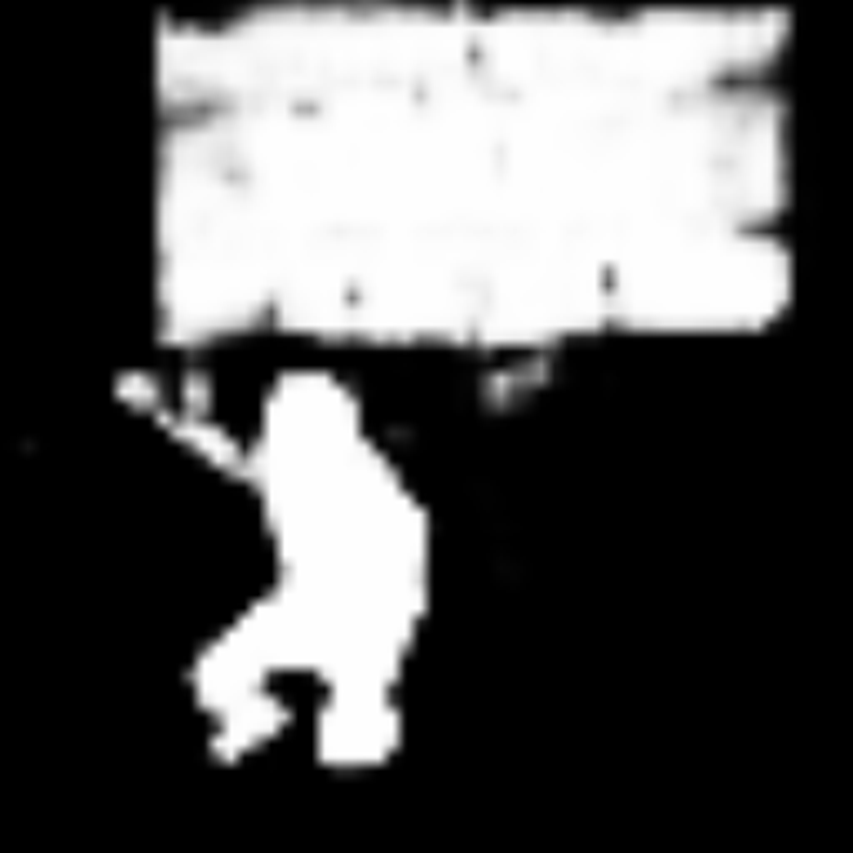}
	\end{subfigure}
	\begin{subfigure}{0.065\textwidth}
		\includegraphics[width=\textwidth]{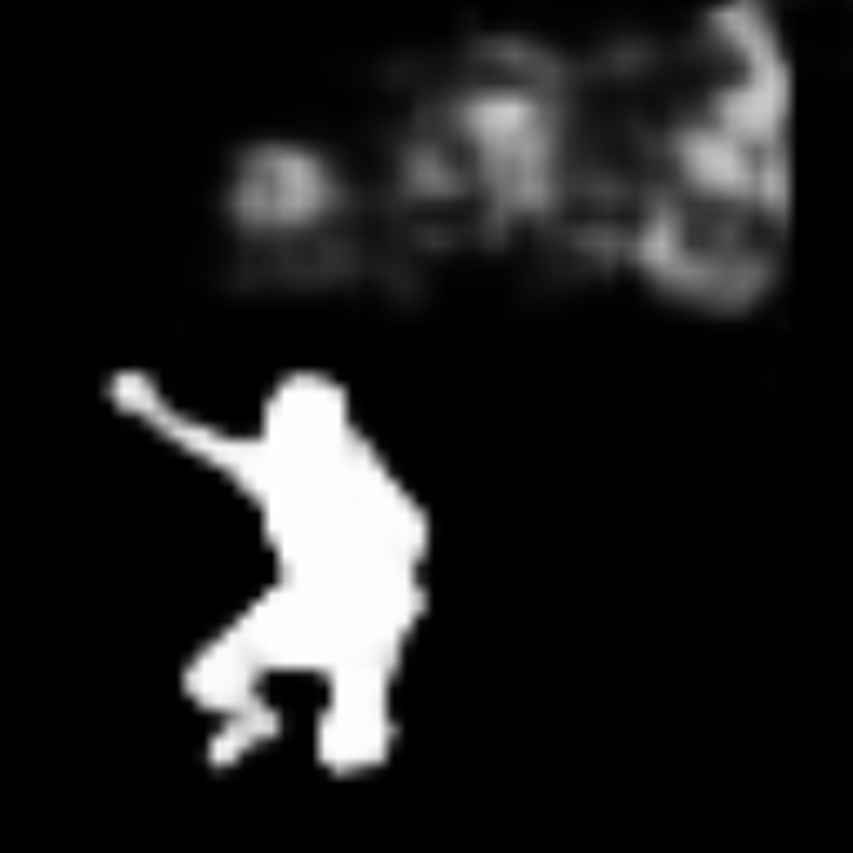}
	\end{subfigure}
    \begin{subfigure}{0.065\textwidth}
		\includegraphics[width=\textwidth]{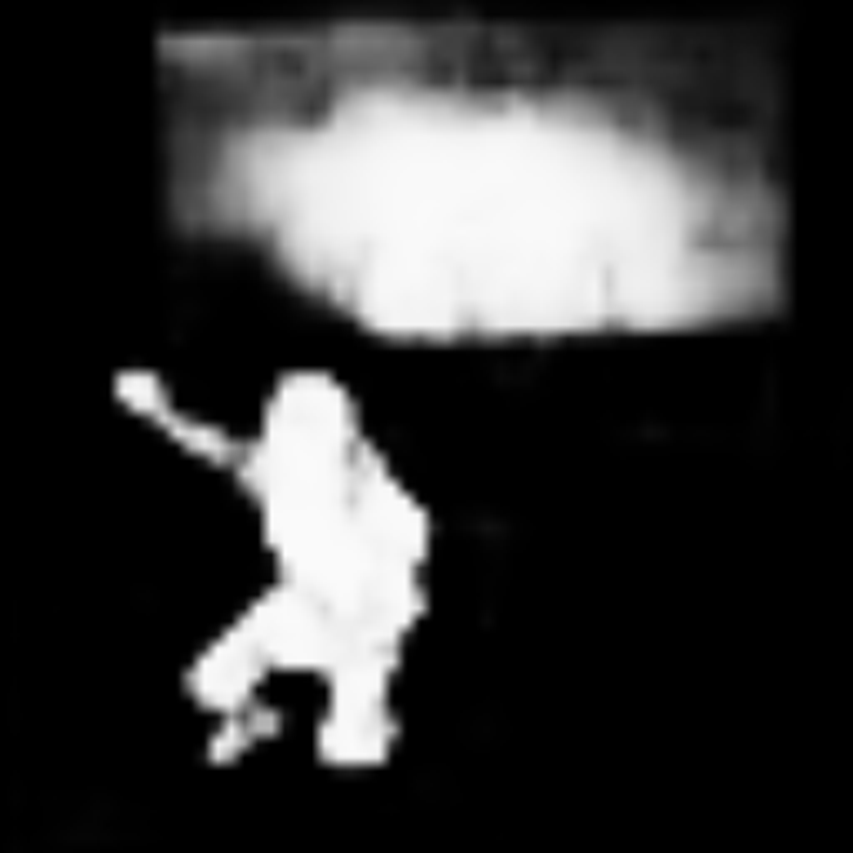}
	\end{subfigure}
	\begin{subfigure}{0.065\textwidth}
		\includegraphics[width=\textwidth]{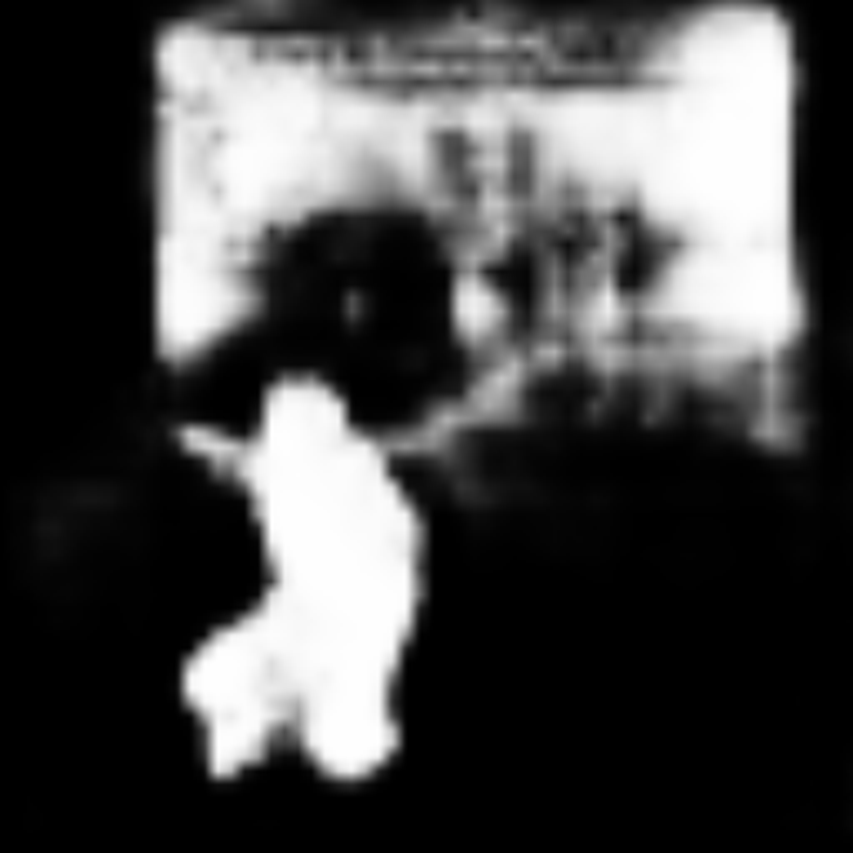}
	\end{subfigure}
	\begin{subfigure}{0.065\textwidth}
		\includegraphics[width=\textwidth]{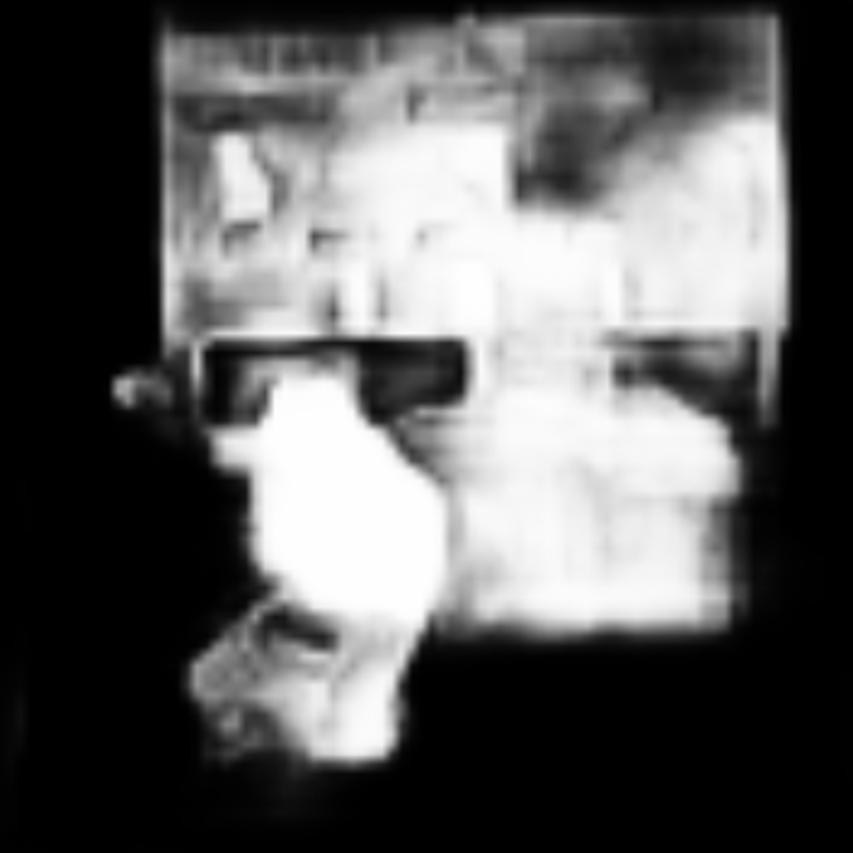}
	\end{subfigure}
	\begin{subfigure}{0.065\textwidth}
		\includegraphics[width=\textwidth]{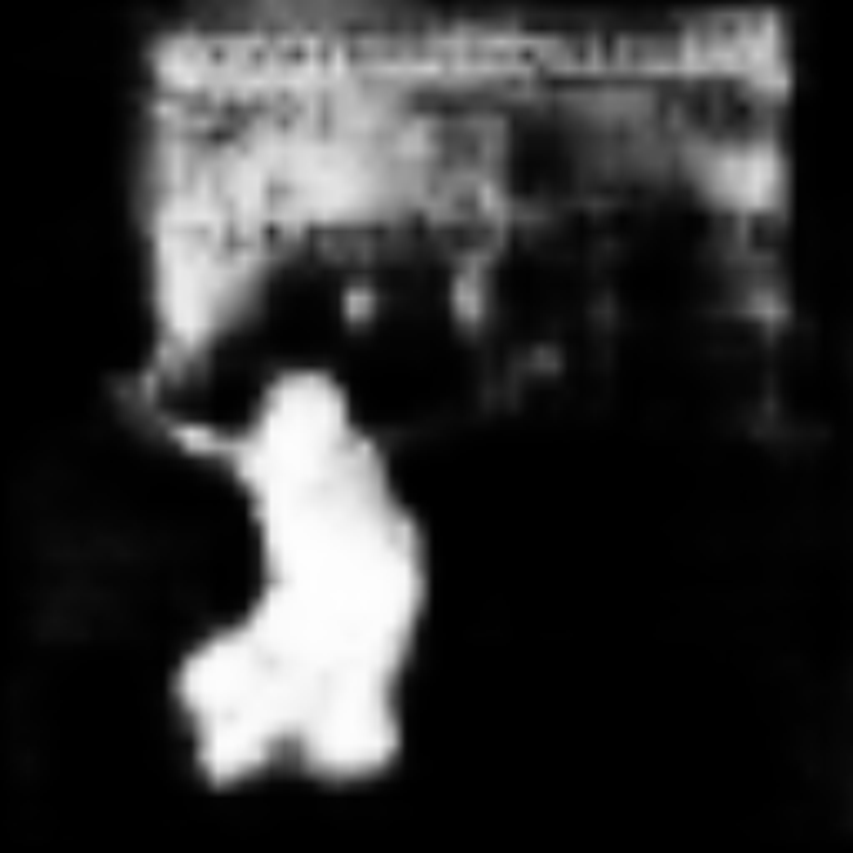}
	\end{subfigure}
	\begin{subfigure}{0.065\textwidth}
		\includegraphics[width=\textwidth]{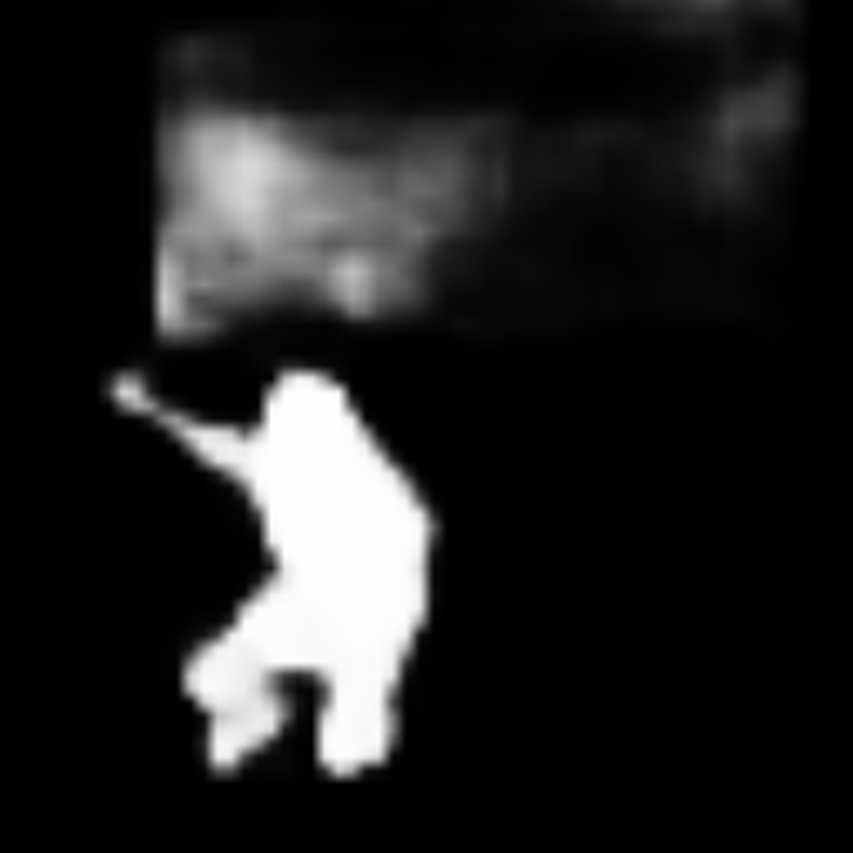}
	\end{subfigure}
	\begin{subfigure}{0.065\textwidth}
		\includegraphics[width=\textwidth]{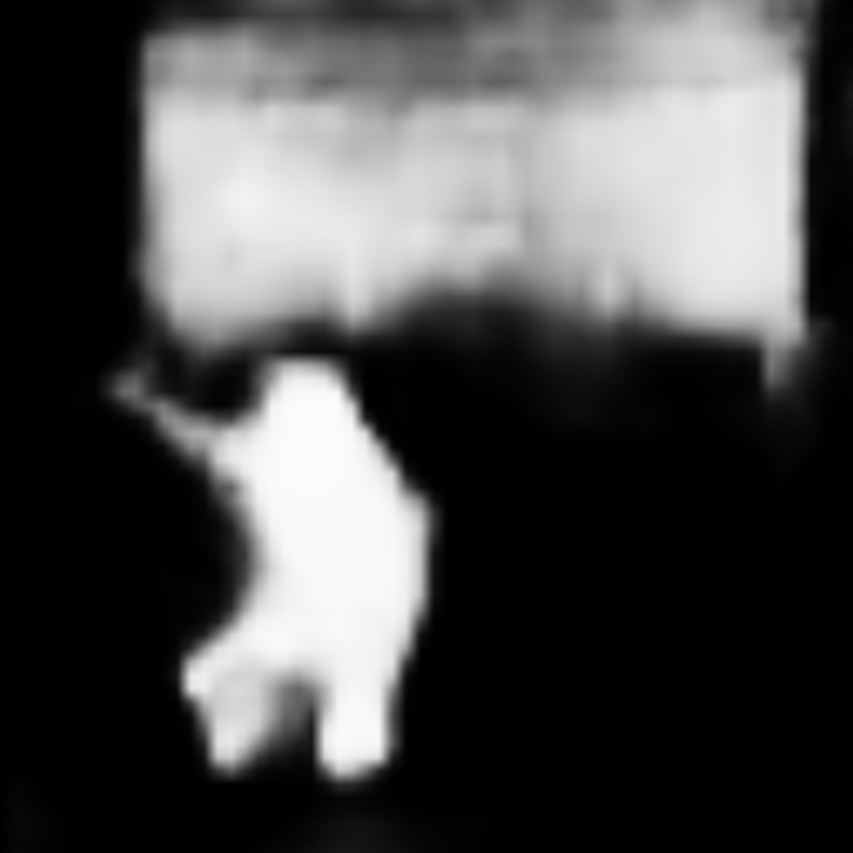}
	\end{subfigure}
	\ \\
	\vspace*{0.5mm}
		\begin{subfigure}{0.065\textwidth} 
		\includegraphics[width=\textwidth]{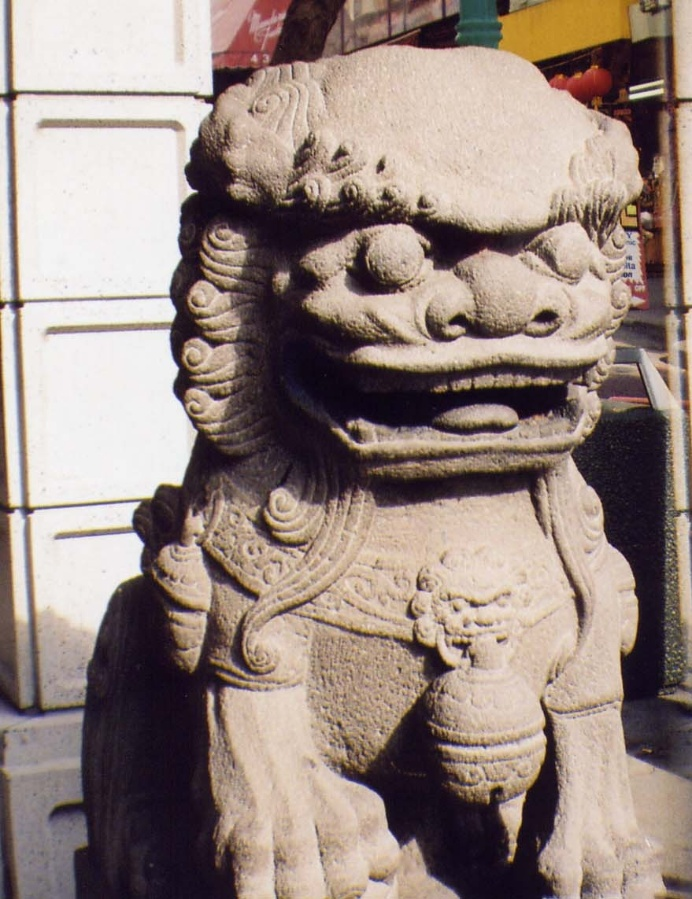}
    \end{subfigure}
	\begin{subfigure}{0.065\textwidth} 
		\includegraphics[width=\textwidth]{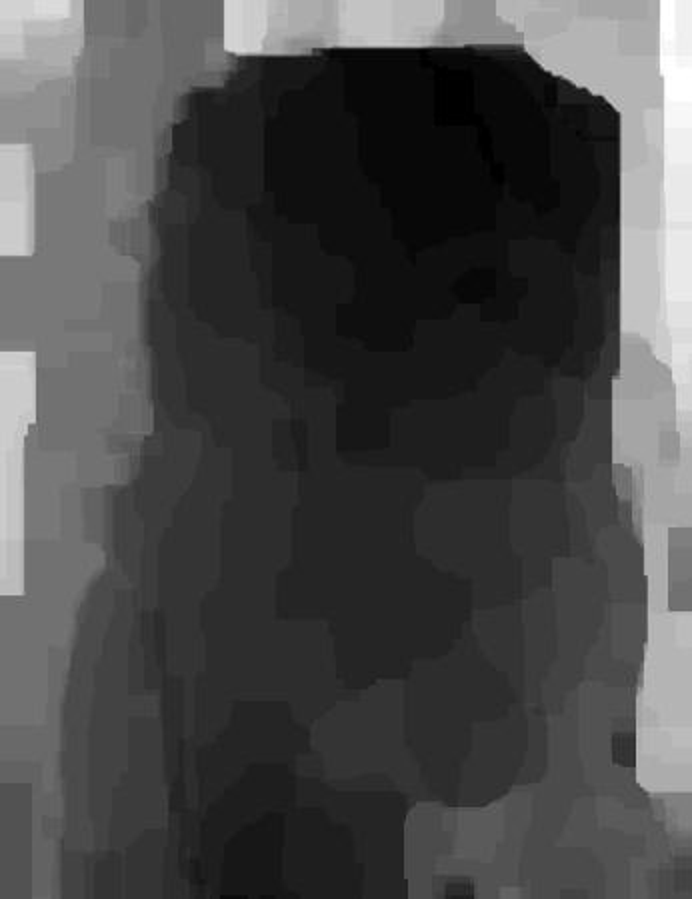}
	\end{subfigure}
	\begin{subfigure}{0.065\textwidth}
		\includegraphics[width=\textwidth]{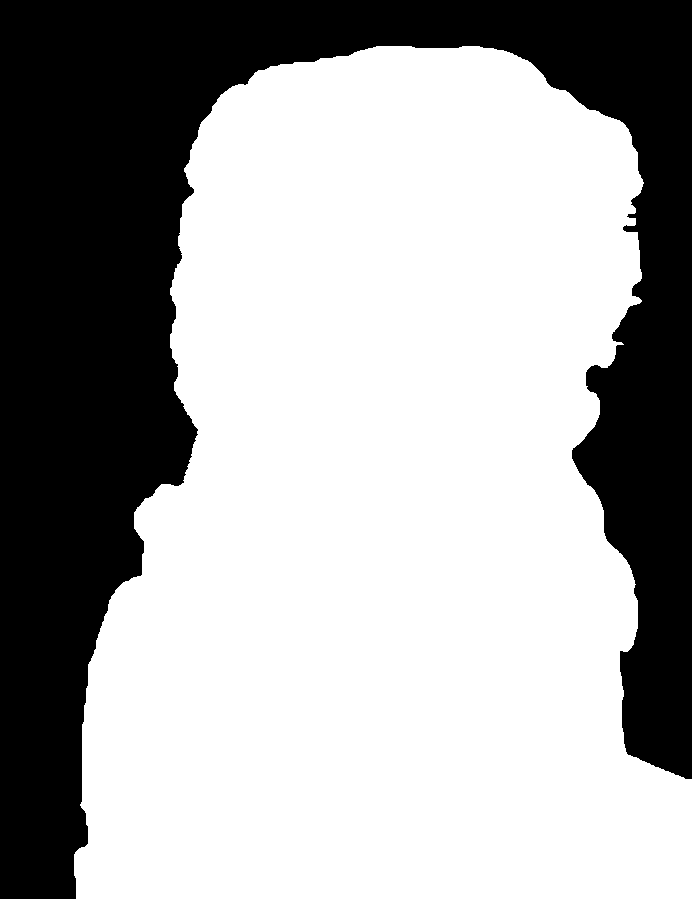}
	\end{subfigure}
    \begin{subfigure}{0.065\textwidth}
		\includegraphics[width=\textwidth]{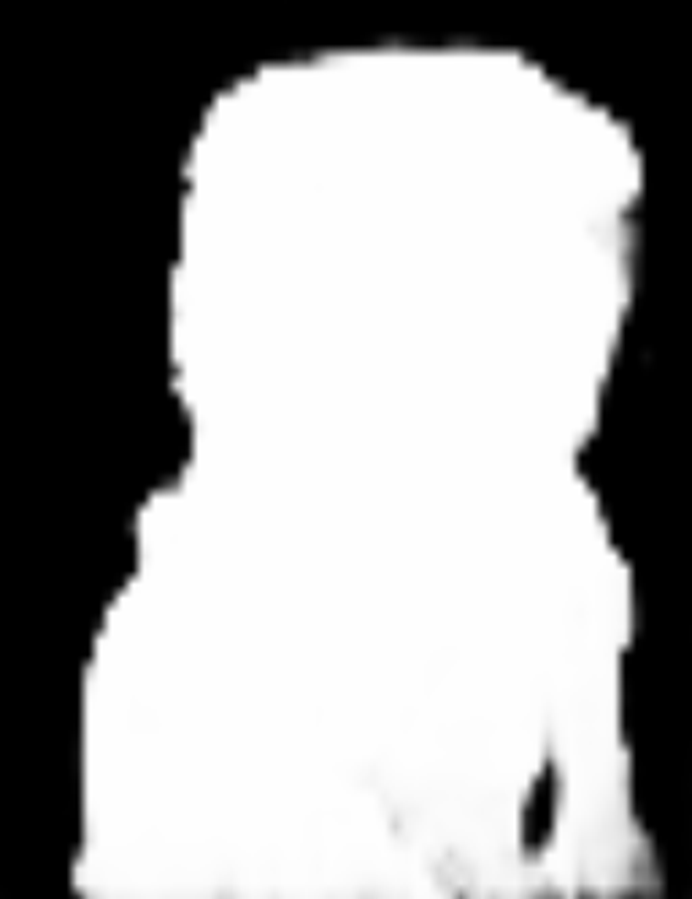}
	\end{subfigure}
    \begin{subfigure}{0.065\textwidth}
		\includegraphics[width=\textwidth]{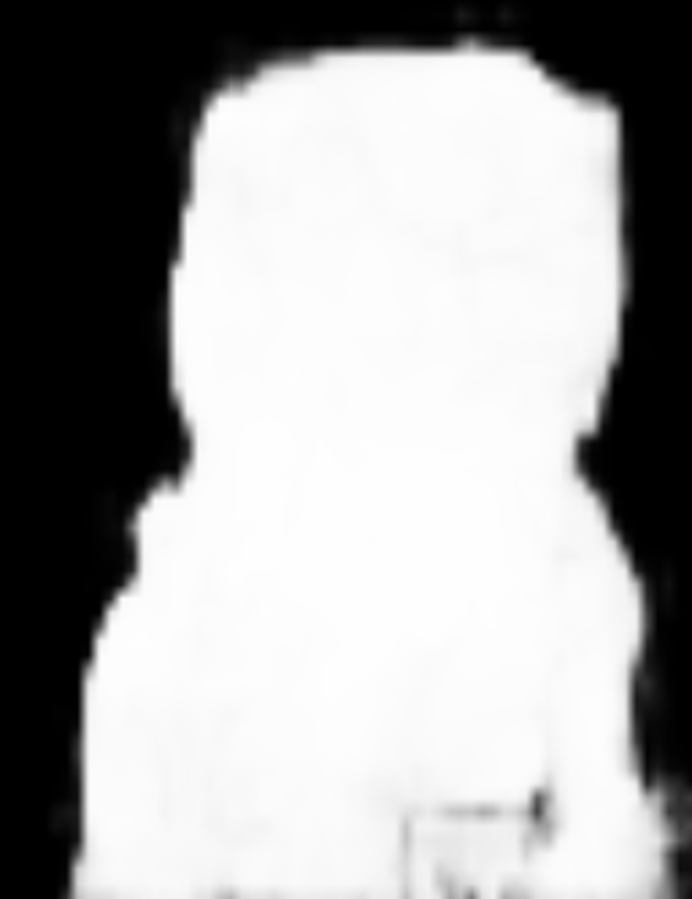}
	\end{subfigure}
    \begin{subfigure}{0.065\textwidth}
		\includegraphics[width=\textwidth]{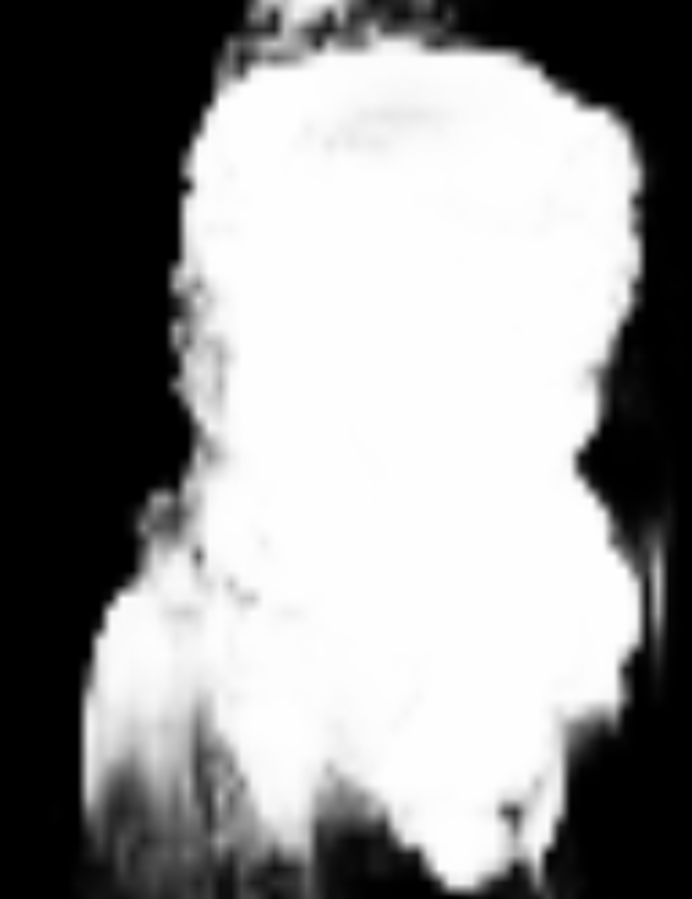}
	\end{subfigure}
	\begin{subfigure}{0.065\textwidth}
		\includegraphics[width=\textwidth]{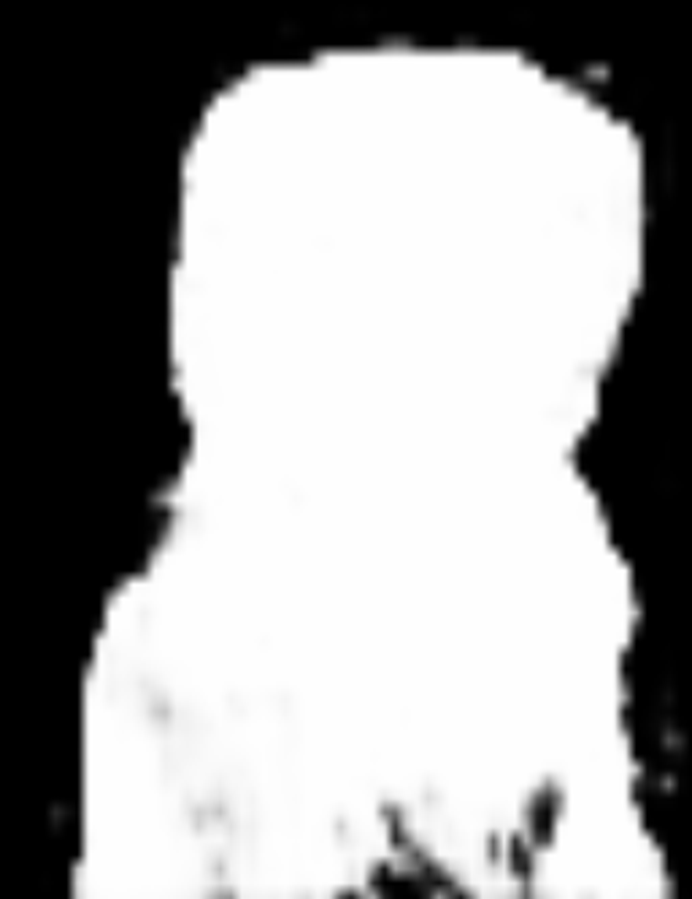}
	\end{subfigure}
	\begin{subfigure}{0.065\textwidth}
		\includegraphics[width=\textwidth]{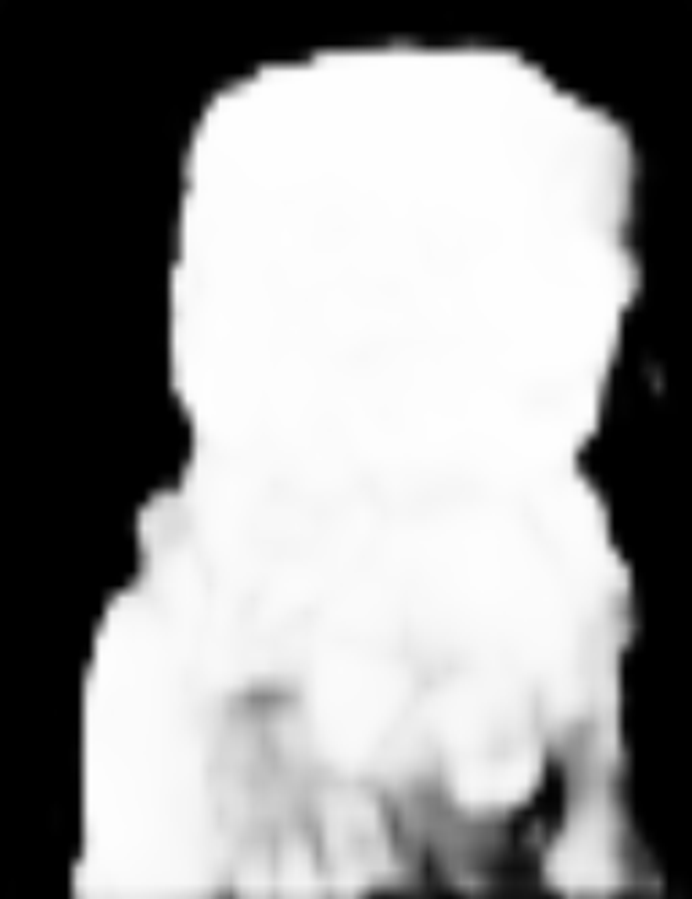}
	\end{subfigure}
    \begin{subfigure}{0.065\textwidth}
		\includegraphics[width=\textwidth]{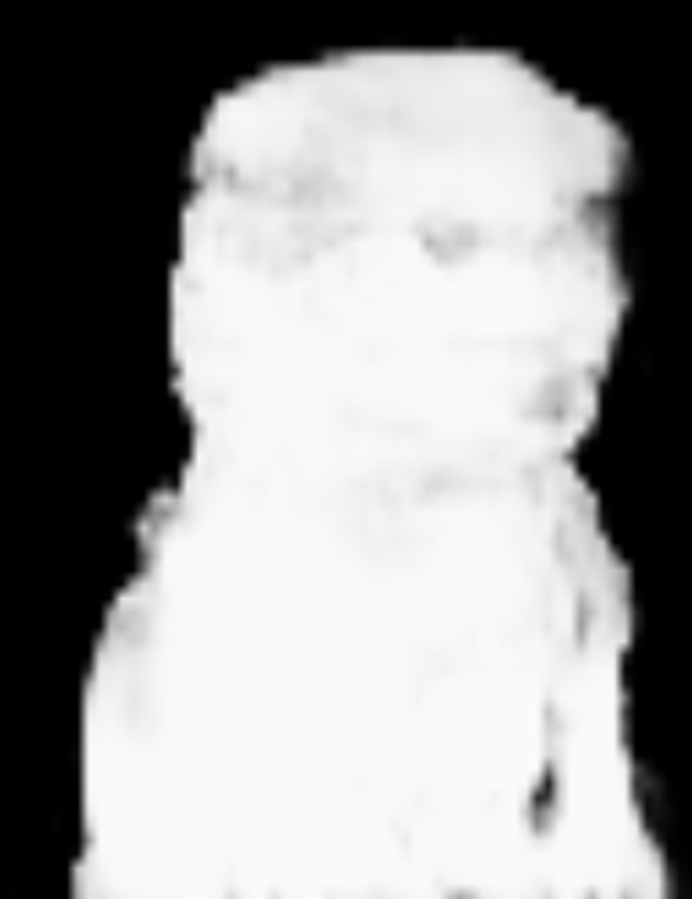}
	\end{subfigure}
	\begin{subfigure}{0.065\textwidth}
		\includegraphics[width=\textwidth]{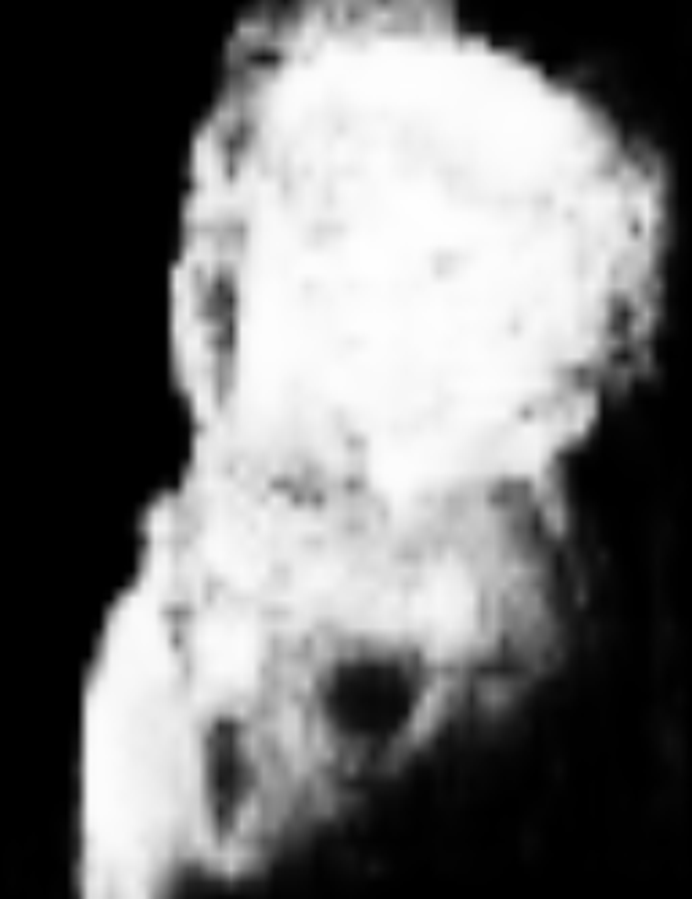}
	\end{subfigure}
	\begin{subfigure}{0.065\textwidth}
		\includegraphics[width=\textwidth]{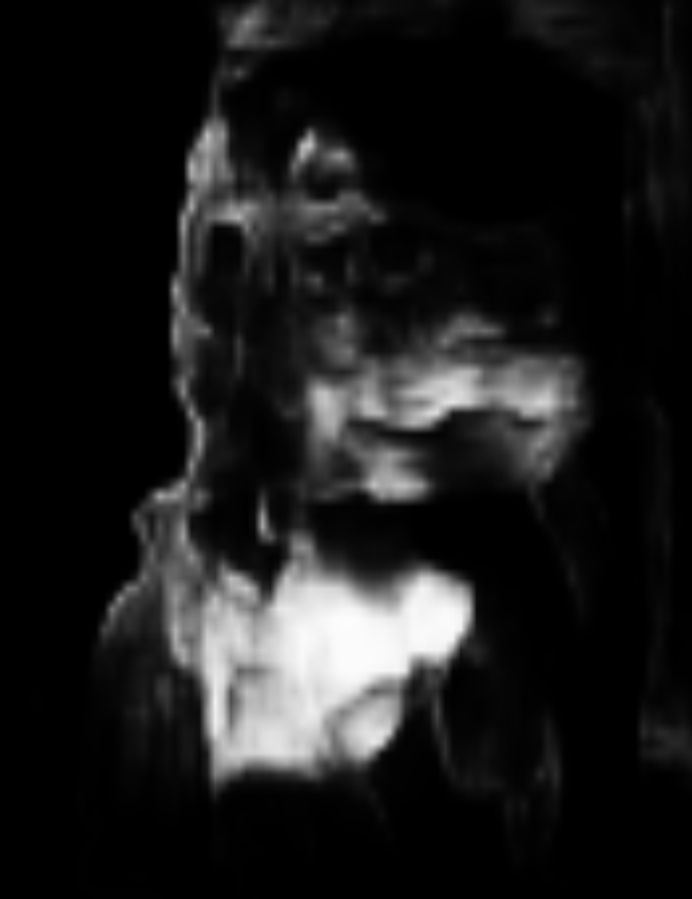}
	\end{subfigure}
	\begin{subfigure}{0.065\textwidth}
		\includegraphics[width=\textwidth]{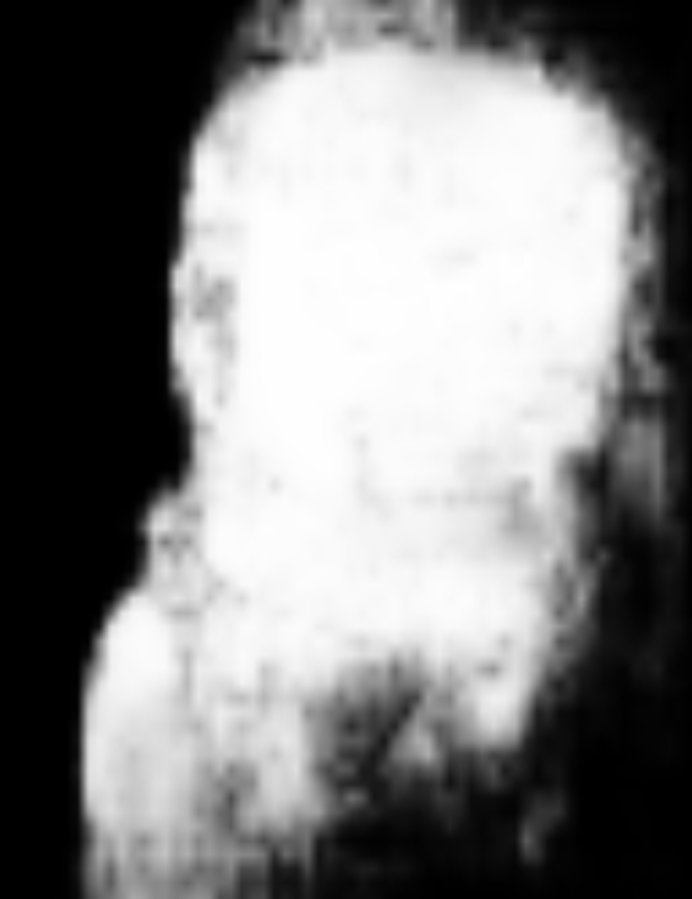}
	\end{subfigure}
	\begin{subfigure}{0.065\textwidth}
		\includegraphics[width=\textwidth]{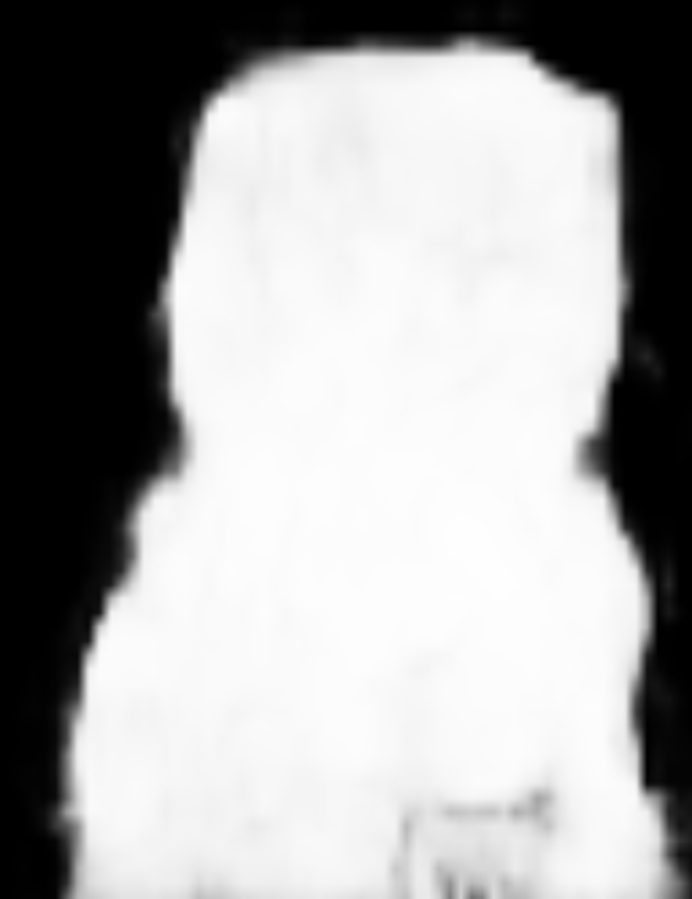}
	\end{subfigure}
	\begin{subfigure}{0.065\textwidth}
		\includegraphics[width=\textwidth]{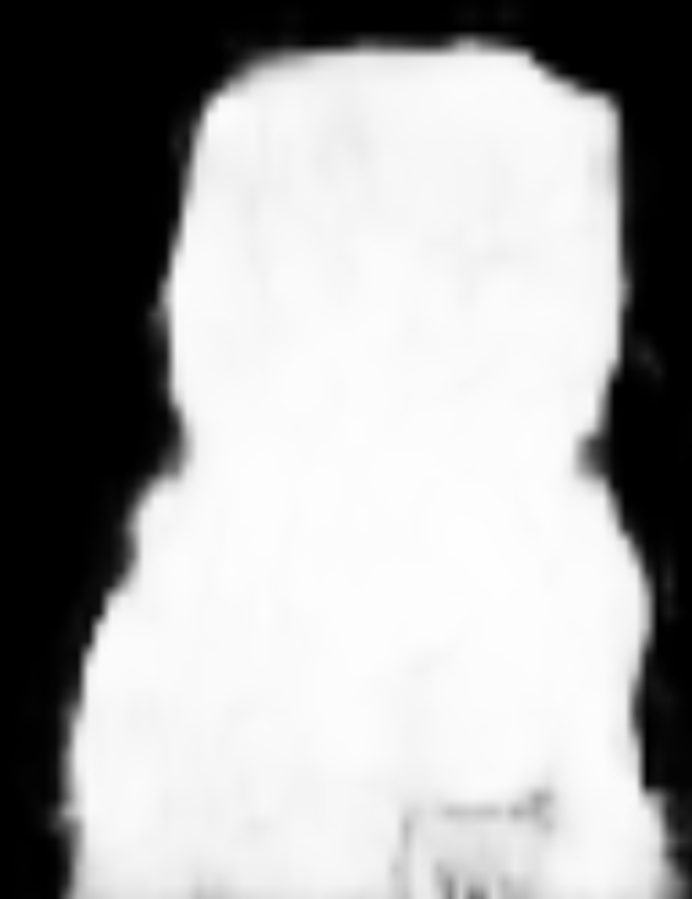}
	\end{subfigure}
	\ \\
	\vspace*{0.5mm}
	\begin{subfigure}{0.065\textwidth} 
		\includegraphics[width=\textwidth]{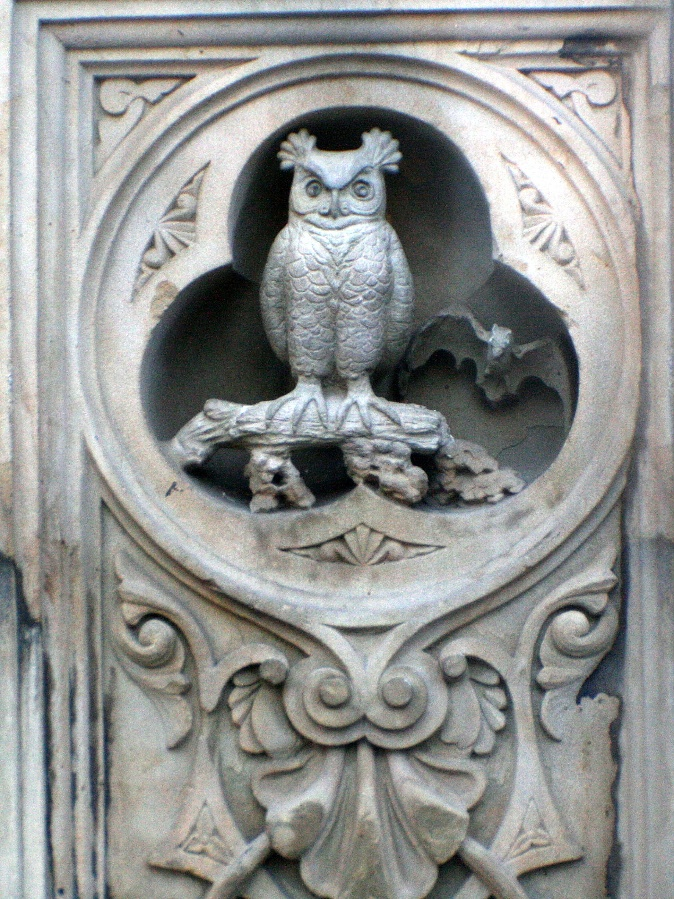}
    \end{subfigure}
	\begin{subfigure}{0.065\textwidth} 
		\includegraphics[width=\textwidth]{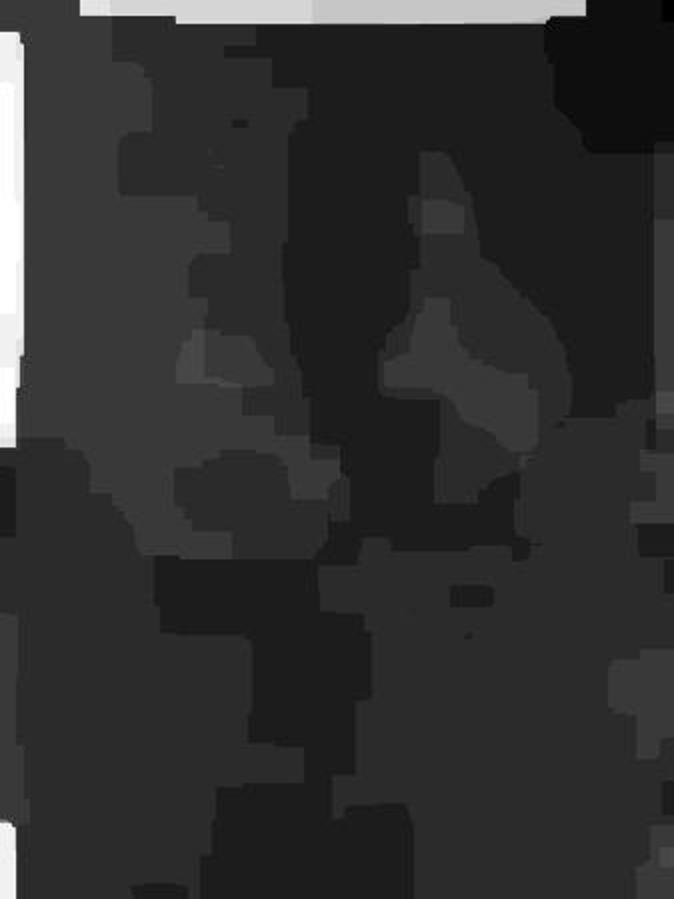}
	\end{subfigure}
	\begin{subfigure}{0.065\textwidth}
		\includegraphics[width=\textwidth]{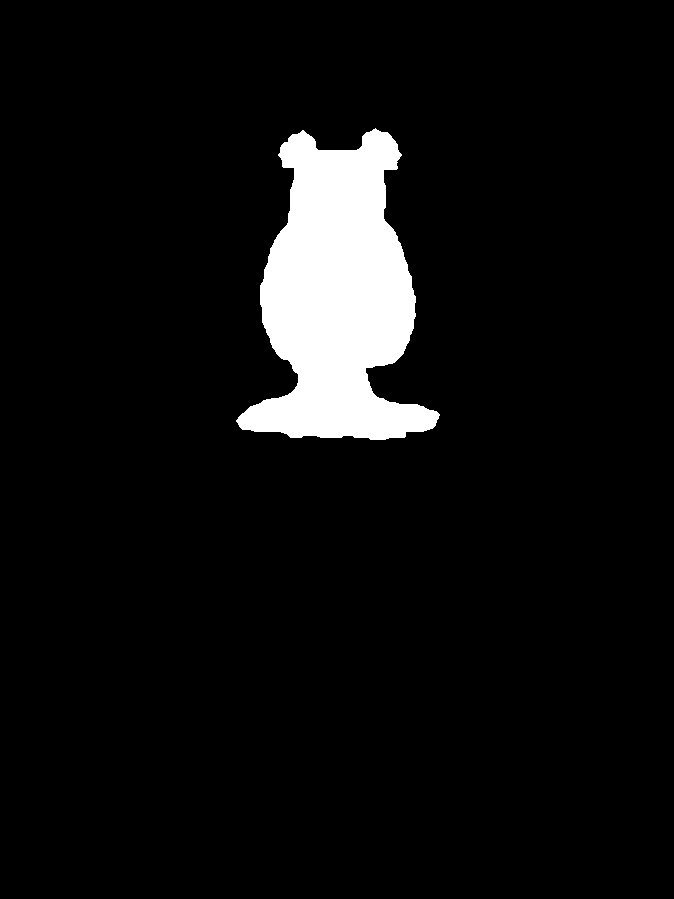}
	\end{subfigure}
    \begin{subfigure}{0.065\textwidth}
		\includegraphics[width=\textwidth]{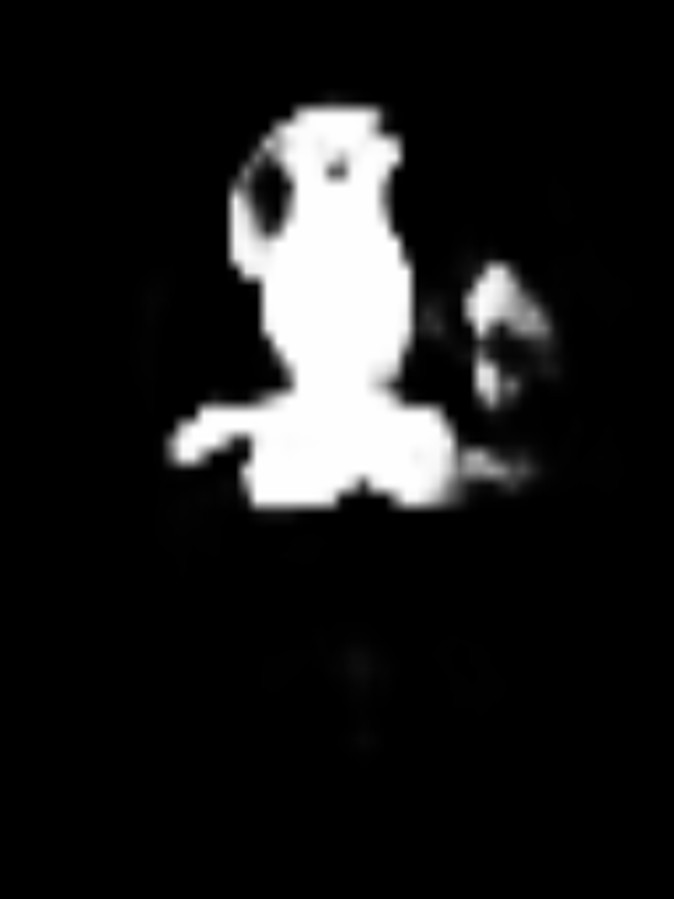}
	\end{subfigure}
    \begin{subfigure}{0.065\textwidth}
		\includegraphics[width=\textwidth]{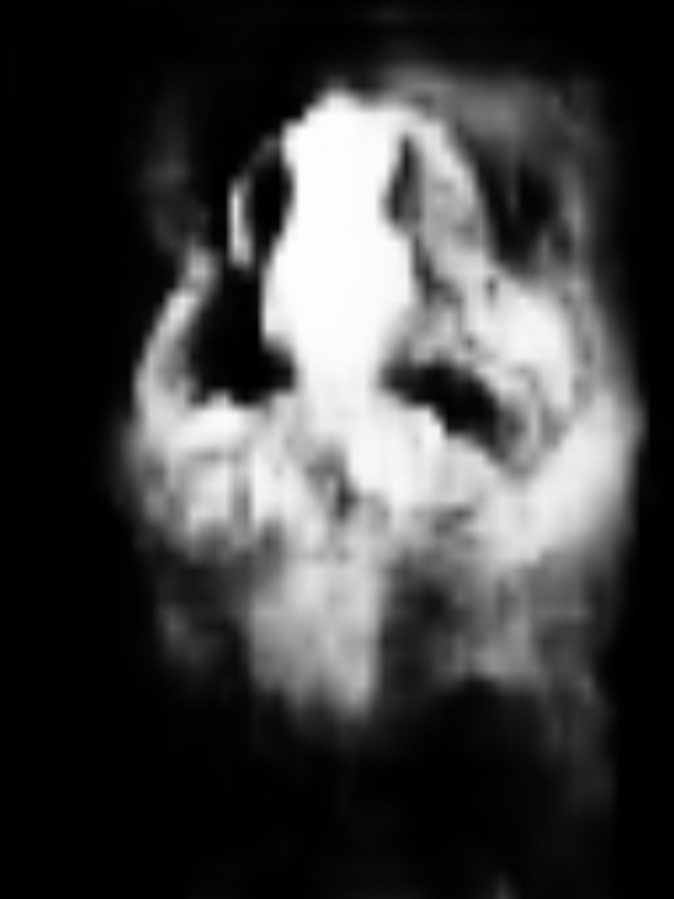}
	\end{subfigure}
    \begin{subfigure}{0.065\textwidth}
		\includegraphics[width=\textwidth]{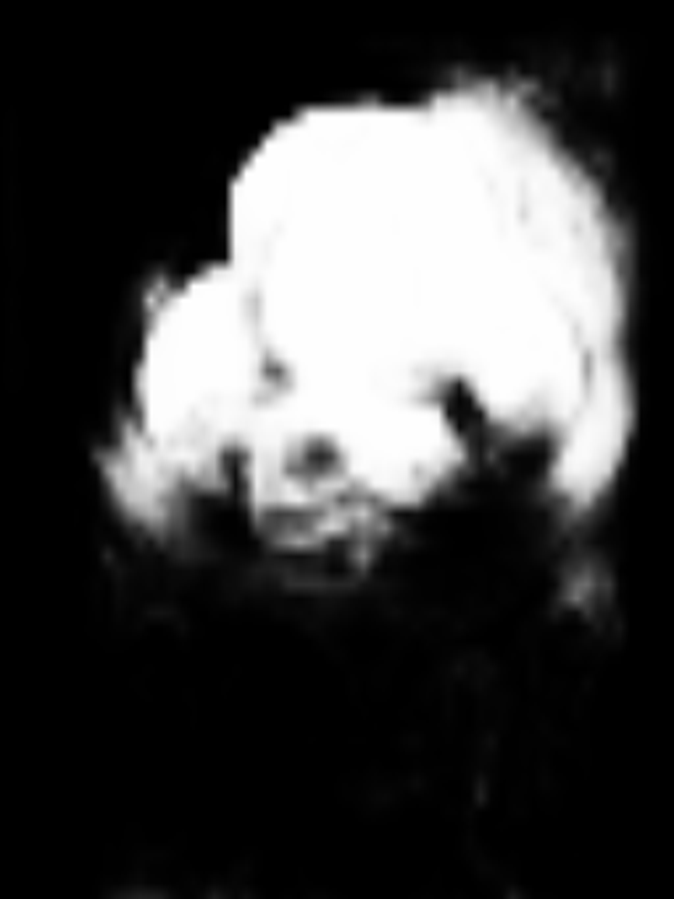}
	\end{subfigure}
	\begin{subfigure}{0.065\textwidth}
		\includegraphics[width=\textwidth]{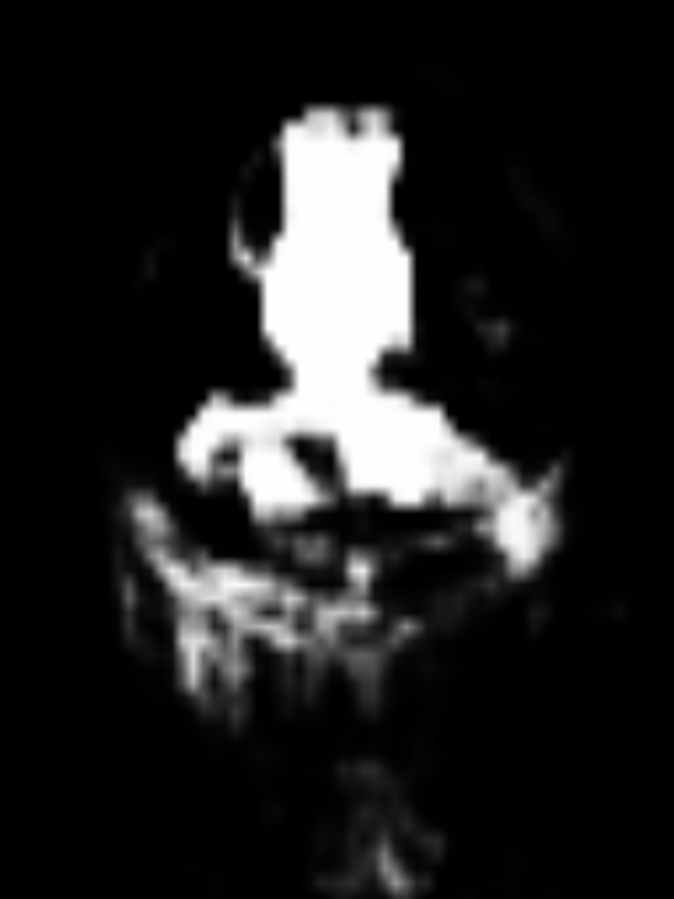}
	\end{subfigure}
	\begin{subfigure}{0.065\textwidth}
		\includegraphics[width=\textwidth]{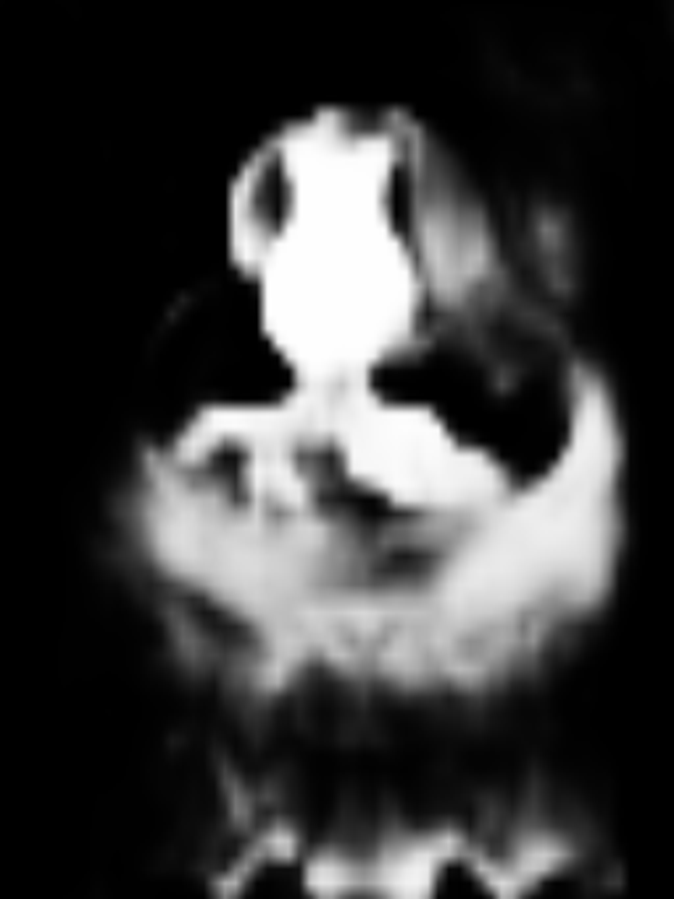}
	\end{subfigure}
    \begin{subfigure}{0.065\textwidth}
		\includegraphics[width=\textwidth]{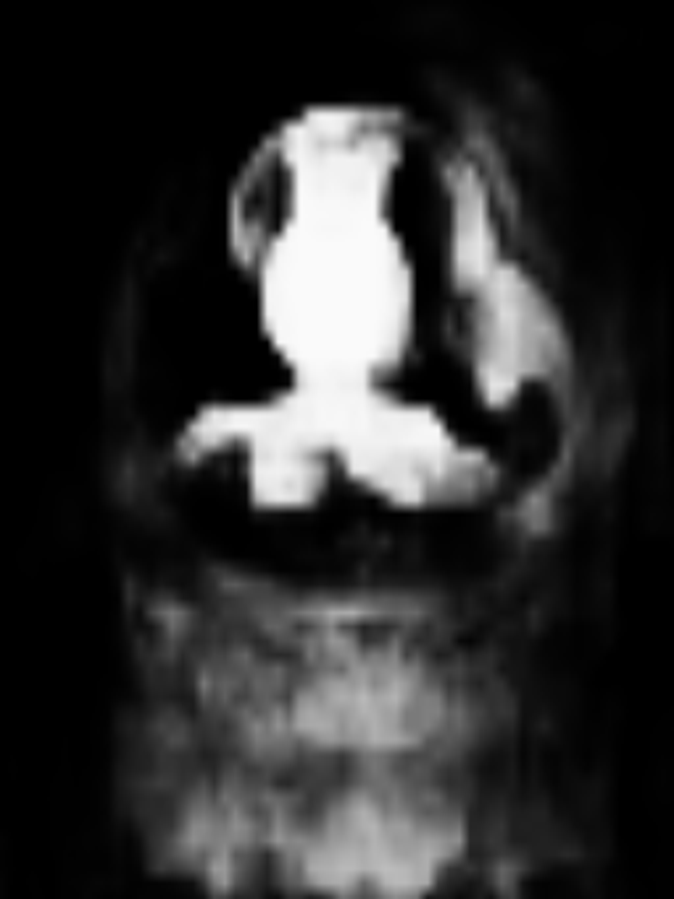}
	\end{subfigure}
	\begin{subfigure}{0.065\textwidth}
		\includegraphics[width=\textwidth]{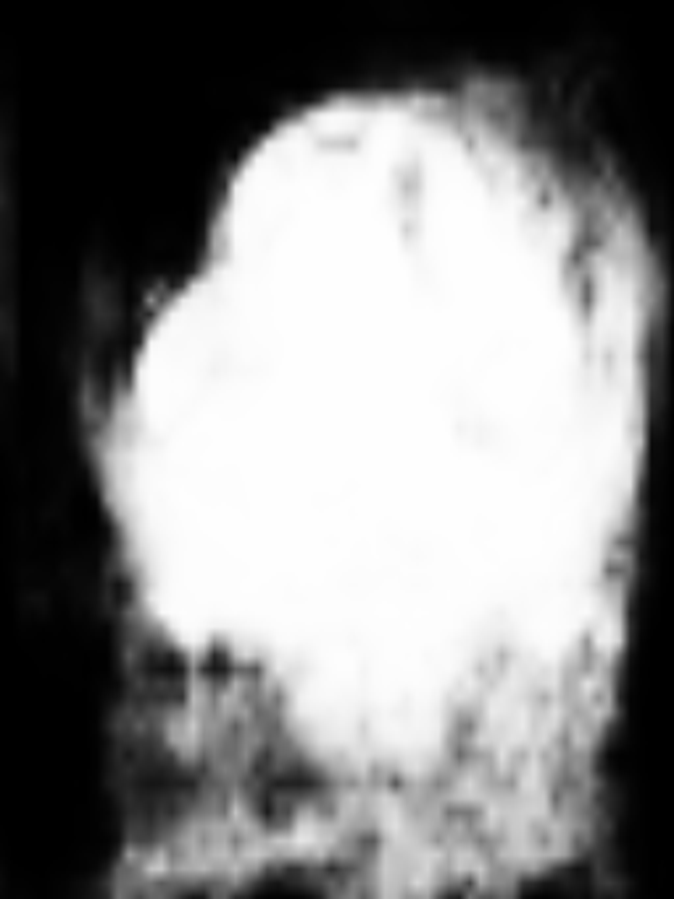}
	\end{subfigure}
	\begin{subfigure}{0.065\textwidth}
		\includegraphics[width=\textwidth]{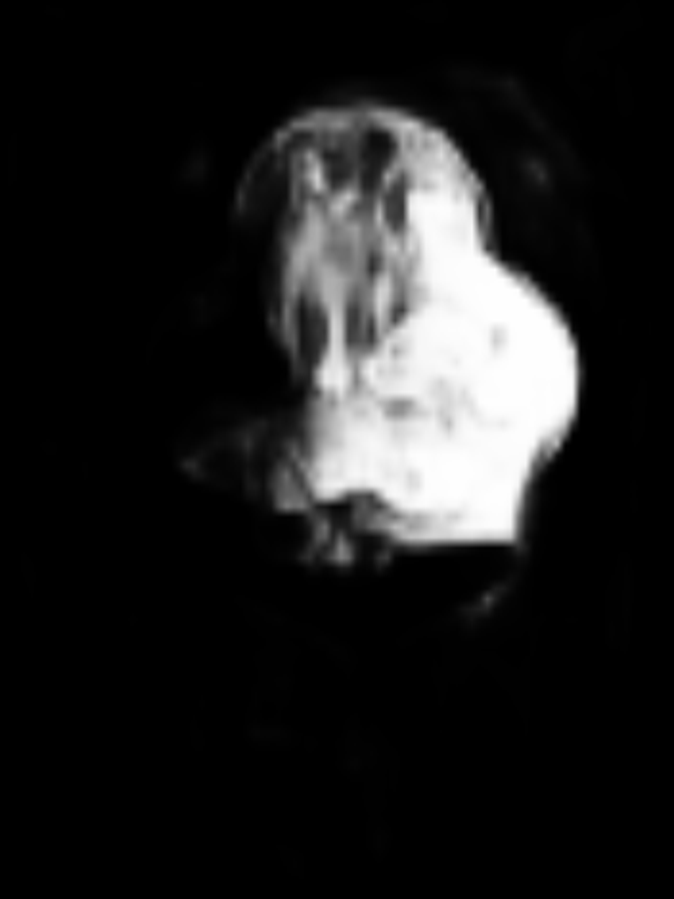}
	\end{subfigure}
	\begin{subfigure}{0.065\textwidth}
		\includegraphics[width=\textwidth]{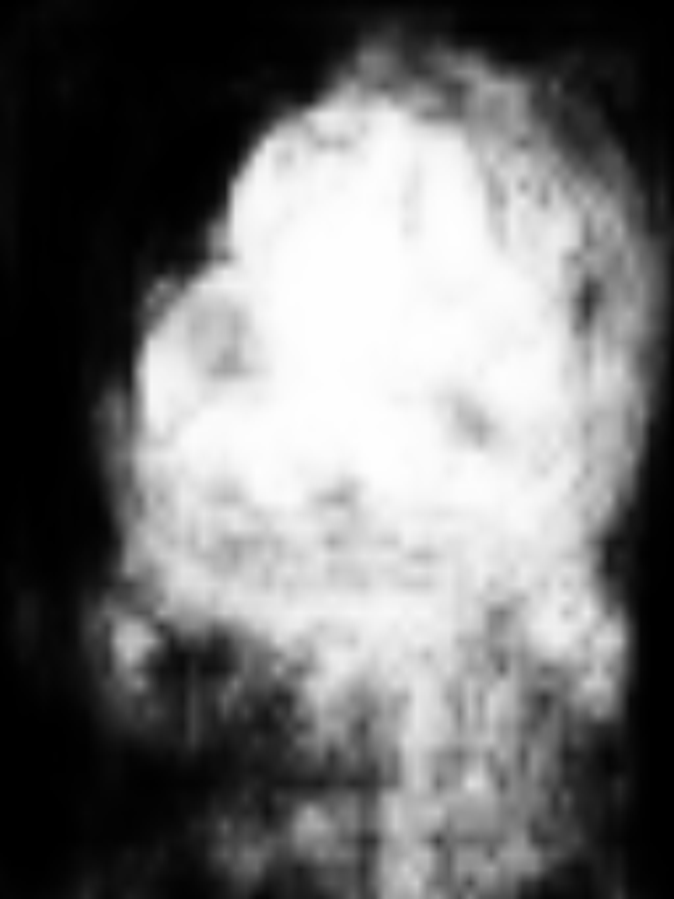}
	\end{subfigure}
	\begin{subfigure}{0.065\textwidth}
		\includegraphics[width=\textwidth]{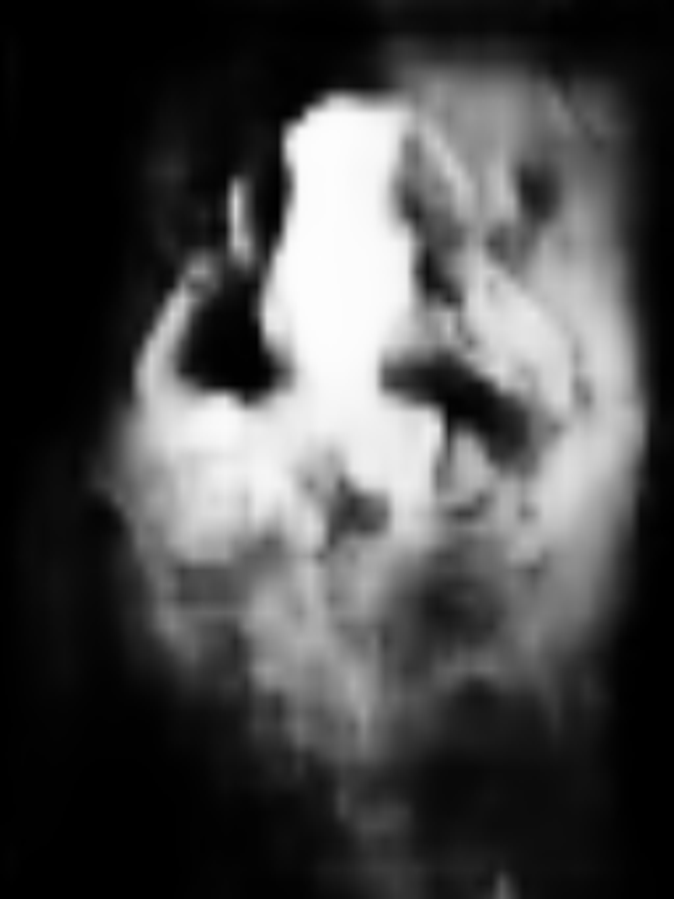}
	\end{subfigure}
	\begin{subfigure}{0.065\textwidth}
		\includegraphics[width=\textwidth]{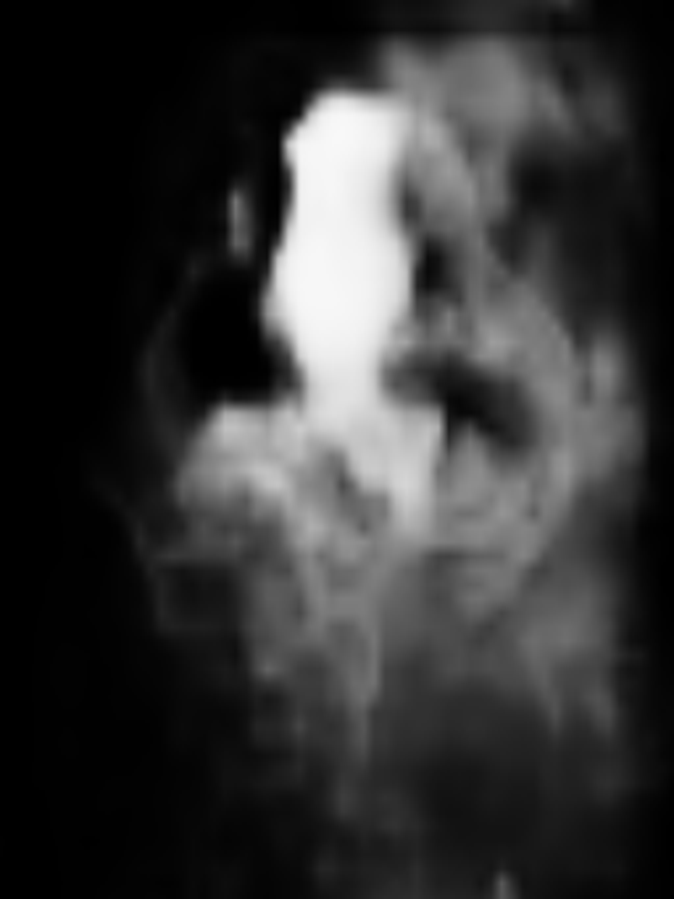}
	\end{subfigure}
	\ \\
	\vspace*{0.5mm}
	\begin{subfigure}{0.065\textwidth} 
		\includegraphics[width=\textwidth]{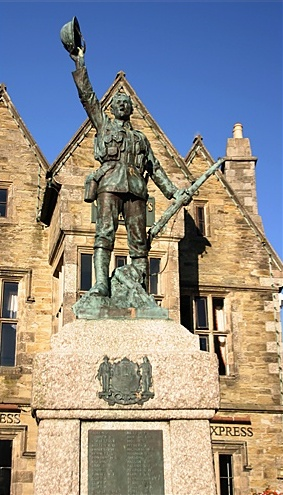}
    \end{subfigure}
	\begin{subfigure}{0.065\textwidth} 
		\includegraphics[width=\textwidth]{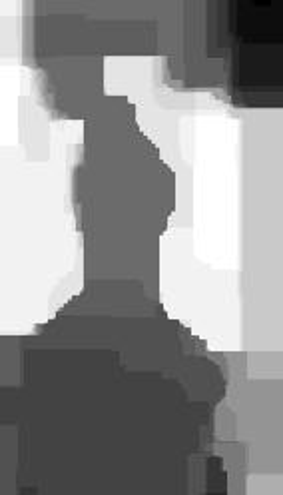}
	\end{subfigure}
	\begin{subfigure}{0.065\textwidth}
		\includegraphics[width=\textwidth]{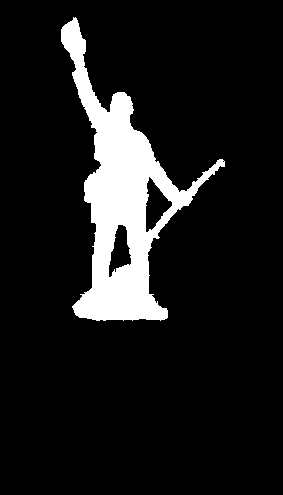}
	\end{subfigure}
    \begin{subfigure}{0.065\textwidth}
		\includegraphics[width=\textwidth]{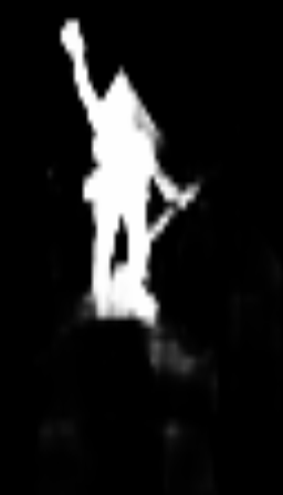}
	\end{subfigure}
    \begin{subfigure}{0.065\textwidth}
		\includegraphics[width=\textwidth]{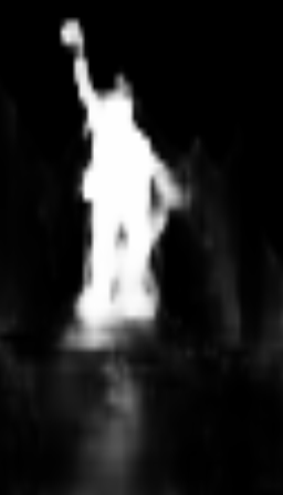}
	\end{subfigure}
    \begin{subfigure}{0.065\textwidth}
		\includegraphics[width=\textwidth]{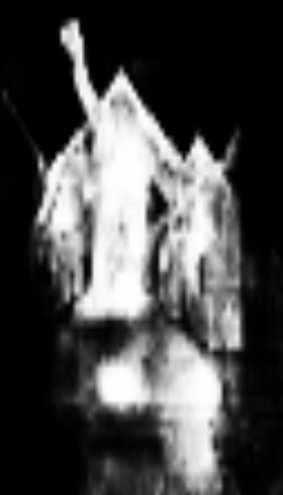}
	\end{subfigure}
	\begin{subfigure}{0.065\textwidth}
		\includegraphics[width=\textwidth]{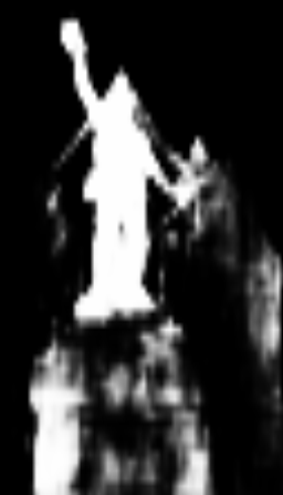}
	\end{subfigure}
	\begin{subfigure}{0.065\textwidth}
		\includegraphics[width=\textwidth]{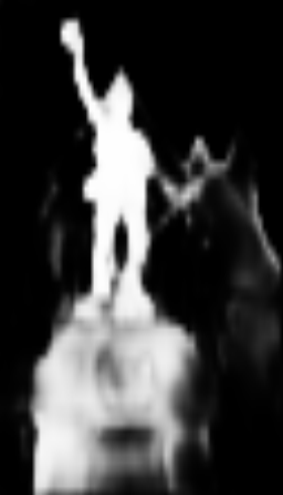}
	\end{subfigure}
    \begin{subfigure}{0.065\textwidth}
		\includegraphics[width=\textwidth]{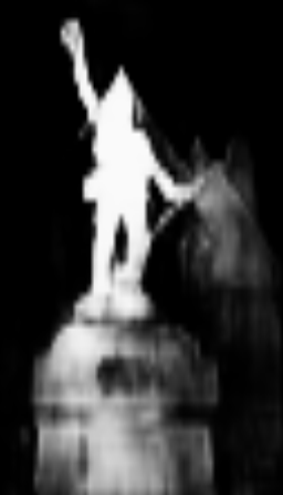}
	\end{subfigure}
	\begin{subfigure}{0.065\textwidth}
		\includegraphics[width=\textwidth]{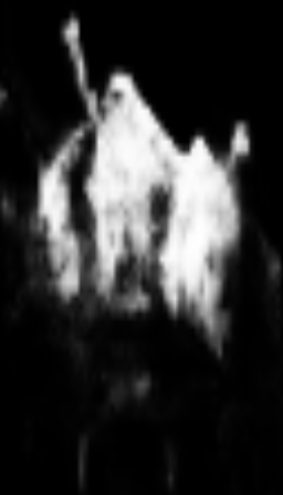}
	\end{subfigure}
	\begin{subfigure}{0.065\textwidth}
		\includegraphics[width=\textwidth]{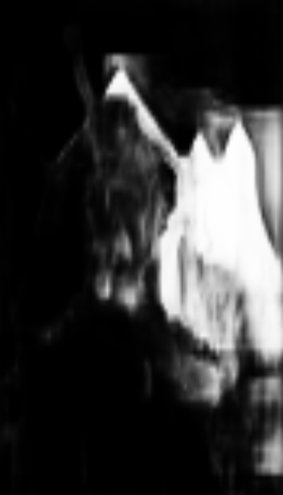}
	\end{subfigure}
	\begin{subfigure}{0.065\textwidth}
		\includegraphics[width=\textwidth]{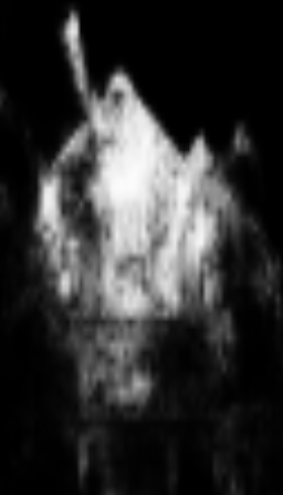}
	\end{subfigure}
	\begin{subfigure}{0.065\textwidth}
		\includegraphics[width=\textwidth]{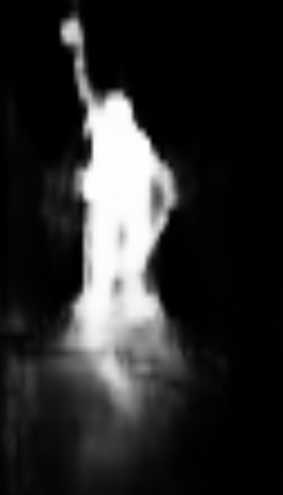}
	\end{subfigure}
	\begin{subfigure}{0.065\textwidth}
		\includegraphics[width=\textwidth]{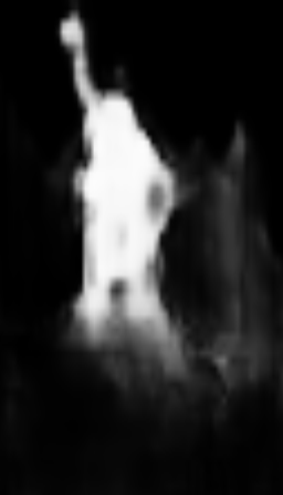}
	\end{subfigure}
	\ \\
	\vspace*{0.5mm}
	\begin{subfigure}{0.065\textwidth} 
		\includegraphics[width=\textwidth]{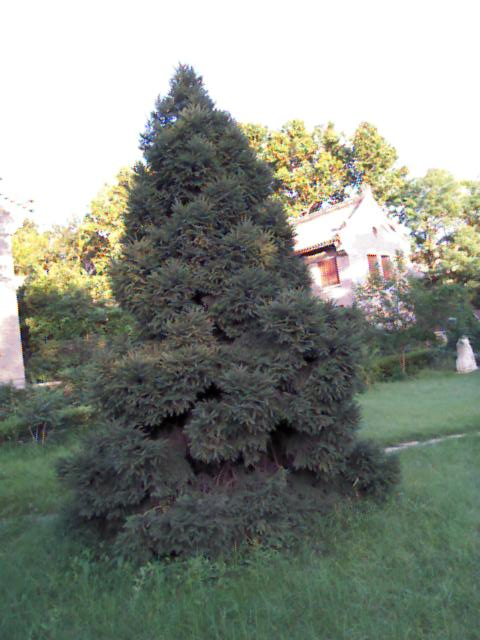}
    \end{subfigure}
	\begin{subfigure}{0.065\textwidth} 
		\includegraphics[width=\textwidth]{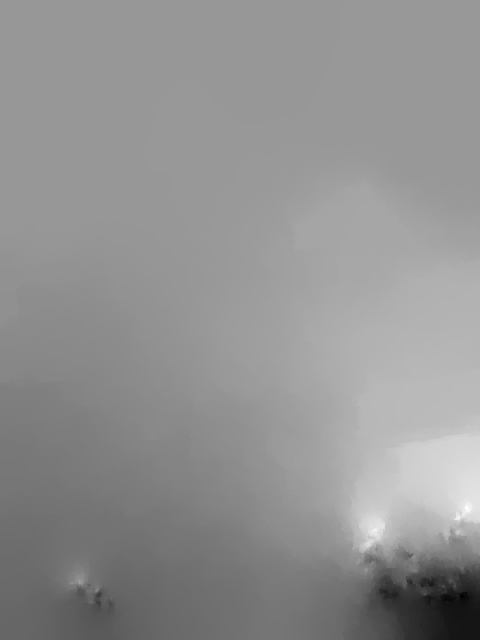}
	\end{subfigure}
	\begin{subfigure}{0.065\textwidth}
		\includegraphics[width=\textwidth]{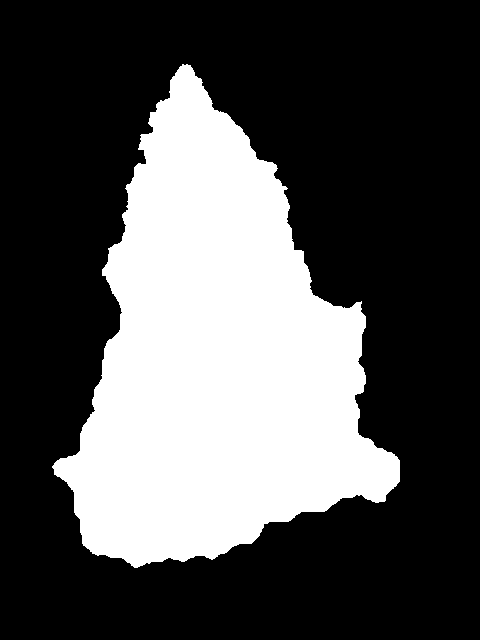}
	\end{subfigure}
    \begin{subfigure}{0.065\textwidth}
		\includegraphics[width=\textwidth]{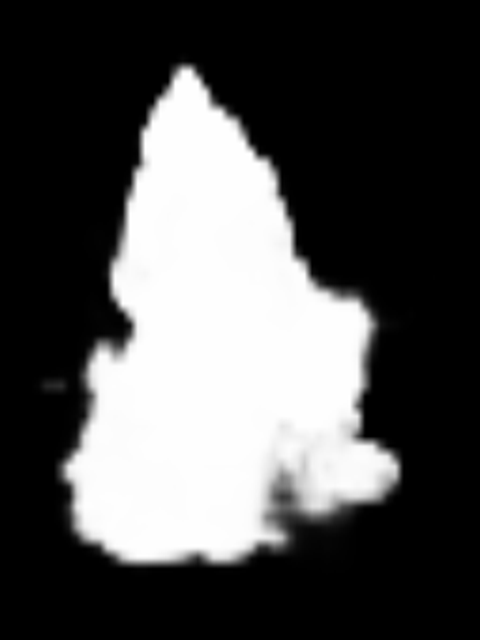}
	\end{subfigure}
    \begin{subfigure}{0.065\textwidth}
		\includegraphics[width=\textwidth]{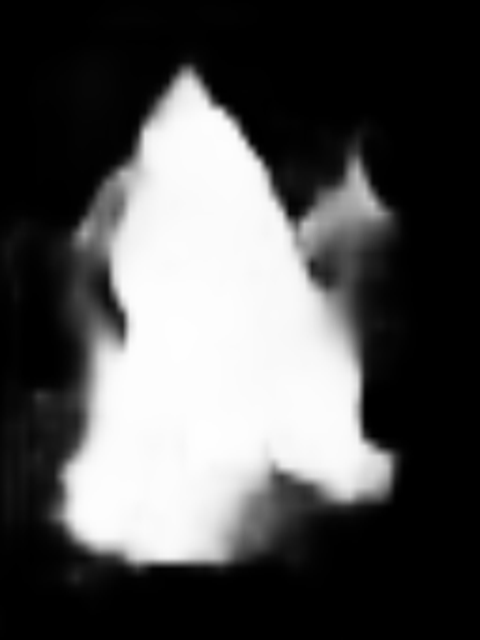}
	\end{subfigure}
    \begin{subfigure}{0.065\textwidth}
		\includegraphics[width=\textwidth]{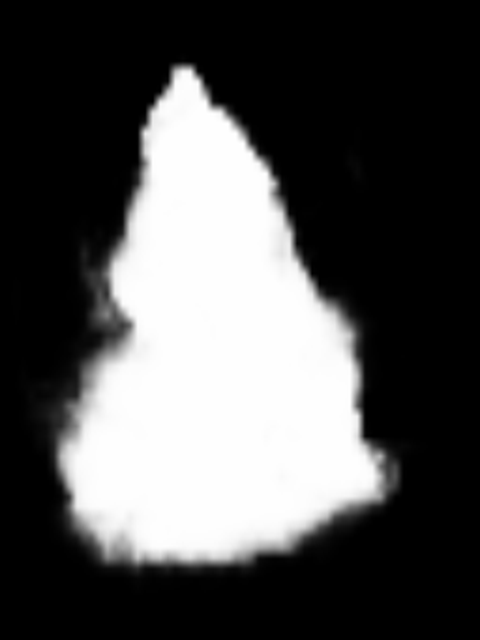}
	\end{subfigure}
	\begin{subfigure}{0.065\textwidth}
		\includegraphics[width=\textwidth]{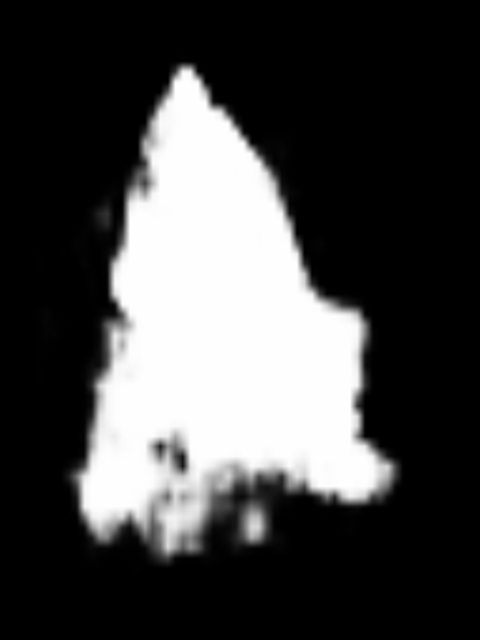}
	\end{subfigure}
	\begin{subfigure}{0.065\textwidth}
		\includegraphics[width=\textwidth]{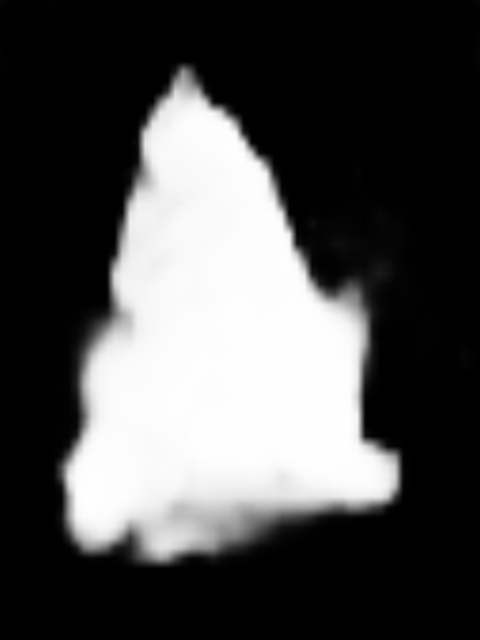}
	\end{subfigure}
    \begin{subfigure}{0.065\textwidth}
		\includegraphics[width=\textwidth]{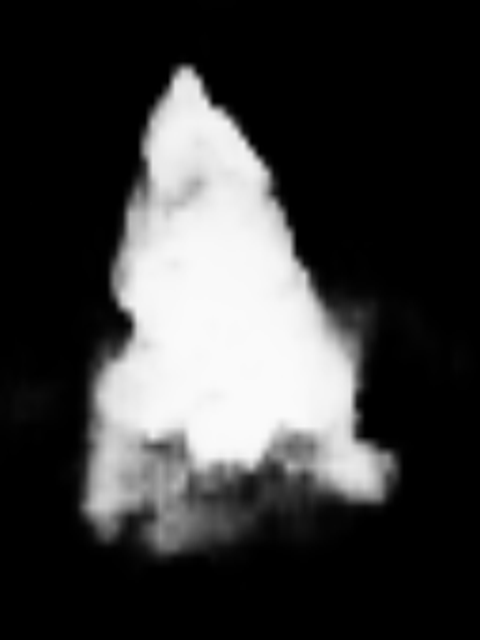}
	\end{subfigure}
	\begin{subfigure}{0.065\textwidth}
		\includegraphics[width=\textwidth]{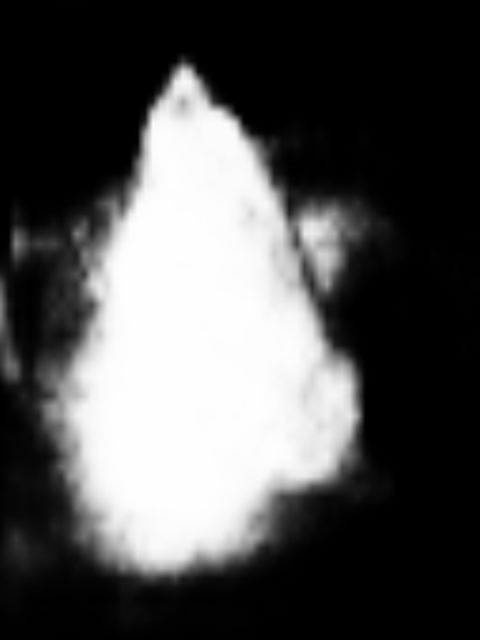}
	\end{subfigure}
	\begin{subfigure}{0.065\textwidth}
		\includegraphics[width=\textwidth]{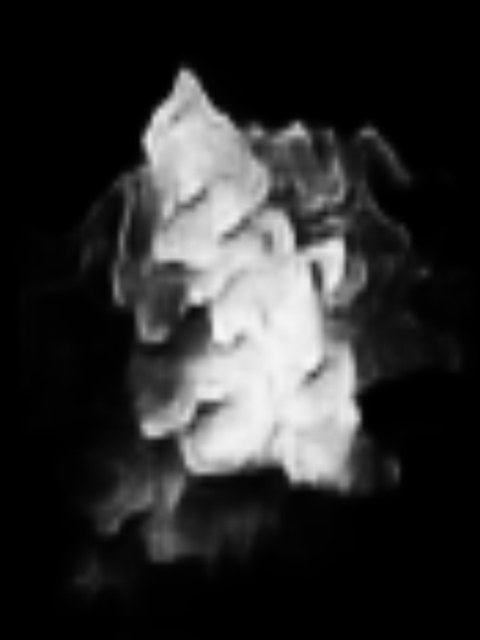}
	\end{subfigure}
	\begin{subfigure}{0.065\textwidth}
		\includegraphics[width=\textwidth]{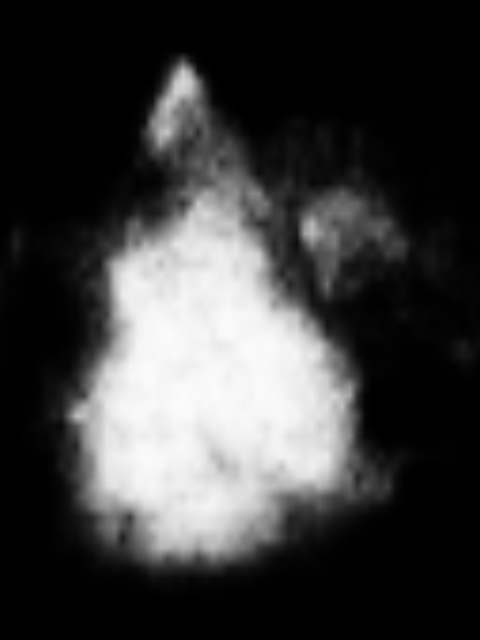}
	\end{subfigure}
	\begin{subfigure}{0.065\textwidth}
		\includegraphics[width=\textwidth]{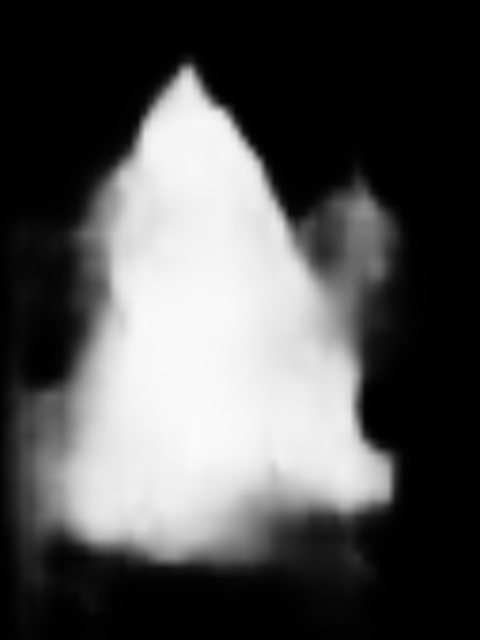}
	\end{subfigure}
	\begin{subfigure}{0.065\textwidth}
		\includegraphics[width=\textwidth]{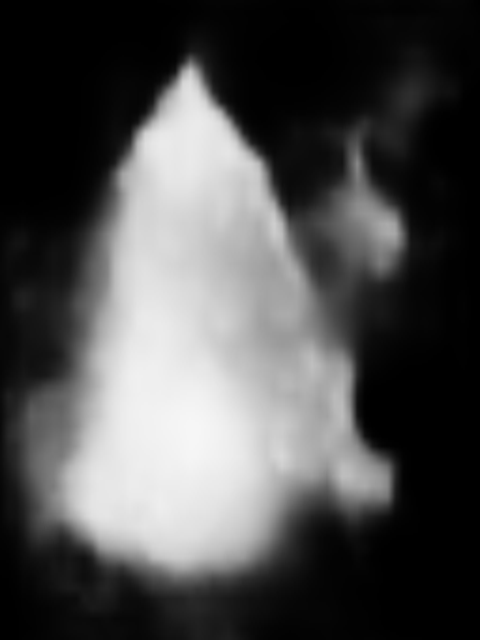}
	\end{subfigure}
	\ \\
	\vspace*{0.5mm}
	\begin{subfigure}{0.065\textwidth} 
		\includegraphics[width=\textwidth]{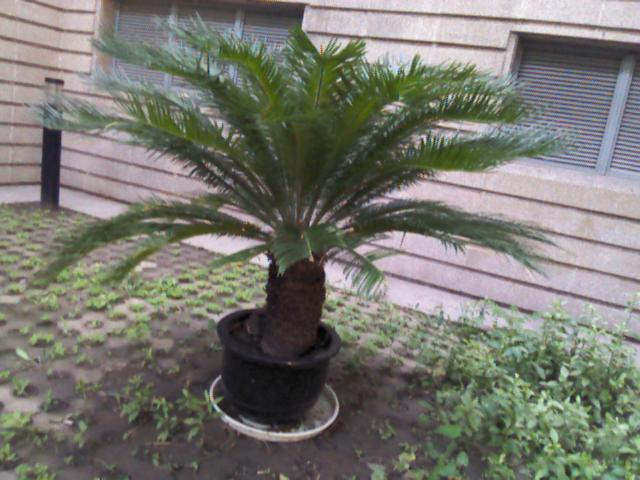}
    \end{subfigure}
	\begin{subfigure}{0.065\textwidth} 
		\includegraphics[width=\textwidth]{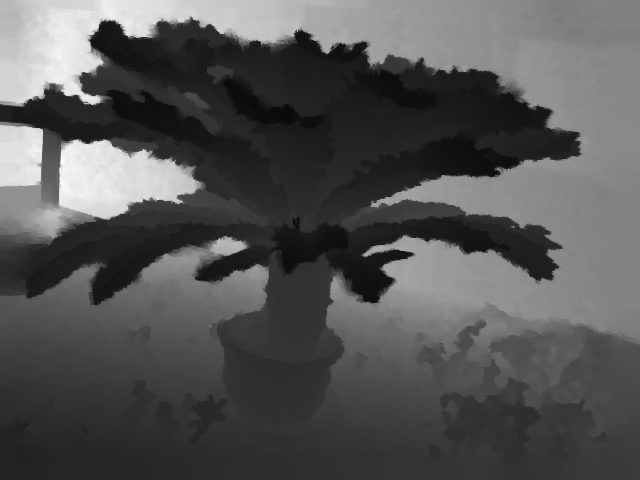}
	\end{subfigure}
	\begin{subfigure}{0.065\textwidth}
		\includegraphics[width=\textwidth]{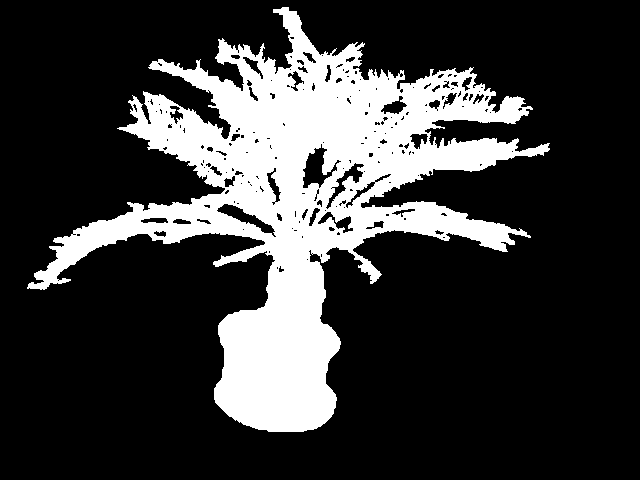}
	\end{subfigure}
    \begin{subfigure}{0.065\textwidth}
		\includegraphics[width=\textwidth]{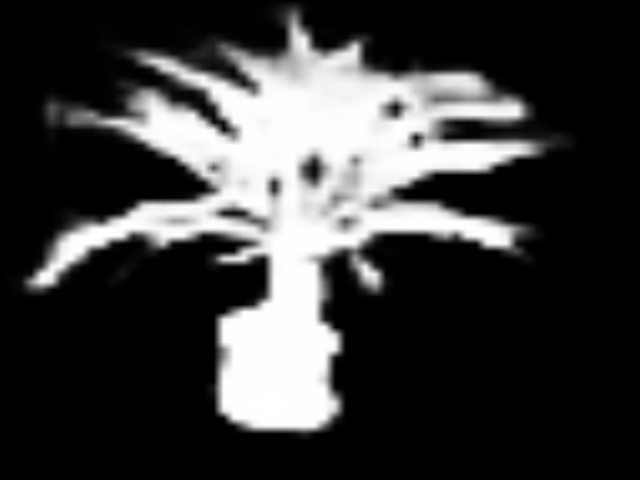}
	\end{subfigure}
    \begin{subfigure}{0.065\textwidth}
		\includegraphics[width=\textwidth]{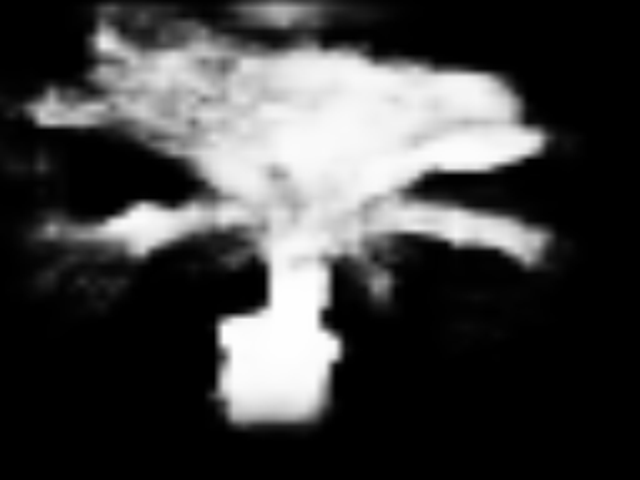}
	\end{subfigure}
    \begin{subfigure}{0.065\textwidth}
		\includegraphics[width=\textwidth]{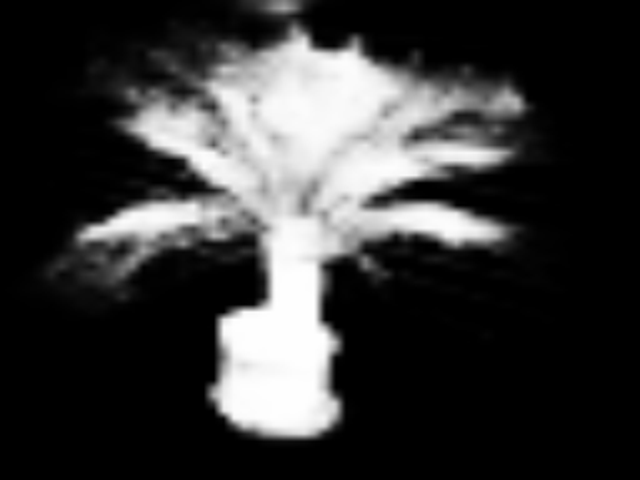}
	\end{subfigure}
	\begin{subfigure}{0.065\textwidth}
		\includegraphics[width=\textwidth]{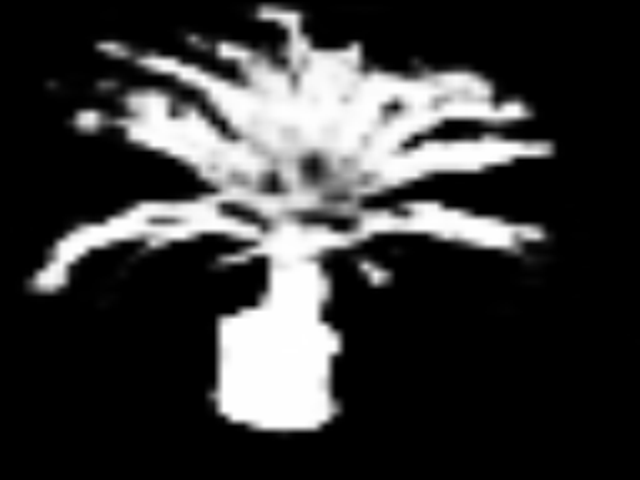}
	\end{subfigure}
	\begin{subfigure}{0.065\textwidth}
		\includegraphics[width=\textwidth]{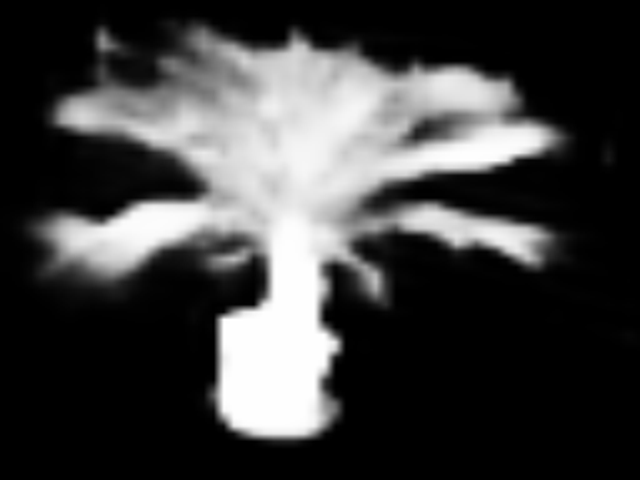}
	\end{subfigure}
    \begin{subfigure}{0.065\textwidth}
		\includegraphics[width=\textwidth]{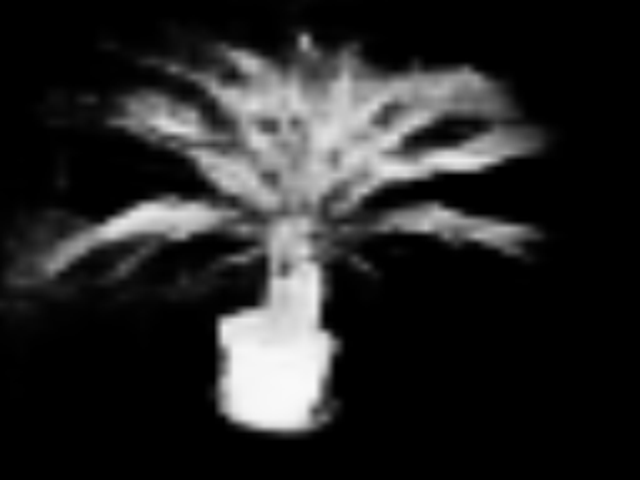}
	\end{subfigure}
	\begin{subfigure}{0.065\textwidth}
		\includegraphics[width=\textwidth]{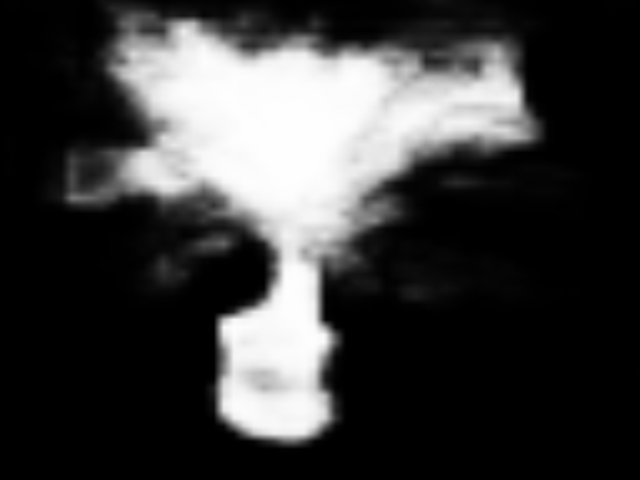}
	\end{subfigure}
	\begin{subfigure}{0.065\textwidth}
		\includegraphics[width=\textwidth]{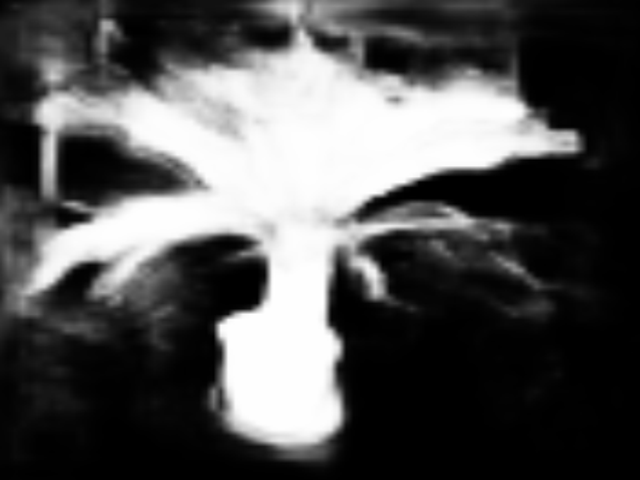}
	\end{subfigure}
	\begin{subfigure}{0.065\textwidth}
		\includegraphics[width=\textwidth]{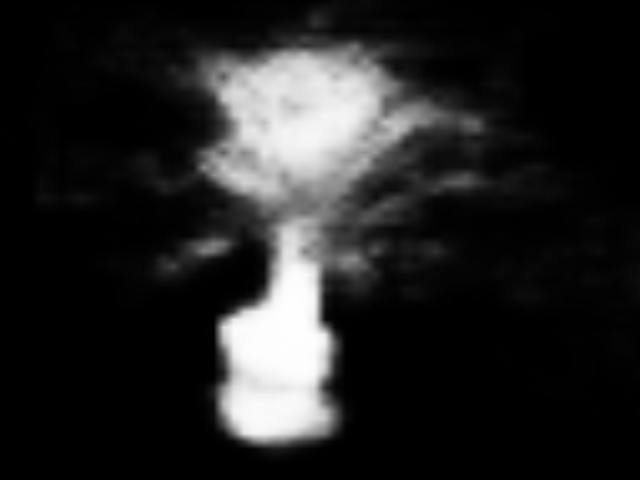}
	\end{subfigure}
	\begin{subfigure}{0.065\textwidth}
		\includegraphics[width=\textwidth]{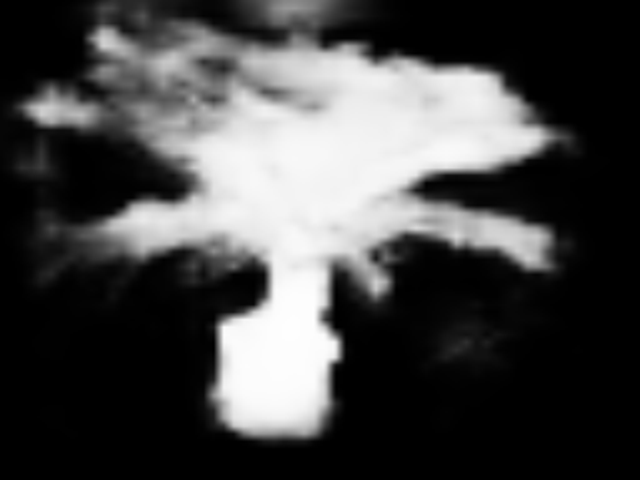}
	\end{subfigure}
	\begin{subfigure}{0.065\textwidth}
		\includegraphics[width=\textwidth]{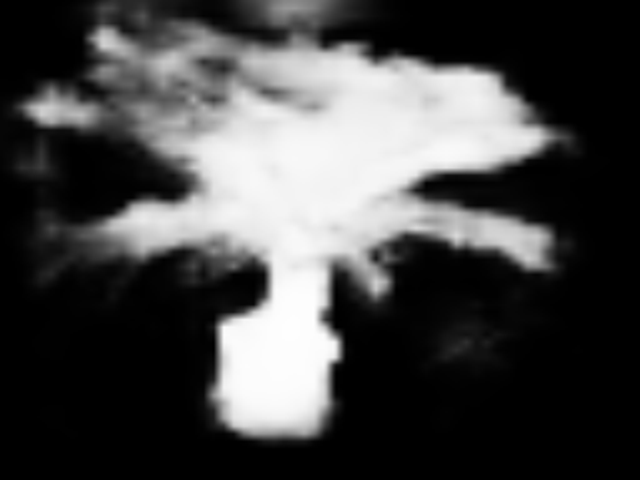}
	\end{subfigure}
	\ \\
	\vspace*{0.5mm}
	\begin{subfigure}{0.065\textwidth} 
		\includegraphics[width=\textwidth]{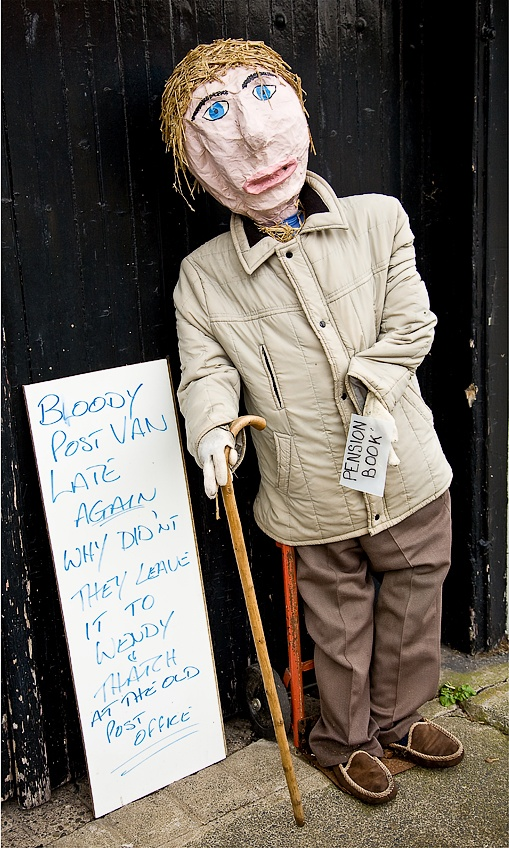}
		\vspace{-5.5mm} \caption{RGB}
    \end{subfigure}
	\begin{subfigure}{0.065\textwidth} 
		\includegraphics[width=\textwidth]{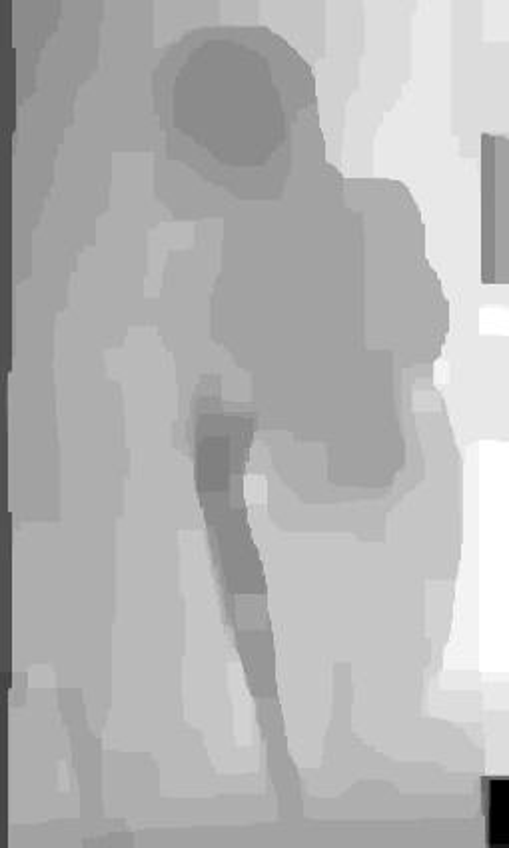}
		\vspace{-5.5mm} \caption{depth}
	\end{subfigure}
	\begin{subfigure}{0.065\textwidth}
		\includegraphics[width=\textwidth]{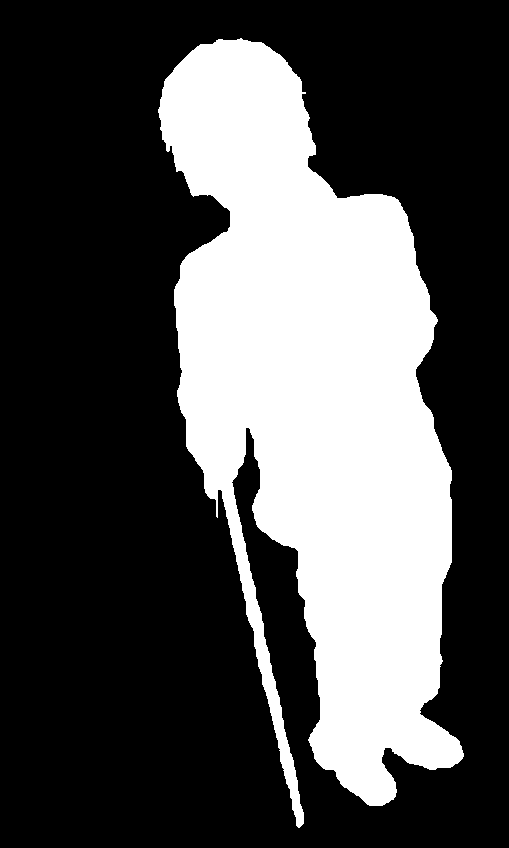}
		\vspace{-5.5mm} \caption{GT}
	\end{subfigure}
    \begin{subfigure}{0.065\textwidth}
		\includegraphics[width=\textwidth]{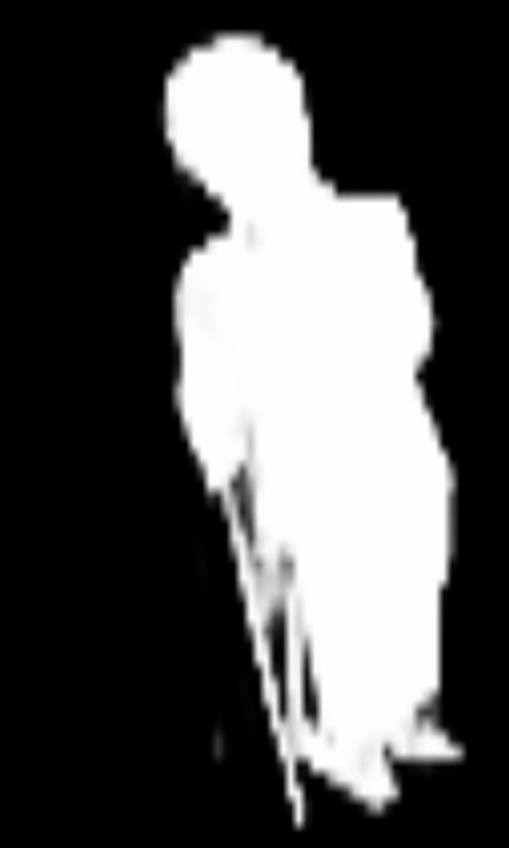}
		\vspace{-5.5mm} \caption{ours}
	\end{subfigure}
    \begin{subfigure}{0.065\textwidth}
		\includegraphics[width=\textwidth]{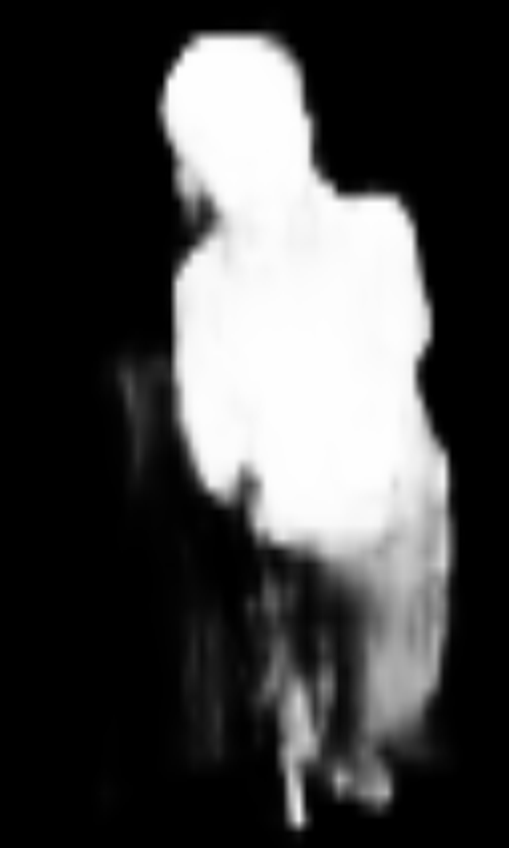}
		\vspace{-5.5mm} \caption{$M_{10}$}
	\end{subfigure}
    \begin{subfigure}{0.065\textwidth}
		\includegraphics[width=\textwidth]{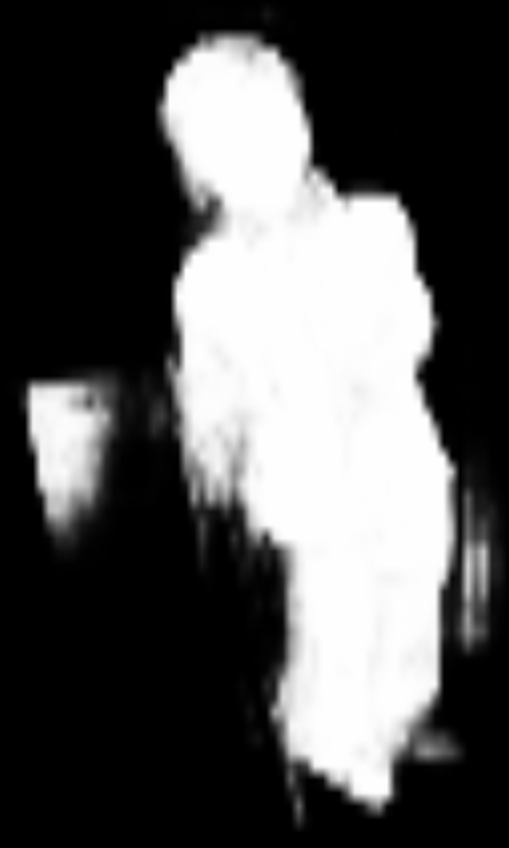}
		\vspace{-5.5mm} \caption{$M_9$}
	\end{subfigure}
	\begin{subfigure}{0.065\textwidth}
		\includegraphics[width=\textwidth]{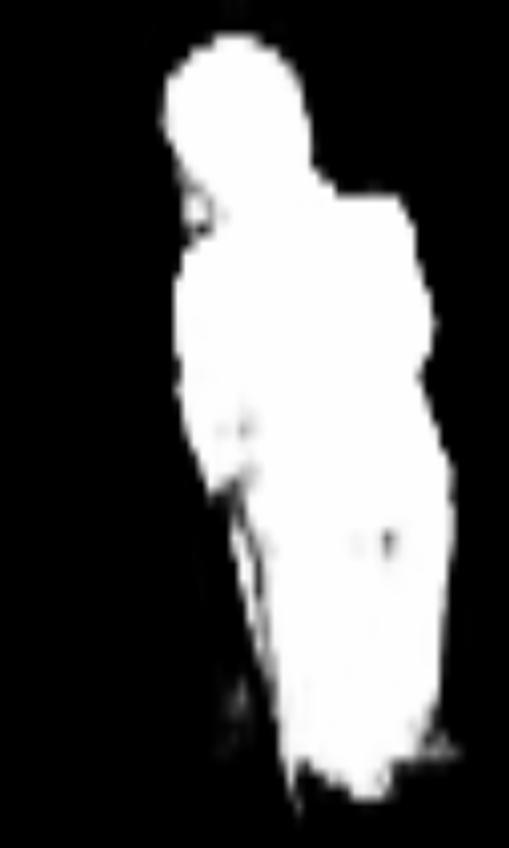}
		\vspace{-5.5mm} \caption{$M_8$}
	\end{subfigure}
	\begin{subfigure}{0.065\textwidth}
		\includegraphics[width=\textwidth]{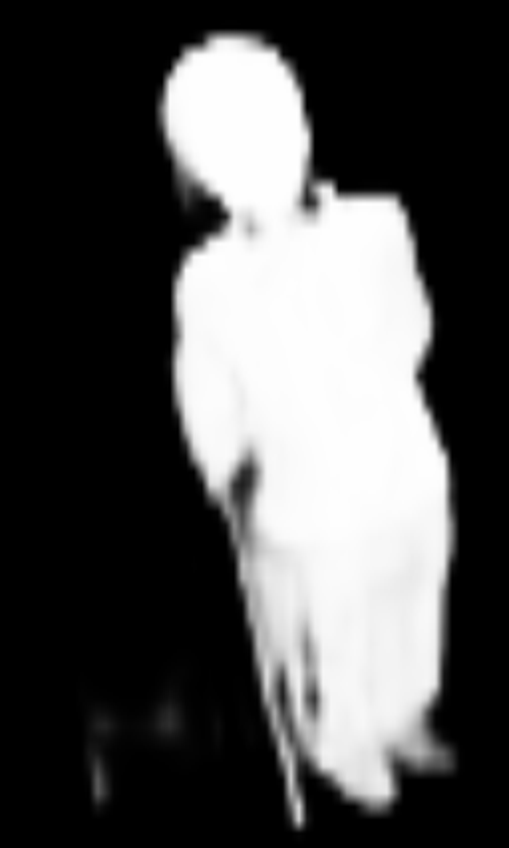}
		\vspace{-5.5mm} \caption{$M_7$}
	\end{subfigure}
    \begin{subfigure}{0.065\textwidth}
		\includegraphics[width=\textwidth]{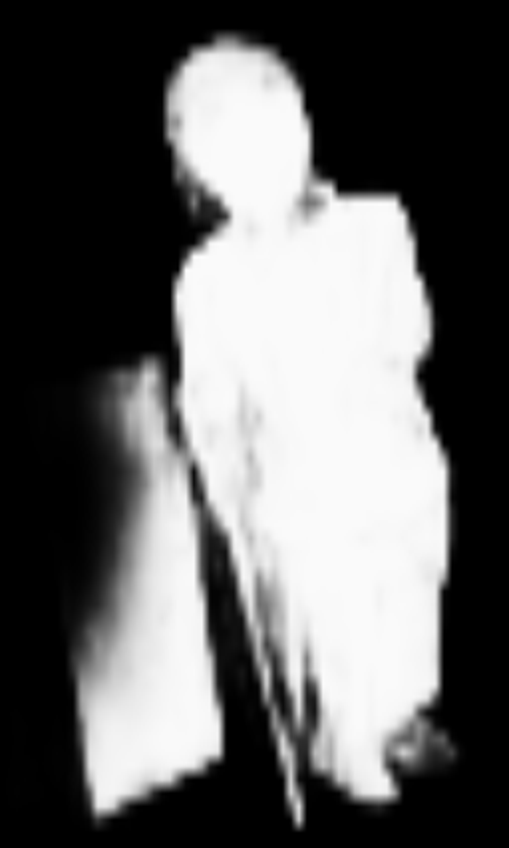}
		\vspace{-5.5mm} \caption{$M_6$}
	\end{subfigure}
	\begin{subfigure}{0.065\textwidth}
		\includegraphics[width=\textwidth]{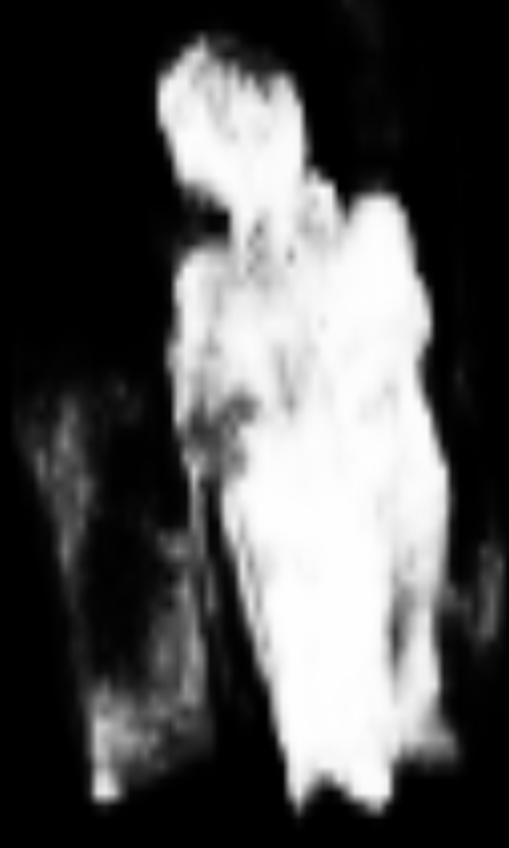}
		\vspace{-5.5mm} \caption{$M_5$}
	\end{subfigure}
	\begin{subfigure}{0.065\textwidth}
		\includegraphics[width=\textwidth]{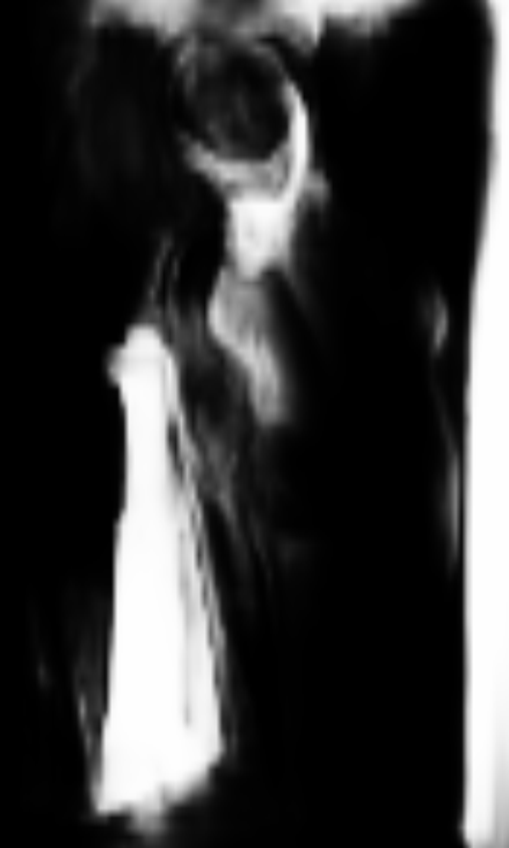}
		\vspace{-5.5mm} \caption{$M_4$}
	\end{subfigure}
	\begin{subfigure}{0.065\textwidth}
		\includegraphics[width=\textwidth]{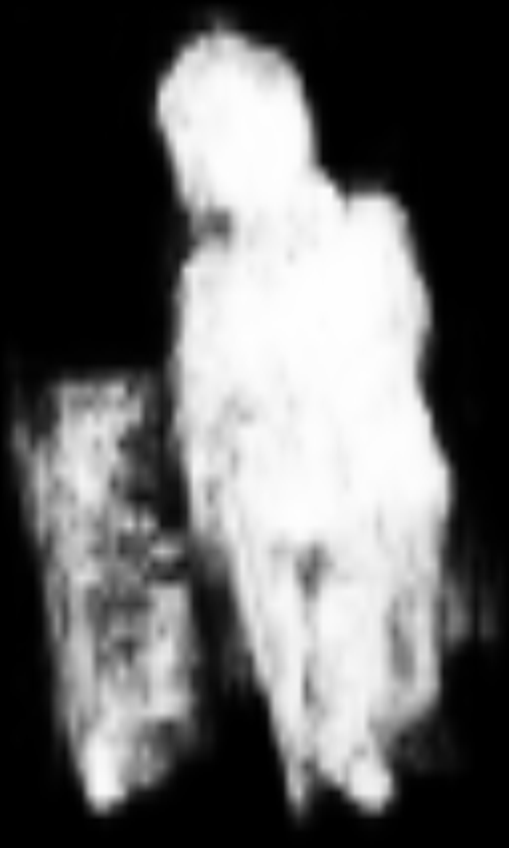}
		\vspace{-5.5mm} \caption{$M_3$}
	\end{subfigure}
	\begin{subfigure}{0.065\textwidth}
		\includegraphics[width=\textwidth]{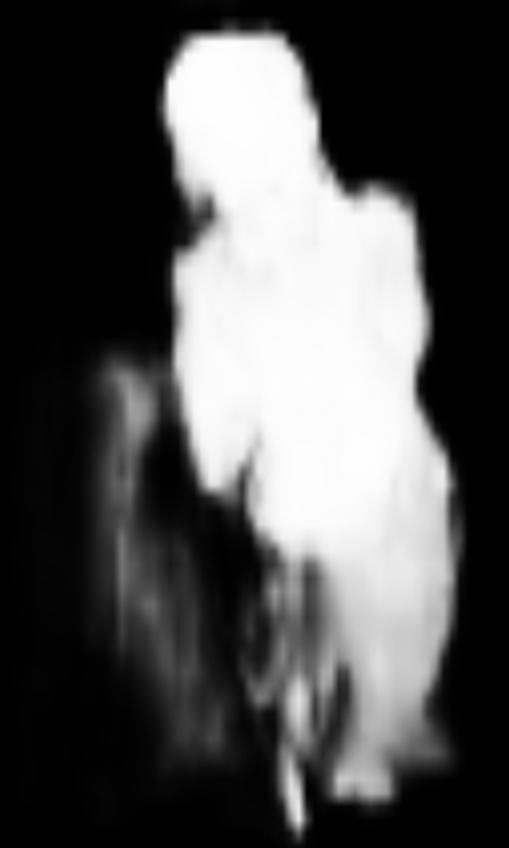}
		\vspace{-5.5mm} \caption{$M_2$}
	\end{subfigure}
	\begin{subfigure}{0.065\textwidth}
		\includegraphics[width=\textwidth]{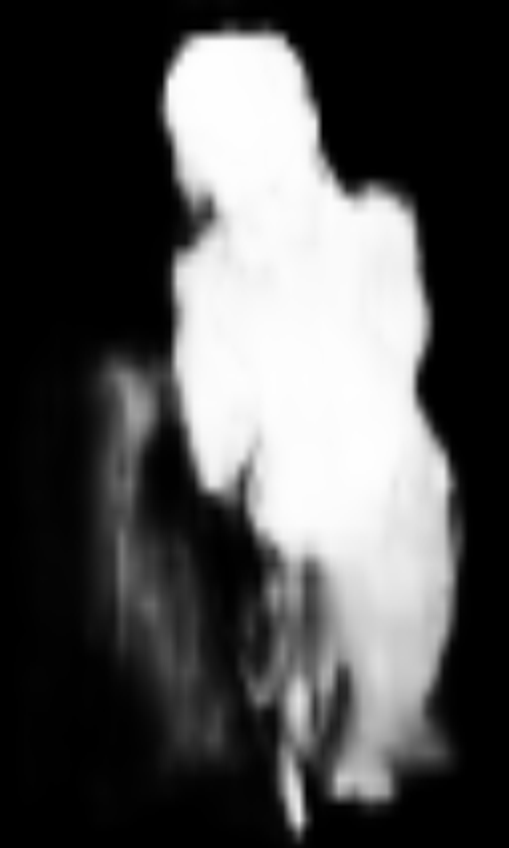}
		\vspace{-5.5mm} \caption{$M_1$}
	\end{subfigure}
	\ \\
	\vspace{-1.0mm}
	\caption{Visual comparison of saliency map results produced by different methods.
            (a) Input RGB image from benchmark datasets;
            (b) Input depth image;
            (c) Ground truths (denoted as 'GT');
            (d)-(n) Saliency maps predicted by our method and ten constructed baseline networks (i.e., $M_{10}$ to $M_1$). Please refer to Section~\ref{subsec:ablation-ayalysis} for the explanation of $M_1$ to $M_{10}$).}
	\label{fig:ablation}
	\vspace{-2.5mm}
\end{figure*}

\vspace{2mm}
\noindent
\textbf{Benchmark Datasets.} We employ seven widely-used benchmark datasets to evaluate our network and state-of-the-art RGB-D saliency detectors. They are
(i) NJU2K~\cite{ju2014depth} (2,000 images),
(ii) NLPR~\cite{peng2014rgbd} (1,000 images),
(iii) STERE~\cite{niu2012leveraging}  (1,000 images),
(iv) RGBD135~\cite{cheng2014depth}  (135 images),
(v)  LFSD~\cite{li2014saliency} (100 images),
(vi) SIP~\cite{fan2020rethinking} (929 images),
(vii) DUTD~\cite{piao2019depth} (1,200 images).
On the DUTD, we follow the setting of~\cite{piao2019depth} and use 800 images for training and 400 images for testing. For the other datasets, we follow recent works~\cite{zhao2019contrast,fu2020jl,fan2020rethinking} to utilize the same 1,500 images from NJU2K and the same $700$ images from NLPR as labeled RGB-D images to train our network for fair comparisons.
For unlabeled RGB images, we utilize 10,553 RGB images from the training set of DUTS~\cite{wang2017learning}.
Although each training image of DUTS has the annotations of the saliency map, we do not use any saliency information of these images when training our dual-semi RGB-D saliency detector.

\vspace{2mm}
\noindent
\textbf{Evaluation metrics.} We adopt four widely-used metrics to quantitatively compare RGB-D saliency detection performance, including S-measure ($S_m$)~\cite{fan2017structure}, F-measure ($F_\beta^{max}$)~\cite{lang2016dual}, E-measure ($E_\phi^{max}$)~\cite{fan2018enhanced}, and Mean Absolute Error ($MAE$)~\cite{perazzi2012saliency}.
In general, a more accurate RGB-D saliency detector shall have a larger $S_m$, a larger $F_\beta^{max}$,  a larger $E_\phi^{max}$, and a smaller $MAE$.

\vspace{1mm}
\noindent
\textbf{Implementation Details.} During training, the backbone is initialize by popular backbones, such as VGG-16~\cite{simonyan2014very}, VGG-19~\cite{simonyan2014very}, ResNet-50~\cite{he2016deep} and HRNet~\cite{sun2019high}, which has been well-trained for the image classification task on the ImageNet~\cite{deng2009imagenet} and other layers are randomly initialized.  Training data is resized to $256 \times 256$ and augmented by random rotation and horizontal flipping. In addition, color jittering is used for the perturbation of unlabeled data. Stochastic gradient descent (SGD) with a momentum of $0.9$ and a weight decay of $0.0005$ is used to optimize the whole network. The learning rate is adjusted by a poly strategy~\cite{liu2015parsenet} with an initial learning rate of $0.001$ and the power of $0.9$. The whole training takes $25$ hours with training batch size $8$ (\ie, $4$ labeled pairs and $4$ unlabeled RGB images) and a maximum iteration of $20,000$ on a single NVIDIA GTX 2080Ti GPU.

\subsection{Comparison with Baselines}
\label{subsec:compare_state_of_the_art}

We evaluate the effectiveness of our network by comparing it against $23$ state-of-the-art RGB-D salient object detectors.
They are LBE~\cite{feng2016local}, DF~\cite{qu2017rgbd}, CTMF~\cite{han2017cnns}, PCF~\cite{chen2018progressively}, 
TANet~\cite{chen2019three}, CPFP~\cite{zhao2019contrast}, DMRA~\cite{piao2019depth}, D$^3$Net~\cite{fan2020rethinking}, SSF~\cite{zhang2020select}, UCNet~\cite{zhang2020uc}, JLDCF~\cite{fu2020jl}, \textcolor{dgreen}{JLDCF[J]~\cite{fu2021siamese}},
HDF-Net~\cite{pang2020hierarchical}, ATSA~\cite{zhang2020asymmetric}, \textcolor{dgreen}{SSDP~\cite{wang2020synergistic}}, \textcolor{dgreen}{DSA$^2$F~\cite{sun2021deep}}, PGA-Net~\cite{chen2020progressively}, DANet~\cite{zhao2020single}, cmMS~\cite{li2020rgb},
Cas-Gnn~\cite{luo2020cascade},   CMWNet~\cite{li2020cross}, CoNet~\cite{ji2020accurate}, and BBS-Net~\cite{fan2020bbs}.
Note that LBE~\cite{feng2016local} utilizes handcrafted features to infer salient objects, while other $22$ methods employ different deep networks to learn discriminative features for RGB-D saliency detection.
To make the comparisons fair, we obtained the saliency maps of all baselines either from the authors or by using their released training models and parameters.

Table~\ref{table:state-of-the-art-part1} reports the $S_m$, $F_\beta^{max}$, $E_\phi^{max}$, and $MAE$ values of our method and baselines on all seven benchmark datasets.
From the results, we find that DDCNN, \ie, our network with only labeled data, also outperforms other compared RGB-D saliency detectors in terms of four evaluation metrics.
By considering unlabeled RGB images, our DS-Net further improves the performance over other saliency detectors on almost all of the seven datasets.
This indicates that our network can more accurately identify salient objects from the input pair of RGB-D data than the compared detectors.
Besides, we also explore the efficiency of semi-supervised learning without any extra RGB images. Specifically, we construct an additional experiment by splitting $3/10$ labeled RGB-D data as the unlabeled RGB images (denoted as ``DDCNN-semi-ourSplit''). As shown in Table~\ref{table:state-of-the-art-part1}, ``DDCNN-semi-ourSplit'' still outperforms DMRA~\cite{piao2019depth}, but fails to suppress the most recent state-of-the-art supervised methods (\ie, BBS-Net~\cite{fan2020bbs} and CoNet~\cite{ji2020accurate}) \textcolor{dgreen}{due to the use of only} $7/10$ labels of RGB-D datasets.
Compared to the best-performing results, our network achieves a $S_m$ improvement of $1.59\%$, a $F_\beta^{max}$ improvement of $1.63\%$, a $E_\phi^{max}$ improvement
of $0.66\%$, and a $MAE$ decrease of $9.89\%$ on the average of seven benchmarks.
 
Fig.~\ref{fig:comparison_real_photos_part1} visually compares the saliency maps produced by our network and the state-of-the-art RGB-D saliency methods.
By observing the different saliency maps, we can conclude that other compared methods in Fig.~\ref{fig:comparison_real_photos_part1} (e)-(m) tend to include non-salient backgrounds or lose salient details in their predicted saliency maps, whereas our DS-Net better detects salient objects from input RGB-D image pairs and our results are more consistent with the ground truths (see Fig.~\ref{fig:comparison_real_photos_part1} (c)).
This indicates that, by leveraging only unlabeled RGB images, our network can suppress non-salient objects and detect more salient pixels than the state-of-the-art RGB-D saliency detectors, which are mainly trained in a supervised manner.

\subsection{Ablation Analysis}
\label{subsec:ablation-ayalysis}

\vspace*{2mm}
\noindent
\textbf{Baseline Network.} We perform ablation experiments to evaluate the effectiveness of the major components in our DS-Net.
We consider ten baselines.
The first seven employ supervised learning for RGB-D saliency detection, using DDCNN as the backbone, without the teacher network.
Specifically, we firstly verify the contributions of fusion between RGB and depth features. ``DDCNN-w/o-DAM''; ``$M_1$'', ``DDCNN-w/o-DGM''; ``$M_2$'' and ``DDCNN-w/o-DIM''; ``$M_3$'' replace DAM, DGM and DIM with a simple feature concatenation operation, respectively.
The fourth baseline (denoted as ``DDCNN-w/o-depth''; ``$M_4$'') removes the depth estimation branch from our DDCNN and does not separate RGB features but directly fuses them with depth features via DAM modules for saliency detection.
The fifth baseline (denoted as ``DDCNN-w/o-reconstr-loss''; ``$M_5$'') removes the reconstruction loss from our DDCNN.
For the other two baselines with supervised learning, one (denoted as ``DDCNN-w/o-pretrain''; ``$M_6$'') directly trains DDCNN on labeled RGB-D data in an end-to-end manner, while the other (``$M_7$'') is our full DDCNN.

The first semi-supervised baseline (denoted as ``DDCNN-semi''; ``$M_8$'') is constructed by directly extending a supervised two-task DDCNN to a semi-supervised learning framework by computing consistency loss on saliency predictions for unlabeled RGB images with their associated pseudo depth maps.
In other words, ``DDCNN-semi'' is equal to removing the consistency loss on the attention maps of our DS-Net.
Furthermore,
we experiment on only labeled data to explore the upper-bound performance of semi-supervised RGB-D saliency detection, where 3/10 labeled RGB-D data are split as the unlabeled RGB images. (denoted as ``DDCNN-semi-ourSplit'; ``$M_9$'').
The last baseline (denoted as ``DDCNN-semi-depthEst''; ``$M_{10}$'') is designed to analyze the quality of estimated depth maps and the contribution of multi-task joint learning. ``$M_{10}$'' firstly utilizes an independent depth estimation model~\cite{lee2018single} to obtain paired RGB-D data for unlabeled RGB images, then trains the DS-Net in a semi-supervised manner.
{\color{dgreen}
Among all the baseline networks, $M_1$ to $M_7$ are trained in a normal supervised manner with labeled RGB-D data. $M_9$ is also trained on labeled RGB-D data only but adopts fewer samples than supervised learning baselines. At last, $M_8$, $M_{10}$ and our DS-Net employ both labeled RGB-D data and unlabeled DUTS images.
}

\vspace*{2mm}
\noindent
\textbf{Quantitative Comparisons.} Table~\ref{tab:ablation1} lists the results of our network and ten baseline networks (i.e., $M_1$ to $M_{10}$) on six benchmark datasets.
%
From the results, we have the following observations:
(i) $M_7$ outperforms $M_1$, $M_2$ and $M_3$  in a large margin, showing that DAM, DGM and DIM have their contributions to the superior performance of our network.
(ii) $M_7$ achieves larger $S_m$ and smaller $MAE$ scores than $M_4$, demonstrating that detecting additional depth maps enables DDCNN to better detect saliency maps from RGB-D image pairs.
(iii) The superior results of $M_7$ over $M_5$ on two evaluation metrics indicates that DDCNN has a more accurate RGB-D saliency detection performance when computing a reconstruction loss on \emph{depth-aware} RGB features and \emph{depth-dispelled} RGB features since the reconstruction loss better separates these two kinds of RGB features. 
(iv) $M_7$ clearly outperforms $M_6$ in terms of $S_m$ and $MAE$ on six benchmark datasets, showing that DDCNN can better identify salient objects by using a pre-trained depth estimation branch for network initialization.
(v) $M_8$ has better $S_m$ and $MAE$ results than $M_7$ on all six benchmark datasets.
This shows that utilizing unlabeled RGB-D image pairs makes the RGB-D saliency detection of DDCNN more accurate, even though their depth maps of input are not the ground truths but are estimated by our DDCNN.
(vi) Although $M_9$ fails to outperform both $M_7$ and $M_8$, it still wins a supervised method DMRA~\cite{piao2019depth}. This result demonstrates that semi-supervised learning improves the capacity of supervised RGB-D saliency detectors and makes a main contribution to our final model.
(vii) Compared to $M_8$, $M_{10}$ degrades performance, quantitatively verifying that our depth estimation branch can produce better depth maps than an independent model. In addition, $M_{10}$ has a competitive result with respect to $M_7$, \textcolor{dgreen}{indirectly proving} that our DIM can merge depth and RGB features adaptively and discard the non-reliable depth pixels with the assistance of the attention mechanism.
(viii) Our DS-Net can more accurately detect salient regions than $M_8$, as indicated by its superior results in two evaluation metrics on six benchmarks.
This further shows that the consistency loss on attention maps contributes to the superior RGB-D saliency detection performance of our network.

\vspace*{2mm}
\noindent
\textbf{Visual comparisons.}
Fig.~\ref{fig:ablation} visually compares the saliency maps predicted by our method and ten baseline networks. Among the results, our method can achieve the most similar saliency maps to the ground truth saliency, demonstrating the effectiveness of our DDCNN design and incorporating unlabeled RGB images into RGB-D saliency detection.

{\color{dgreen}
\subsection{Discussion}

\begin{table*}[!t]
	\setlength\tabcolsep{2pt}
	\caption{Quantitative results for depth estimation on NJU2K and NLPR datasets. }
	\label{tab:depth-comparisons}
    \resizebox{1.0\textwidth}{!}{
        \begin{tabular}{c|cccc|cccc|cccc}
        \toprule
        \multirow{2}{*}{\textbf{Datasets}}& \multicolumn{4}{c|}{\textbf{Our DDCNN}}& \multicolumn{4}{c|}{\textbf{SG-Loss [2]}}& \multicolumn{4}{c}{\textbf{DBE [3]}} \\
        & $MAE\downarrow$& $RMSE\downarrow$& $iMAE\downarrow$& $iRMSE\downarrow$& $MAE\downarrow$& $RMSE\downarrow$& $iMAE\downarrow$& $iRMSE\downarrow$& $MAE\downarrow$& $RMSE\downarrow$& $iMAE\downarrow$& $iRMSE\downarrow$ \\
        \midrule
        \midrule
        NJU2K& 253.71& 961.12& 1.02& 3.21& 249.32& 955.19& 1.01& 3.15& 288.64& 974.36& 1.55& 3.91 \\

        NLPR& 244.68& 954.34& 1.01& 3.10& 239.16& 943.66& 0.97& 3.05& 285.31& 969.78& 1.31& 3.77 \\
        
        \bottomrule
        \end{tabular}
    }
\end{table*}
\begin{figure}[!t]
	\centering
    \vspace*{0.5mm}
	\begin{subfigure}{0.085\textwidth}
	    \includegraphics[width=\textwidth]{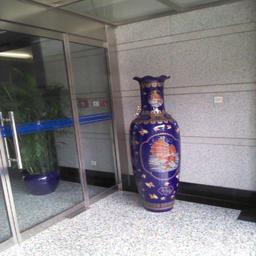}
	\end{subfigure}
	\begin{subfigure}{0.085\textwidth}
        \includegraphics[width=\textwidth]{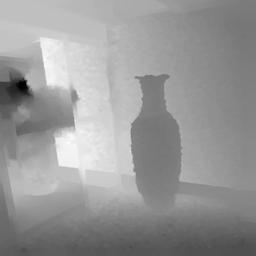}
	\end{subfigure}
    \begin{subfigure}{0.085\textwidth}
		\includegraphics[width=\textwidth]{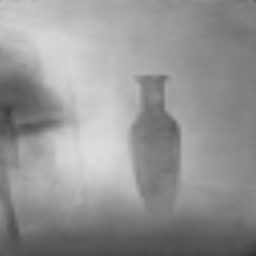}
	\end{subfigure}
    \begin{subfigure}{0.085\textwidth}
		\includegraphics[width=\textwidth]{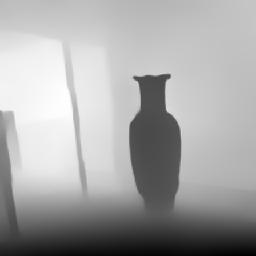}
	\end{subfigure}
    \begin{subfigure}{0.085\textwidth}
		\includegraphics[width=\textwidth]{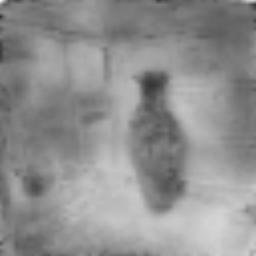}
	\end{subfigure}
	\ \\
    \vspace*{0.5mm}
	\begin{subfigure}{0.085\textwidth}
	    \includegraphics[width=\textwidth]{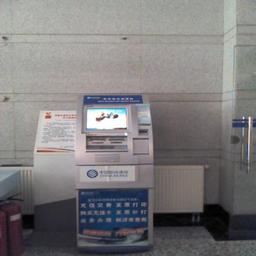}
	\end{subfigure}
	\begin{subfigure}{0.085\textwidth}
        \includegraphics[width=\textwidth]{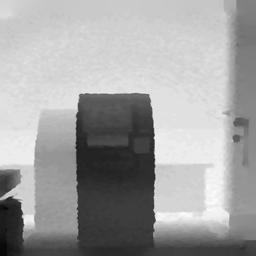}
	\end{subfigure}
    \begin{subfigure}{0.085\textwidth}
		\includegraphics[width=\textwidth]{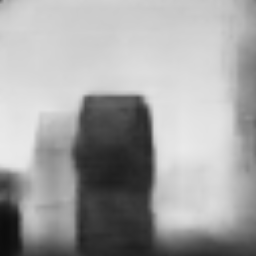}
	\end{subfigure}
    \begin{subfigure}{0.085\textwidth}
		\includegraphics[width=\textwidth]{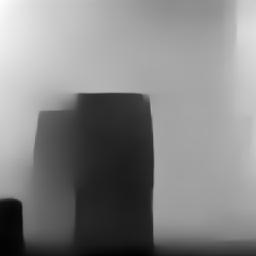}
	\end{subfigure}
    \begin{subfigure}{0.085\textwidth}
		\includegraphics[width=\textwidth]{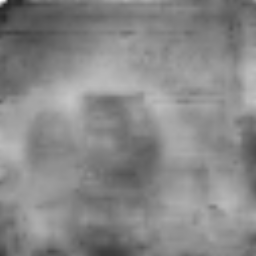}
	\end{subfigure}
	\ \\
   \vspace*{0.5mm}
   \begin{subfigure}{0.085\textwidth}
   	    \includegraphics[width=\textwidth]{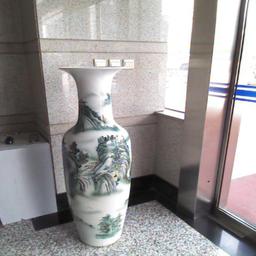}
        \vspace{-5.5mm} \caption{\footnotesize{RGB}}
   \end{subfigure}
   \begin{subfigure}{0.085\textwidth}
   	    \includegraphics[width=\textwidth]{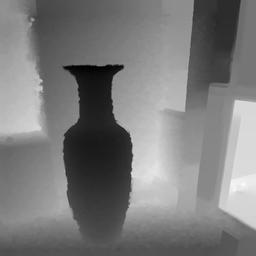}
        \vspace{-5.5mm} \caption{\footnotesize{depth}}
   \end{subfigure}
   \begin{subfigure}{0.085\textwidth}
   	    \includegraphics[width=\textwidth]{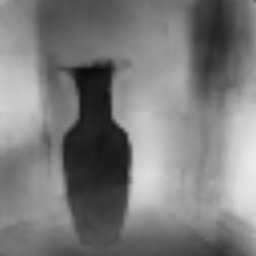}
        \vspace{-5.5mm} \caption{\footnotesize{ours}}
   \end{subfigure}
    \begin{subfigure}{0.085\textwidth}
		\includegraphics[width=\textwidth]{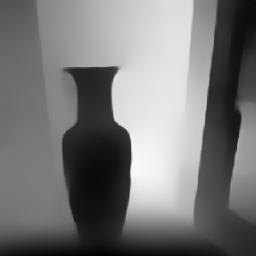}
		\vspace{-5.5mm} \caption{\footnotesize{sg-loss}}
	\end{subfigure}
    \begin{subfigure}{0.085\textwidth}
		\includegraphics[width=\textwidth]{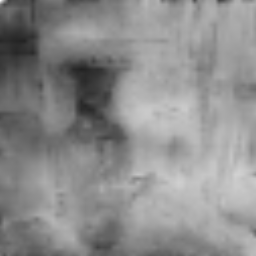}
        \vspace{-5.5mm} \caption{\footnotesize{DBE}}
	\end{subfigure}
	\caption{\textcolor{dgreen}{Qualitative results of our depth estimation branch.}
	}
	\label{fig:depth-comparisons}
	\vskip -10pt
\end{figure}

\vspace*{2mm}
\noindent
\textbf{Evaluation of the depth maps.}
The depth estimation branch in our work is to align the labeled RGB-D data and unlabeled RGB images as inputs and does not limit the choice of the specific method.
Although we only implement a small branch with pretty simple architecture, it is still essential to evaluate the quality of depth maps generated by the branch.
Specifically, we train the SG-Loss~\cite{xian2020structure} and DBE~\cite{lee2018single} with the same setting as the depth estimation branch of our DDCNN.
We adopt four widely used metrics, including Mean Average Error (MAE), Root Mean Square Error (RMSE), and their inverses.
Although only the depth maps for unlabeled DUTS images are fed into our DS-Net, they have no available paired ground truth depth maps for evaluation.
Thus, we report the results on the unseen testing set of NJU2K and NLPR benchmarks.
Quantitative and qualitative results are summarized in Table~\ref{tab:depth-comparisons} and Fig.~\ref{fig:depth-comparisons}.
From the results, although our depth estimation branch is devised with a pretty simple decoder, it still outperforms the DBE in a significant margin and achieves comparable results against the more recent depth estimation model (\ie SG-Loss).
These results show that our feature disentangling can extract depth-related features correctly and ensure the quality of pseudo depth maps for unlabeled RGB images.
What is more, the powerful feature encoder also regularizes the RGB-D saliency detection to make up for the lack of perfect depth quality.

\vspace*{2mm}
\noindent
\textbf{Effects of the hyperparameters.}
We conduct a group of experiments to select the best configurations for hyperparameters $\alpha$ in Eq.~\ref{eq:supervised_loss}, $\gamma$ in Eq.~\ref{eq:consistency_loss} and $\beta_1$, $\beta_2$ and $\lambda_{max}$ in Eq.~\ref{eq:total_loss}. Table~\ref{tab:hyperparameters-lambda} and Fig.~\ref{fig:hyperparameters} summarize the results quantitatively and qualitatively.
\begin{table*}[!t]
	\setlength\tabcolsep{2pt}
    \caption {Quantitative results of our method with different configurations of hyperparameters $\beta_1$ and $\beta_2$.}
    \label{tab:hyperparameters-lambda}	
    \resizebox{1.0\textwidth}{!}{
        \begin{tabular}{c|cccc|cccc|cccc|cccc}
        \toprule
        \multirow{2}{*}{\textbf{Datasets}}& \multicolumn{4}{c|}{\textbf{$\beta_1=1.0, \beta_2=1.0$}}& \multicolumn{4}{c|}{\textbf{$\beta_1=1.0, \beta_2=0.01$}}& \multicolumn{4}{c|}{\textbf{$\beta_1=0.01, \beta_2=1.0$}}& \multicolumn{4}{c}{\textbf{$\beta_1=0.01, \beta_2=10.0$}} \\
        & $S_m \; \uparrow$& $F_\beta^{max}\uparrow$& $E_\phi^{max}\uparrow$& $MAE\downarrow$& $S_m \; \uparrow$& $F_\beta^{max}\uparrow$& $E_\phi^{max}\uparrow$& $MAE\downarrow$& $S_m \; \uparrow$& $F_\beta^{max}\uparrow$& $E_\phi^{max}\uparrow$& $MAE\downarrow$& $S_m \; \uparrow$& $F_\beta^{max}\uparrow$& $E_\phi^{max}\uparrow$& $MAE\downarrow$ \\
        \midrule
        \midrule
        NJU2k& 0.929& 0.943& 0.946& 0.036& 0.936& 0.951& 0.952& 0.032& \textbf{0.950}& \textbf{0.965}& \textbf{0.966}& \textbf{0.024}& 0.943& 0.958& 0.959& 0.028 \\
        NLPR& 0.933& 0.932& 0.949& 0.024& 0.939& 0.938& 0.956& 0.022& \textbf{0.952}& \textbf{0.953}& \textbf{0.970}& \textbf{0.018}& 0.945& 0.946& 0.963& 0.020 \\
        STERE& 0.891& 0.892& 0.920& 0.052& 0.898& 0.900& 0.929& 0.047& \textbf{0.914}& \textbf{0.915}& \textbf{0.947}& \textbf{0.037}& 0.906& 0.907& 0.938& 0.042 \\
        RGBD135& 0.905& 0.904& 0.930& 0.033& 0.915& 0.914& 0.940& 0.029& \textbf{0.936}& \textbf{0.933}& \textbf{0.961}& \textbf{0.021}& 0.926& 0.923& 0.951& 0.025 \\
        LFSD& 0.840& 0.849& 0.864& 0.092& 0.852& 0.861& 0.878& 0.083& \textbf{0.878}& \textbf{0.885}& \textbf{0.905}& \textbf{0.064}& 0.865& 0.873& 0.891& 0.073 \\
        SIP& 0.846& 0.869& 0.889& 0.073& 0.860& 0.884& 0.904& 0.066& \textbf{0.886}& \textbf{0.915}& \textbf{0.933}& \textbf{0.051}& 0.873& 0.899& 0.919& 0.058 \\  
        
        \bottomrule
        \end{tabular}
    }
\end{table*}
\begin{figure}[t]
	\centering

	\begin{subfigure}[t]{0.49\columnwidth}
    \includegraphics[width=\textwidth]{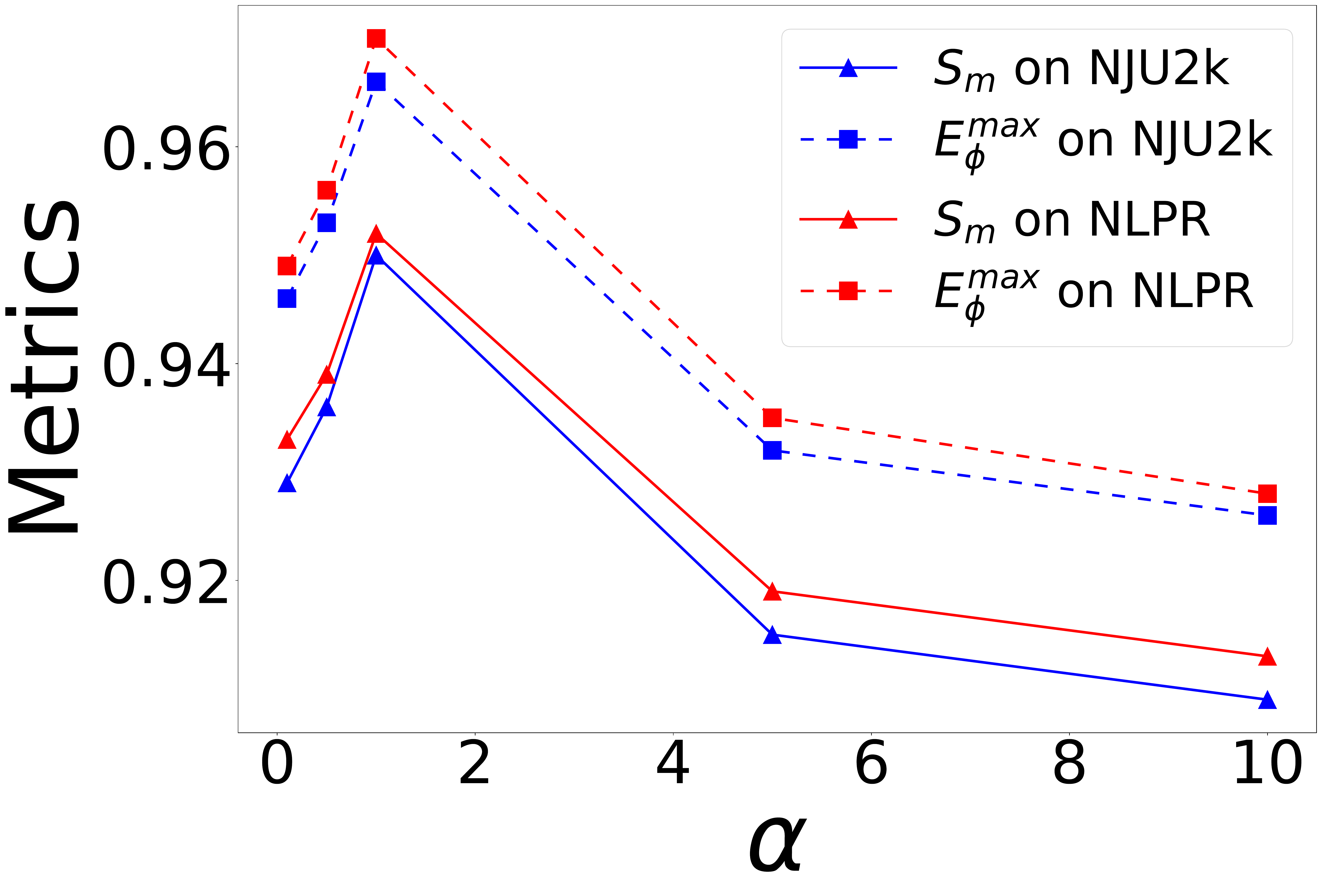}
	\end{subfigure}
	\hfill
	\begin{subfigure}[t]{0.49\columnwidth}
    \includegraphics[width=\textwidth]{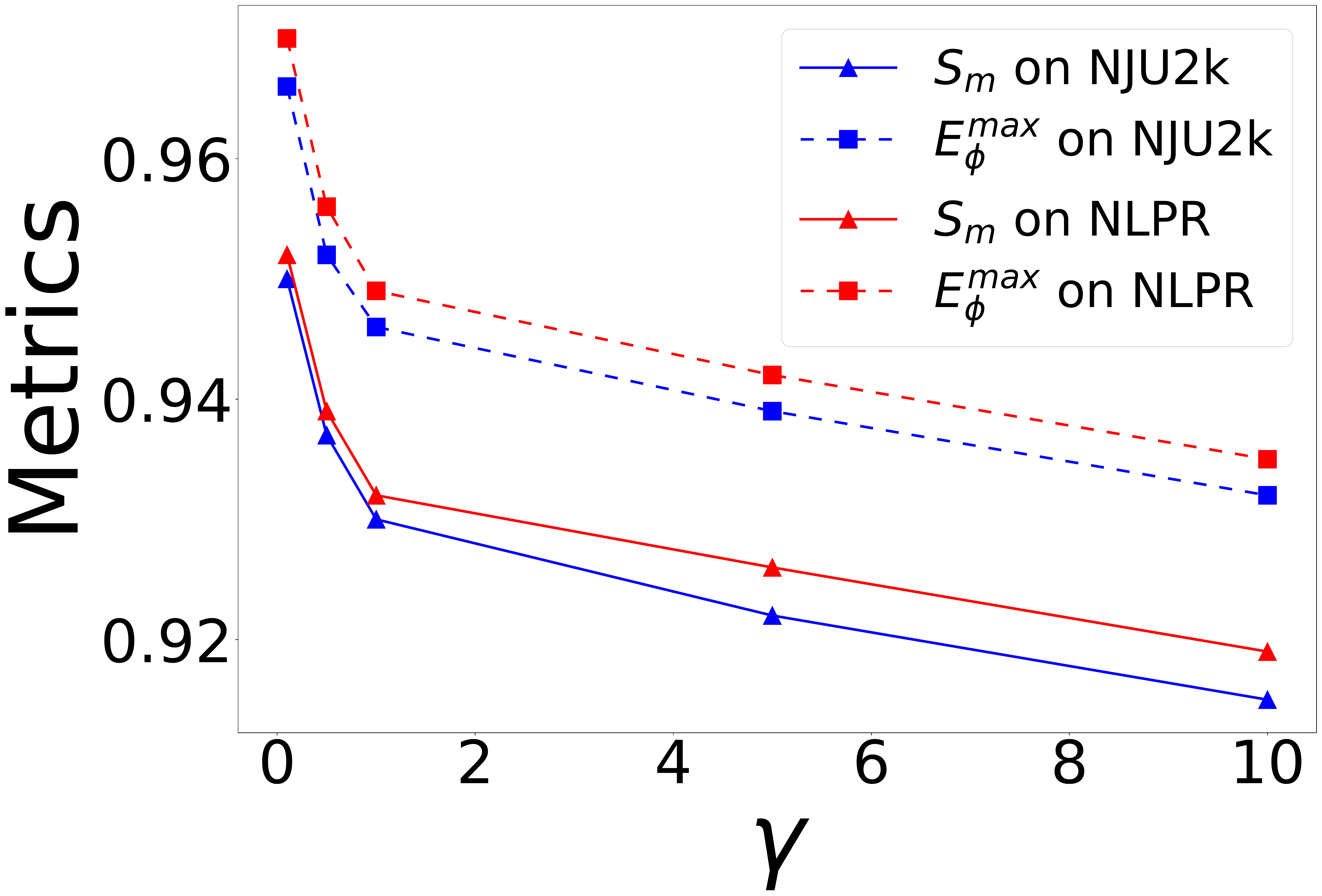}
	\end{subfigure}
	\\
	\begin{subfigure}[t]{0.49\columnwidth}
    \includegraphics[width=\textwidth]{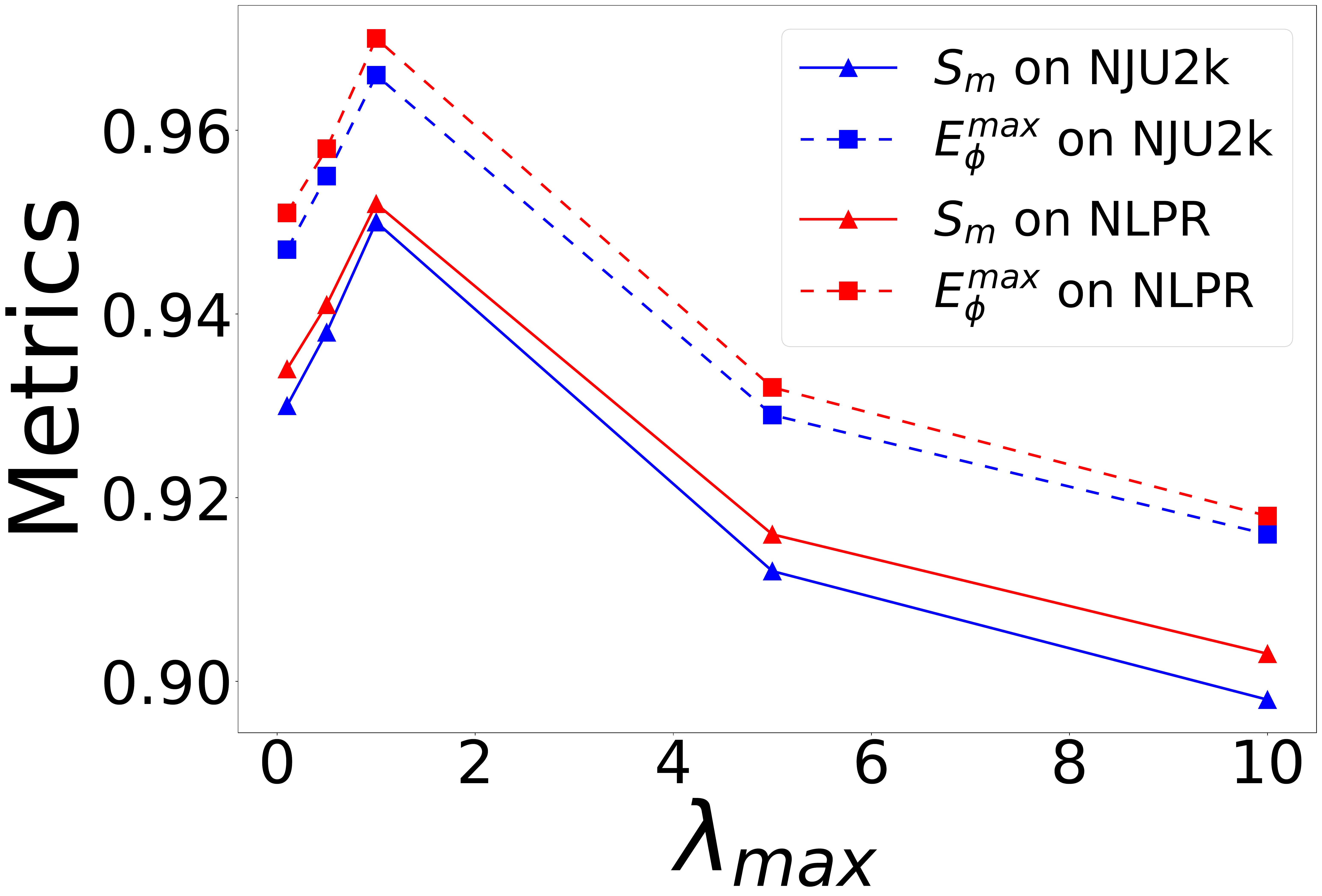}
    \label{subfig:hyperparameters-lambdamax}
	\end{subfigure}
	\hfill
	\begin{subfigure}[t]{0.49\columnwidth}
    \includegraphics[width=\textwidth]{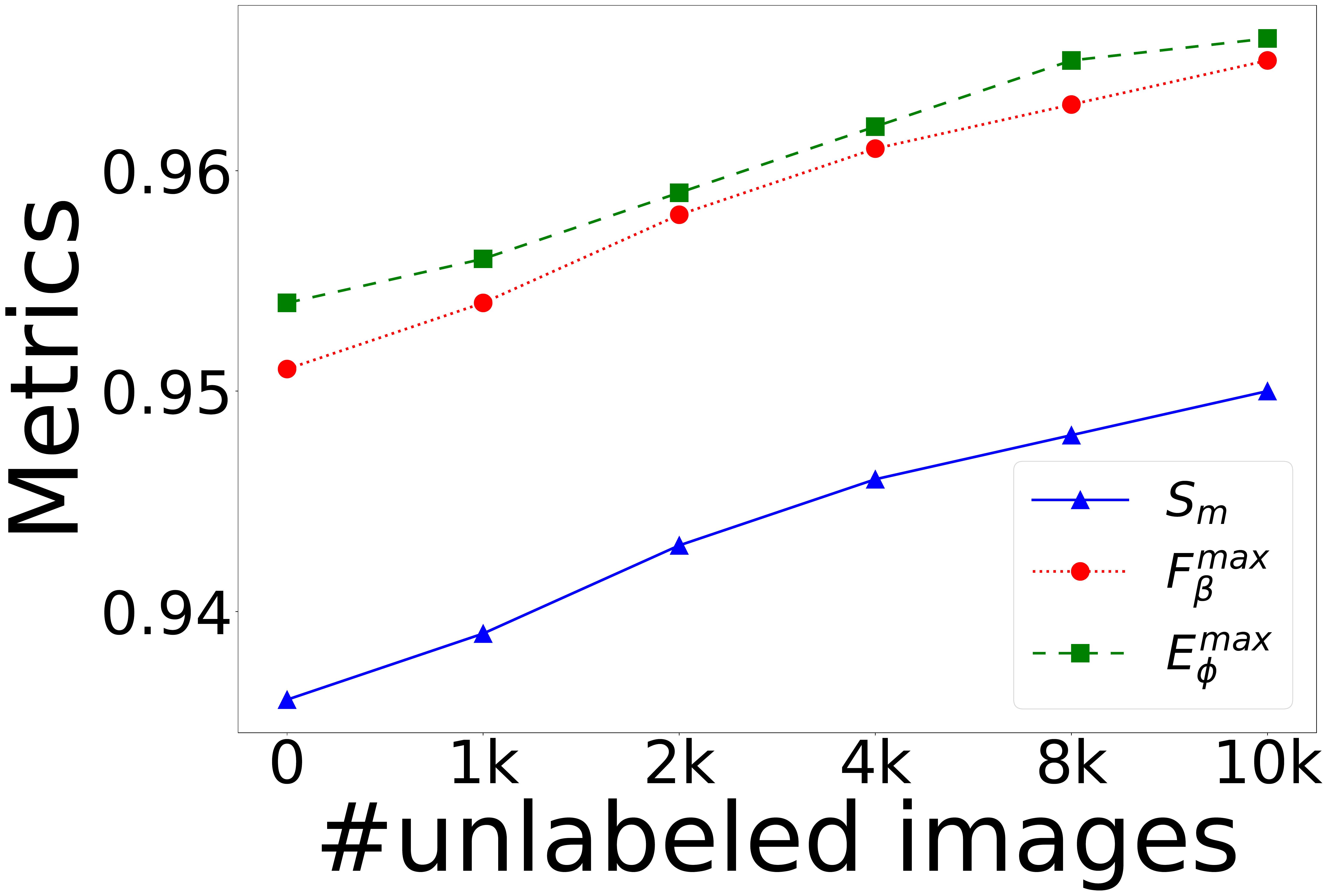}
	\end{subfigure}
    \caption{ Illustration of performance (in $S_m$ and $E_\phi^{max}$) of our model with different hyperparameter configurations and numbers of unlabeled images.}
    \label{fig:hyperparameters}
    \vspace{-5mm}
\end{figure}
\begin{table*}[!t]
	\setlength\tabcolsep{1pt}
    \caption {Quantitative results of our method and baseline networks on NJU2K~\cite{ju2014depth}, NLPR~\cite{peng2014rgbd}, STERE~\cite{niu2012leveraging}, RGBD135~\cite{cheng2014depth}, LFSD~\cite{li2014saliency}, and SIP~\cite{fan2020rethinking}.}
    \label{tab:discussions}
    \resizebox{1.0\textwidth}{!}{
    \begin{tabular}{c|cc|cc|cc|cc|cc|cc}
    \toprule
    \multirow{2}{*}{\textbf{Name}} &
    \multicolumn{2}{c|}{\textbf{NJU2K~\cite{ju2014depth}}} & \multicolumn{2}{c|}{\textbf{NLPR~\cite{peng2014rgbd}}} & \multicolumn{2}{c|}{\textbf{STERE~\cite{niu2012leveraging}}} &
    \multicolumn{2}{c|}{\textbf{RGBD135~\cite{cheng2014depth}}} & \multicolumn{2}{c|}{\textbf{LFSD~\cite{li2014saliency}}} & \multicolumn{2}{c}{\textbf{SIP~\cite{fan2020rethinking}}}  \\

    & $S_m\uparrow$ &$MAE\downarrow$
    & $S_m\uparrow$ &$MAE\downarrow$
    & $S_m\uparrow$ &$MAE\downarrow$
    & $S_m\uparrow$ &$MAE\downarrow$
    & $S_m\uparrow$ &$MAE\downarrow$
    & $S_m\uparrow$ &$MAE\downarrow$ \\
    
    \midrule
    \midrule
    
    ``DDCNN-depthDispelledOnly''& 0.870& 0.071& 0.878& 0.041& 0.825& 0.096& 0.818& 0.066& 0.731& 0.171& 0.735& 0.136 \\
    ``DDCNN-depthAwareOnly''& 0.850& 0.083& 0.858& 0.047& 0.802& 0.111& 0.788& 0.078& 0.693& 0.199& 0.696& 0.158 \\
    \textbf{our DDCNN}& \textbf{0.936}& \textbf{0.033}& \textbf{0.939}& \textbf{0.024}& \textbf{0.907}& \textbf{0.040}& \textbf{0.925}& \textbf{0.025}& \textbf{0.862}& \textbf{0.075}& \textbf{0.881}& \textbf{0.052} \\

    \hline
    
    ``DDCNN-semi-linear''& 0.867& 0.073& 0.874& 0.042& 0.821& 0.098& 0.813& 0.068& 0.724& 0.176& 0.729& 0.140 \\
    \textbf{our DDCNN-semi}& \textbf{0.941}& \textbf{0.025}& \textbf{0.942}& \textbf{0.020}& \textbf{0.911}& \textbf{0.038}& \textbf{0.934}& \textbf{0.024}& \textbf{0.869}& \textbf{0.071}& \textbf{0.883}& \textbf{0.051} \\
    
    \hline
    
    ``DS-Net-w/o-DAM''& 0.902& 0.053& 0.906& 0.032& 0.860& 0.073& 0.865& 0.048& 0.789& 0.129& 0.794& 0.103 \\
    ``DS-Net-w/o-DGM''& 0.909& 0.049& 0.913& 0.030& 0.868& 0.068& 0.875& 0.044& 0.801& 0.120& 0.807& 0.095 \\
    ``DS-Net-w/o-DIM''& 0.916& 0.045& 0.920& 0.028& 0.876& 0.063& 0.885& 0.041& 0.814& 0.110& 0.820& 0.088 \\
    ``DS-Net-linear''& 0.901& 0.053& 0.906& 0.032& 0.860& 0.073& 0.865& 0.048& 0.789& 0.129& 0.794& 0.103 \\
    \textbf{our DS-Net} & \textbf{0.950}& \textbf{0.024}& \textbf{0.952}& \textbf{0.018}& \textbf{0.914}& \textbf{0.037}& \textbf{0.936}& \textbf{0.021}& \textbf{0.878}& \textbf{0.064}& \textbf{0.886}& \textbf{0.051} \\    
    
    \bottomrule
    \end{tabular}
    }
\end{table*}

From the results, the configuration with $\alpha=1.0$, $\gamma=0.1$, $\beta_1=0.01$ and $\beta_2=1.0$ and $\lambda_{max}=1.0$ achieves the best results, which is consistent with our empirical setting obtained after a few warm-up epochs. The reason behind this is that the best configuration may be the one maintaining the best balance for the order of magnitudes during the whole training process.

\vspace*{2mm}
\noindent
\textbf{Role of the depth-dispelled features.}
In this paper, $R_i^s$ is designed to predict the RGB-D saliency. However, because of the possible non-equivalent mapping of RGB features disentangling~\cite{liu2018exploring, hsieh2018learning, ren2021interpreting}, we cannot guarantee that all the helpful information for saliency prediction is encoded into only $R_i^s$. 
Consequently, while predicting the final RGB-D saliency, we merge $R_i^s$, $R_i^d$, and corresponding depth features to utilize as much RGB information as possible.
In addition, to further verify the role of $R_i^s$, we conduct a group of experiments where RGB-D saliency is predicted by fusing $R_i^s$ (or $R_i^d$) and depth features with our DAM (or DGM) only, which are denoted as ``DDCNN-depthDispelledOnly'' and ``DDCNN-depthAwareOnly'', respectively. 
Table~\ref{tab:discussions} shows the quantitative results.
From the results, our DDCNN significantly outperforms the baselines with the depth-dispelled features or depth-aware features only. It indicates that predicting RGB-D saliency with depth-dispelled features can recover the RGB information as completely as possible and contributes to improving saliency detection accuracy.

\vspace*{2mm}
\noindent
\textbf{Role of DIM in the semi-supervised learning setting.}
To further explore how the different components work under the semi-supervised learning setting, we construct three additional baselines denoted as ``DS-Net-w/o-DAM'', ``DS-Net-w/o-DGM'', and ``DS-Net-w/o-DIM'' by replacing DAM, DGM, and DIM of our DS-Net with a simple feature concatenation operation, respectively.
Quantitative results are summarized in Table~\ref{tab:discussions}.
From the results, our DS-Net outperforms ``DS-Net-w/o-DAM'', ``DS-Net-w/o-DGM'' and ``DS-Net-w/o-DIM'' by a substantial margin, demonstrating the necessity of DAM, DGM, and DIM in the semi-supervised learning setting.
In addition, ``DS-Net-w/o-DAM'', ``DS-Net-w/o-DGM'' and ``DS-Net-w/o-DIM'' win the corresponding supervised baselines (\ie ``DDCNN-w/o-DAM ($M_1$)'', ``DDCNN-w/o-DGM ($M_2$)'' and ``DDCNN-w/o-DIM ($M_3$)'') in terms of six benchmark datasets, showing that the relatively weak baselines can also benefit from our semi-supervised framework.

\vspace*{2mm}
\noindent
\textbf{Accuracy of the attention maps.}
Attention consistency is computed at multi-scale features from different convolutional layers.
This contributes progressive and globally coherent guidance to semi-supervised learning, which ensures that each step of feature learning in our framework is relatively correct.
To further demonstrate the role of consistency loss on intermediate attention maps and evaluate their accuracy, we report our accuracy of the attention maps by using a linear evaluation protocol.
Specifically, we extract the attention maps with four scales from the different layers of ``DDCNN-semi($M_8$)'' and our DS-Net, respectively,  and train a logistic regression model to predict the final saliency with the resolution of $64\times64$, $32\times32$, $16\times16$ and $8\times8$.
The training and testing split is consistent with the setting of the whole network. 
Table~\ref{tab:discussions} quantitatively summarizes the results of linear evaluation with the resolution of $64\times64$.
From the results, although ``DS-Net-linear'' degrades performance w.r.t. the normal DS-Net due to freezing the attention maps and most layers of the network, it outperforms the baseline ``DDCNN-semi-linear'' by a considerable margin.
This demonstrates that the consistency loss on attention maps significantly improves the accuracy of the corresponding attention maps and contributes to the final saliency predictions.

\begin{figure}[!t]
	\centering
    \vspace*{0.5mm}
	\begin{subfigure}{0.1\textwidth}
	    \includegraphics[width=\textwidth]{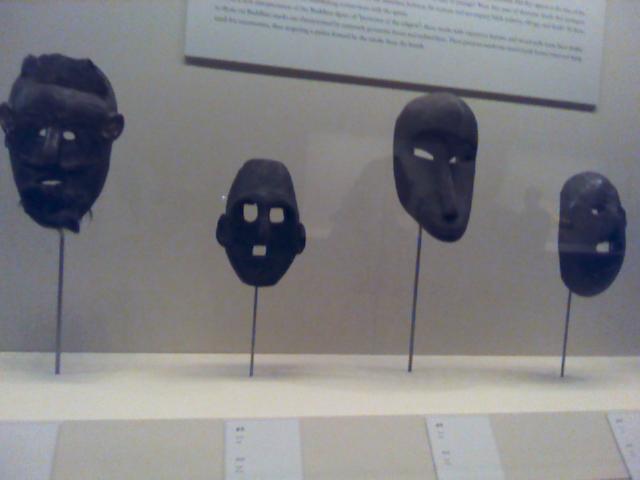}
	\end{subfigure}
	\begin{subfigure}{0.1\textwidth}
        \includegraphics[width=\textwidth]{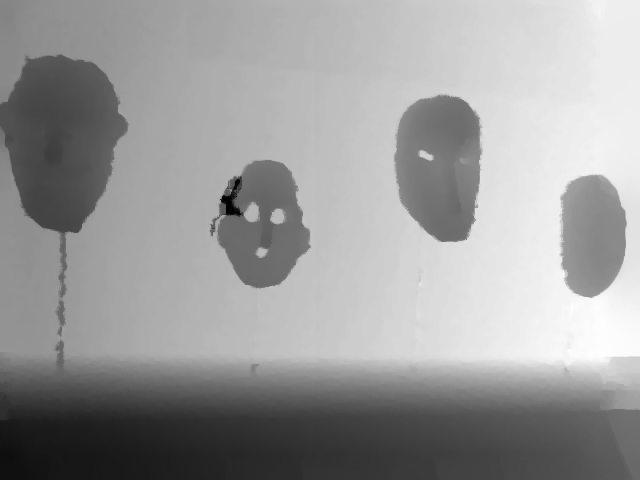}
	\end{subfigure}
    \begin{subfigure}{0.1\textwidth}
		\includegraphics[width=\textwidth]{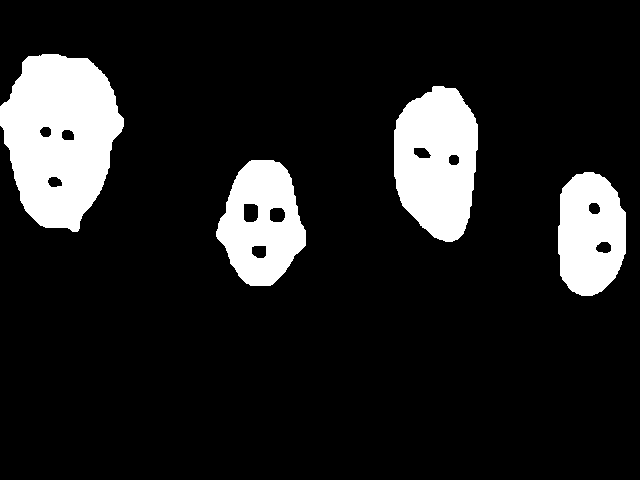}
	\end{subfigure}
    \begin{subfigure}{0.1\textwidth}
		\includegraphics[width=\textwidth]{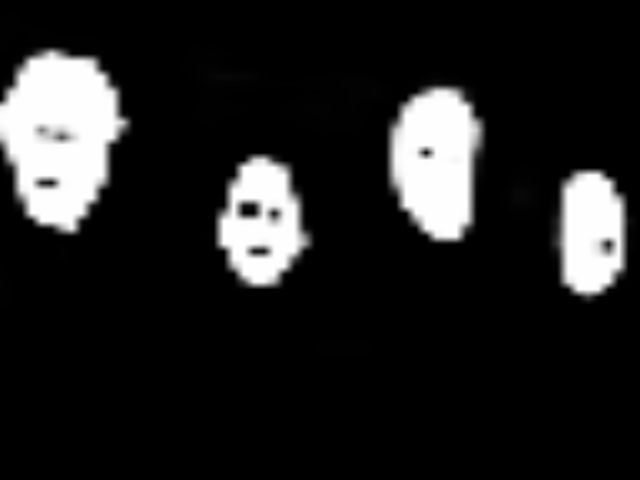}
	\end{subfigure}
	\ \\
    \vspace*{0.5mm}
	\begin{subfigure}{0.1\textwidth}
	    \includegraphics[width=\textwidth]{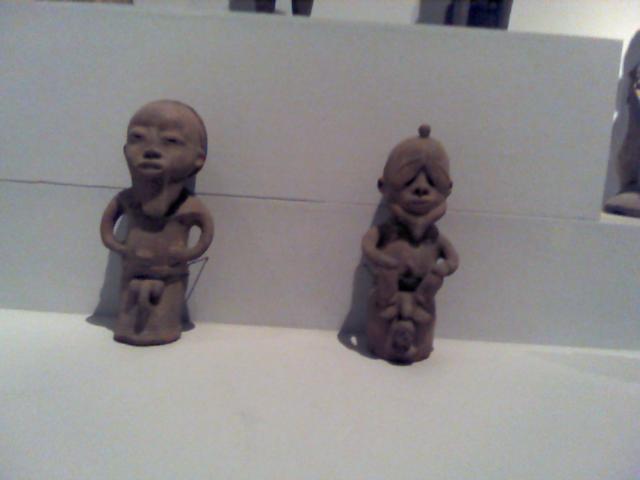}
	\end{subfigure}
	\begin{subfigure}{0.1\textwidth}
        \includegraphics[width=\textwidth]{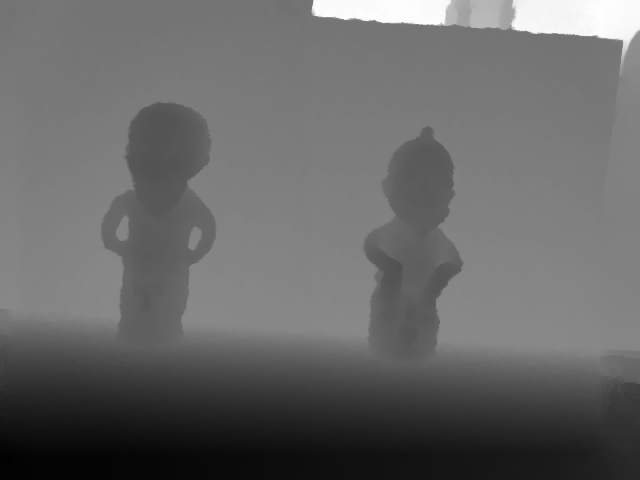}
	\end{subfigure}
    \begin{subfigure}{0.1\textwidth}
		\includegraphics[width=\textwidth]{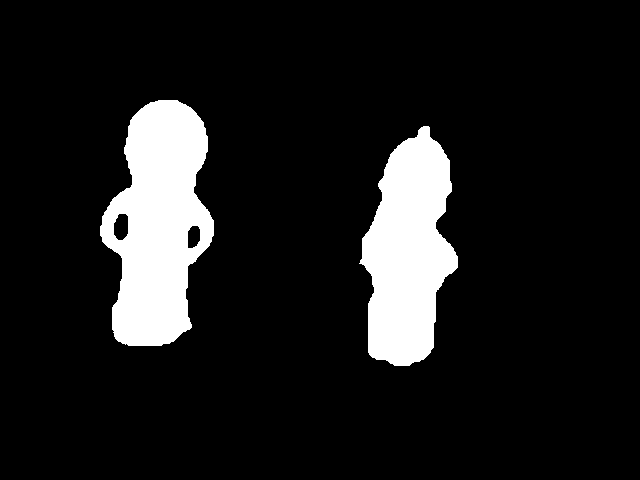}
	\end{subfigure}
    \begin{subfigure}{0.1\textwidth}
		\includegraphics[width=\textwidth]{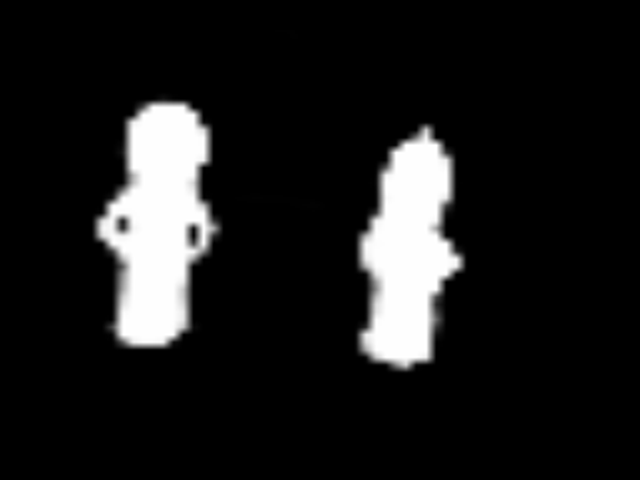}
	\end{subfigure}
	\ \\
   \vspace*{0.5mm}
   \begin{subfigure}{0.1\textwidth}
   	    \includegraphics[width=\textwidth]{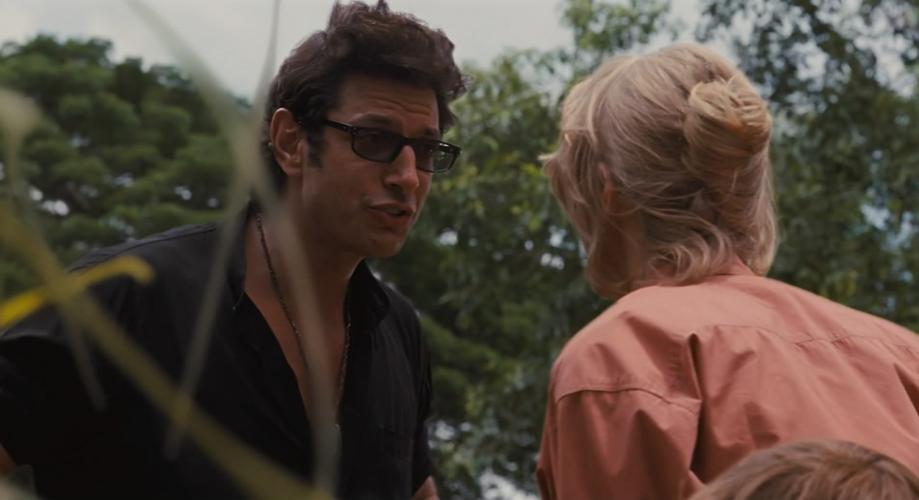}
        \vspace{-5.5mm} \caption{\footnotesize{RGB}}
   \end{subfigure}
   \begin{subfigure}{0.1\textwidth}
   	    \includegraphics[width=\textwidth]{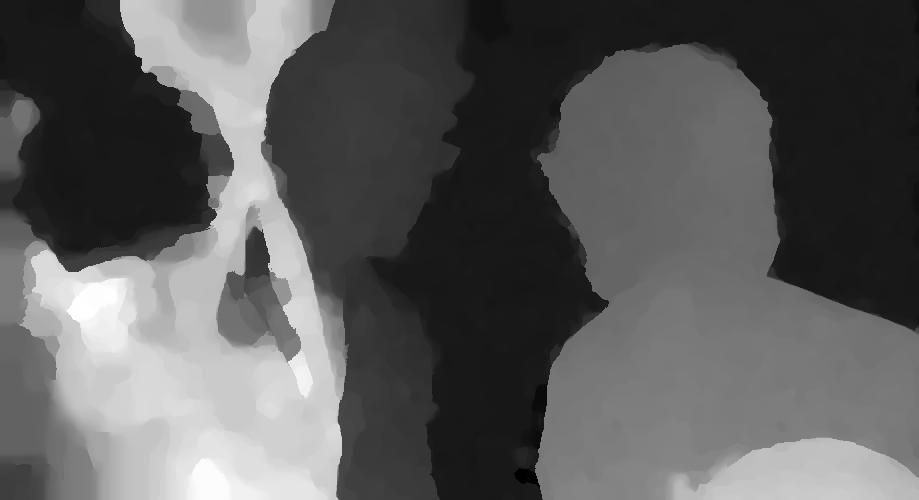}
        \vspace{-5.5mm} \caption{\footnotesize{depth}}
   \end{subfigure}
   \begin{subfigure}{0.1\textwidth}
   	    \includegraphics[width=\textwidth]{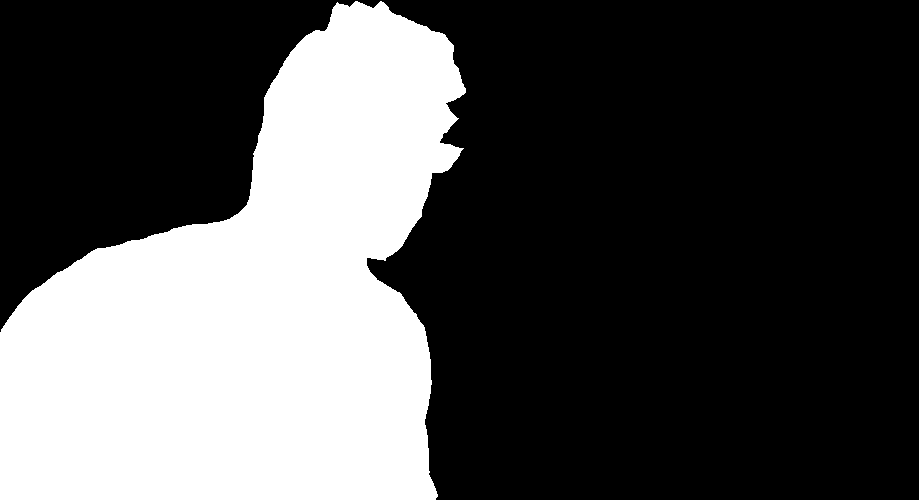}
        \vspace{-5.5mm} \caption{\footnotesize{GT}}
   \end{subfigure}
   \begin{subfigure}{0.1\textwidth}
   	    \includegraphics[width=\textwidth]{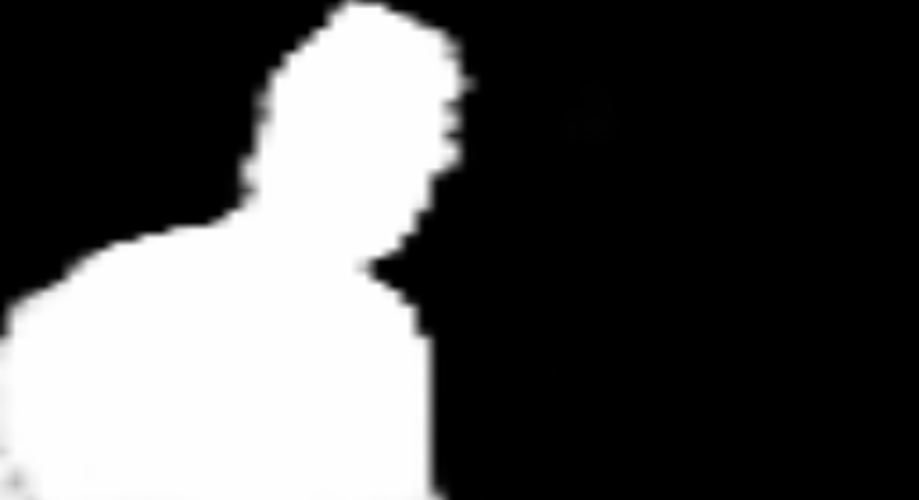}
        \vspace{-5.5mm} \caption{\footnotesize{ours}}
   \end{subfigure}

	\caption{Samples with multiple objects.
	(a) Input RGB image from benchmark datasets;
	(b) Input depth image;
	(c) Ground truths (denoted as ’GT’); and 
	(d) Our results. }
	\label{fig:challenging}
	\vskip -10pt
\end{figure}

\vspace*{2mm}
\noindent
\textbf{Effects of the number of unlabeled images.}
In general, a model will gain a stronger generalization ability when fed into more samples.
To further explore the relationship between the model performance and the number of unlabeled samples, we conduct five additional ablation experiments using our DS-Net on 1k, 2k, 4k, 8k, and 10k (10,553) unlabeled RGB images, respectively.
All of the samples are randomly sampled from the DUTS dataset.
Fig.~\ref{fig:hyperparameters} presents the effect of the number of unlabeled samples qualitatively.
From the results, on the one hand, the model achieves higher and higher metric performance with the increase of the number of unlabeled samples, demonstrating that semi-supervised learning can benefit from more training data significantly.
On the other hand, the performance gain decreases as more unlabeled data are fed into the model, indicating that the model has an upper-bound performance no matter how much data from the same domain is employed.
This indicates a research direction worth exploring, which is to improve further the generalization ability of the model by using out-of-domain data.

\vspace*{2mm}
\noindent
\textbf{Samples with multiple objects.}
To study the upper-bound performance of our model and find the capability boundary of it, we delineate several samples with multiple objects.
These samples are extremely challenging for existing state-of-the-art RGB-D saliency detectors, even for a human annotator.
We carefully selected a few in Fig.~\ref{fig:challenging}.
Apparently, our model can detect salient objects in these complicated samples accurately.
This indicates that our model can handle samples with different levels of difficulty and achieve superior overall performance.

}

\section{Conclusion}
\label{sec:conclusion}

In this paper, we have presented a Dual-Semi RGB-D Salient Object Detection Network (DS-Net) by leveraging unlabeled RGB images to assist the RGB-D SOD task in a semi-supervised manner. 
A depth decoupling convolutional neural network (DDCNN) has been proposed to jointly estimate the pseudo depth maps for RGB images and predict saliency maps for paired RGB-D images. Our DDCNN disentangles the features from RGB images into depth-aware features and depth-dispelled features, which enables the network to identify the latent features specific to each modality.
Experimental results on seven benchmark datasets have demonstrated the effectiveness of our method. 
We have also shown that the semi-supervised manner could further improve the performance of RGB-D SOD, even using an RGB image with the pseudo depth map.

\section*{Acknowledgement}
This work has been supported in part by the National Key Research and Development Program of China (2018AAA0101900), Zhejiang NSF (LR21F020004), Chinese Knowledge Center of Engineering Science and Technology (CKCEST), Hikvision-Zhejiang University Joint Research Center, National Natural Science Foundation of China (Grant No. 61902275) and The Hong Kong Polytechnic University under Grant P0030419, Grant P0030929, and Grant P0035358.

{\small
	\bibliographystyle{IEEEtran}
	\bibliography{egbib}

\begin{thebibliography}{10}
\providecommand{\url}[1]{#1}
\csname url@samestyle\endcsname
\providecommand{\newblock}{\relax}
\providecommand{\bibinfo}[2]{#2}
\providecommand{\BIBentrySTDinterwordspacing}{\spaceskip=0pt\relax}
\providecommand{\BIBentryALTinterwordstretchfactor}{4}
\providecommand{\BIBentryALTinterwordspacing}{\spaceskip=\fontdimen2\font plus
\BIBentryALTinterwordstretchfactor\fontdimen3\font minus
  \fontdimen4\font\relax}
\providecommand{\BIBforeignlanguage}[2]{{%
\expandafter\ifx\csname l@#1\endcsname\relax
\typeout{** WARNING: IEEEtran.bst: No hyphenation pattern has been}%
\typeout{** loaded for the language `#1'. Using the pattern for}%
\typeout{** the default language instead.}%
\else
\language=\csname l@#1\endcsname
\fi
#2}}
\providecommand{\BIBdecl}{\relax}
\BIBdecl

\bibitem{cong2018review}
R.~Cong, J.~Lei, H.~Fu, M.-M. Cheng, W.~Lin, and Q.~Huang, ``Review of visual
  saliency detection with comprehensive information,'' \emph{IEEE Transactions
  on Circuits and Systems for Video Technology}, 2018.

\bibitem{wang2019salient}
W.~{Wang}, Q.~{Lai}, H.~{Fu}, J.~{Shen}, H.~{Ling}, and R.~{Yang}, ``Salient
  object detection in the deep learning era: An in-depth survey,'' \emph{IEEE
  Transactions on Pattern Analysis and Machine Intelligence}, 2021.

\bibitem{fan2020rethinking}
D.-P. Fan, Z.~Lin, Z.~Zhang, M.~Zhu, and M.-M. Cheng, ``Rethinking {RGB-D}
  salient object detection: Models, data sets, and large-scale benchmarks,''
  \emph{IEEE Transactions on Neural Networks and Learning Systems}, 2020.

\bibitem{feng2016local}
D.~Feng, N.~Barnes, S.~You, and C.~McCarthy, ``Local background enclosure for
  {RGB-D} salient object detection,'' in \emph{CVPR}, 2016, pp. 2343--2350.

\bibitem{song2017depth}
H.~Song, Z.~Liu, H.~Du, G.~Sun, O.~Le~Meur, and T.~Ren, ``Depth-aware salient
  object detection and segmentation via multiscale discriminative saliency
  fusion and bootstrap learning,'' \emph{IEEE Transactions on Image
  Processing}, vol.~26, no.~9, pp. 4204--4216, 2017.

\bibitem{cong2017co}
R.~Cong, J.~Lei, H.~Fu, Q.~Huang, X.~Cao, and C.~Hou, ``Co-saliency detection
  for {RGBD} images based on multi-constraint feature matching and cross label
  propagation,'' \emph{IEEE Transactions on Image Processing}, vol.~27, no.~2,
  pp. 568--579, 2017.

\bibitem{cong2019going}
R.~Cong, J.~Lei, H.~Fu, J.~Hou, Q.~Huang, and S.~Kwong, ``Going from {RGB} to
  {RGBD} saliency: A depth-guided transformation model,'' \emph{IEEE
  Transactions on Cybernetics}, 2019.

\bibitem{chen2018progressively}
H.~Chen and Y.~Li, ``Progressively complementarity-aware fusion network for
  {RGB-D} salient object detection,'' in \emph{CVPR}, 2018, pp. 3051--3060.

\bibitem{zhao2019contrast}
J.-X. Zhao, Y.~Cao, D.-P. Fan, M.-M. Cheng, X.-Y. Li, and L.~Zhang, ``Contrast
  prior and fluid pyramid integration for {RGBD} salient object detection,'' in
  \emph{CVPR}, 2019, pp. 3927--3936.

\bibitem{piao2019depth}
Y.~Piao, W.~Ji, J.~Li, M.~Zhang, and H.~Lu, ``Depth-induced multi-scale
  recurrent attention network for saliency detection,'' in \emph{ICCV}, 2019,
  pp. 7254--7263.

\bibitem{fu2020jl}
K.~Fu, D.-P. Fan, G.-P. Ji, and Q.~Zhao, ``{JL-DCF}: Joint learning and
  densely-cooperative fusion framework for {RGB-D} salient object detection,''
  in \emph{CVPR}, 2020, pp. 3052--3062.

\bibitem{zhang2020uc}
J.~Zhang, D.-P. Fan, Y.~Dai, S.~Anwar, F.~S. Saleh, T.~Zhang, and N.~Barnes,
  ``Uc-net: uncertainty inspired {RGB-D} saliency detection via conditional
  variational autoencoders,'' in \emph{CVPR}, 2020, pp. 8582--8591.

\bibitem{zhang2020select}
M.~Zhang, W.~Ren, Y.~Piao, Z.~Rong, and H.~Lu, ``Select, supplement and focus
  for {RGB-D} saliency detection,'' in \emph{CVPR}, 2020, pp. 3472--3481.

\bibitem{liu2020learning}
N.~Liu, N.~Zhang, and J.~Han, ``Learning selective self-mutual attention for
  {RGB-D} saliency detection,'' in \emph{CVPR}, 2020, pp. 13\,756--13\,765.

\bibitem{Li2020_TC}
C.~Li, R.~Cong, S.~Kwong, J.~Hou, H.~Fu, G.~Zhu, D.~Zhang, and Q.~Huang,
  ``{ASIF-Net: Attention Steered Interweave Fusion Network for RGB-D Salient
  Object Detection},'' \emph{IEEE Transactions on Cybernetics}, pp. 1--13,
  2020.

\bibitem{liu2019salient}
Z.~Liu, S.~Shi, Q.~Duan, W.~Zhang, and P.~Zhao, ``Salient object detection for
  {RGB-D} image by single stream recurrent convolution neural network,''
  \emph{Neurocomputing}, vol. 363, pp. 46--57, 2019.

\bibitem{huang2018rgbd}
P.~Huang, C.-H. Shen, and H.-F. Hsiao, ``{RGBD} salient object detection using
  spatially coherent deep learning framework,'' in \emph{IEEE 23rd
  International Conference on Digital Signal Processing (DSP)}.\hskip 1em plus
  0.5em minus 0.4em\relax IEEE, 2018, pp. 1--5.

\bibitem{han2017cnns}
J.~Han, H.~Chen, N.~Liu, C.~Yan, and X.~Li, ``Cnns-based {RGB-D} saliency
  detection via cross-view transfer and multiview fusion,'' \emph{IEEE
  transactions on cybernetics}, vol.~48, no.~11, pp. 3171--3183, 2017.

\bibitem{wang2019adaptive}
N.~Wang and X.~Gong, ``Adaptive fusion for {RGB-D} salient object detection,''
  \emph{IEEE Access}, vol.~7, pp. 55\,277--55\,284, 2019.

\bibitem{li2020cross}
G.~Li, Z.~Liu, L.~Ye, Y.~Wang, and H.~Ling, ``Cross-modal weighting network for
  {RGB-D} salient object detection,'' \emph{ECCV}, 2020.

\bibitem{chen2020multi}
Z.~Chen, L.~Zhu, L.~Wan, S.~Wang, W.~Feng, and P.-A. Heng, ``A multi-task mean
  teacher for semi-supervised shadow detection,'' in \emph{CVPR}, 2020, pp.
  5611--5620.

\bibitem{InfNet2020}
D.-P. Fan, T.~Zhou, G.-P. Ji, Y.~Zhou, G.~Chen, H.~Fu, J.~Shen, and L.~Shao,
  ``{Inf-Net: Automatic COVID-19 Lung Infection Segmentation From CT Images},''
  \emph{IEEE Transactions on Medical Imaging}, vol.~39, no.~8, pp. 2626--2637,
  aug 2020.

\bibitem{laine2016temporal}
S.~Laine and T.~Aila, ``Temporal ensembling for semi-supervised learning,'' in
  \emph{International Conference On Learning Representations}, 2017.

\bibitem{tarvainen2017mean}
A.~Tarvainen and H.~Valpola, ``Mean teachers are better role models:
  Weight-averaged consistency targets improve semi-supervised deep learning
  results,'' in \emph{NIPS}, 2017, pp. 1195--1204.

\bibitem{chen2017deeplab}
L.-C. Chen, G.~Papandreou, I.~Kokkinos, K.~Murphy, and A.~L. Yuille,
  ``{DeepLab}: Semantic image segmentation with deep convolutional nets, atrous
  convolution, and fully connected crfs,'' \emph{IEEE Transactions on Pattern
  Analysis and Machine Intelligence}, vol.~40, no.~4, pp. 834--848, 2017.

\bibitem{woo2018cbam}
S.~Woo, J.~Park, J.-Y. Lee, and I.~So~Kweon, ``Cbam: Convolutional block
  attention module,'' in \emph{Proceedings of the European conference on
  computer vision (ECCV)}, 2018, pp. 3--19.

\bibitem{qu2017rgbd}
L.~Qu, S.~He, J.~Zhang, J.~Tian, Y.~Tang, and Q.~Yang, ``{RGBD} salient object
  detection via deep fusion,'' \emph{IEEE Transactions on Image Processing},
  vol.~26, no.~5, pp. 2274--2285, 2017.

\bibitem{chen2019three}
H.~Chen and Y.~Li, ``Three-stream attention-aware network for {RGB-D} salient
  object detection,'' \emph{IEEE Transactions on Image Processing}, vol.~28,
  no.~6, pp. 2825--2835, 2019.

\bibitem{fu2021siamese}
K.~Fu, D.-P. Fan, G.-P. Ji, Q.~Zhao, J.~Shen, and C.~Zhu, ``Siamese network for
  rgb-d salient object detection and beyond,'' \emph{IEEE Transactions on
  Pattern Analysis and Machine Intelligence}, 2021.

\bibitem{pang2020hierarchical}
Y.~Pang, L.~Zhang, X.~Zhao, and H.~Lu, ``Hierarchical dynamic filtering network
  for {RGB-D} salient object detection,'' \emph{ECCV}, 2020.

\bibitem{chen2020progressively}
S.~Chen and Y.~Fu, ``Progressively guided alternate refinement network for
  {RGB-D} salient object detection,'' \emph{ECCV}, 2020.

\bibitem{zhao2020single}
X.~Zhao, L.~Zhang, Y.~Pang, H.~Lu, and L.~Zhang, ``A single stream network for
  robust and real-time {RGB-D} salient object detection,'' \emph{ECCV}, 2020.

\bibitem{li2020rgb}
C.~Li, R.~Cong, Y.~Piao, Q.~Xu, and C.~C. Loy, ``{RGB-D} salient object
  detection with cross-modality modulation and selection,'' \emph{ECCV}, 2020.

\bibitem{luo2020cascade}
A.~Luo, X.~Li, F.~Yang, Z.~Jiao, H.~Cheng, and S.~Lyu, ``Cascade graph neural
  networks for {RGB-D} salient object detection,'' in \emph{ECCV}.\hskip 1em
  plus 0.5em minus 0.4em\relax Springer, 2020, pp. 346--364.

\bibitem{zhang2020asymmetric}
M.~Zhang, S.~X. Fei, J.~Liu, S.~Xu, Y.~Piao, and H.~Lu, ``Asymmetric two-stream
  architecture for accurate {RGB-D} saliency detection,'' in \emph{ECCV}, 2020.

\bibitem{wang2020synergistic}
Y.~Wang, Y.~Li, J.~H. Elder, R.~Wu, H.~Lu, and L.~Zhang, ``Synergistic saliency
  and depth prediction for rgb-d saliency detection,'' in \emph{Proceedings of
  the Asian Conference on Computer Vision}, 2020.

\bibitem{sun2021deep}
P.~Sun, W.~Zhang, H.~Wang, S.~Li, and X.~Li, ``Deep rgb-d saliency detection
  with depth-sensitive attention and automatic multi-modal fusion,'' in
  \emph{Proceedings of the IEEE/CVF Conference on Computer Vision and Pattern
  Recognition}, 2021, pp. 1407--1417.

\bibitem{fan2020bbs}
D.-P. Fan, Y.~Zhai, A.~Borji, J.~Yang, and L.~Shao, ``{BBS-Net}: {RGB-D}
  salient object detection with a bifurcated backbone strategy network,'' in
  \emph{ECCV}.\hskip 1em plus 0.5em minus 0.4em\relax Springer, 2020, pp.
  275--292.

\bibitem{ji2020accurate}
W.~Ji, J.~Li, M.~Zhang, Y.~Piao, and H.~Lu, ``Accurate {RGB-D} salient object
  detection via collaborative learning,'' \emph{ECCV}, 2020.

\bibitem{ju2014depth}
R.~Ju, L.~Ge, W.~Geng, T.~Ren, and G.~Wu, ``Depth saliency based on anisotropic
  center-surround difference,'' in \emph{IEEE international conference on image
  processing (ICIP)}.\hskip 1em plus 0.5em minus 0.4em\relax IEEE, 2014, pp.
  1115--1119.

\bibitem{peng2014rgbd}
H.~Peng, B.~Li, W.~Xiong, W.~Hu, and R.~Ji, ``{RGBD} salient object detection:
  a benchmark and algorithms,'' in \emph{ECCV}.\hskip 1em plus 0.5em minus
  0.4em\relax Springer, 2014, pp. 92--109.

\bibitem{niu2012leveraging}
Y.~Niu, Y.~Geng, X.~Li, and F.~Liu, ``Leveraging stereopsis for saliency
  analysis,'' in \emph{CVPR}, 2012, pp. 454--461.

\bibitem{cheng2014depth}
Y.~Cheng, H.~Fu, X.~Wei, J.~Xiao, and X.~Cao, ``Depth enhanced saliency
  detection method,'' in \emph{Proceedings of international conference on
  internet multimedia computing and service}, 2014, pp. 23--27.

\bibitem{li2014saliency}
N.~Li, J.~Ye, Y.~Ji, H.~Ling, and J.~Yu, ``Saliency detection on light field,''
  in \emph{CVPR}, 2014, pp. 2806--2813.

\bibitem{wang2017learning}
L.~Wang, H.~Lu, Y.~Wang, M.~Feng, D.~Wang, B.~Yin, and X.~Ruan, ``Learning to
  detect salient objects with image-level supervision,'' in \emph{CVPR}, 2017,
  pp. 136--145.

\bibitem{fan2017structure}
D.-P. Fan, M.-M. Cheng, Y.~Liu, T.~Li, and A.~Borji, ``Structure-measure: A new
  way to evaluate foreground maps,'' in \emph{ICCV}, 2017, pp. 4548--4557.

\bibitem{lang2016dual}
C.~Lang, J.~Feng, S.~Feng, J.~Wang, and S.~Yan, ``Dual low-rank pursuit:
  Learning salient features for saliency detection,'' \emph{IEEE Transactions
  on Neural Networks and Learning Systems}, vol.~27, no.~6, pp. 1190--1200,
  2016.

\bibitem{fan2018enhanced}
D.-P. Fan, C.~Gong, Y.~Cao, B.~Ren, M.-M. Cheng, and A.~Borji,
  ``Enhanced-alignment measure for binary foreground map evaluation,'' in
  \emph{IJCAI}, 2018, pp. 698--704.

\bibitem{perazzi2012saliency}
F.~Perazzi, P.~Kr{\"a}henb{\"u}hl, Y.~Pritch, and A.~Hornung, ``Saliency
  filters: Contrast based filtering for salient region detection,'' in
  \emph{CVPR}, 2012, pp. 733--740.

\bibitem{simonyan2014very}
K.~Simonyan and A.~Zisserman, ``Very deep convolutional networks for
  large-scale image recognition,'' \emph{arXiv preprint arXiv:1409.1556}, 2014.

\bibitem{he2016deep}
K.~He, X.~Zhang, S.~Ren, and J.~Sun, ``Deep residual learning for image
  recognition,'' in \emph{CVPR}, 2016, pp. 770--778.

\bibitem{sun2019high}
K.~Sun, Y.~Zhao, B.~Jiang, T.~Cheng, B.~Xiao, D.~Liu, Y.~Mu, X.~Wang, W.~Liu,
  and J.~Wang, ``High-resolution representations for labeling pixels and
  regions,'' \emph{arXiv preprint arXiv:1904.04514}, 2019.

\bibitem{deng2009imagenet}
J.~Deng, W.~Dong, R.~Socher, L.-J. Li, K.~Li, and L.~Fei-Fei, ``Image{N}et: A
  large-scale hierarchical image database,'' in \emph{CVPR}, 2009, pp.
  248--255.

\bibitem{liu2015parsenet}
W.~Liu, A.~Rabinovich, and A.~C. Berg, ``Parsenet: Looking wider to see
  better,'' \emph{arXiv preprint arXiv:1506.04579}, 2015.

\bibitem{lee2018single}
J.-H. Lee, M.~Heo, K.-R. Kim, and C.-S. Kim, ``Single-image depth estimation
  based on fourier domain analysis,'' in \emph{Proceedings of the IEEE
  Conference on Computer Vision and Pattern Recognition}, 2018, pp. 330--339.

\bibitem{xian2020structure}
K.~Xian, J.~Zhang, O.~Wang, L.~Mai, Z.~Lin, and Z.~Cao, ``Structure-guided
  ranking loss for single image depth prediction,'' in \emph{Proceedings of the
  IEEE/CVF Conference on Computer Vision and Pattern Recognition}, 2020, pp.
  611--620.

\bibitem{liu2018exploring}
Y.~Liu, F.~Wei, J.~Shao, L.~Sheng, J.~Yan, and X.~Wang, ``Exploring
  disentangled feature representation beyond face identification,'' in
  \emph{Proceedings of the IEEE Conference on Computer Vision and Pattern
  Recognition}, 2018, pp. 2080--2089.

\bibitem{hsieh2018learning}
J.-T. Hsieh, B.~Liu, D.-A. Huang, L.~Fei-Fei, and J.~C. Niebles, ``Learning to
  decompose and disentangle representations for video prediction,'' in
  \emph{Proceedings of the 32nd International Conference on Neural Information
  Processing Systems}, 2018, pp. 515--524.

\bibitem{ren2021interpreting}
J.~Ren, M.~Li, Z.~Liu, and Q.~Zhang, ``Interpreting and disentangling feature
  components of various complexity from dnns,'' in \emph{International
  Conference on Machine Learning}.\hskip 1em plus 0.5em minus 0.4em\relax PMLR,
  2021, pp. 8971--8981.

\end{thebibliography}
}

\end{document}